\def\equationautorefname~#1\null{Equation (#1)\null}
\def\sectionautorefname~#1\null{Section #1\null}
\def\subsectionautorefname~#1\null{Section #1\null}
\def\subsubsectionautorefname~#1\null{Section #1\null}
\def\figureautorefname~#1\null{Figure #1\null}
\documentclass[twocolumn]{aastex63}
\usepackage{amsmath}
\usepackage{xparse}
\usepackage{hyperref}
\usepackage{longtable}
\usepackage[flushleft]{threeparttable}
\usepackage[titletoc,title]{appendix}
\usepackage{bm}

\newcommand*{\blue}{\textcolor{black}} 
\newcommand*{\ajn}{\textcolor{black}} 
\newcommand*{\ergs}{erg s$^{-1}$ } 
\newcommand*{\mum}{$\mu$m} 
\received{}
\revised{}
\accepted{April 8 2026}
\graphicspath{{./}{figures/}}

\begin{document}

\title{Infrared spectral signatures of light r-process elements in kilonovae} 

\correspondingauthor{Anders Jerkstrand}
\email{anders.jerkstrand@astro.su.se}

\author[0000-0001-8005-4030]{Anders Jerkstrand}
\affiliation{The Oskar Klein Centre, Department of Astronomy, Stockholm University, AlbaNova, SE-10691 Stockholm, Sweden}
\author[0000-0001-5255-0782]{Quentin Pognan}
\affiliation{Max Planck Institute for Gravitational Physics (Albert Einstein Institute), Am Mühlenberg 1, Potsdam-Golm, 14476, Germany}
\author[0000-0001-6595-2238]{Smaranika Banerjee}
\affiliation{Institut f\"ur Kernphysik, Technische Universit\"at Darmstadt, Schlossgartenstr.~2, 64289 Darmstadt, Germany}
\affiliation{The Oskar Klein Centre, Department of Astronomy, Stockholm University, AlbaNova, SE-10691 Stockholm, Sweden}
\author[0000-0002-9604-1434]{N. C. Sterling}
\affiliation{University of West Georgia, 1601 Maple Street, Carrollton, GA 30118, USA}
\author[0000-0002-6224-3492]{Jon Grumer}
\affiliation{Theoretical Astrophysics, Department of Physics and Astronomy, Uppsala University, Box 516, SE-751 20 Uppsala, Sweden}
\author[0009-0005-3148-513X]{Niamh Ferguson}
\affiliation{Department of Physics, University of Strathclyde, Glasgow G4 0NG, UK}
\author{Keith Butler}
\affiliation{Institut für Astronomie und Astrophysik, Scheinerstr. 1, 81679 München, Germany}
\author[0000-0002-8094-6108]{James Gillanders}
\affiliation{Astrophysics sub-Department, Department of Physics, University of Oxford, Keble Road, Oxford OX1 3RH, UK}
\author[0000-0002-8229-1731]{Stephen Smartt}
\affiliation{Astrophysics sub-Department, Department of Physics, University of Oxford, Keble Road, Oxford OX1 3RH, UK}
\author[0000-0003-4443-6984]{Kyohei Kawaguchi}
\affiliation{Max Planck Institute for Gravitational Physics (Albert Einstein Institute), Am Mühlenberg 1, Potsdam-Golm, 14476, Germany}
\affiliation{Center of Gravitational Physics and Quantum Information,
 Yukawa Institute for Theoretical Physics, 
Kyoto University, Kyoto, 606-8502, Japan}
\author[0000-0001-9688-2332
]{Blanka Vilagos}
\affiliation{The Oskar Klein Centre, Department of Astronomy, Stockholm University, AlbaNova, SE-10691 Stockholm, Sweden}
\begin{abstract}
A central question regarding neutron star mergers is whether they are able to produce all the r-process elements, from first to third peak. We here study theoretical infrared signatures of first-peak elements with spectral synthesis modelling. By combining state-of-the-art NLTE physics with new radiative and collisional data for these elements, we identify several promising diagnostic lines from Ge, As, Se, Br, Kr and Zr. The models give self-consistent line luminosities and indicate specific features that probe emission volumes at early phases ($\sim$10d), the product of ion mass and electron density in late phases ($\gtrsim$75d), and in some cases direct ionic masses at intermediate phases. Emission by [Se~I] 5.03 \mum\ + [Se~III] 4.55 \mum\ is the only one from the first r-process peak that could explain the Spitzer photometry of AT2017gfo. However, the models show consistently that with a Kr/Te and Se/Te ratio following the solar r-process pattern, Kr + Se emission is dominant over Te for the blend at 2.1 \mum\ observed in both AT2017gfo and AT2023vfi. The somewhat better line profile fit with [Te~III] may suggest that both AT2017gfo and AT2023vfi had a strongly sub-solar production of the light r-process elements. An alternative scenario could be that Kr + Se in an asymmetric morphological distribution generates the feature. Further JWST spectral observations holds promise to determine the light r-process production of kilonovae, and in particular whether the light elements are made in a slow disk wind or in a fast proto-NS wind. We identify specific needs for further atomic data for $Z=31-40$ elements.
\end{abstract}


\section{Introduction} \label{sec:intro}

The James Webb Space Telescope (JWST) is currently revolutionising many branches of astronomy, opening up the near infrared (NIR) and mid infrared (MIR) regimes which can now be studied with excellent sensitivity and spectral resolution. For explosive transients, such as supernovae (SNe) and kilonovae (KNe), the infrared contains rich signatures of atomic transitions that are often less affected by uncertain physical conditions than optical lines. While the Spitzer Space Telescope revealed much information about Type II SNe \citep{Kotak2006,Kotak2009}, other SN types were not observed. The first kilonova, AT2017gfo, was photometrically detected in the Spitzer warm phase \citep{Kasliwal2022}. However, it is JWST, and in a few years the Extremely Large Telescope (ELT), that will give us detailed spectral information about these transients in the infrared.

The importance of the infrared regime for trans-iron elements has been established in studies of planetary nebulae (PNe). The first detection of neutron-capture elements in a PN, or any type of nebula, was the detection of optical Kr and Xe lines by \citet{Pequignot1994}. The near-infrared range subsequently yielded detections of [Kr III] 2.199 $\mu$m and [Se IV] 2.287 $\mu$m \citep{Dinerstein2001a}, [Zn IV] 3.625 $\mu$m \citep{Dinerstein2001b}, [Rb IV] 1.5973 $\mu$m, [Cd IV] 1.7204 $\mu$m, [Ge VI] 2.1930 $\mu$m \citep{Sterling2016}, [Br V] 1.6429 $\mu$m, [Te III] 2.1019 $\mu$m \citep{Madonna2018}, and [Rb III] 1.356 $\mu$m \citep{Dinerstein2021}. These detections prompted work to develop the needed atomic data for these elements to carry out abundance analyses \citep[e.g.,][]{Sterling2011,SterlingWitthoeft2011,Sterling2015,McLaughlin2019,Macaluso2019}.
%
The post-diffusion phase ($t \gtrsim 1$ week for kilonovae) gives unique information about explosive transients as the innermost layers, typically containing most of the ejecta mass, become visible as the nebula becomes more transparent. With radiative transfer playing a smaller role, and Doppler broadening being smaller for the inner layers, spectral signatures emerge more cleanly than in the diffusion phase. At the same time, modelling of the physical conditions becomes more challenging, as the necessary abandonment of Local Thermodynamic Equilibrium (LTE) for full non-LTE (NLTE) collisional radiative modelling  brings about sensitivity to a large variety of collisional and radiative processes \citep[see e.g.][for a review]{LRCA}, and requires an extensive computational machinery. For kilonovae, comprehensive NLTE modelling work has been done by \citet[][simplified NLTE in the optically thin limit]{Hotokezaka2021,Hotokezaka2022,Hotokezaka2023} and by \citet[][full NLTE with radiative transfer]{Pognan2022a,Pognan2022b, Pognan2023, Pognan2025}. The impact of recombination rates on NLTE ionization and spectra was studied by \citet{Banerjee2025}, and a study of how NLTE ionization effects depend on density and temperature at early phases was carried out by \citet{Brethauer2026}. \citet{Tarumi2023} and \citet{Sneppen2024} have studied NLTE effects on He line formation.

What is looked for in nebular analysis are signatures that, based on inspection of the line formation mechanisms, can be established as diagnostics with good robustness (i.e., not too sensitive to physical conditions or uncertain physics or parameters). In this endeavor, the infrared is a particularly potent regime. Under typical nebular plasma conditions, IR lines often originate from forbidden transitions within the ground term of a given ion, and with the relatively small excitation energies ($T_{exc} \lesssim 2000 \left(\lambda/2\ \mu m\right)^{-1}$ K), the line luminosities can be in a quite temperature-insensitive regime (if $T \gtrsim T_{exc}$). Further, as transitions between fine structure states of a given term  tend to have small A-values and large collision strengths, the critical densities for them are quite low and they can be close to LTE for quite long times, giving no or weak sensitivity to the electron density. Finally, line opacity generally declines with wavelength, so there are no or weak radiative transfer effects. What results from these properties is that IR lines are excellent diagnostics of ionic masses (optically thin case) and/or emission volume (optically thick case) \citep{Moseley1989, Haas1990, Wooden1993, Jerkstrand2012,Jerkstrand2017}.

These benign properties of the infrared regime give expectations of strong scientific return with JWST for explosive transients. Extending the observable limit from up to 2.5 $\mu$m (accessible from the ground) to 30 $\mu$m, and with excellent sensitivity, JWST is expected to be able to reveal key spectral signatures of KNe and other transients. It has already shown its value by the observations of infrared emission of the presumed kilonova AT2023vfi at a distance of 292 Mpc \citep{Levan2024}.

With these prospects, it is imperative to improve our theoretical understanding of kilonova infrared emission, and to develop model predictions for a variety of compact object merger types. This topic was opened up by \citet{Hotokezaka2021}, who studied nebular spectral emission by neodymium, including in the IR. \citet{Hotokezaka2022} produced theoretical predictions of IR emission from solar composition ejecta with fixed parameterised physical conditions (temperature and ionization), in a limited NLTE approach (thermal collisions and spontaneous radiative decay included). The strongest emission was obtained for [Se III] 5.74 $\mu$m, whereas the Spitzer 4.5 and 3.6 $\mu$m bands were found to be dominated by [Se~III] 4.55 and Br + As lines, respectively.

From this exploratory modelling, the next step is modelling with calculation of physical conditions, with some consideration of the density structure of the ejecta. This is the main goal of this paper. We have presented the optical (and some near-infrared) properties of 1D full NLTE spectral models in \citet{Pognan2022a,Pognan2023} - here the focus is on the near- and mid-infrared emission of models with a similar setup. With the importance of accurate atomic physics, we have adopted an approach to initially limit the focus to the infrared emission from light r-process elements only, $Z \leq 40$ (with one exception, Te ($Z=52$), due to its possible identification in observed spectra), carrying out an extensive atomic data compilation and calibration effort for these elements. Complementary information from varying ejecta parameters and composition are presented in \citet{Pognan2026}.

\section{Modelling setup} 

\begin{figure}[htb]
\includegraphics[width=1\linewidth]{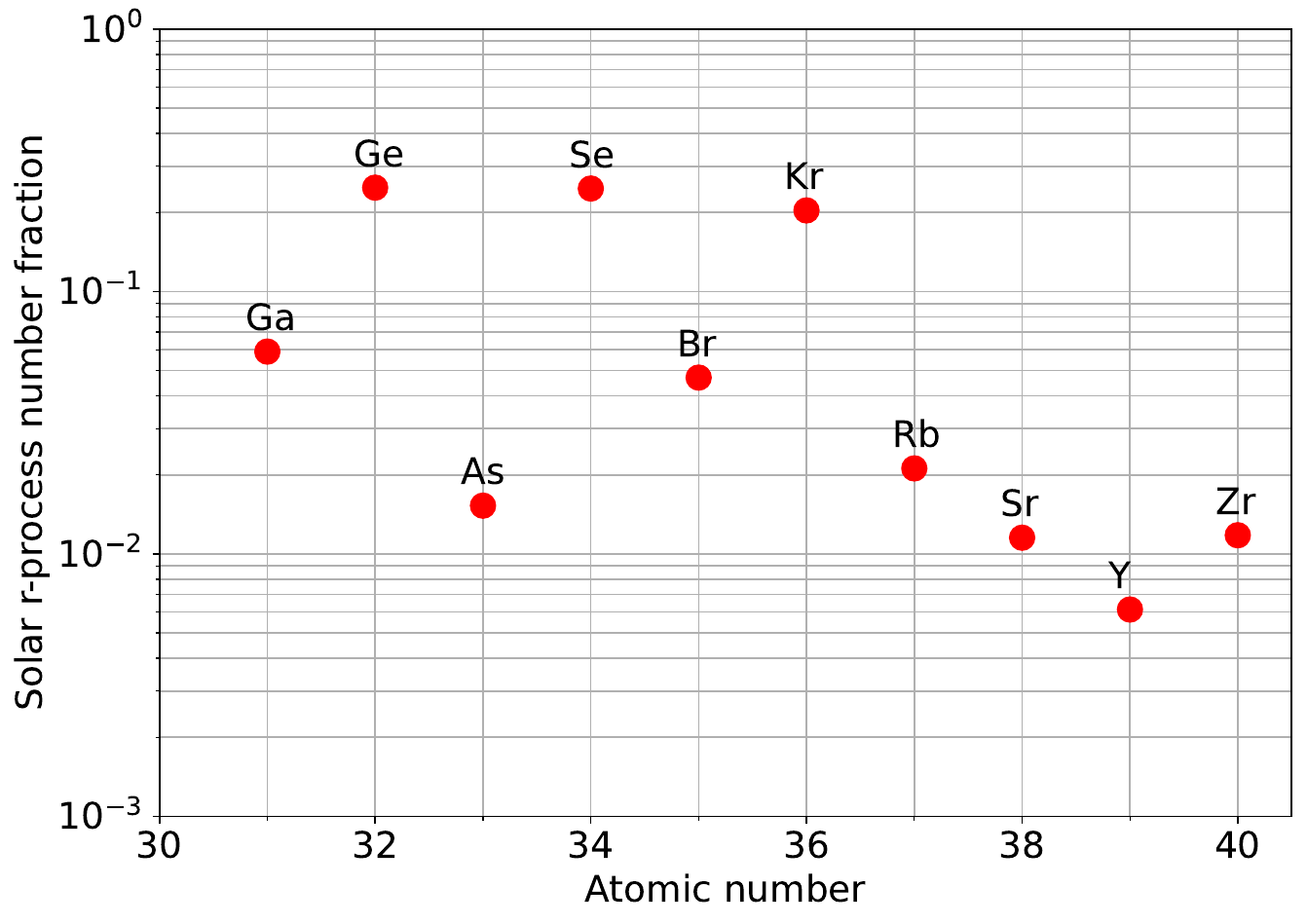}
\caption{The solar ``r-process'' (i.e., non s-process) number abundances in the $Z=31-40$ range, according to the \citet{Prantzos2020} estimation. The value for Te ($Z=52$, not plotted) is 2.3\%.}
\label{fig:abundances}
\end{figure}

The goal of the study is to establish an orientation of the nebular near and mid-infrared signatures of light r-process elements in KN ejecta, being aided by self-consistent physical NLTE modelling. We use the \texttt{SUMO} code \citep{Jerkstrand11,Jerkstrand2012}, adopting a 1D approach where our base-line model is similar to the models in \citet{Pognan2023}. Because the light r-process elements ($31 \leq Z \leq 40$) make up about 80\% of the solar r-process composition by mass \citep[and 87\% by number, Fig. \ref{fig:abundances},][]{Prantzos2020}, it is plausible that the ejecta of many KNe are dominated by these elements. The ejecta mass of AT2017gfo has been estimated to be about 0.05 $M_\odot$ \citep[e.g.][]{Smartt2017,Villar17,Tanaka17,Wollaeger2018,Vieria2025}, which is the value we use for the model ejecta here.

Light r-process elements are made in significant amounts in regions where the electron fraction $Y_e$ (number of protons relative to number of nucleons) is raised to $\gtrsim 0.3$ due to positron capture and neutrino irradiation. In current simulations, dynamic, secular (disk outflow), and hypermassive neutron star (HMNS) polar wind components can all have parts with such conditions \citep[e.g.][]{Wanajo2014,Goriely2015,Fujibayashi2020b,Just2023,Curtis2023}. However, only the disk outflow component reaches, in current simulations, the kind of masses inferred for AT2017gfo \citep[and one may also note that ejecta mass was likely yet larger in AT2023vfi, ][]{Rastinejad2025}. It is therefore reasonable to associate the baseline model with such ejecta. 
The disk outflow can achieve high $Y_e$ values ($\gtrsim 0.2$) due to positron capture \citep[e.g.][]{Fernandez2013, Fujibayashi:2020qda, Just2022b} or even values as high as 0.4-0.5 via neutrino irradiation by either a hypermassive neutron star \citep{Metzger2014} or the disk itself \citep{Just2015,Just2023}.

The density profiles of various high-$Y_e$ components can vary, both between simulations and within a simulation with position angle. The analytic limit for a disk wind is a $-2$ power law, whereas other components can be steeper, roughly described by $-3$ or $-4$ power laws \citep[e.g.][]{Kawaguchi2021,Just2022b,Fernandez2024}. We use a -3 power law throughout as a value that lies relatively close to all scenarios.

The mean velocity of a disk outflow is significantly lower than for dynamic ejecta or HMNS outflows, although with magnetohydrodynamic effects the difference is somewhat reduced \citep{Kiuchi2023}. We choose an inner boundary velocity $v_{in}=0.02c$ to obtain mean velocities ($\langle\beta=v/c\rangle=0.078$ for a -3 power law) roughly consistent with disk outflow simulations \citep[e.g., $\langle\beta\rangle\sim0.05$ in the simulations of][]{Just2023}, but a higher value $v_{in}=0.08c$ (giving $\langle\beta\rangle=0.13$ for a -3 power law) in an alternative model which would correspond closer to the other components \citep[e.g., $\beta=0.1-0.2$ for the proto-NS wind in the simulations of][]{Just2023}. The ejecta are terminated at $v_{out}=0.2c$, with material faster than this playing a subdominant role at nebular phases. 

The abundances are set to the solar r-process pattern for $Z=31-40$ and $Z=52$, as estimated by \citet{Prantzos2020}, plus 1\% abundance each (by mass) of Fe, Ni and Zn. Although the study is focused on the $Z \leq 40$ elements, Te ($Z=52$) was added due to its possible identification in both AT2017gfo \citep{Hotokezaka2023} and AT2023vfi \citep{Levan2024}. Inspection of the detailed abundance patterns in the simulations of \citet{Kawaguchi2021,Just2023, Fujibayashi2023a,Fernandez2024}, shows that the $Z=34-40$ pattern is relatively close to the solar one for most simulations. For $Z=31-33$, there is, however, more variation and simulations tend to produce lower abundances than solar.
\citet{Goriely1999} assesses the solar r-process abundances of $Z=31$ (Ga) as very uncertain, with potentially a value much lower than the standard one.

All the simulations mentioned above give abundances of $Z=26, 28, 30$ being the highest in the $Z=20-30$ range, and of order 1\% of the $Z=31-40$ total abundance, which we adopt. By using the solar r-process abundance distribution the model is made as generic as possible, but still in broad agreement with KN simulations, although not tied to any particular one.

The zoning is done with logarithmic 10\% steps in velocity, resulting in of order 20 zones (depending on $v_{in}$ and $v_{out}$). This is an increase compared to the 5 zones used in \citet{Pognan2023}, which better spatially resolves the physical conditions and radiative transfer. 

Radioactive powering varies significantly with composition \citep[e.g.][]{Wanajo2014} and also with uncertain nuclear physics for fixed $Y_e$ \citep{Barnes2021}. Kilonovae have multiple ejecta components, with different $Y_e$ and different morphologies. For this reason, we allow a certain freedom for the power deposition, that is intended to bracket the range of conditions in KNe. The baseline powering, ``low power'', has the decay power  of a $Y_e=0.35$ composition \citep{Wanajo2014}, whereas ``high power'' multiplies this by a factor $f$. This scenario could represent e.g. a $Z=30-40$ ejecta component exposed to radioactive decay particles from another more neutron-rich composition in the ejecta. We choose an approach where we boost the power in the high-velocity model with a factor $f=3$. This allows us to probe the likely span of ionization states, with the low-velocity, low-power model giving a quite neutral ejecta, whereas the high-velocity, high-power one gives high ionization (both low density and high power work to increase ionization). In this way we can bracket the range of possible lines from the studied elements. A study of the impact of $f$ is provided in Appendix \ref{sec:sens_f}.

Thermalization is computed in the same way as in previous papers, with time-delay accounted for using the analytic methods of \citet{Kasen2019} and \citet{Waxman2019}, but specific flows into heating and ionization channels (non-thermal excitation is ignored) using the Spencer-Fano routine of \texttt{SUMO}.    

The baseline atomic data is the \texttt{FAC} calculation set presented in \citet{Pognan2023}. However, we have here also carried out extensive calibrations of energy levels, wavelengths, and A-values relevant for IR lines, and also added in new recombination and collision rates, with details outlined in Appendix \ref{sec:atomicdata}. Of particular importance are the new dielectronic recombination rates for light r-process elements computed by \citet{Banerjee2025}. For collision strengths, computed values are used for several ions. When the values are unknown, we make a separate treatment between forbidden fine-structure lines (most important for this study) and others, because the fine-structure transitions often have significantly larger collision strengths than higher-energy ones. For the first group, we fitted the formula $\Upsilon = \mbox{const} \times g_u g_l$ \citep{Axelrod1980}, where $g_u$ and $g_l$ are the statistical weights of upper and lower levels, to all the calculated fine-structure data for the included elements (see appendix), obtaining a best-fit coefficient of $\rm{const}=0.41$. For the others, we used the Axelrod value derived for iron ($\rm{const}=0.004$) boosted by a factor 10 to bring better agreement with recent calculations for trans-iron elements \citep[][]{Bromley2023,Mulholland2024,Mulholland2025,Dougan2025,McCann2025}. 

We study two models, A and B, with properties summarized in Table \ref{table:models}. We compute the models at 10d, 40d and 80d, in the steady-state approximation \citep[see][for an in-depth study of this approximation]{Pognan2022a}, at wavelengths from 400 Å to 30 $\mu$m.

\begin{table*}[htb]
\centering
\begin{tabular}{cccc}
\hline
Name & Composition & Deposition & $v_{in}$ \\ 
\hline
A & Solar for $Z=31-40$,52 + $26, 28, 30$ at 1\% & Low ($f=1$) & 0.02  \\
B &       Same as A                                             & High ($f=3$) & 0.08 
\\
\hline
\end{tabular}
\caption{Summary of the models studied. Both models have $v_{out}=0.2c$ and $-3$ power law density profiles.}
\label{table:models}
\end{table*}

\section{Data} 
We will make comparisons of the computed models to observations of AT2017gfo and AT2023vfi.

We will make use of the +10.4d spectrum of AT2017gfo presented by \citet{Pian2017}; specifically, we use the reduced version of the ENGRAVE data release.\footnote{\url{http://www.engrave-eso.org/AT2017gfo-Data-Release}} The spectrum has been corrected for redshift with $z=0.009843$ (luminosity distance $D_L = 40$ Mpc) and (Milky Way) extinction with $E(B-V)=0.11$ mag \citep{Hjorth2017, Smartt2017}. We smooth the spectrum using a 20-bin boxcar.

\citet{Kasliwal2022} reported on the detection of AT2017gfo in the Spitzer 4.5 $\mu$m band at +43d and at +74d \citep[see also][for an alternative data reduction]{Villar2018}. The Spitzer observations also gave upper limits in the 3.6 $\mu$m band. The reported 4.5 $\mu$m band flux at +43d is $F_\nu = 6.43 \times 10^{-29}$ erg s$^{-1}$ cm$^{-2}$ Hz$^{-1}$, which corresponds to
%
\begin{multline}
\blue{F_\lambda =  F_\nu \frac{c}{\lambda^2}} =
\blue{= 9.5 \times 10^{-12}\  \mbox{erg s}^{-1} \mbox{cm}^{-2}  \mbox{cm}^{-1}} \\
\blue{= 9.5 \times 10^{-20}\  \mbox{erg s}^{-1} \mbox{cm}^{-2}  \mbox{Å}^{-1}.} 
\end{multline}
This is equivalent to $F_\lambda \times \lambda_{\mu m} = 4.3 \times 10^{-19}\ \mbox{erg s}^{-1} \mbox{cm}^{-2}  \mbox{Å}^{-1}$ ($\lambda_{\mu m}=4.5$). We control our conversion by taking
\begin{multline}
L_{band} = F_\lambda \times \Delta \lambda_{band} \times 4 \pi d^2 \\ 
\approx 9.5 \times 10^{-20}\ \mbox{erg s}^{-1} \mbox{cm}^{-2}\mbox{Å}^{-1} \times 10^4 \mbox{Å} \times 4 \pi \times \\ 
\left(40 \times 10^6 \times 3.08 \times 10^{18}\mbox{cm}\right)^2
\blue{= 1.8 \times 10^{38}\ \mbox{erg s}^{-1}}, 
\end{multline}
%
similar to the value $2 \times 10^{38}$ erg s$^{-1}$ reported in \citet{Hotokezaka2022}.
%

The 3.6 $\mu$m-band upper limit magnitude at +43d of 23.21 corresponds to a flux limit of $F_\nu=1.9 \times 10^{-29}$ erg s$^{-1}$ cm$^{-2}$ Hz$^{-1}$ \footnote{\href{https://lweb.cfa.harvard.edu/~  dfabricant/huchra/ay145/mags.html}{https://lweb.cfa.harvard.edu/\textasciitilde  dfabricant/huchra/ay145/mags.html}}, or equivalently $F_\lambda \lambda_{\mu m} = 1.6 \times 10^{-19}\  \mbox{erg s}^{-1} \mbox{cm}^{-2}  \mbox{Å}^{-1}$.
For the +74d data, the corresponding values in the 4.5 $\mu$m band are $F_\nu = 1.04 \times 10^{-29}$ erg s$^{-1}$ cm$^{-2}$ Hz$^{-1}$, corresponding to $F_\lambda \lambda_{\mu m} = 6.9 \times 10^{-20}\ \mbox{erg s}^{-1} \mbox{cm}^{-2}  \mbox{Å}^{-1}$. The 3.6 $\mu$m band upper limit of 23.05 corresponds to $F_\lambda \lambda_{\mu m} = 1.8 \times 10^{-19}\ \mbox{erg s}^{-1} \mbox{cm}^{-2}  \mbox{Å}^{-1}$.

We use the JWST spectra of GRB 230307A/AT2023vfi, taken at +29d and +61d \citep{Levan2024}, utilizing the reduction of \citet{Gillanders2025} as our standard set\footnote{The reduced spectra were downloaded from 
\href{https://ora.ox.ac.uk/objects/uuid:5032f338-aff0-4089-9700-03dc5c965113}{https://ora.ox.ac.uk/objects/uuid:5032f338-aff0-4089-9700-03dc5c965113}.}. 
Different reductions are similar at +29d but significantly different at +61d, for the latter epoch we plot both for completeness. The +29d spectrum was smoothed with a 3-bin boxcar, and the +61d spectrum was  smoothed by a 5-bin boxcar.  
The spectra were corrected for redshift with $z=0.0646$ (luminosity distance 292 Mpc), and (Milky Way) extinction with $E(B-V)=0.076$ mag \citep{Levan2024}. We plot the spectra redward of 1.9\,\mum, as they are dominated by the GRB afterglow blueward of this \citep[see][]{Gillanders2023_AT2023vfi, Levan2024, Gillanders2025}, see also \citet{Yang2024} for complementary afterglow observations and modelling.

\section{Analytic limits of IR line formation} 
Before we proceed to study the model spectra, we review the various analytic limits of nebular line formation \citep[see also][]{Jerkstrand2017}.

\subsection{Optically thick, LTE limit}
In this limit, the 1-zone line luminosity is
\begin{multline}
L = V \times n_u A  \beta_S h \nu = V \frac{n_u A h \nu}{n_l A \lambda^3 \frac{1}{8\pi} \frac{g_u}{g_l} t \left(1- \frac{g_l}{g_u}\frac{n_u}{n_l}\right)} \\
= V \frac{8 \pi h \nu}{\lambda^3} t^{-1} \frac{e^{-h\nu/kT}}{\left(1- e^{-h\nu/kT}\right)} = \frac{4\pi V}{ct} \lambda B_{\lambda} (T),
\label{eq:thicklte}
\end{multline}
where $V$ is the volume, $n_u$ and $n_l$ are the number densities of the upper and lower states, $A$ is the Einstein coefficient for spontaneous emission, $\beta_S$ is the Sobolev escape probability, $g_u$ and $g_l$ are the statistical weights of the upper and lower levels, $t$ is time, and the other symbols have their usual meaning. If $h\nu \ll kT$, then $ e^{-h\nu/kT} \approx 1 - h\nu/kT$, giving
\begin{equation}
\blue{L \approx \frac{8 \pi k}{\lambda^3 t} V T.}
\label{eq:LTEthick}
\end{equation}

\subsection{Strong NLTE limit}
Assuming that collisional excitation from $n_l$ dominates the inflow to $n_u$, but the electron density $n_e$ is low enough that deexcitation is always radiative, we have
\begin{equation}
\blue{n_u A \beta_S = n_l n_e \frac{8.6 \times 10^{-6}}{\sqrt{T}}\frac{1}{g_l}\Upsilon(T) e^{-h\nu/kT}},
\end{equation}
where $\Upsilon$ is the effective collision strength, usually weakly dependent on temperature. Then
\begin{equation}
L = V \frac{hc}{\lambda}n_l n_e \frac{8.6 \times 10^{-6}}{\sqrt{T}} \frac{1}{g_l}\Upsilon e^{-h\nu/kT}.
\end{equation}
Taking $n_l \approx n_{ion}$, and using $M_{ion} = V n_{ion} \mu m_p$, where $\mu$ is the atomic weight of the ion in a.m.u., this becomes
\begin{equation}
L \approx \frac{8.6 \times 10^{-6} hc}{\lambda g_l \mu m_p} T^{-1/2}e^{-h \nu/kT} M_{ion} n_e \Upsilon.
\label{eq:strongnlte}
\end{equation}
If $h\nu \ll kT$, then the exponential factor approaches unity. Because $\Upsilon$-factors have weak $T$-dependencies, the temperature dependency is then approximately $\propto T^{-1/2}$.

\subsection{\blue{Optically thin, LTE limit}}
This limit yields
\begin{equation} 
L = V n_l \frac{g_u}{g_l} e^{-h\nu/kT} A h \nu = \frac{hc}{\lambda \mu m_p} \frac{g_u}{g_l} e^{-h\nu/kT} M_{ion} A.
\label{eq:thinlte}
\end{equation}
If $h\nu \ll kT$, then the exponential factor approaches unity, so there is no temperature-dependency.

\subsection{Summary}
Thus, line formation can occur in three different regimes. For $T \gg T_{exc}$ (often fulfilled for MIR lines), the optically thick LTE regimes probes $V \times T$ (Eq. \ref{eq:LTEthick}), the strong NLTE regime probes $M_{ion} \times n_e/\sqrt{T}$ (Eq. \ref{eq:strongnlte}), and the optically thin LTE regime probes $M_{ion}$  (Eq. \ref{eq:thinlte}). Even if the luminosity is generated over a region with varying physical conditions, the total luminosity may still robustly reveal information on masses or volumes. For example, if formation is in the third regime, it does not matter how $T$ and $n_e$ vary over the region. If formation is in the first regime, the emission volume can be inferred no matter what the $n_e$ distribution is (and if a good average temperature can be assessed).

Note that while lines with $\lambda < hc/kT$ are more sensitive to temperature than the Eq. \ref{eq:LTEthick}, \ref{eq:strongnlte}, \ref{eq:thinlte} limits, ratios of such lines with similar excitation energies can still give abundances ratios to good accuracy.

\begin{figure*}[htb]
\centering
\includegraphics[width=0.6\linewidth]{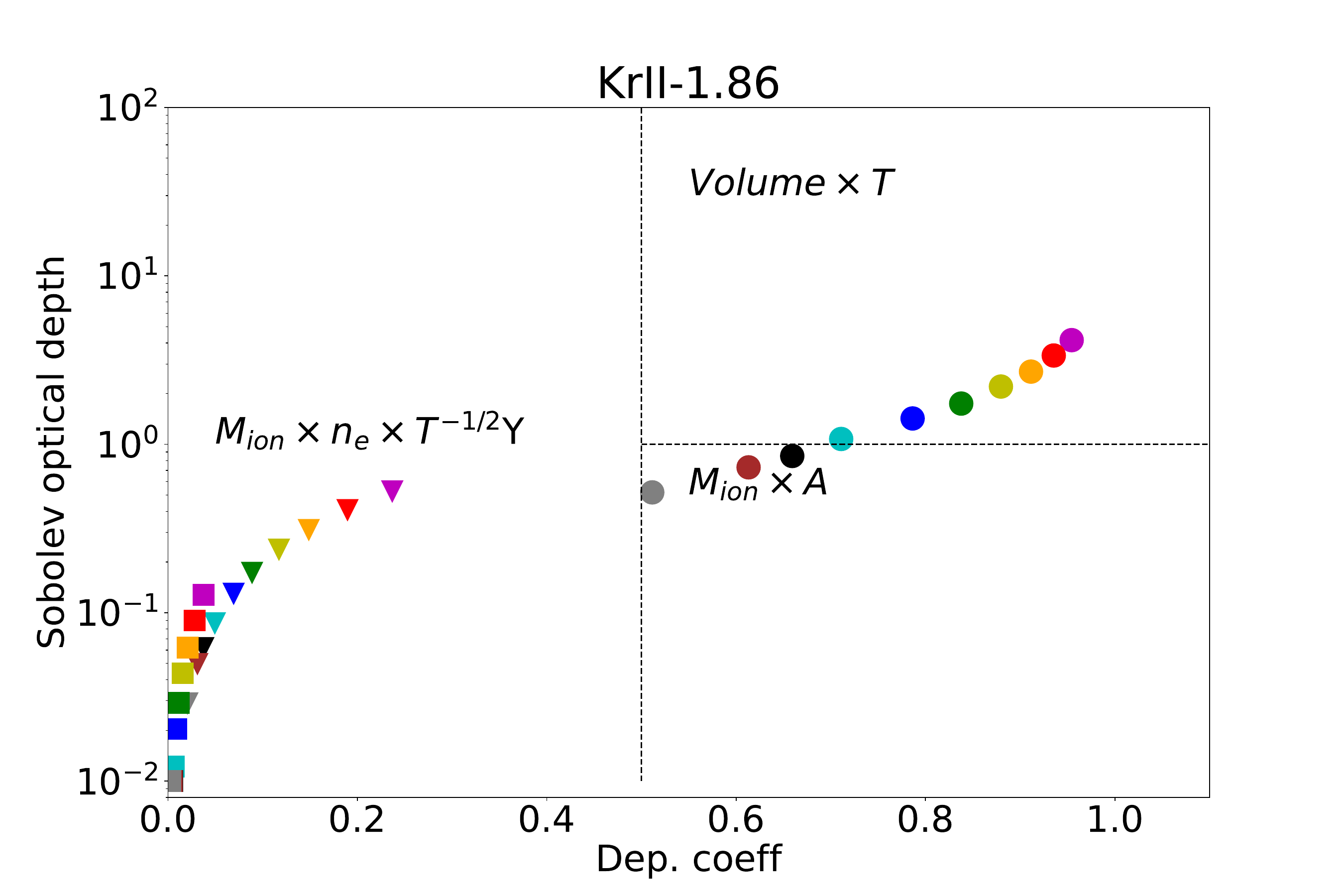}
\includegraphics[width=0.6\linewidth]{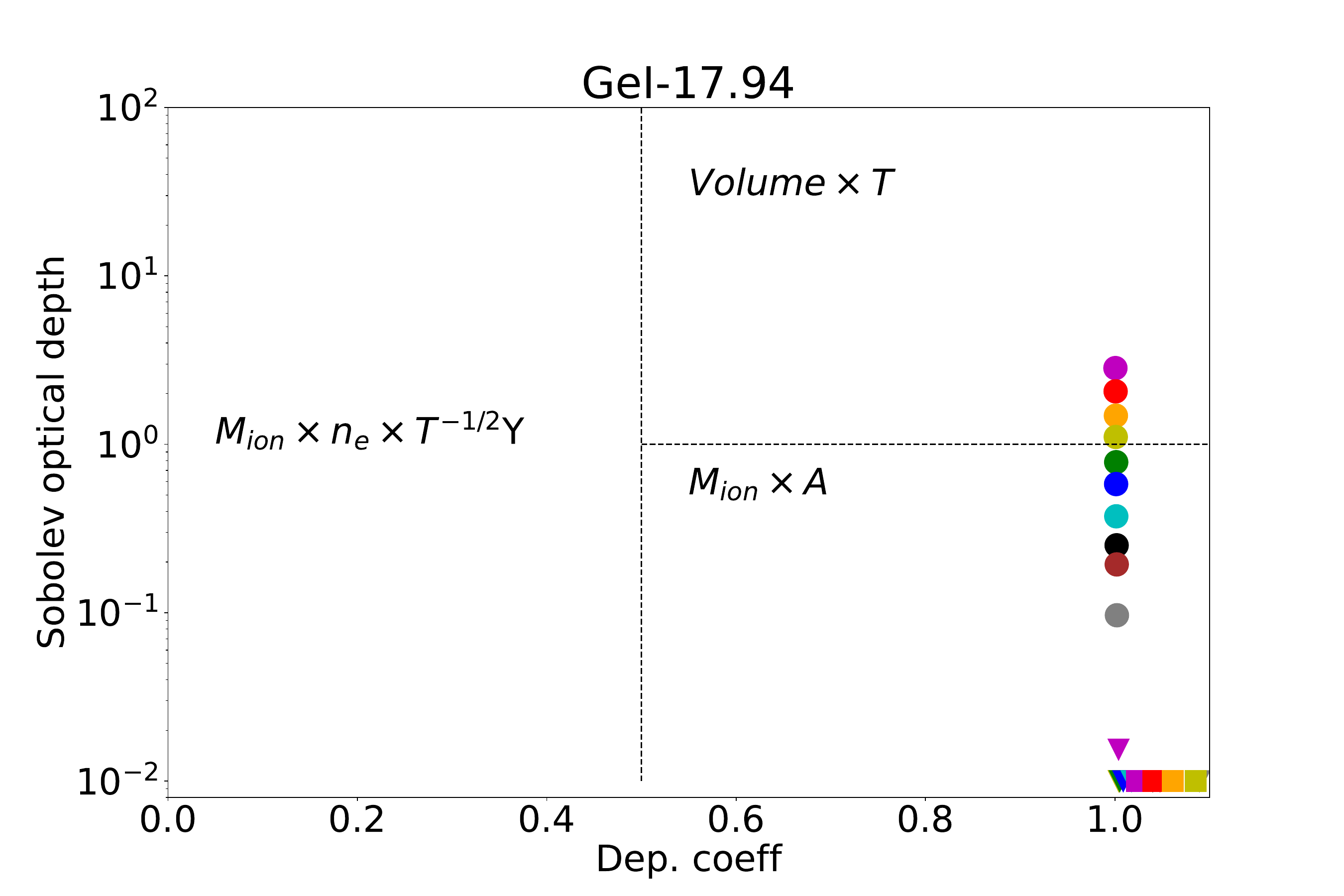}
\caption{Formation regimes for [Kr~II] 1.86 \mum\ (top) and [Ge~I] 17.94 \mum\ (bottom), in model A at 10d (circles), 40d (triangles) and 80d (squares). The x-axis shows departure coefficient for the upper level, and y-axis the Sobolev optical depth for the transition. Each color corresponds to a zone, with purple the innermost zone. Zone 1 has $v\sim 0.02$c, zone 10 has $v\sim0.06c$. The stated temperature-dependencies are for the limit $T \gg T_{exc}$.}
\label{fig:regimes-zonespec}
\end{figure*}

Figure \ref{fig:regimes-zonespec} plots the position in the departure coefficient (level population relative to its value for the case of an atom in LTE) - optical depth plane for the [Kr II] 1.86 \mum\ and [Ge I] 17.94 \mum\ lines, in model A at 10d, as examples. For each epoch, the values in the innermost 10 zones are plotted, which are the ones with the most energy deposition, and therefore will be dominant for line formation. Dividing lines between the different line formation limits discussed above are also drawn.

[Kr II] 1.86 \mum\ is at 10d mostly in the upper right corner, which means its luminosity probes volume but not mass. At 40d and 80d, on the other hand, it has moved to the left side domain, which means its luminosity probes the product of Kr II mass and electron number density (and $T^{-1/2}$). Thus, by the time the Kr II mass can be diagnosed, the regime is already where there is dependency on $n_e$ - the line never (cleanly) passes through the $M_{ion} A$ regime (bottom right).

[Ge I] 17.94 \mum\ (bottom panel) behaves differently. It straddles the upper right and bottom right regimes at 10d, indicating hybrid formation that needs detailed modelling for interpretation. Then, at 40d and 80d the line is fully in the bottom right corner, which means direct probing of $M_{ion}$. As the departure coefficients are close to unity (LTE), the line is far away from any $n_e$ sensitivity which gives good robustness for mass inferences.

The accuracy to which an ionic mass can be inferred in the most favorable case; optically thin LTE and $T_{exc} \ll T$ depends in theory on the accuracy of the A-value of the transition only. These are known to expected accuracies of a few percent for the low-lying M1/E2 transitions giving the dominant IR lines. In practice, the accuracy will be more limited by uncertainty in the exact value of $\exp{\left(-T_{exc}/T\right)}$, which e.g. is 0.9 if $T_{exc}/T=0.1$ but 0.74 if $T_{exc}/T = 0.3$. Uncertainty in the measured line luminosity, coming from uncertainties in distance, flux calibration, and blending with other lines, will in many cases also be influential for which accuracy masses can be determined to.

By studying the spatial and temporal variation of these formation regimes, it is possible to identify which lines give what diagnostic information at which epochs - valuable for developing high-return observation strategies.

\section{Results} 
\subsection{Model A, evolution of physical conditions}

Over the modelled time span (10-80d), model A has a radioactive powering (after thermalization) ranging from $4.6\times 10^{39}$ \ergs at 10d to $9.1\times 10^{38}$ \ergs\ at 80d, which is a factor 5.1 decline. This is a significantly smaller power decline than a canonical $t^{-1.3}$ decay (and declining thermalization efficiency would steepen further) which would give a factor $\gtrsim$15. The reason is that at this $Y_e$ a few isotopes dominate the power output rather than a large ensemble. The power for $Y_e \sim 0.35$ is lower than for low-$Y_e$ compositions at 10d (by a factor 3-4), but higher at 80d (by a factor $\sim$2) \citep{Wanajo2014}.

\begin{figure}[htb]
\centering
\includegraphics[width=1\linewidth]{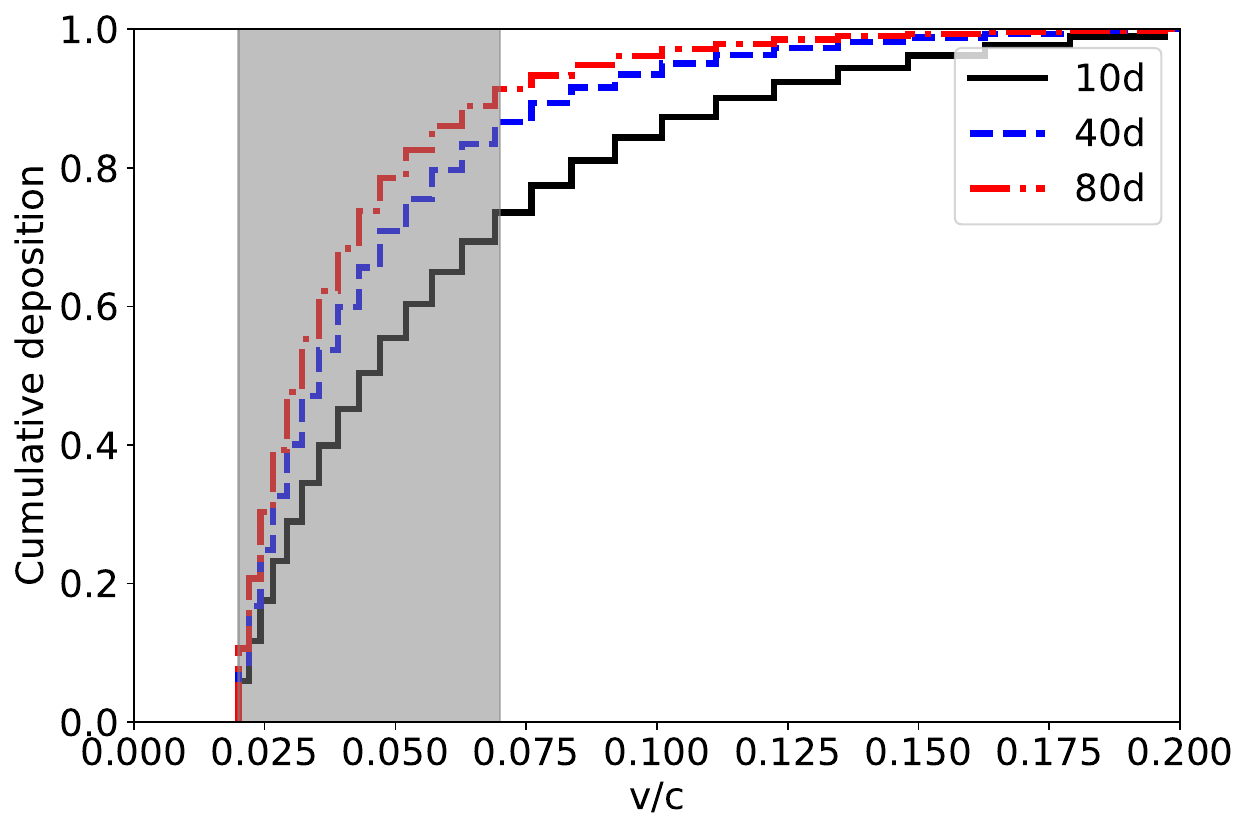}
\includegraphics[width=1\linewidth]{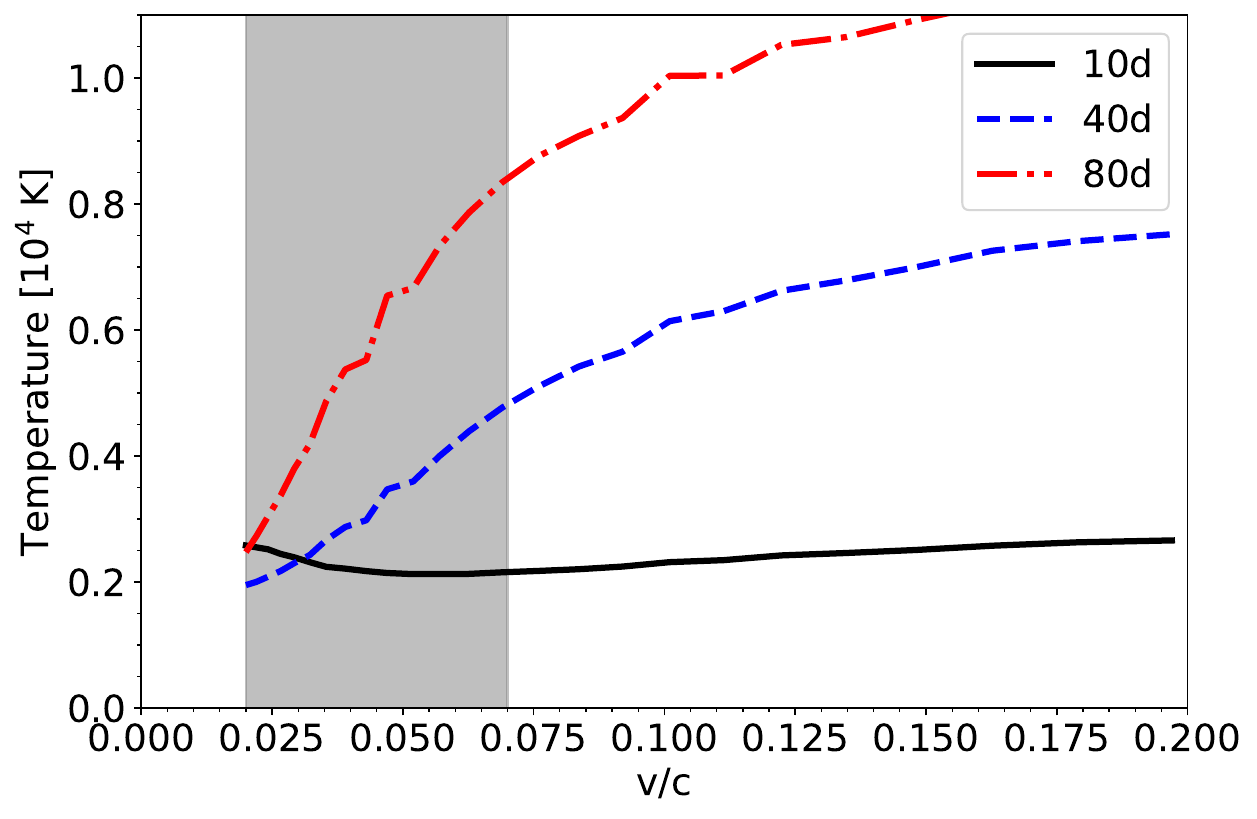}
\includegraphics[width=1\linewidth]{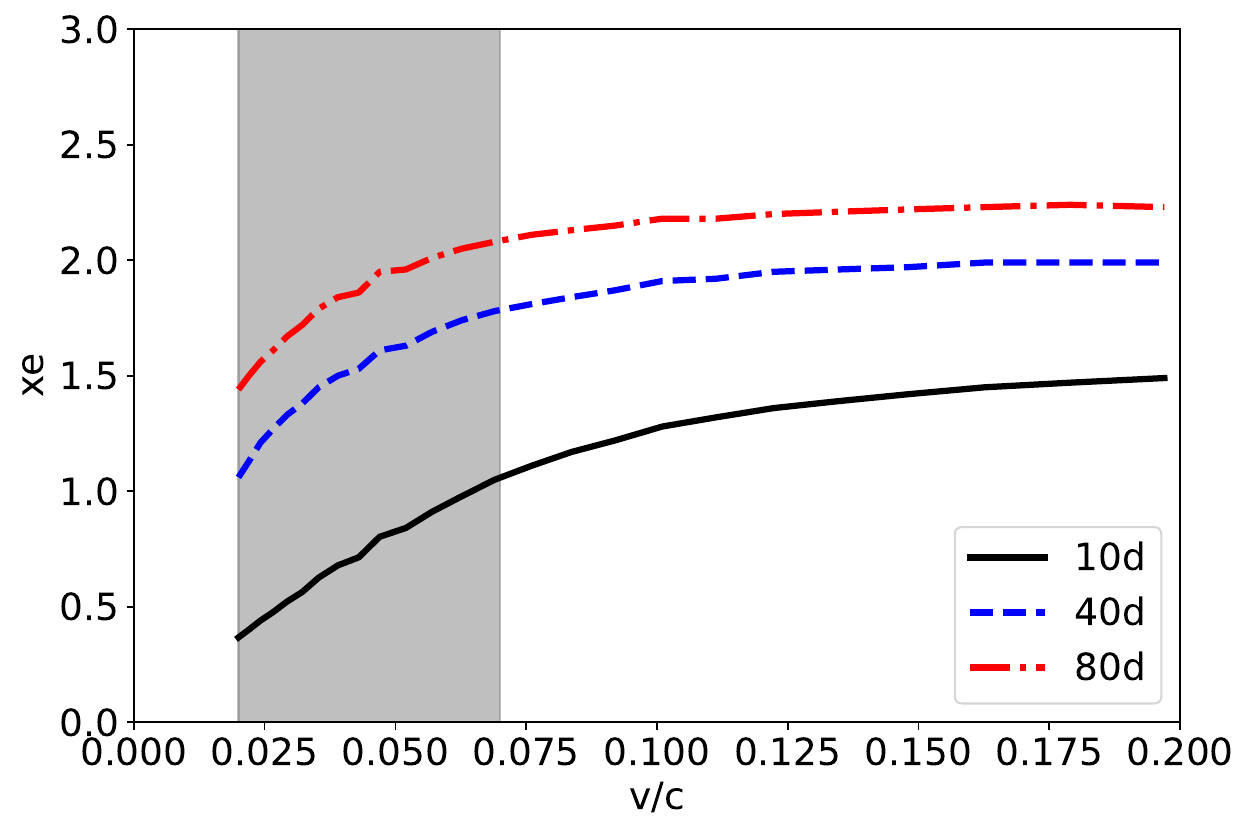}
\caption{Physical conditions in model A; cumulative radioactive deposition (top), electron temperature (middle), electron fraction (bottom), at 10d (black, solid), 40d (blue, dashed) and 80d (red, dash-dotted). The $\gtrsim$80\% deposition region is marked gray.}
\label{fig:Alow-phys}
\end{figure}

Figure \ref{fig:Alow-phys} (top) shows the cumulative radioactive deposition versus velocity in model A. Some $\gtrsim$80\% of the deposition occurs in the layers inside 0.07c, corresponding to the inner $\sim$10 shells of the total 25. The central region dominance can be seen to increase with time, as thermalization inefficiency increasingly sets in for the outer layers. This effect will lead to a narrowing of emission lines with time, and hence cleaner, less blended spectra at late nebular epochs.

Figure \ref{fig:Alow-phys} (middle) shows the temperature versus velocity at the three epochs. At 10d the ejecta are close to isothermal at around 2300 K in the line-forming region ($v/c \lesssim 0.07$). With time there is then a general trend of increasing temperatures, as radioactive powering decreases more slowly than the collisional cooling efficiency \citep{Hotokezaka2021,Pognan2022a}, although the innermost few zones break this trend and become somewhat cooler between 10d and 40d (changes in ionization and reduced photon trapping can also affect the temperature). In contrast to at 10d, there is at later epochs a quite steep temperature gradient ($\sim2000-5000$ K over $v/c=0.02-0.07$ at 40d and $\sim2500-8000$ K at 80d). The general temperature scale in the line-forming region indicates that emission lines with $\lambda \gtrsim hc/\left(k\times 2000~\mbox{K}\right) \gtrsim\ 7\ \mu$m will at all epochs be quite insensitive to the detailed temperature (proportional to $T^\alpha$, where $\alpha=-1/2,0$, 1, depending on regime, as opposed to exponential dependency if $\lambda < hc/kT$), making mid-infrared lines uniquely powerful probes.

Figure \ref{fig:Alow-phys} (bottom) shows the ionization (electron fraction $x_e$) versus velocity. In the line forming region ($\lesssim 0.07c$) the characteristic value is $x_e\sim$0.7 at 10d (neutrals and singly ionized species dominate), increasing to $x_e \sim 1.5$ at 40d (singly and doubly ionized species dominate) and $x_e\sim$1.8 at 80d (doubly ionized species dominate). While $x_e$ does increase with velocity coordinate, this effect only slightly dampens the electron density profile from the model density profile ($n_{nuclei} \propto v^{-3}$), roughly to a $n_e \propto v^{-2}$ form in the inner layers. As such, the electron density varies by about an order of magnitude over the line-forming region ($\sim 0.02-0.07$c). The variation of $T$ and $n_e$ is important information for assessing single-zone approaches. Ionization structures of selected elements are plotted in Figs. \ref{fig:modelA-low-10d-ionstructure}-\ref{fig:modelA-low-80d-ionstructure}. 

\begin{figure*}[htb]
\centering
\includegraphics[width=0.7\linewidth]{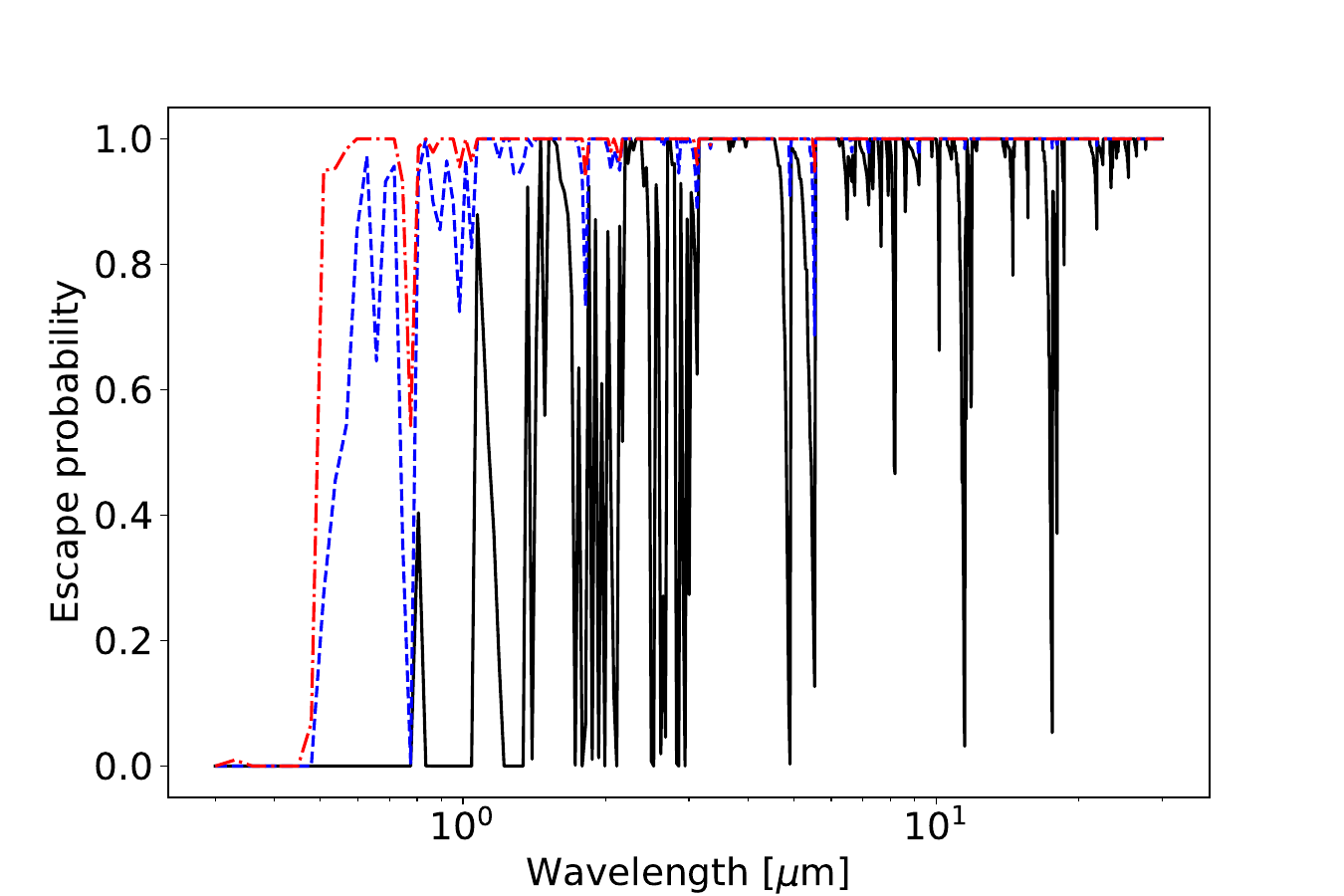}
\caption{\blue{Escape probability for a photon emitted at the centre of the nebula at the modelled epochs (10d; black, solid, 40d; blue, dashed, 80d;  red, dash-dotted) for model A.}}
\label{fig:pesc-multi}
\end{figure*}

Figure \ref{fig:pesc-multi} shows the escape probability for a photon emitted at the centre of the nebula versus wavelength. At 10d, the optical range is completely opaque, and much of the NIR as well, although certain escape windows exist there, where the line opacity is sufficiently low. The MIR range is optically thin for the most part at 10d although some lines remain optically thick, causing absorption ranges. By 40d the NIR range has largely opened up, and at 80d conditions are mostly optically thin down to $\sim$5000 Å, although a few lines remain optically thick in the innermost zones, e.g. Rb I 7802, 7949 Å. 

\subsection{Spectrum of model A, 10d} 
\begin{figure*}[htb]
\centering
\includegraphics[width=0.72\linewidth]{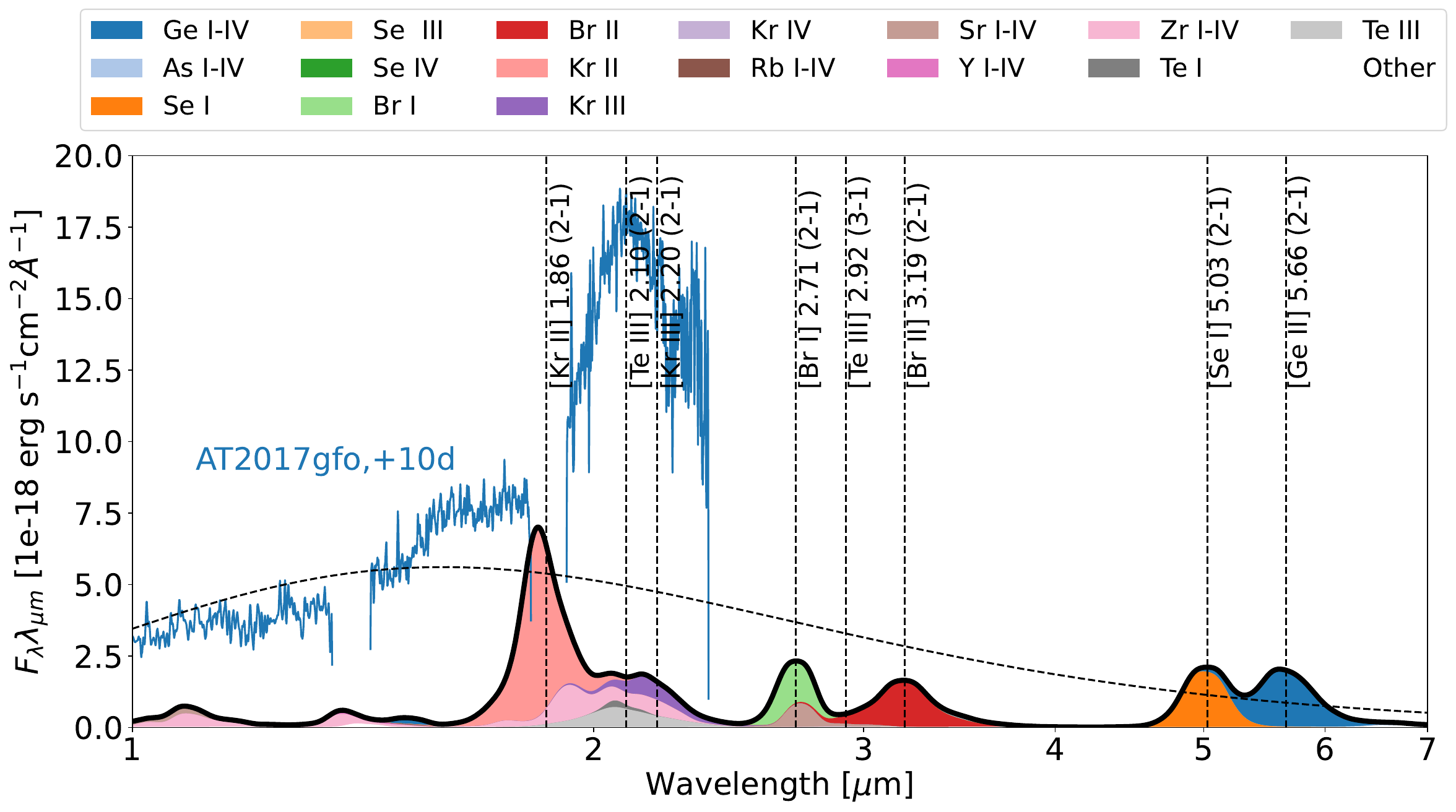}
\includegraphics[width=0.72\linewidth]{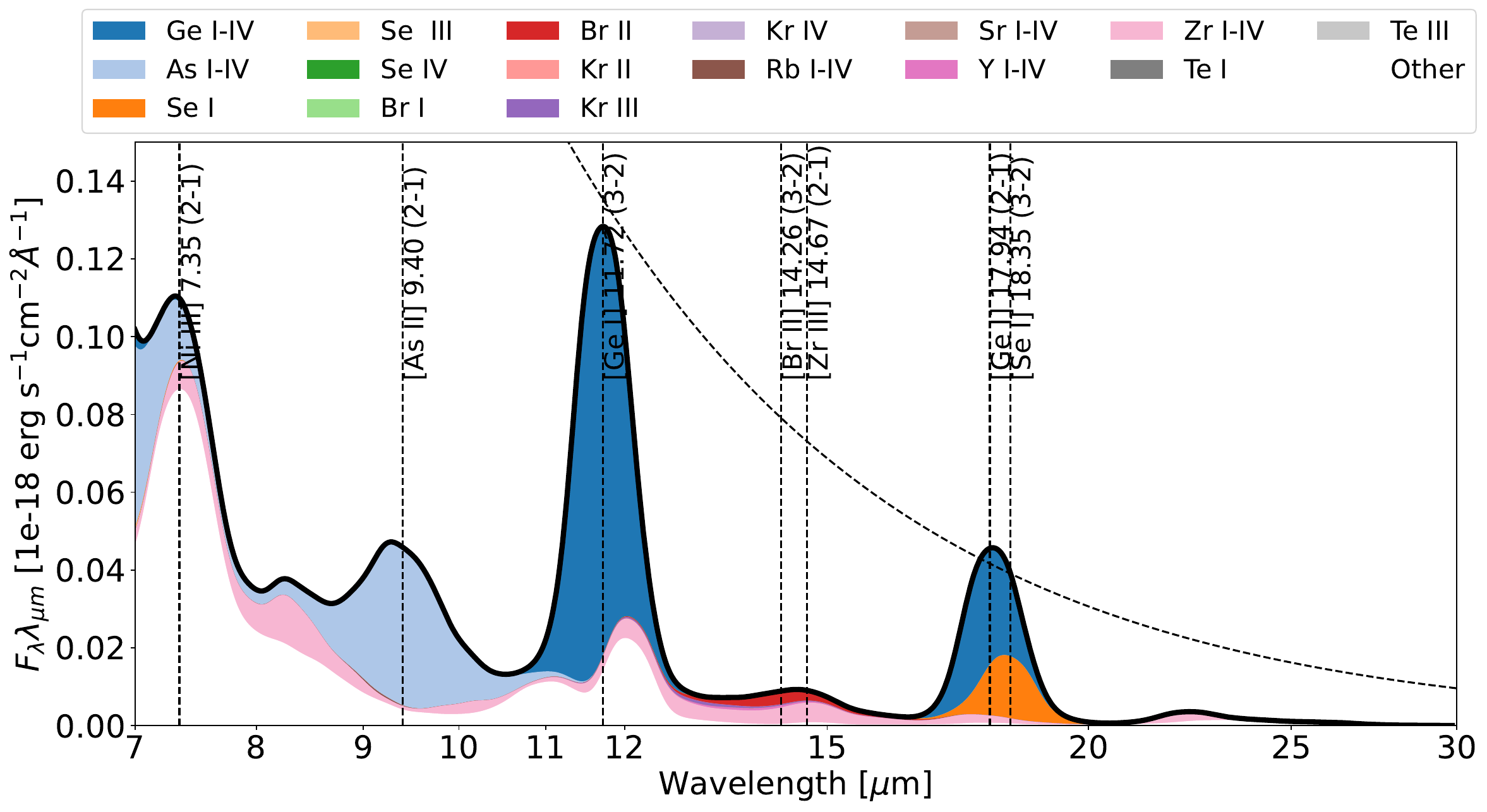}
\caption{Spectrum of model A at 10d. The observed (ground-based) spectrum of AT2017gfo at +10.4d \citep{Pian2017} is plotted in blue. Overplotted is also a 2300 K blackbody (black, dashed) - lines forming in the optically thick LTE regime have peak fluxes at or close to this curve.
}
\label{fig:spec-Alow-10d}
\end{figure*}

Figure \ref{fig:spec-Alow-10d} shows the model spectrum at 10d. Line emission is dominated by [Kr II] 1.86 \mum, [Br~I] 2.71 \mum, [Br~II] 3.19 \mum, [Se~I] 5.03 \mum, [Ge~II] 5.66 \mum, [Ge~I] 11.72 \mum, [Ge~ I] 17.94 \mum\, [As~II] 9.40 \mum, and [Se~I] 18.35 \mum. There is also some emission in the allowed Sr I transitions between 5s4d($^3$D) to 5s5p($^3$P$^{\rm o}$) at 2.60-3.01 \mum. Emission is also seen for [Ni~III] 7.35 \mum\ and [Zr~III] 14.67 \mum\ (some blending with [Br~II] 14.26 \mum).

\begin{figure}[htb]
\centering
\includegraphics[width=1\linewidth]{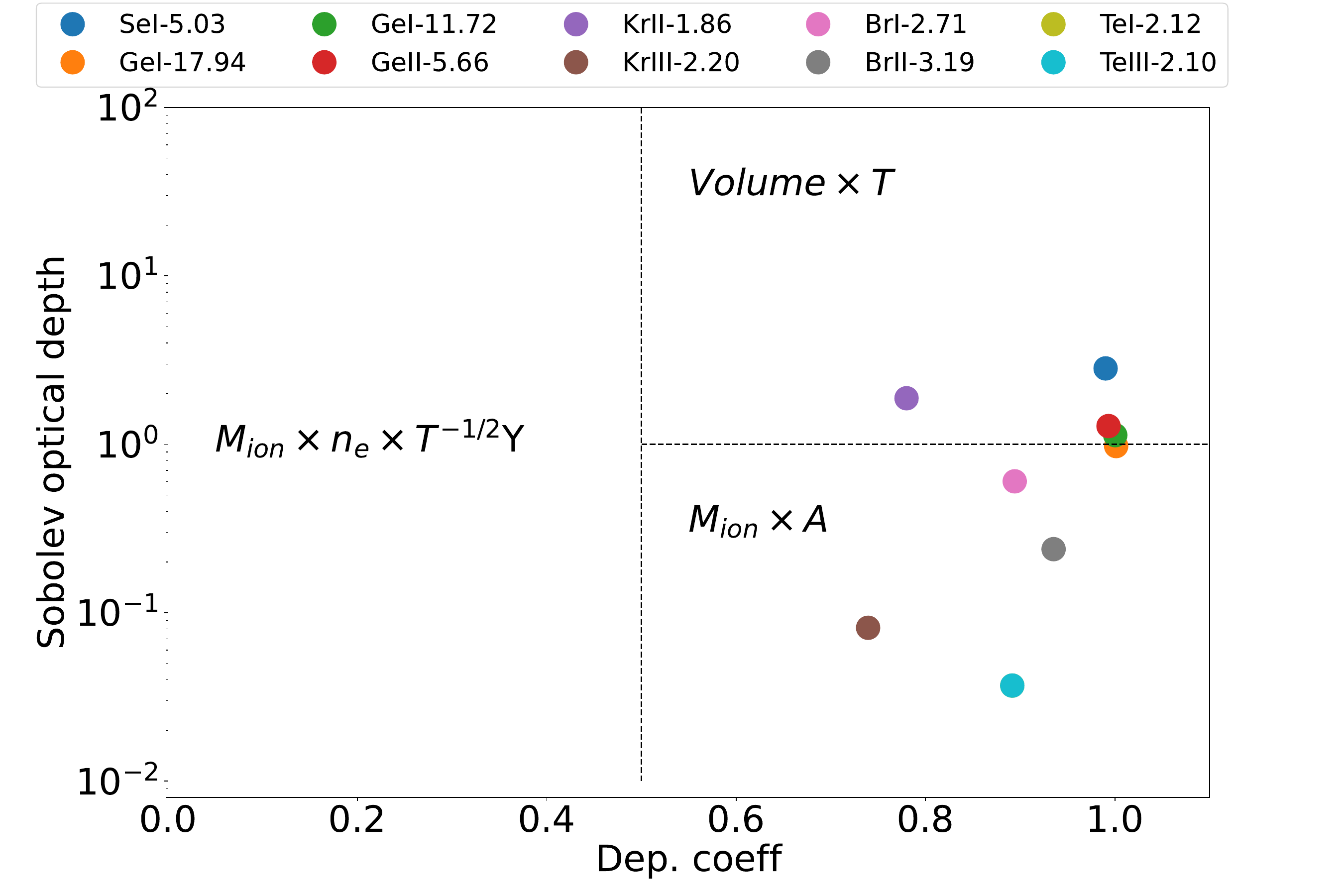}\\
\includegraphics[width=1\linewidth]{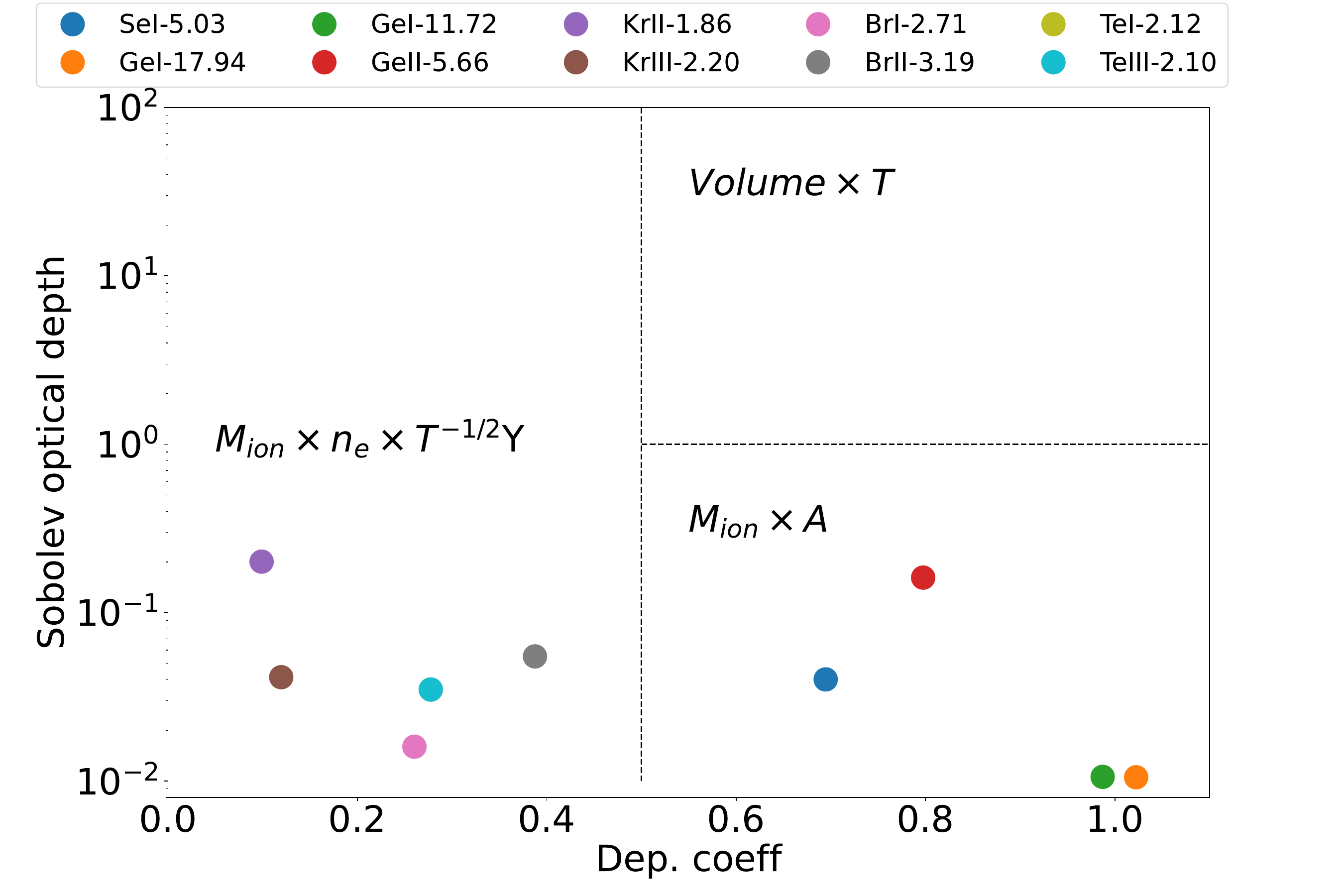}\\
\includegraphics[width=1\linewidth]{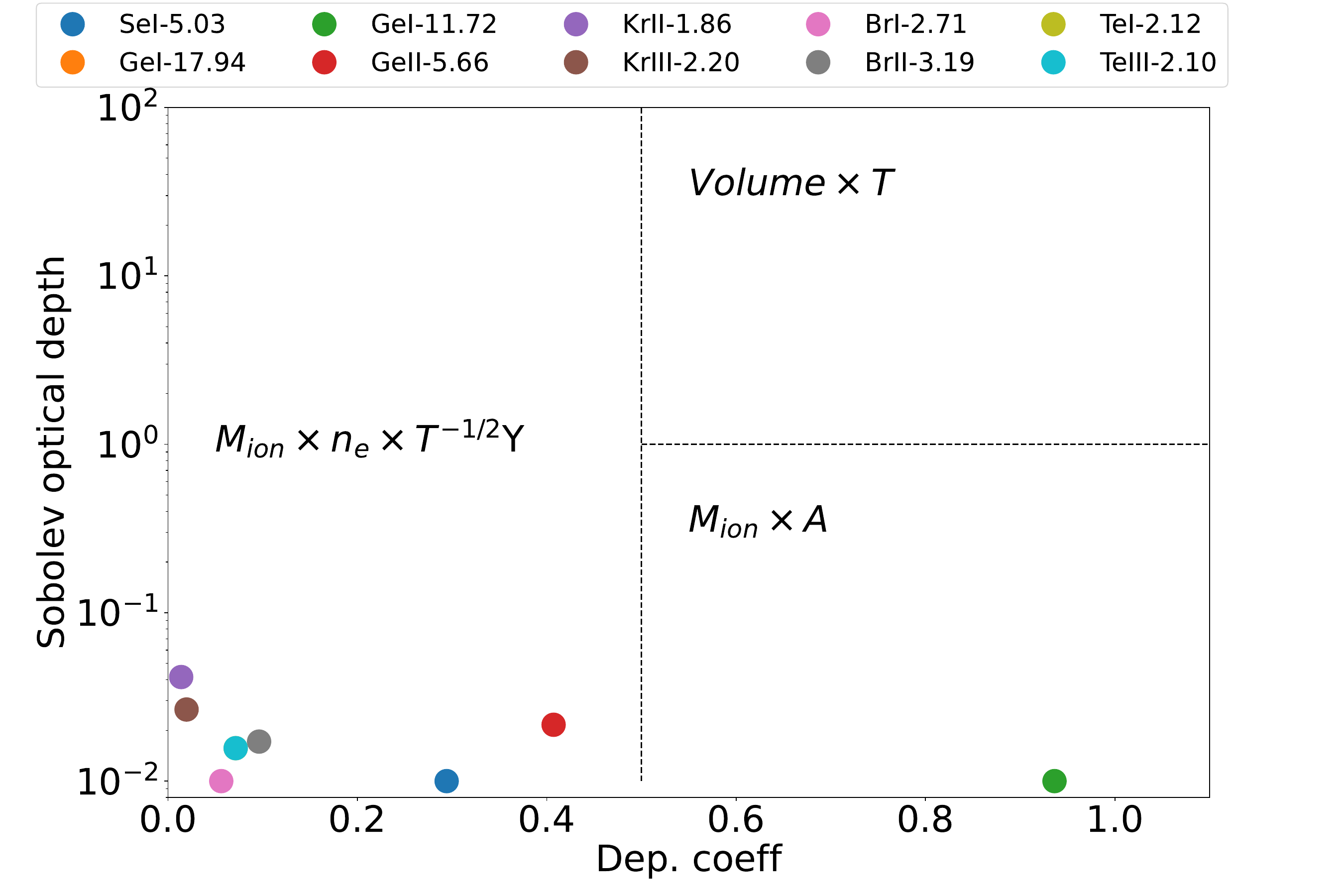}
\caption{Formation regimes in model A at 10d (top), 40d (middle) and 80d (bottom). Each point represents the zone-averaged value over the innermost eight shells. Optical depths below $10^{-2}$ are plotted at this floor value. The stated temperature dependencies are for the limit $T \gg T_{exc}$.}
\label{fig:regimes}
\end{figure}

\blue{Figure \ref{fig:regimes} (upper left) shows the zone-averaged formation regimes of the main lines at 10d. We can see that LTE is a quite good approximation, with all departure coefficients $\gtrsim 0.7$. We can also see that many of the lines are still optically thick. In those cases the specific ion masses are not important; each line samples a blackbody (at characteristic line-forming region temperature of $\sim$2300 K) and only emission volume and temperature affect the line flux (Eq. \ref{eq:thicklte}).}

\blue{At 2300 K, a blackbody peaks at $hc/\left(4.96 kT\right) = 1.3 \ \mu$m. The closest optically thick LTE  line to this is [Kr~II] 1.86 $\mu$m, which therefore becomes the strongest line in the spectrum. Other optically thick lines then follow on the black-body curve with peak luminosities in rough agreement with the theoretical limit (optically thick LTE case). Some variation occurs as the volume over which a given transition is optically thick and in LTE can vary. Radiative transfer effects still operate for $\lambda \lesssim 2\ \mu$m (Fig. \ref{fig:pesc-multi}) which there complicates this simple picture.} 

Comparison to AT2017gfo at +10.4d shows that AT2017gfo was brighter by a factor several in the NIR compared to the model. As described above, the $Y_e \approx 0.35$ composition has significantly less decay power than lower-$Y_e$ compositions at this epoch, and the discrepancy is possibly explained by the presence of such low-$Y_e$ material. 

Looking at spectral details, the model gives distinct emission lines in the NIR from Zr~I 1.37 \mum\ (this may be [Rb~III] 1.36 \mum\ emission scattering in optically thick Zr I lines), and [Kr~II] 1.86 \mum. These lines unfortunately lie in regions of strong atmospheric absorption, and hence the reconstructed flux of \citet{Pian2017} is uncertain at these wavelengths. If one takes the reconstructed fluxes at face value, the lack of these features in the observations around 10d might be due to several reasons. One is opacity provided by heavier elements ($Z>40$ elements are not included in the model, except Te); Fig. \ref{fig:pesc-multi} shows that even this light composition is largely optically thick in the NIR at 10d, and heavier elements would further enhance the opacity. Another is cooling by heavier elements (this would require these elements to be co-spatial). Another is that AT2017gfo has significantly less $Z=30-40$ elements than in this model (solar). A final possibility is that the velocity distribution in this model is unrepresentative. However, definitive conclusions cannot be drawn from the ground-based spectra of AT2017gfo. JWST observations of KNe, such as those for AT2023vfi \citep{Levan2024}, are needed to accurately measure the strengths of these Zr I/[Rb III], and [Kr II] features.

Irrespective of which, or combination of which factors, one valuable conclusion is that none of these $Z=31-40$ elements provide a compelling candidate line to explain the strong observed 2.1 \mum\ feature. [Kr~III] 2.20 \mum\ would be the line closest in wavelength, but it is relatively weak in the model, with most Kr being in the I and II states. There is also a Zr~I line at 2.24 \mum\ (0.02c separation) that is optically thick out to $\sim$0.05c, absorbing some of the [Kr~III] 2.20 \mum\ emission. The tellurium feature is a blend of [Te~I] 2.10 \mum\ and [Te~III] 2.10 \mum\ in this model. Its luminosity is substantially below the observed value. \ajn{The Te I lines becomes stronger than the Te III line when ionization is low/moderate, and can give signatures also in core-collapse supernova models \citep{Ricigliano2025}.}

\paragraph{Kr discussion.} [Kr~III] 2.20 $\mu$m is one of only three lines (the others being [Te III] 2.10 \mum\ and [Br II] 3.19 \mum) that directly probe ion mass at this epoch (Fig. \ref{fig:regimes}). It is however quite weak, as only about 1\% of the Kr in the model is in the III state. Dividing Eq. \ref{eq:thinlte} with Eq. \ref{eq:thicklte}, and using $A=2.0$ s$^{-1}$, $M_{\mathrm{Kr\ III}}=10^{-4}\ M_\odot$, $v=0.05c$, gives $L_{\rm LTE,thin}/L_{\rm LTE,thick} \sim 0.1$.

Let us consider ways to bring the [Kr III] line up towards observed levels for the 2.1 $\mu$m feature. If a different ionization structure would increase the Kr III abundance by a factor 10, the line would move into the optically thick regime, for the same overall density, and line luminosity would go up by a factor $\sim$10, all else remaining similar, according to the above ratio. Changing the volume would not easily achieve this, as a smaller volume would reduce the LTE, optically thick regime luminosity, and a larger volume would not achieve optical thickness (and would also move the regime towards NLTE). Increasing power would be more promising. It would lead to more Kr III, and a higher temperature. But, as long as a significant Kr II abundance exists, [Kr II] 1.86 $\mu$m would also strengthen. It would seem difficult to achieve a strong [Kr III] line but a weak [Kr II] line (with the caveat that most of the [Kr II] spectral feature lies is blocked by atmospheric absorption), unless the doubly ionized state becomes dominant to the singly ionized state; in the model the opposite relation holds, and calculated  recombination rates for the Kr ions are in use \citep[even though DR calculations for the quite low temperatures here carry uncertainties, e.g.][]{Ferland2003,Sterling2015}. 

\subsection{Spectrum of model A, 40d}
\begin{figure*}[htb]
\centering
\includegraphics[width=0.72\linewidth]{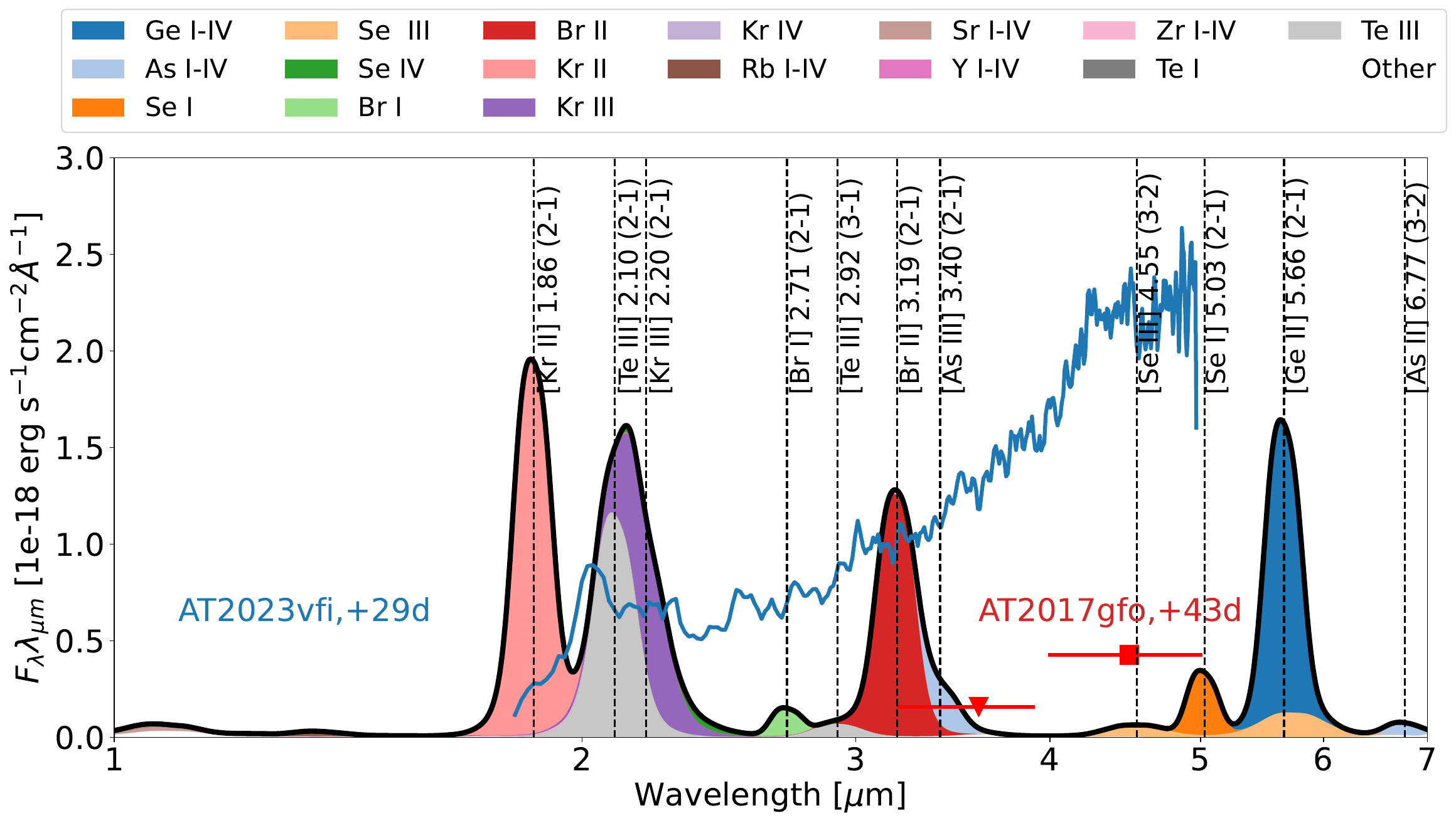}
\includegraphics[width=0.72\linewidth]{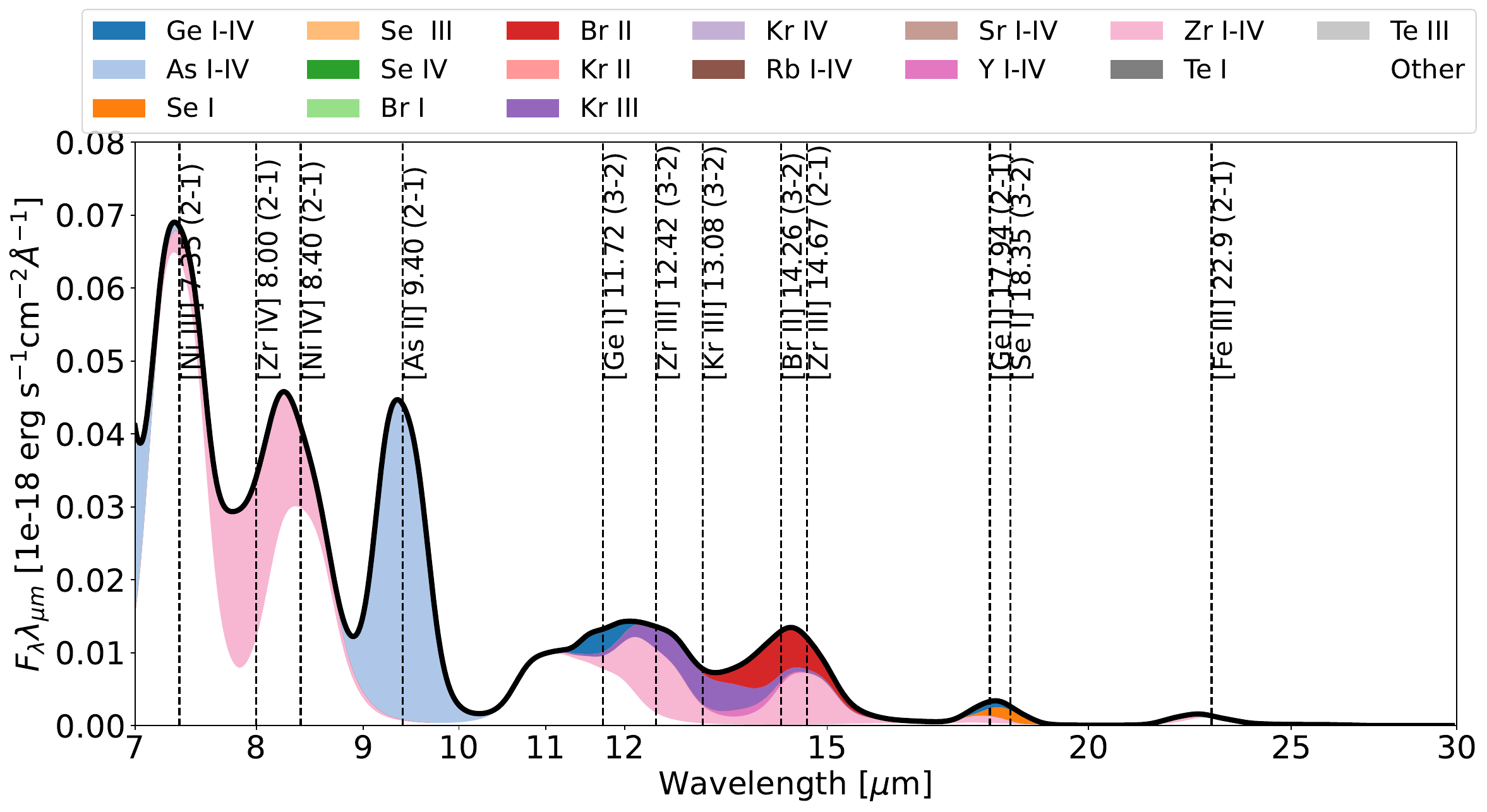}
\caption{Spectrum of model A at 40d. Observed photometry of AT2017gfo at +43d and observed spectra of AT2023vfi at +29d are also shown. The model and AT2023vfi spectra have been scaled to a distance of 40 Mpc (distance of AT2017gfo).}
\label{fig:spec-Alow-40d}
\end{figure*}

\blue{The spectrum at 40d is shown in Fig. \ref{fig:spec-Alow-40d}. The neutral germanium lines have now significantly weakened, as ionization has increased. The iron-group elements give relatively strong emission between $10-13$ $\mu$m, and also by [Ni II] 7.35 $\mu$m, 
whereas [Br II] and [Zr III] create a blend at around 14 $\mu$m. Clear detection potential is predicted for Ni, As, Br, Kr, Se, and Ge.}

All strong lines are now optically thin (Fig. \ref{fig:regimes}). The long-wavelength lines are still close to LTE, whereas the shorter-wavelength ones are now in the NLTE regime - it is however still the same lines that dominate as at 10d. The somewhat higher ionization leads to the [Kr III]/[Kr II], [Br II]/[Br I], and [Ge II]/[Ge I] line ratios now being higher.

The model predicts the best targets for direct ion mass determinations (optically thin, close to LTE, and weak $T$ sensitivity) to be [Ge I] 17.94 $\mu$m, [Ge I] 11.72 $\mu$m, [Ge II] 5.66 $\mu$m, and [Se I] 5.03 \mum. The Ge I lines are however quite weak and blended, but the [Ge II] and [Se I] lines are strong. Neutral Ge makes up 1-10\% of the Ge in the line-forming region, and singly ionized Ge $\gtrsim$60\%. Therefore joint observations of the Ge I and Ge II emission lines holds promise to get a good constraint on the total Ge mass. Selenium has a substantial Se II abundance ($\gtrsim 50$\%), which means observations of the I and III lines (Se II gives no strong lines) still need a significant correction factor to estimate the total Se mass.

The strongest line in the model is, as at 10d, [Kr~II] 1.86 $\mu$m. There is no sign of this line in the +29d JWST spectrum of AT2023vfi. At this epoch, there is also no possibility to shift model flux over to the [Kr~III] 2.20 $\mu$m line (by a higher ionization state); it is already too strong in the model compared to observations. While at 10d radiative transfer effects at these wavelengths are still present to some degree, at 40d they are minor \citep[at least from the elements included here, and models including also heavier elements show little absorption by these,][]{Pognan2026}; the model predicts none that could absorb [Kr~II] 1.86 $\mu$m or [Kr~III] 2.20 $\mu$m emission.

It would seem difficult to resolve this ``Kr issue'' with either an ionization effect (only small fractions reside in I and IV states) or a radiative transfer effect (epoch is too late). Apart from the option of there being little or no Kr in the ejecta, the remaining option is a lower temperature. If the temperature were 1500 K instead of the roughly 3000 K in the model, the [Kr II] 1.86 $\mu$m luminosity would go down a factor 13. This might happen if there are many other cooling lines from heavier elements, not included in the model, as observations of AT2023vfi may suggest with its smooth spectrum. A counterpoint to this is that the power in model A is quite low, clearly lower than what is needed for AT2017gfo at +10d (by a factor $\sim$5) and for AT2023vfi at +29d (factor $\sim$2) - but heavier elements will bring in higher power, acting to raise the temperature.

The Te emission is at this epoch almost exclusively due to [Te III] 2.10 \mum. The luminosity is, as at 10d, similar to the [Kr II] + [Kr III] emission. The observed feature in AT2023vfi has a peak close to 2 \mum, not in clean agreement with any line.

\subsection{Spectrum of model A, 80d}

\begin{figure*}[htb]
\centering
\includegraphics[width=0.72\linewidth]{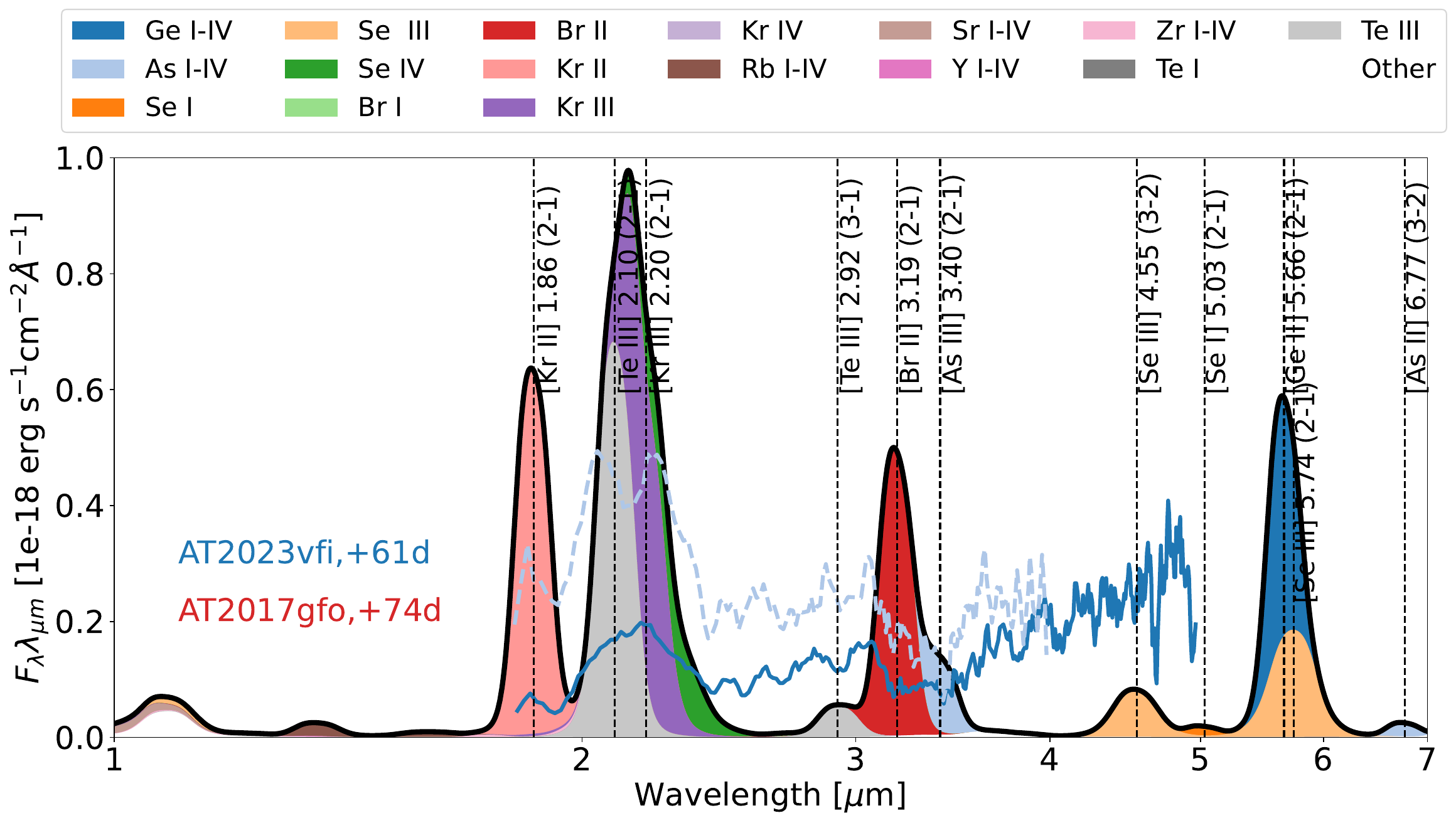}
\includegraphics[width=0.72\linewidth]{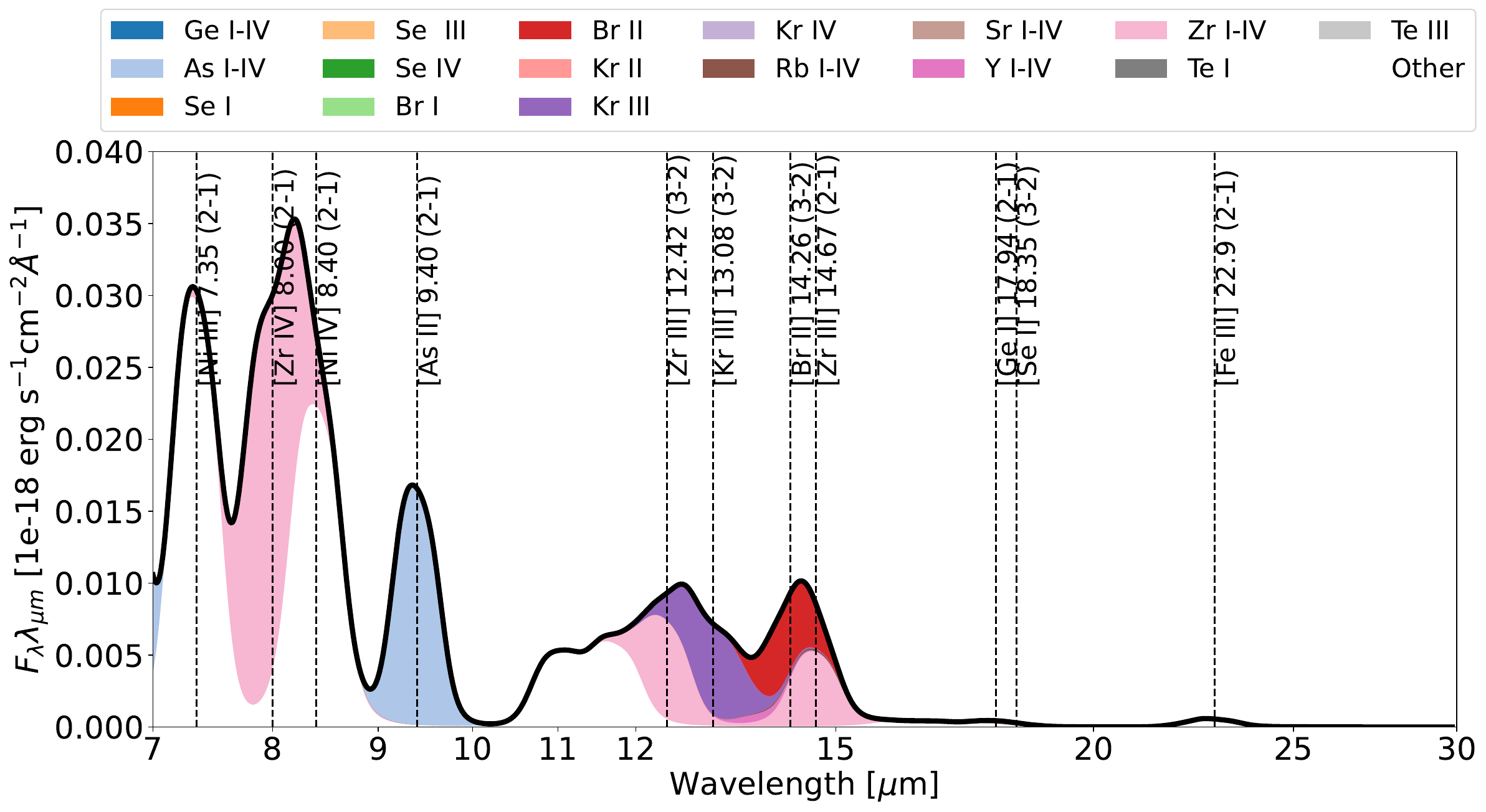}
\caption{\blue{Spectrum of model A at 80d. Observed photometry of AT2017gfo at +74d and observed spectra of AT2023vfi at +61d are also shown; our standard reduction set \citep{Gillanders2025} is plotted solid, we plot also the reduction of \cite{Levan2024} (dashed) for completeness. The model and AT2023vfi spectra have been scaled to a distance of 40 Mpc (distance of AT2017gfo).}}
\label{fig:spec-Alow-80d}
\end{figure*}

The spectrum at 80d is shown in Fig. \ref{fig:spec-Alow-80d}. The ejecta are now somewhat hotter (characteristic temperature in the line-forming region of $\sim$5000 K) and more ionized (characteristic $x_e$ of $\sim$1.8), than at 40d. The $1-10$ $\mu$m region is qualitatively similar as at +40d, although the line luminosities are down by a factor $\sim$2. The neutral lines have now mostly disappeared. In the $10-30$ $\mu$m range, the Ge I lines have disappeared, while some of the other features remain at similar luminosities as at 40d. The iron-group cluster at $10-13$ $\mu$m has strengthened further, and it blends into a complex of lines from Ge I, Zr III, Kr III, and Br II extending up to 15 $\mu$m. 

The +74d Spitzer 4.5 \mum\ observation of AT2017gfo is matched reasonably well by the combination of [Se III] 4.55 \mum\ and [Se I] 5.03 \mum\ emission in the model. As at 40d, the 3.6 \mum\ upper limit is at or somewhat below the model prediction - with [Br II] 3.19 \mum\  tending to give a bit too much luminosity into the band.

For AT2023vfi, the predicted [Br II] 3.19 \mum\ line is not, as at 40d, observed,  and the issue with [Kr II] 1.86 \mum\ is also still there. The observed feature seen close to 2.1 \mum\ lies somewhat more towards the red at +61d, and would match with a relatively flat-topped [Kr III] 2.20 \mum\  line - in model A the line is however too narrow. Both the [Kr III] 2.20 \mum\ and [Te III] 2.10 \mum\ model lines have similar luminosities as the AT2023vfi feature.

\subsection{Model B - evolution of physical conditions}
\begin{figure}[htb]
\centering
\includegraphics[width=1\linewidth]{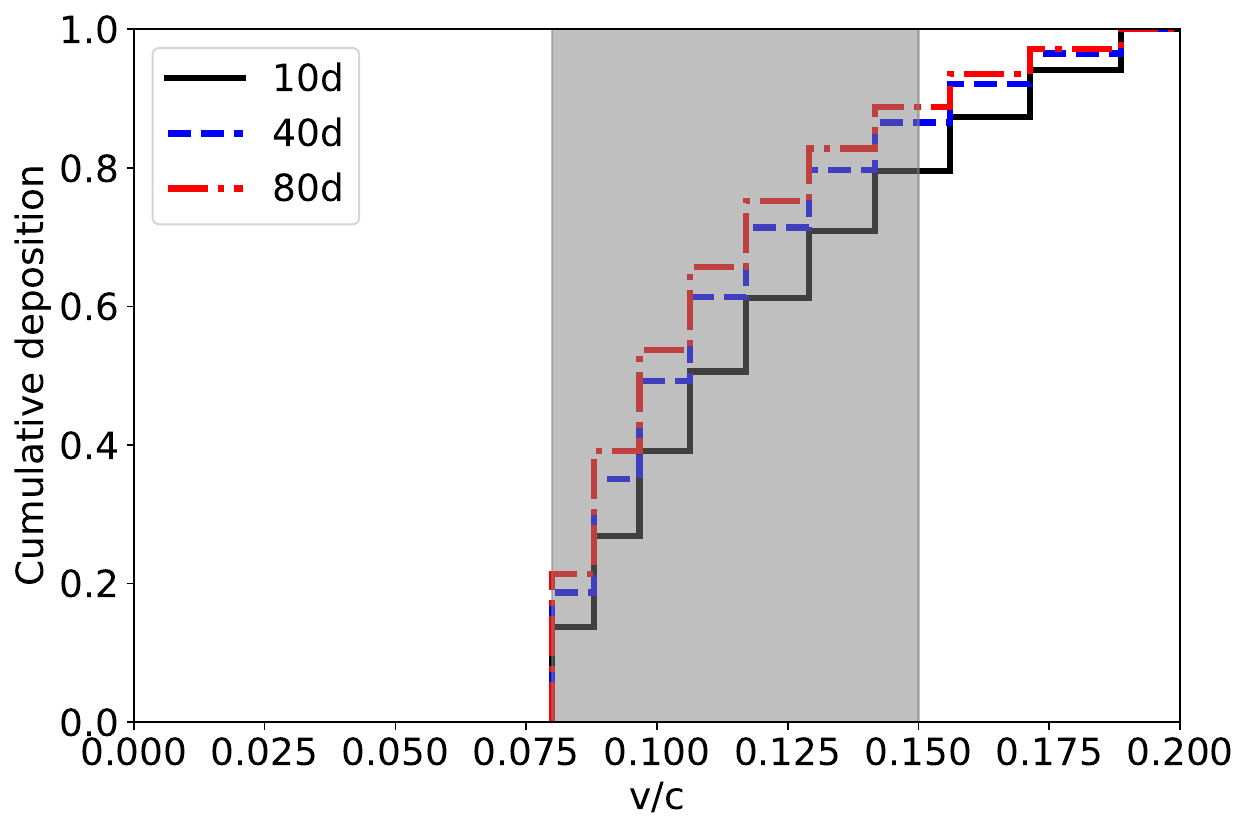}
\includegraphics[width=1\linewidth]{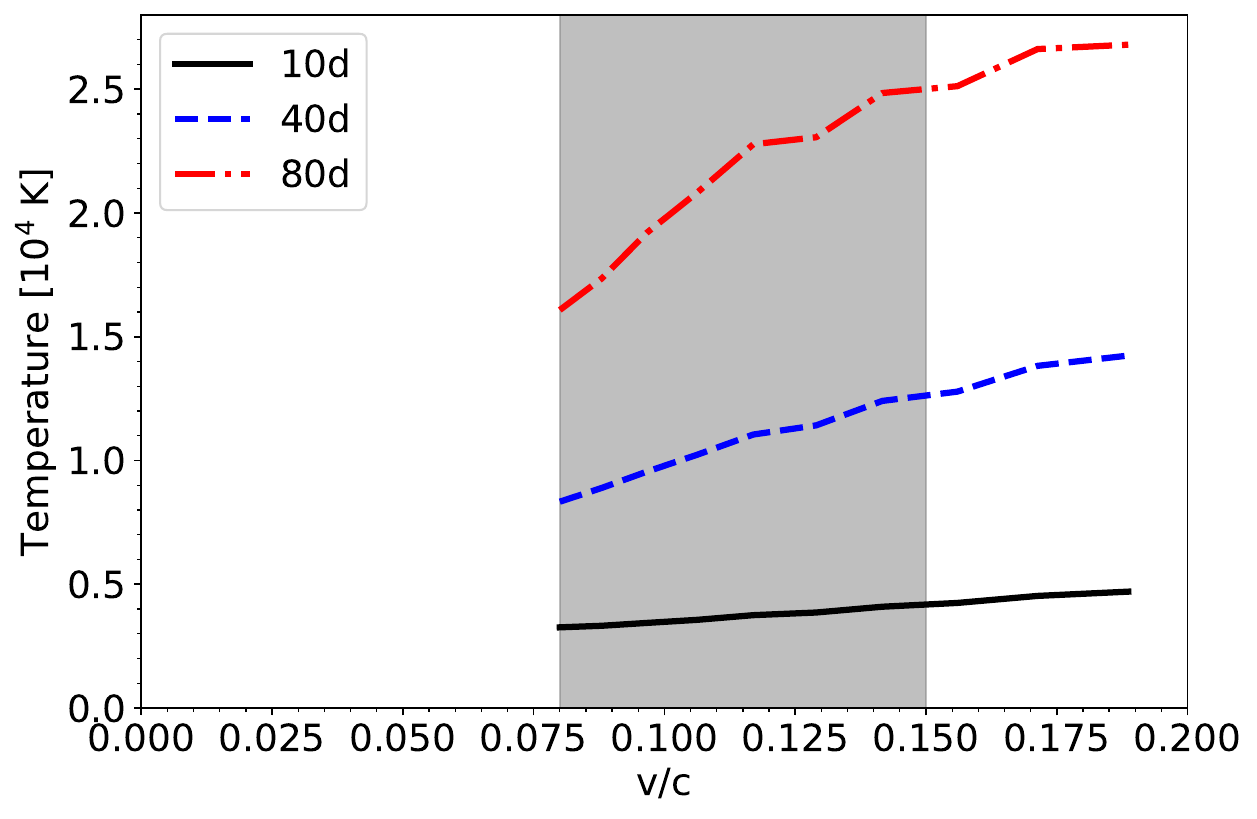}
\includegraphics[width=1\linewidth]{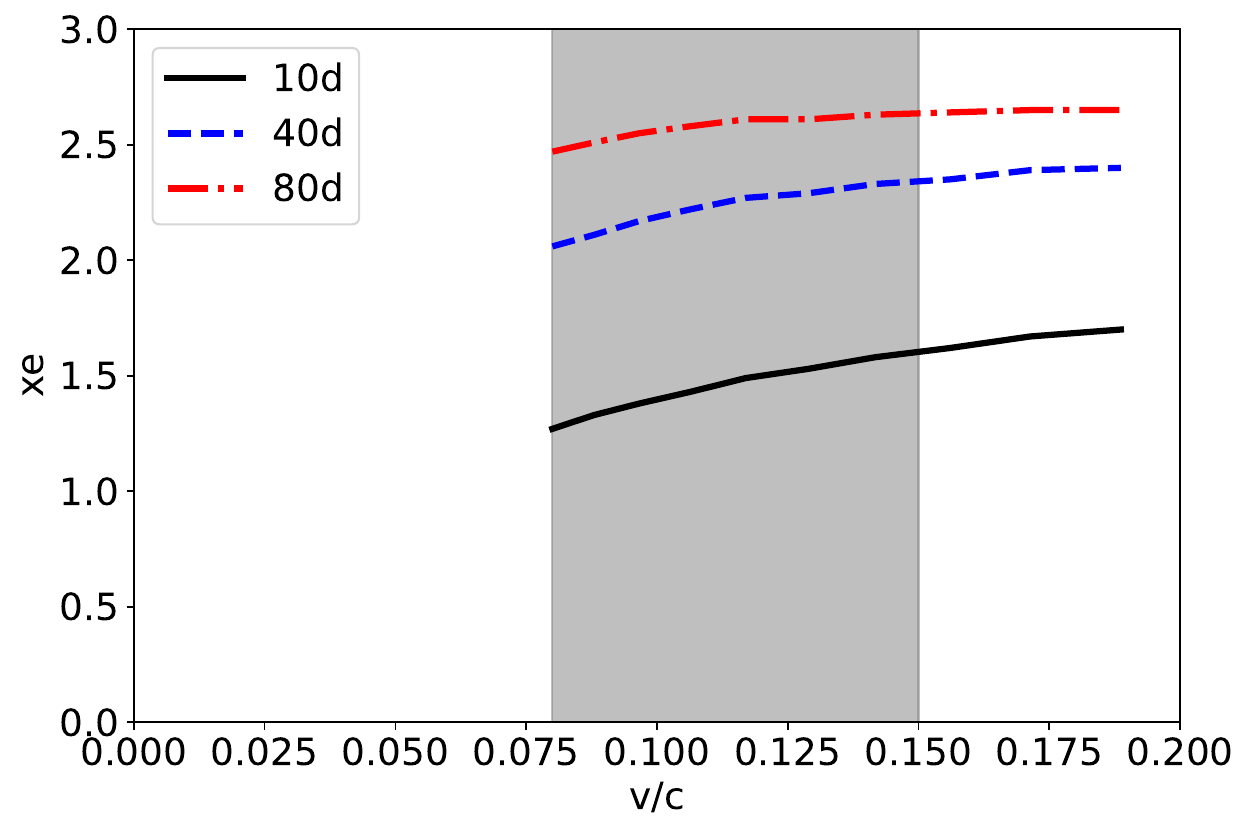}
\caption{Physical conditions in model B; cumulative radioactive deposition (top), electron temperature (middle), electron fraction (bottom), at 10d (black, solid), 40d (blue, dashed) and 80d (red, dash-dotted). The $\gtrsim$80\% deposition region is marked gray.}
\label{fig:Bhigh-physical}
\end{figure}

The physical conditions in model B are plotted in Fig. \ref{fig:Bhigh-physical}. Model B reaches 80\% deposition at $v/c \sim 0.15$, thus emission lines will emerge a factor 2-3 broader than in model A. In terms of power levels, the factor $\sim$3 decay boost combines with lower thermalization efficiencies (due to the lower densities) to give deposition ratios, relative to model A, of 2.2, 1.3 and 0.9 at 10d, 40d and 80d, respectively.

The lower densities, and the higher deposition (except at 80d), lead to higher temperatures, especially at 40d and 80d when model B reaches characteristic temperatures of 10,000 and 20,000 K, much higher than model A ($\sim$3000 and $\sim$6000 K, respectively). This means that more IR lines now come into the regime $T_{exc} \ll T$. Ionization is also higher, with $x_e$ values up by factors $\sim$1.5.

Ionization structures of selected elements are plotted in Figs. \ref{fig:modelB-high-10d-ionstructure}-\ref{fig:modelB-high-80d-ionstructure}. At 40d and 80d in this model, the triply-ionized state is the dominant ion for some elements. If the 4-times ionized state is significantly populated, this can lead to an artificially large abundance in the IV charge state in our models, which do not include higher ionization states. This is likely to be the case when the fraction of the IV state exceeds $\sim$2/3, as is the case for Ge, As, Kr, and Zr. Of these species, only [Zr IV] has a detectable line (at 8.00 \mum) in the wavelength range considered, and thus the flux of this line in Figures \ref{fig:spec-Bhigh-40d} and \ref{fig:spec-Bhigh-80d} should be considered an upper limit. No such effect is found in model A-low at any epoch.

We also looked at possible IR lines from (not included) V stages; neither Ge V, As V, Se V, Rb V, Sr V, Y V, Zr V or Te V have any potential strong IR lines. Ga V has its 2-1 (fine-structure) line at 2.79 \mum, Kr V has its 2-1 transition at 2.67 \mum, and 3-2 at 1.36 \mum. Br V has its 2-1 transition at 1.64 \mum.

\subsection{Spectrum of model B - 10d}
\begin{figure*}[htb]
\centering
\includegraphics[width=0.72\linewidth]{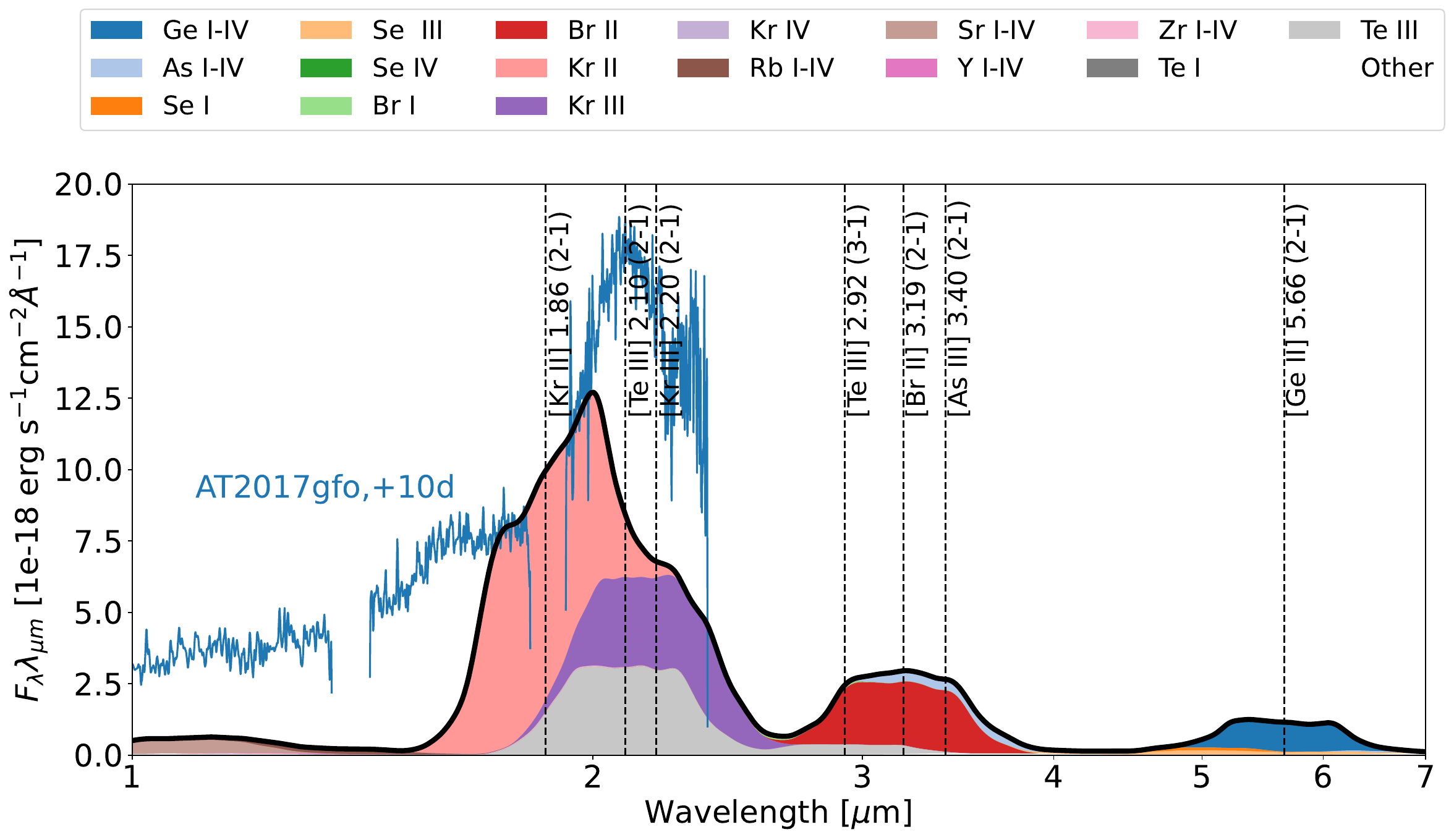}
\includegraphics[width=0.72\linewidth]{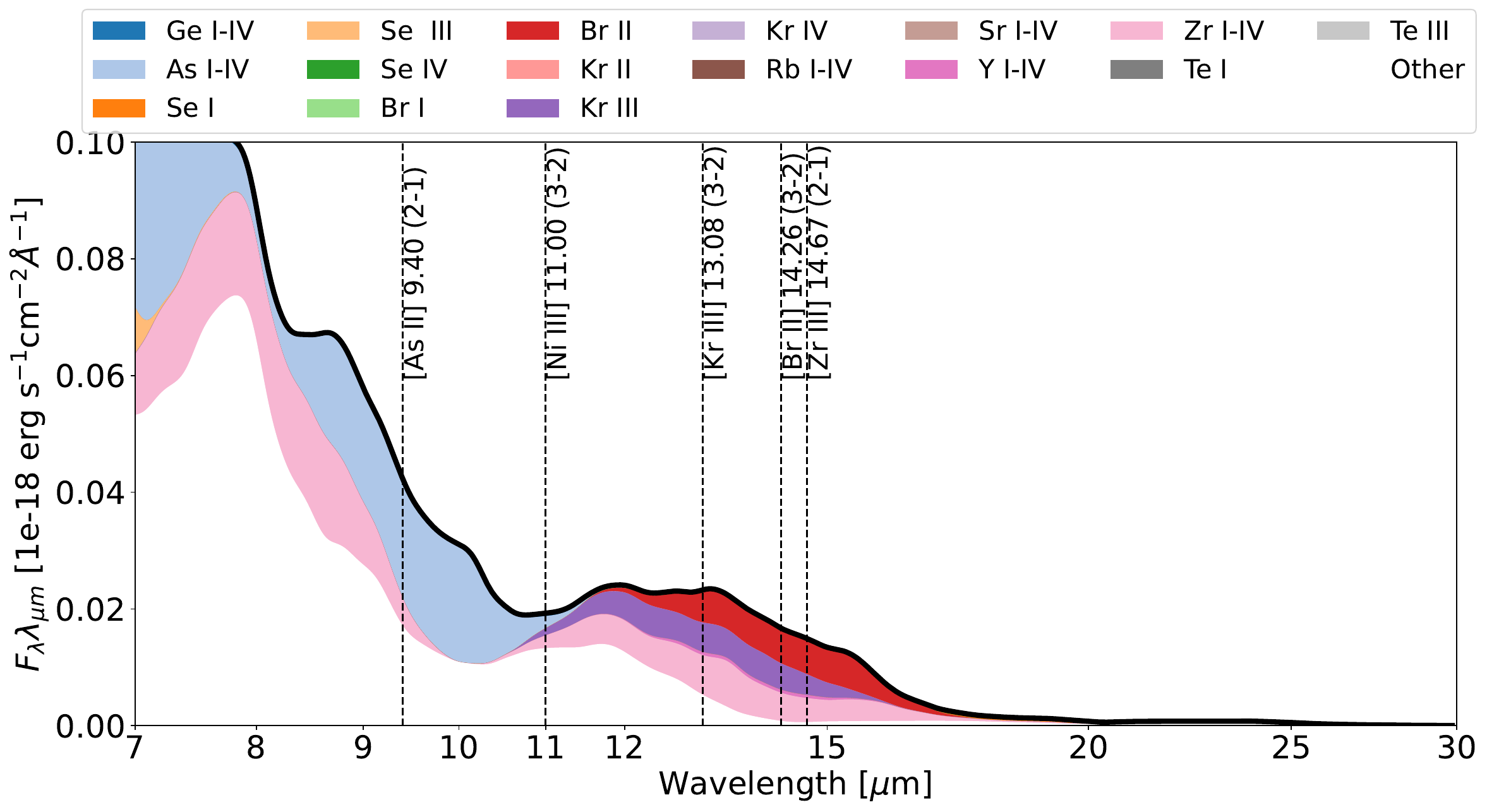}
\caption{\blue{Spectrum of model B at 10d, compared to the same data as in Fig. \ref{fig:spec-Alow-10d}.}}
\label{fig:spec-Bhigh-10d}
\end{figure*}

The spectrum of model B at 10d is shown in Figure \ref{fig:spec-Bhigh-10d}. [Kr II] 1.86 \mum\ is, as in model A, the strongest line. [Kr~III] 2.20 \mum\ is significantly stronger than in model A, which together with the [Te~III] 2.10 \mum\ contribution moves the blend towards the red, giving a peak at $\sim$2 \mum. The Kr lines are now formally consistent with the AT2017gfo spectrum, but the shape of the triple-blend feature is also not in particularly good agreement with the observed feature, which is quite well described by a Gaussian centred at 2.1 \mum\ \citep[which would correspond to a Gaussian emissivity distribution if by a single line,][]{Jerkstrand2017}. Considering the 7.4d-10.4d evolution, \citet{Gillanders2024} favor a fit consisting of two Gaussians centred at 2.05 \mum\ and 2.14 \mum, respectively, with $v_{\rm FWHM}=0.1c$. It is of interest that these wavelengths would match [Te~III] 2.10 \mum\ and [Kr~III] 2.20 \mum\ if both have a blueshift of $\sim 0.05/2.1 = 0.024c$. Such a blueshift could arise either due to an intrinsic ejecta asymmetry, or due to radiative transfer effects \citep{Jerkstrand2015}. 

At longer wavelengths, [Br~II] 3.19 \mum\ and [Ge~II] 5.66 \mum\ are distinct, as in model A. The neutral lines, [Br~I] 2.71 \mum\ and [Se~I] 5.03 \mum\ are now however mostly gone, as is the NIR contribution by zirconium (much of it is Zr I in model A, and neutral abundances are generally lower in model B).

Beyond 10 \mum, the neutral Ge lines in model A are gone, and instead [Kr~III] 13.08 \mum\ and [Br~II] 14.26 \mum\ have emerged - the flux levels are however low and there is severe blending throughout. This model predicts observations between 1-7 \mum\ to be most rewarding (in terms of lines least affected by blending), and Br II and Ge II appear promising for detection at $\sim$10d irrespective of the velocity of the light r-process component (strong lines predicted by both models A and B). Inspection of the line formation regimes (Fig. \ref{fig:regimes-B-high}) shows that Br II and Ge II form in the optically thin LTE regime, thus with this type of ejecta the ionic masses are possible to directly probe at 10d. The fractions of Br and Ge in the singly ionized states are quite high in the model (20-50\% for Br and $\sim$80\% for Ge, Fig. \ref{fig:modelB-high-10d-ionstructure}), so these ionic masses are in turn useful tracers of the total element masses.

\subsection{Spectrum of model B - 40d}
\begin{figure*}[htb]
\centering
\includegraphics[width=0.72\linewidth]{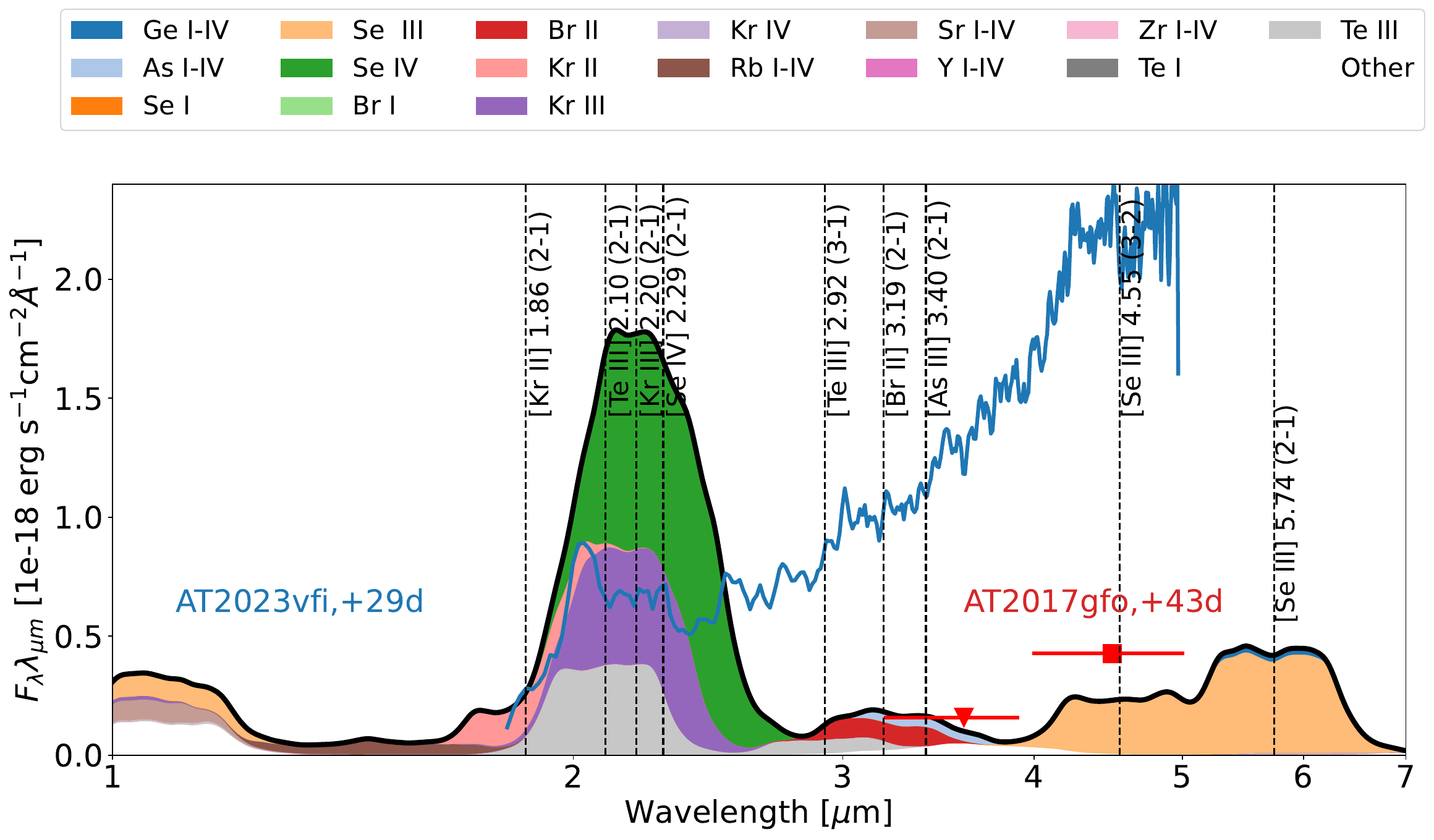}
\includegraphics[width=0.72\linewidth]{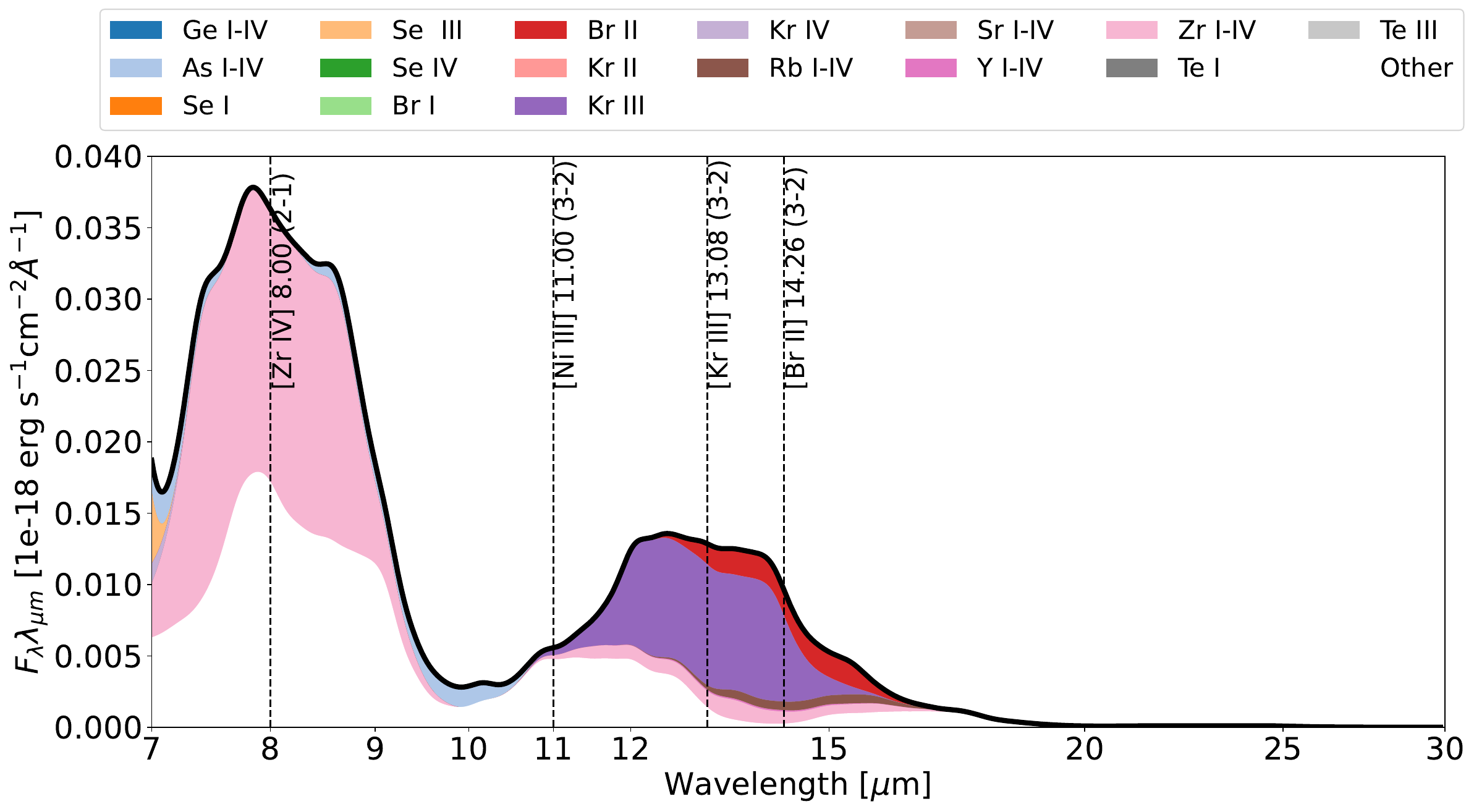}
\caption{\blue{Spectrum of model B at 40d, compared to the same data as in Fig. \ref{fig:spec-Alow-40d}.}}
\label{fig:spec-Bhigh-40d}
\end{figure*}

The spectrum of model B at 40d is shown in Figure \ref{fig:spec-Bhigh-40d}. The narrow [Kr~II] 1.86 \mum\ line in model A has here morphed into a broader feature. The [Kr~II] 1.86 \mum\ luminosity is now subdominant to [Kr~III] 2.20 \mum\, and in addition [Se~IV] 2.29 \mum\ has emerged, with even more luminosity than the [Kr~III] line. Together with a relatively modest [Te~III] 2.10 \mum\ contribution, the 2.1 feature is therefore now a four-way blend. The model feature is in reasonably good agreement with the observed feature in AT2023vfi, both in terms of luminosity and (roughly) line profile - although the detailed interpretation depends on what one assumes about the underlying (quasi)-continuum.

Neutral Br and Ge lines are weak, while [Br~II] 3.19 \mum\ is present, with some blending of [As III] 3.40 \mum. Both [Se~I] 5.03 \mum\ and [Ge~II] 5.66 \mum\ have disappeared and given way to [Se~III] 4.55 \mum\ and [Se~III] 5.74 \mum. What one may note is that in both a low and high velocity model, it is always selenium that would produce the type of flux observed with Spitzer for AT2017gfo - neutral selenium in the first case and doubly ionized selenium in the second case. At longer wavelengths, [Zr IV] 8.00 \mum\ and [Kr~III] 13.08 \mum\ are the dominant lines. 

\begin{figure}[htb]
\centering
\includegraphics[width=1\linewidth]{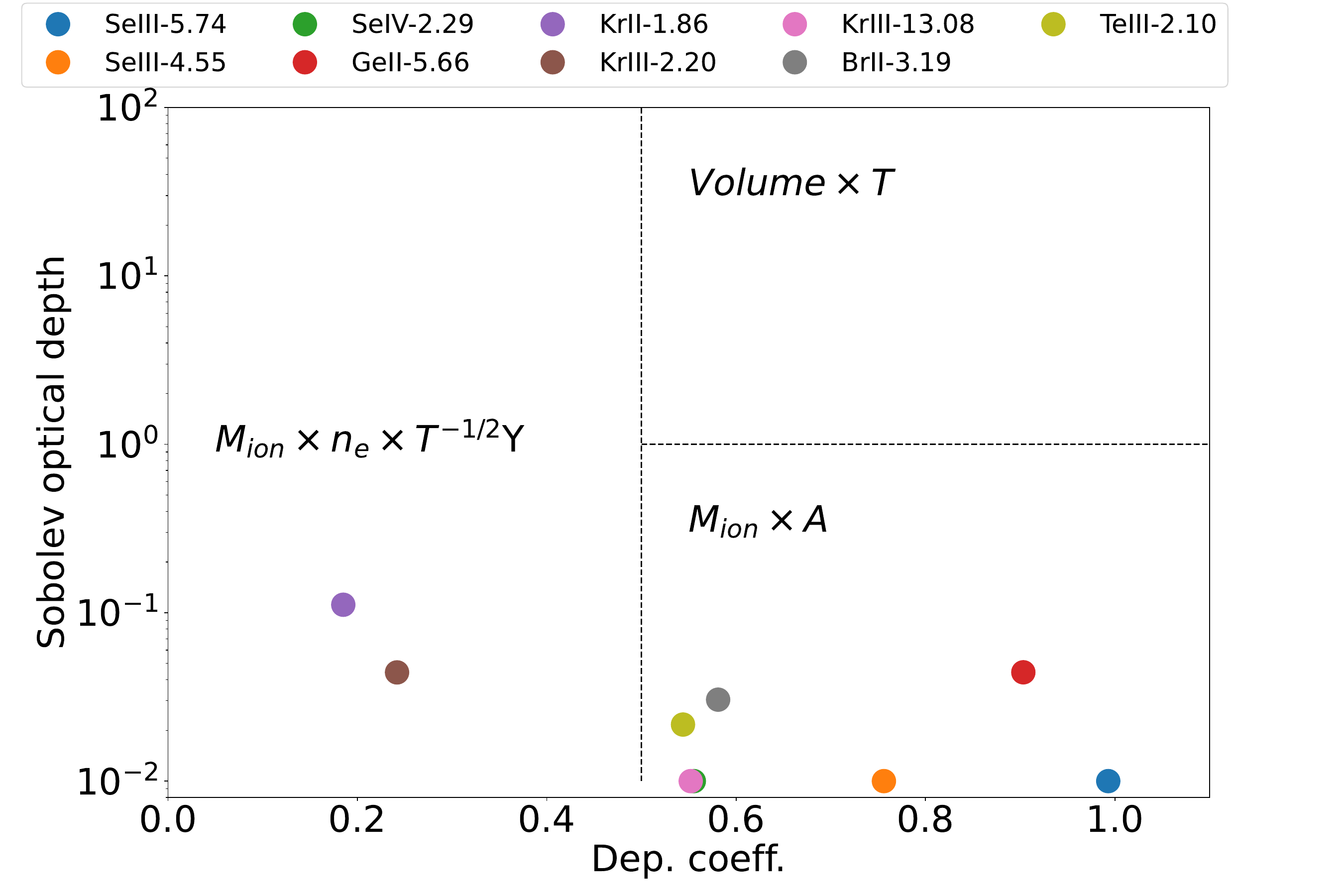}\\
\includegraphics[width=1\linewidth]{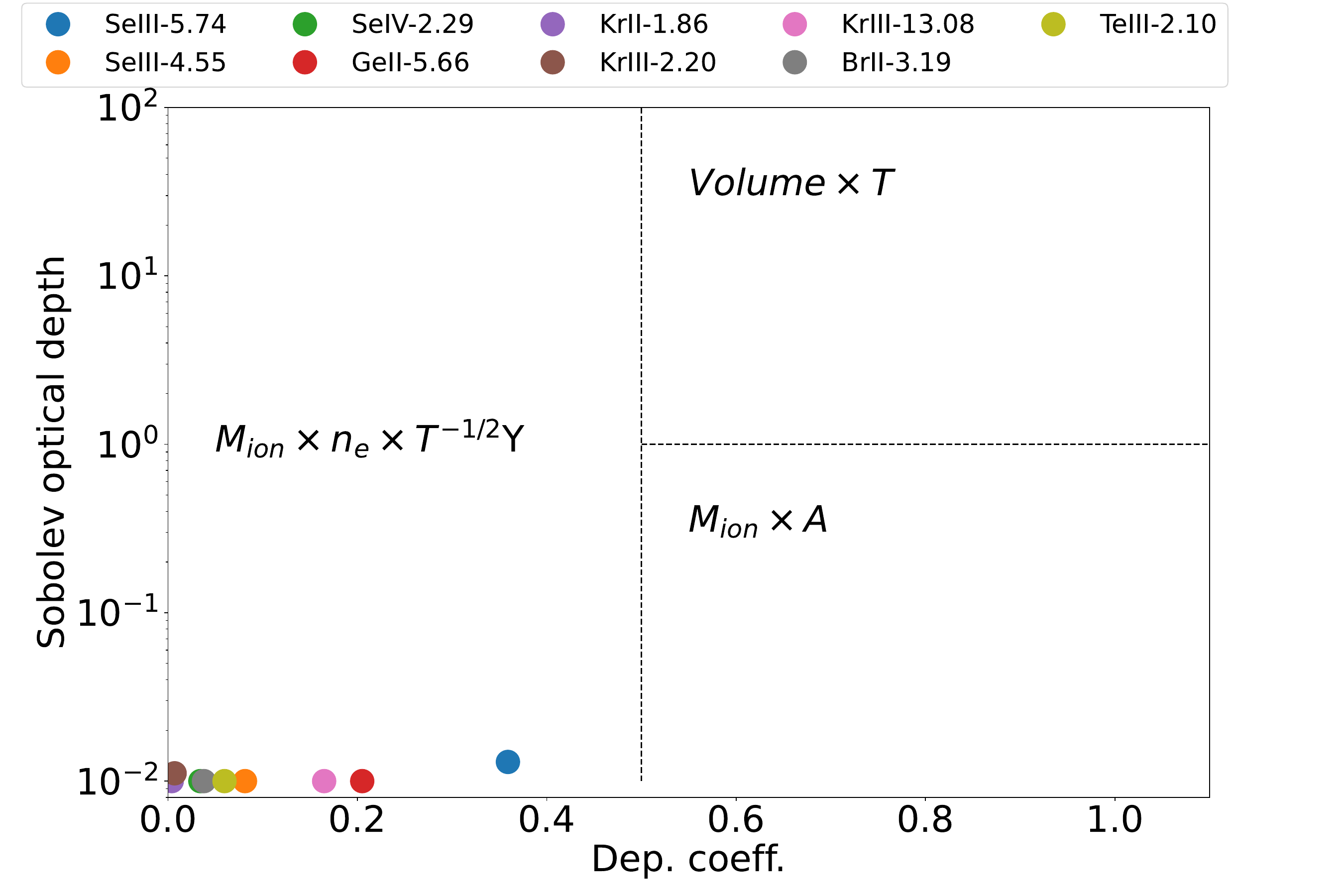}\\
\includegraphics[width=1\linewidth]{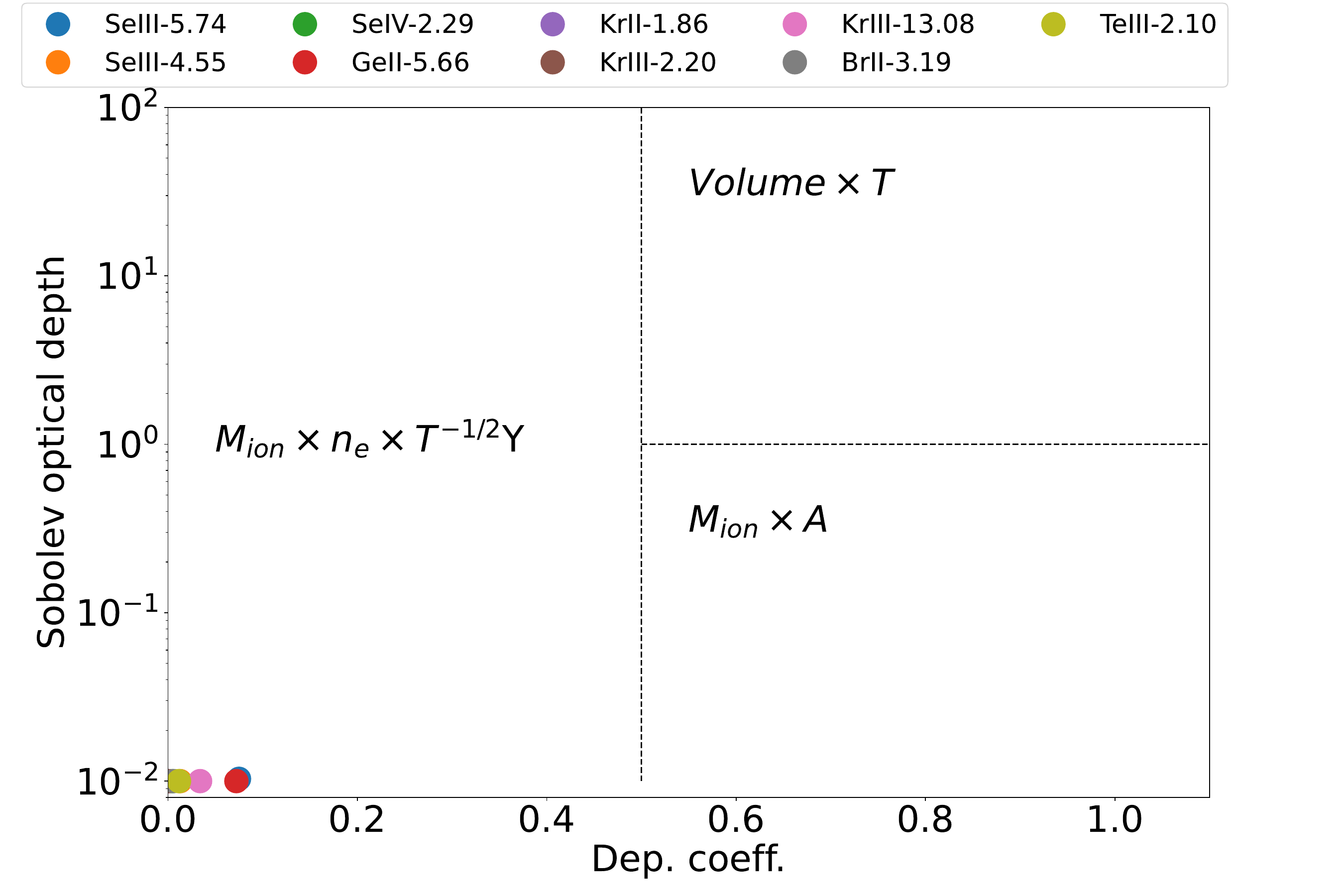}
\caption{\blue{Formation regimes in model B at 10d (top), 40d (middle) and 80d (bottom). Each point represents the zone-averaged values over the innermost eight shells. Optical depths below $10^{-2}$ are plotted at this floor value. The stated temperature dependencies are for the limit $T \gg T_{exc}$.}}
\label{fig:regimes-B-high}
\end{figure}

All lines are now well into the low density limit (``strong NLTE'') in the line formation plane (Fig. \ref{fig:regimes-B-high}) - thus A-values play no role for the line luminosities, which instead are directly proportional to collision strengths. In addition, as the temperature is $\gtrsim$$10,000$ K, relative IR line luminosities are temperature insensitive and differ mainly by the combination of ionic masses and collision strengths. The models make use of detailed collision strengths for the Kr II, Kr III, Se III, Se IV, Te III and Zr IV transitions but not for Br II lines.

\subsection{Spectrum of model B - 80d}
\begin{figure*}[htb]
\centering
\includegraphics[width=0.72\linewidth]{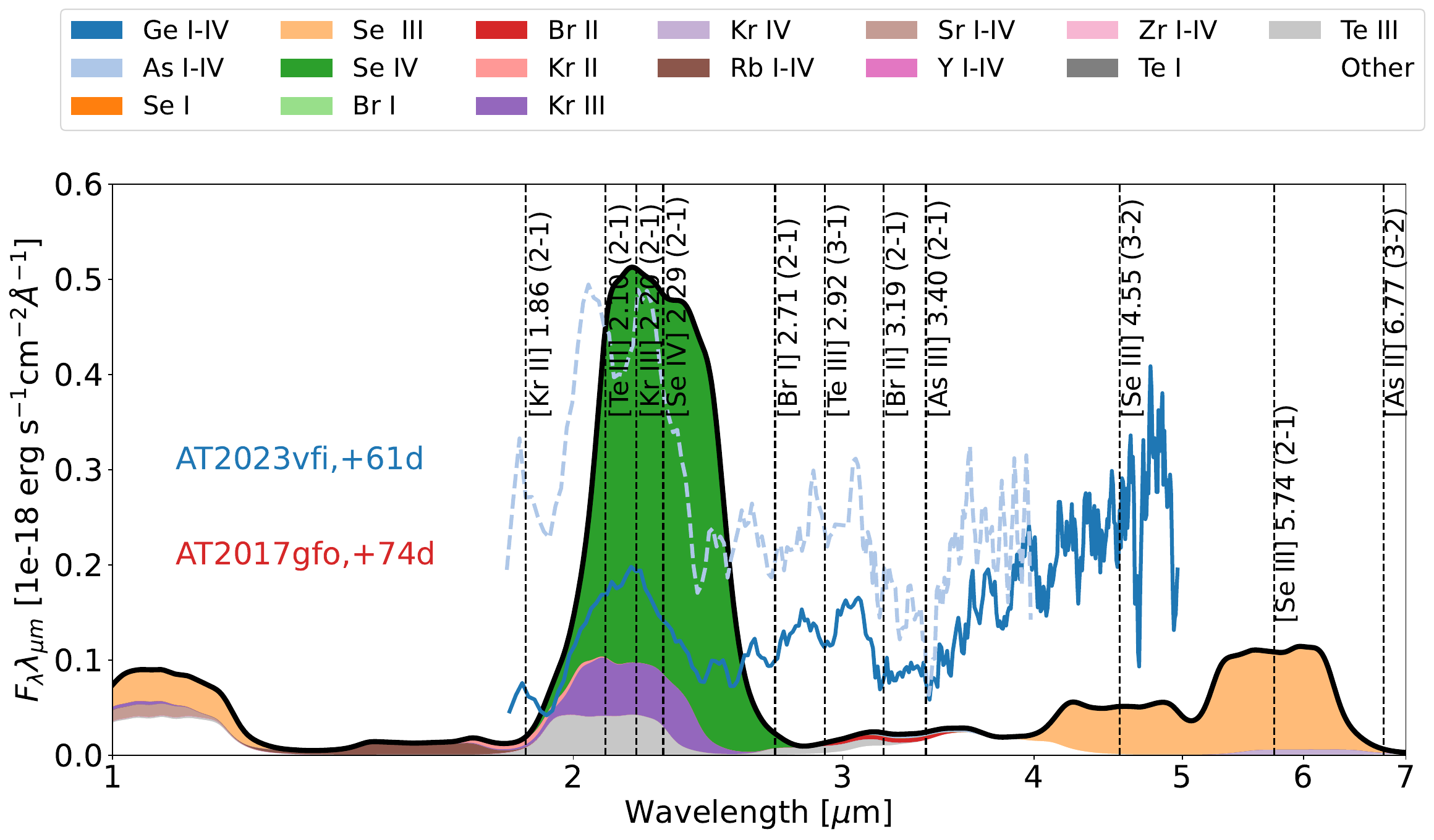}
\includegraphics[width=0.72\linewidth]{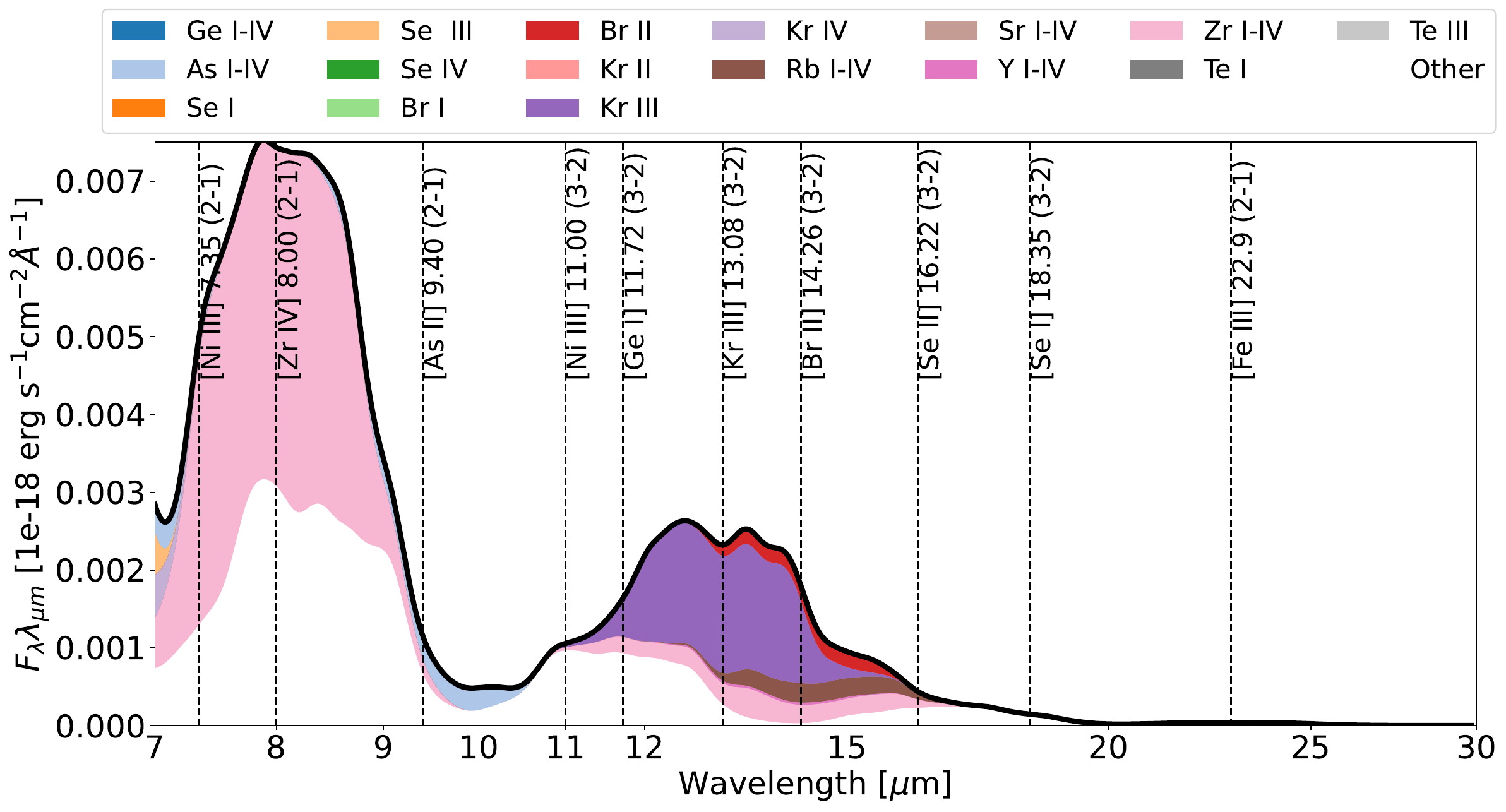}
\caption{\blue{Spectrum of model B at 80d, compared to the same data as in Fig. \ref{fig:spec-Alow-80d}.}}
\label{fig:spec-Bhigh-80d}
\end{figure*}

The spectrum of model B at 80d is shown in Figure \ref{fig:spec-Bhigh-80d}. Changes to +40d are quantitative but not qualitative. [Se~IV] 2.29 \mum\ now contributes the majority of the flux in the 2.1 \mum\ blend, with [Kr~III] 2.20 \mum\ having second strongest contribution. As at 40d, a solar ratio between Te and the lighter elements would mean Te does not dominate the feature. The [Se~III] 4.55 \mum\ line remains in reasonable agreement with the Spitzer observations - spectral observations with JWST of the 5-7 \mum\ region hold potential to confirm or reject Se III based on the 5.74 \mum\ line.

Evolution beyond 10 \mum\ is minor - the same lines are seen as at 40d and they are formed in the same regime, the model therefore predicts limited additional value of observing at the long wavelengths again if data at $\sim$40d already exists.

\subsection{Summary of predicted signatures}

Table \ref{table:predictions} summarizes predicted signatures from the $Z=31-40$ elements in the two models. Six of these ten elements are predicted to have detectable lines; Ge, As, Se, Br, Kr and Zr. The other four elements - Ga, Rb, Sr, Y - give no strong IR lines in the models, at any epoch.

This table can be used to design a JWST observation strategy depending on target element. For example, detection of Ge is predicted to require one MIRI spectrum at around 10d - (unblended) Ge I lines should emerge if ejecta conditions are similar to model A, and Ge II lines if they are similar to conditions in model B. Analogously, a NIRSpec spectrum at 10-40d holds potential to reveal Br I and Br II for model A conditions and Br II (and perhaps Br IV) for model B conditions.

\begin{table}
\centering
\begin{tabular}{ccc}
\hline
Ion  & Line (\mum)  & Detectable in \\
\hline
Ge I & 11.72 & A10\\
    &  17.94 & A10\\
Ge II & 5.66 & A10-80, B10\\
\hline
As II & 6.77 & A40-80,\\
      & 9.40 & A10-80, (B10)\\
As III & 3.40 & A40-80, (B40-B80)\\
\hline
Se I & 5.03 & A10-80 \\
Se III & 4.55 & A40-80, B40-80 \\
       & 5.74 & (A40-80), B40-80\\
Se IV & 2.29 & (B40), B80 \\
\hline
Br I & 2.71 & A10-40 \\
Br II & 3.19 & A10-80, B10-40 \\
      & 14.26 & (A10-80), (B10-40) \\
Br IV & 3.34 & (B40-80) \\
      & 3.81 & (B40-80) \\ 
\hline
Kr II & 1.86 & A10-80, B10-80\\
Kr III & 2.20 & (A10-80), (B10-80) \\
\hline
Zr I & 14.91 & (A40-80)\\
Zr II  & 22.28 & A10\\
Zr III & 12.42 & (A10-80), (B10)\\
Zr IV & 8.00 & (B40-80)\\
\hline
\end{tabular}
\caption{Summary of predicted IR lines. Lines that suffer significant blending are marked in parenthesis.}
\label{table:predictions}
\end{table}

\section{Discussion} 
From the considered model set, krypton stands out as an element giving strong predicted lines that are not clearly seen in the observations of AT2017gfo and AT2023vfi. We therefore look in some detail into the formation of these lines, and what constraints may be put on krypton abundances.

In the optically thin LTE limit, the line ratio of [Kr~III] 2.20 \mum\ and [Te~III] 2.10 is given by
\begin{multline}
\frac{L(\rm Kr\ III)}{L(\rm Te\ III)} =\\
\frac{x_{\rm Kr}}{x_{\rm Te}} \frac{y_{\rm Kr,3}}{y_{\rm Te,3}} \frac{g_{\rm u,KrIII}/Z_{\rm KrIII}(T)}{g_{\rm u, TeIII}/Z_{\rm TeIII}(T)}\frac{\rm A_{KrIII}}{\rm A_{TeIII}} \frac{h\nu_{\rm KrIII}}{h\nu_{\rm TeIII}} e^{\frac{302 K}{T}} \\
\approx 2.7 \frac{y_{\rm Kr,3}}{y_{\rm Te,3}}.
\end{multline}
where $x_{El}$ denotes the number fraction of the element and $y_{El,n}$ denotes the fraction of that element in ionization state $n$, with $n=1$ representing the neutral atom. The last equation uses the solar value of $x_{\rm Kr}/x_{\rm Te}= 8.7$ \citep{Lodders2003}, and, because the temperature is much higher than 302 K, the last term has been put to unity. It also takes the partition function ratio to be $Z_{\rm KrIII}/ Z_{\rm TeIII}\approx 4$ (it varies between 2-5 over $T=0-10,000$ K).

Past some time ($\gtrsim$40d in model B), the lines both form in the low-density NLTE regime. Their luminosity ratio is then, approximating most of the ions to be in their ground states,
\begin{equation}
\frac{L(\rm Kr\ III)}{L(\rm Te\ III)} = \frac{x_{\rm Kr}}{x_{\rm Te}} \frac{y_{\rm Kr,3}}{y_{\rm Te,3}} \frac{\Upsilon_{\rm Kr\ III}/g_{\rm u,Kr\ III}}{\Upsilon_{\rm Te\ III}/g_{\rm u,Te\ III}}\frac{h\nu_{\rm KrIII}}{h\nu_{\rm TeIII}}e^{\frac{302 K}{T}},
\end{equation}
Inserting the known values of the collision strengths \citep{Schoning1997,Madonna2018}, and again putting the temperature factor to unity, this becomes
\begin{equation}
\frac{L(\rm Kr\ III)}{L(\rm Te\ III)} \approx 4.9 \frac{y_{\rm Kr,3}}{y_{\rm Te,3}}.
\end{equation}
Thus, for a solar Kr/Te ratio the Te III line can dominate the Kr III one only if the fraction of Te in the doubly ionized state is at least a factor $3-5$ larger than the fraction of Kr in the doubly ionized state.

Kr I and Kr II have higher ionization potentials (14.0 and 24.4 eV, respectively) than Te I and Te II (9.0 and 18.6 eV, respectively). All else equal, this would imply that more Kr stays in neutral and singly ionized states compared to Te. However, recombination rates also affect this. For Kr we use calculated rates \citep{Sterling2011}, so there is limited uncertainty from this direction (although DR calculations are challenging due to uncertain positions of doubly excited states). While for Te only the Te III to Te II recombination stage has published rates \citep{Singh2025}, this does not directly affect the question of how to achieve a lower $y_{\rm Kr,3}/y_{\rm Te,3}$ ratio (to allow Te III to explain the 2.1 \mum\ feature at a solar Kr/Te ratio), as in the models here $y_{\rm Te,3}$ is already large, $\gtrsim 0.3$.

In model A, the characteristic $y_{\rm Kr,3}$ value in the line-forming region is $\sim$1\% at 10d, growing to $\sim$10\% at 40d and $\sim$30\% at 80d (Figs. \ref{fig:modelA-low-10d-ionstructure}-\ref{fig:modelA-low-80d-ionstructure}). The corresponding $y_{\rm Te,3}$ values are $\sim$5\%, 50\%, and 70\%. Thus, at 40d and 80d, the $y_{\rm Te,3}$ value is close to its maximum (unity), and to make a significantly weaker [Kr~III] contribution would mean reducing $y_{\rm Kr,3}$. But $y_{\rm Kr,2}$ is already high, and would, for a movement toward a less ionized solution, become yet higher, further overproducing the [Kr~II] 1.86 \mum\ line.

In model B, both $y_{\rm Kr,3}$ and $y_{\rm Te,3}$ are high at all epochs, $\gtrsim 40\%$ (Figs \ref{fig:modelB-high-10d-ionstructure}-\ref{fig:modelB-high-80d-ionstructure}). At 40d and 80d, the neutral and singly ionized states are subdominant and Kr and Te are mainly in the doubly and triply ionized states. But again, because the krypton recombination rates are known, it is not straightforward to reduce the $y_{\rm Kr,3}$ value.

An analogous analysis of [Kr~II] 1.86 \mum\ gives, in the optically thin LTE phase, using $Z_{\rm KrII}/Z_{\rm TeIII} \approx 2$ (range $1-4$ over $0-10,000$ K),
\begin{equation}
\frac{L(\rm Kr\ II)}{L(\rm Te\ III)} \approx 5.8 \frac{y_{\rm Kr,2}}{y_{\rm Te,3}} e^{\frac{-884 K}{T}},
\end{equation}
and in the low-density limit
\begin{equation}
\frac{L(\rm Kr\ II)}{L(\rm Te\ III)} \approx 0.81 \frac{y_{\rm Kr,2}}{y_{\rm Te,3}}e^{\frac{-884 K}{T}}.
\end{equation}
Both model A and B give $\frac{y_{\rm Kr,2}}{y_{\rm Te,3}} \gtrsim$ 1 (an exception is model B at 80d where the ratio is lower). Thus, a solar Kr/Te ratio yields a [Kr II] 1.86 \mum\ line comparable in strength to [Te III] 2.20 \mum, as $T \gtrsim 884$ K.

Thus, if an observed feature at 2.1 \mum\ is \emph{assumed} to be caused mainly by Te III - both models A and B, and this analysis, show that this is impossible with a solar (r-process component) Kr/Te ratio (at least in the ejecta component giving rise to the [Te~III] 2.10 \mum\ line); the Kr II + Kr III luminosity is always larger than that of Te III for plausible physical conditions.  
Krypton gives uniquely constraining information because both Kr II and Kr III have strong 2-1 lines in this spectral region. The general picture of a subsolar production of light r-process elements is also reinforced by the fact that quite strong predicted [Br~II] 3.19 \mum\ is not observed in AT2023vfi, with model A giving a particularly clear disagreement. A possible interpretation is that Kr and Br abundances are both significantly subsolar compared to Te.

What if Kr and Te exist in different regions, with different physical conditions? The excitation temperatures are between 6500-7700 K for these three lines. At face value, this means that also relatively small zone temperature differences could drive big differences in relative luminosities. But - if the ejecta mass is $\sim$0.05 $M_\odot$ and the 31-40 elements are produced in solar ratios, there must be strong Kr lines produced - as we show here equal or stronger to observed features and/or non-detections in AT2017gfo and AT2023vfi. Thus - one could boost Te by postulating that it exists in another, hotter and/or more ionized zone - but such a model would not fit the data as the Kr + Te + Se complex would be much too strong. 

If subsolar production of Br and Kr is the case in the first two observed KNe, important consequences follow for the question of the origin of the light r-process elements. Alternative sources for r-process nucleosynthesis - neutrino-driven winds in CCSNe, collapsar disks, and jets and magnetorotational SNe, in many studies appear promising for light r-process production \citep[e.g.][]{Travaglio2004,Cowan2021,Arcones2023}.

At low metallicity, the weak s-process is not yet effective in massive stars, and therefore can their heavy element ejection be ascribed to explosive nucleosynthesis. Germanium shows no correlation with europium (Eu) in low-metallicity stars \citep{Cowan2005}, suggesting two different sources. At the same time, Ge correlates well with Fe, which suggests a coproduction. But already Se and Sr show close to solar r-process abundances with respect to heavier elements in several metal-poor stars \citep{Roederer2012,Roederer2022}. These results suggest the main r-process (presumably from kilonovae with perhaps a contribution by collapsars) is capable in making the solar r-process pattern from $Z=34$.

These results appear to be in good agreement with several current KN nucleosynthesis models, which achieve patterns close to solar from $Z=34$, but subsolar for $Z=31-33$. There is however some tension with nucleosynthesis yield models for CCSNe \citep{Sukhbold2016,Wanajo2018}, which tend to give close to solar production all the way up to $Z\sim38$ (i.e., implicating that production of these elements is small for sources producing the  $Z\geq 39$ elements).

Two interesting hypothesis tests, with consideration of what the models in this paper show, now present themselves. One is that the Ge r-process component (estimated as 36\% by \citet{Prantzos2020}) comes mainly from CCSNe. This is already a quite convincing scenario from CCSN theory, NSM theory, and low-metallicity star observations - but detections or constraining upper limits on Ge emission directly from r-process sources are needed to fully test this. The models here show that the hypothesis is testable with MIR observations of KNe, e.g. of the [Ge~II] 5.66 \mum\ line.

The other hypothesis is that the r-process components of Se, Br and Kr (estimated at 61\%, 74\%, and 61\% by \citet{Prantzos2020}) come mainly from KNe. There is tension here between CCSN theory, NSM theory, and metal-poor r-process enriched stars - and there is also a certain degree of tension within the KN analysis presented here. Selenium, at solar abundance, is able to roughly match  the Spitzer photometry of AT2017gfo. On the other hand, the lack of observed Br and Kr lines suggests significantly subsolar Br and Kr production. Further JWST data, and yet more accurate models, hold clear promise to settle this question. \ajn{There is also a close relation to the question of strontium production by KNe \citep{Watson2019,Domoto2021,Perego2022,Tarumi2023}, as well as possible identifications of Y \citep{Sneppen2023} and/or Rb \citep{Pognan2023}.}

With the complex ejecta structures and physical formation processes, the interpretation of kilonova spectra is still in its infancy. Significant work has been done in the past few years to identify possible signatures, in particular in the infrared where radiative transfer effects and line blending tend to be less severe. \citet{Hotokezaka2023} used a single-zone emissivity model where the composition is the solar one, but starting at $Z=38$ (i.e., the abundance of Kr and other $Z<38$ elements is zero). The physical conditions were fixed to ionization structure $\rm I-II-III-IV =0.25, 0.4, 0.25, 0.1$ for all elements, and temperature to $T=2000$ K. The model has ejecta mass 0.05 $M_\odot$ and uniform density with $v_{max}=0.07c$ (as such the characteristic line widths are somewhere between models A and B here).

The [Te III] 2.10 \mum\ luminosity in model A here at 10d is significantly below the one of \citet{Hotokezaka2023}.
There are two main reasons for this. First, the Te mass fraction in the models here is 3.8\% (the solar value), whereas in the \citet{Hotokezaka2023} model it is 11\%; with the same total ejecta mass of 0.05 $M_\odot$ this means a factor $\sim$3 lower Te mass in our case. Second, our lower value for $y_{\rm Te,3}$ of $\sim$0.05 in model A, compared to the assumed 0.25 in \citet{Hotokezaka2023}, combines with the 3 times lower Te mass to give a Te III mass ratio of $\sim$15. Both in model A here, and in the model of \citet{Hotokezaka2023}, electron densities are larger than $n_{e,crit}$, so the line is formed close to LTE and only ion mass and temperature affect the predicted flux. The typical temperature in model A is around 2300 K, and while this seems close to the assumed 2000 K in \citet{Hotokezaka2023}, it leads to a factor 1.5 higher emissivity through the Boltzmann factor (entering both in LTE and NLTE limits), lowering the difference to a factor $\sim$10.

When the line does not form in LTE, the situation is even more complex. In model B, the much larger volume of the line-forming region leads to lower electron densities (factor $\sim$30), and the [Te~III] 2.10 \mum\ line forms in the transition regime between LTE and NLTE at 10d (Fig. \ref{fig:regimes-B-high}). The luminosity ratio of NLTE vs LTE emission is $L(\mbox{NLTE})/L(\mbox{LTE}) = g_l /g_u \left(n_e /n_{e,crit}\right)$, so the emission per Te III ion is lower in NLTE (as then $n_e < n_{e,crit}$), all else similar. However, a higher temperature ($\sim$4000 K) significantly brings up the Boltzmann factor (factor 3.1 from 2300 K and factor 5.1 from 2000 K), and now $y_{\rm Te,3}$ has increased to $\gtrsim 0.5$, together giving a factor 35 increase for LTE emission compared to model A, and a factor 10 increase compared to \citet{Hotokezaka2023}. But going into NLTE damps this to factors of $\sim$3.5 and $\sim$1. Finally, our lower Te mass gives a luminosity of $\sim$1/2 of the one of \citet{Hotokezaka2023}.

The above analysis illustrates how the luminosity of a line can vary quite dramatically with density, composition, and energy deposition, which determine the temperature and ionization, in a model. In model A, the density to energy deposition ratio is too high to allow for sufficient Te in the doubly ionized state. In model B, it is too low, but with a lesser discrepancy.

\subsection{Complementary information from \citet{Pognan2026}} 
In \citet{Pognan2026}, a \texttt{SUMO} model grid is constructed where ejecta mass and composition (held fixed here) is varied. This grid gives complementary information about how lines can change over ejecta parameter space. One result is that for the case of no dynamic ejecta component ($f_{dyn}=0$, the composition most similar to the one used here), the spectrum is not strongly sensitive to ejecta mass, in the sense that Se I + Se III, Br I, Br II, Ni III lines are dominant for all cases at 10d. The Ni III lines are stronger than here because the abundance is higher (about 7\% compared to 1\% here).  The grid demonstrates that as long as mixing with heavier matter is limited (e.g., a full solar composition would have $f_{dyn} \lesssim 0.1$), the light r-process elements still fully dominate the IR spectra - thus removing a major source of uncertainty regarding how generic results from this paper (using models with only light elements) may be taken.

Some more sensitivity is seen around the 2 \mum\ region. The result here, that $L(\mbox{[Kr II]}\ 1.86\ \mu m + \mbox{[Kr III]}\ 2.20\ \mu m) \gtrsim L(\mbox{[Te III]}\ 2.10\ \mu m)$ for a roughly solar Kr/Te abundances is not recovered in the \citet{Pognan2026} models, despite having a ratio of $\sim$1.5 solar. The main reason for this is that the \citet{Pognan2026} grid uses a modified Axelrod approximation for the Kr II and Kr III collision strengths; these are over a factor 10 too low compared to the calculated values used here. The line forms in NLTE in the \cite{Pognan2026} models which gives direct proportionality to the collision strengths and thus a factor 10 too low Kr II and Kr III luminosities. Similarly, the [Se IV] collision strength estimate (0.16) is too low compared to the calculated 7.2 used here, leading to no or weak [Se~IV] 2.29 $\mu$m line in the \citet{Pognan2026} grid.

At the same time, the models here lack the $Z > 40$ elements (except Te) that also partially contribute to the 2.1 $\mu$m bump in the \citet{Pognan2026} models. A full understanding of the formation of the 2.1 \mum\ bump in AT2017gfo and AT2023vfi is clearly challenging. There are several ways such a broad bump could be generated - from a single feature dominating
(for which one can directly infer the velocity of the emitting region) to blends of two
or more features (where such extractions are more difficult). 

\section{Conclusions and future work}
We have studied the infrared spectral signatures of light r-process elements ($Z\leq 40$, plus tellurium, $Z=52$) in kilonovae, over the epochs 10-80d using 1D NLTE models. We conclude the following.

\begin{itemize}

\item From the $Z=31-40$ range, distinct signatures in the MIR are predicted for [Ge~I] 11.72 \mum, [Ge~I] 17.94 \mum, [Ge~II] 5.66 \mum, [As~II] 6.77 \mum, [As~II] 9.40 \mum, [As~III] 3.40 \mum, [Se~I] 5.03 \mum, [Se~I] 18.35 \mum, [Se~III] 4.55 \mum, [Se~III] 5.74 \mum, [Se~IV] 2.29 \mum, [Br~I] 2.71 \mum, [Br~II] 3.19 \mum, [Br~II] 14.26 \mum, [Kr~II] 1.86 \mum, [Kr~III] 2.20 \mum, [Kr~III] 13.08 \mum, and [Zr IV] 8.00 \mum. In the models explored here, low-velocity ejecta (associated e.g. with a disk outflow) generate lines mainly from neutral, singly and doubly ionized species, whereas high-velocity ejecta (associated e.g. with a PNS component) generate lines mainly from singly, doubly, and triply ionized species.

\item Lines evolve over time from an initially optically thick LTE regime (probing emission volume) to an optically thin NLTE regime (probing the product of ionic mass and electron density). Some lines pass through an optically thin LTE regime (directly probing ionic mass). Longer wavelength lines are less sensitive to temperature and can therefore more robustly be used to infer emission volumes and/or ionic masses.

\item Comparison to AT2017gfo Spitzer photometry, from $Z \leq 40$ candidates only [Se I] 5.03 \mum\ and [Se III] 4.55 \mum\ produce significant luminosity in the 4.5 \mum\ band, and model predictions are in rough agreement with the observed band fluxes. Following the proposal of [Se III] 4.55 \mum\ using models with parameterized physical conditions \citep{Hotokezaka2022}, these are the first self-consistent models successfully explaining this data (using a Se mass of $0.014$ $M_\odot$). In the 3.6 \mum\ band, [Br II] 3.19 \mum\ and [As III] 3.40 \mum\ are the dominant contributors, with fluxes at or below the upper limits. 

\item Comparison to AT2023vfi JWST spectra, a light composition-only component cannot produce the smooth observed spectrum, instead giving discrete emission lines with gaps between with low  flux. This may suggest that heavier elements are present that either do much of the cooling, and/or reprocess the light element emission.

\item For a low-velocity, light-composition ejecta, [Kr~II] 1.86 \mum\ and [Kr~III] 2.20 \mum\ produce narrow, well-separated lines, inconsistent with observations of both AT2017gfo and AT2023vfi. In a higher velocity model these features partially overlap giving a bright feature centred close to 2.1 \mum, as observed in both AT2017gfo and AT2023vfi. While the detailed line profile in AT2017gfo still fits better with a single line centred at 2.1 \mum\ \citep[with Te III 2.10 \mum\ the leading candidate,][]{Hotokezaka2023}, the situation is less clear for AT2023vfi. A solar Br/Te ratio leads to predicted strong lines from [Br~I] 2.71 \mum\ and [Br~II] 3.19 \mum, which are not observed in AT2023vfi. The summary picture suggests that both AT2017gfo and AT2023vfi ejected significantly less light ($Z \leq 36$) r-process material than a solar composition. This is of interest in the context of various challenges with KNe being the main source of first r-process peak elements.
\end{itemize}

The results in this paper provide indications of subsolar production of light r-process elements in KNe. However, to more firmly establish this, progress is needed along at least three fronts. 

The first is more atomic data for NLTE modelling. 
While the atomic data situation has significantly improved for the $Z=31-40$ range in the last years \citep[e.g.][]{Banerjee2025,Singh2025,McCann2025}, now enabling nebular emission modelling for both planetary nebulae and kilonovae with quite good accuracy, certain important data is still missing. This includes collision strengths for Ge I and II, As II and III, Se I and II, Br I, II and III, as well as recombination rates for Ge and As ions. For yet heavier elements, this kind of data is still mostly missing.

The second is improved NLTE spectral models taking into consideration deviations from spherical symmetry \citep[being done already for the LTE phase, e.g.][]{Bulla2019,Collins2023}, the multi-component nature of KNe, and doing thermalization and other microphysics to higher level of detail. While MIR lines are quite insensitive to temperature, NIR lines are more so, and there is always a direct sensitivity to ionization state. We put focus here on ratios of nearby lines of similar ions, but individual line luminosities are sensitive to physical conditions.

The third is observations of more KNe in the infrared - we establish here the unique diagnostic possibilities of this range, but the data so far is very limited. Progress requires JWST observations with unique r-process signatures predicted all along the $1-30$ \mum\ range.

\appendix 

\section{Atomic data updates}
\label{sec:atomicdata}
The dielectronic recombination rates calculated for Se, Rb, Sr, Y, and Zr by \citet{Banerjee2025} and Kr \citep{Sterling2011} were used. 
The dielectronic and radiative recombination rates for Br 
and the radiative recombination rates for Rb, from Kerlin et al. (in prep.), were also incorporated.

As outlined in more detail below, some energy levels in our \texttt{FAC} base-set \citep{Pognan2023} were corrected to experimental values or, lacking that, to more accurate theoretical ones. In such calibrations, lines connecting to the level had A-values rescaled by a factor $\left(\lambda_{old}/\lambda_{new}\right)^3$ \citep{Hotokezaka2021}. The numbering of levels in the descriptions below refer to the energy-ordering in the \texttt{FAC} model atoms \citep[using a spectroscopic
configuration-interaction (SCI) method,][]{Pognan2023}.

\subsection{Ga I}
\noindent The energy of level 2 ($^2$P$^{\rm o}_{3/2}$) was corrected to NIST \citep{Shirai2007}. The Ga I isoelectronic sequence (Ga I, Ge II, As III, Se IV,..) has a single p-shell valence electron, giving a doublet ground term $^2$P with $J=3/2$ or $J=1/2$ depending on the spin of the electron. A potential IR line is therefore $^2$P$^{\rm o}_{3/2}-^2$P$^{\rm o}_{1/2}$ (2-1, 12.10 $\mu$m, $A=4.6 \times 10^{-3}$\ s$^{-1})$.

\subsection{Ga II}
\noindent The first excited state in Ga II lies high at 47,367 cm$^{-1}$, no calibrations were made. The Ga II isoelectronic sequence (Zn I, Ga II, Ge III, As IV, ...) has a 4s$^2$ ground configuration with no fine structure splitting.

\subsection{Ga III}
\noindent The first excited state in Ga III lies high at at 65,169 cm$^{-1}$, no calibrations were made. The Ga III isoelectronic sequence (Cu I, Zn II, Ga III, Ge IV, ...) has a single-electron (4s)  ground configuration.

\subsection{Ga IV}
\noindent The first excited state in Ga IV lies high at 149,512 cm$^{-1}$, no calibrations were made. The Ga IV isoelectronic sequence (Cu II, Zn III, Ga IV,...\footnote{Ni I breaks pattern with a different ground term, 3d$^8$($^3$F)4s$^2$}.) has a 3d$^{10}(^1\rm S)$  ground configuration with no splitting.

\subsection{Ge I}
\noindent The energies for levels 2-5 ($^3$P$_1$, $^3$P$_2$, $^1$D$_2$, $^1$S$_0$) were corrected to NIST values \citep{Sugar1993}. Theoretical A-values for transitions between these first 5 levels are provided by \citet[][Hartree-plus-statistical-exchange and Hartree-Fock with relativistic corrections]{Biemont1986-2}. The Ge I isoelectronic sequence (Ge I, As II, Se III, Br IV, ..) has two p-shell electrons. In the ground term the spins are aligned (so $2S+1=3$); and splitting occurs by three allowed values for the orbital magnetic number giving $J=2,1,0$. Transitions with $\Delta J=1$ have significantly higher A-values than $\Delta J=2$, giving two main lines from the ground term. Potential IR lines are $^3$P$_1-^3$P$_0$ (2-1, 17.94 \mum, $A=3.3\times 10^{-3}$\ s$^{-1})$, $^3$P$_2-^3$P$_1$ (3-2, 11.72 \mum, $A=8.0\times 10^{-3}$\ s$^{-1})$, and (trans-term) $^1$D$_2-^3$P$_0$ (4-3, 1.75 \mum, $A=0.098$\ s$^{-1})$.

\subsection{Ge II}
\noindent The energy of level 2 ($^2$P$^{\rm o}_{3/2}$) was corrected to NIST \citep{Sugar1993}. A potential IR line is $^2$P$^{\rm o}_{3/2}-^2$P$^{\rm o}_{1/2}$ (2-1, 5.66 $\mu$m, $A=0.050$ s$^{-1}$).

\subsection{Ge III}
\noindent The first excited state in Ge III lies high at 61,733 cm$^{-1}$, no calibrations were made.

\subsection{Ge IV}
\noindent The first excited state in Ge IV lies high at 81,311 cm$^{-1}$, no calibrations were made.

\subsection{As I} 
\noindent The energies of levels 2-5 ($^2$D$^{\rm o}_{3/2}$, $^2$D$^{\rm o}_{5/2}$, $^2$P$^{\rm o}_{1/2}$, $^2$P$^{\rm o}_{3/2}$) were corrected to NIST values \citep{Moore1971}. 
Theoretical A-values for transitions between the first 5 levels are provided by \citet[][their table 6]{Biemont1986-2}. The As I isoelectronic sequence (As I, Se II, Br III, Kr IV, ..) has a 4p$^3$$(^4\rm S)$ ground configuration with no splitting. IR lines can only be generated from the second term, 4p$^3(^2\rm D^o)$. For As I the $^2$D$^{\rm o}_{5/2}-^2$D$^{\rm o}_{3/2}$ transition is however beyond JWST range at 31 \mum, and with a low A-value.

\subsection{As II} 
\noindent The energies of levels 2-5 ($^3$P$_1$, $^3$P$_2$, $^1$D$_2$, $^1$S$_0$) were corrected to NIST values \citep{Moore1971}. Theoretical A-values for transitions between the first 5 levels are provided by \citet{Biemont1986-2}. 
Potential IR lines include $^3$P$_1-^3$P$_0$ (2-1, 9.40 $\mu$m, $A=0.022$ s$^{-1}$) and $^3$P$_2-^3$P$_1$ (3-2, 6.77 $\mu$m, $A=0.043$ s$^{-1}$). 

\subsection{As III}
\noindent  The energy of level 2 ($^2$P$^{\rm o}_{3/2}$) was corrected to the NIST value \citep{Moore1971}. A potential IR line is $^2$P$^{\rm o}_{3/2}-^2$P$^{\rm o}_{1/2}$ (2-1, 3.40 $\mu$m, $A=0.23$ s$^{-1}$.)

\subsection{As IV}
\noindent The first excited state in As IV lies high at 75,812 cm$^{-1}$, no calibrations were made.

\subsection{Se I}
\noindent The energies of levels 2-5 ($^3$P$_1$, $^3$P$_0$, $^1$D$_2$, $^1$S$_0$) were corrected to NIST values \citep{Moore1971}. Theoretical A-values for transitions between the first 5 levels are provided by \citet[][Hartree-Fock with relativistic corrections, their Table 8]{Biemont1986-1}. The Se I isoelectronic sequence (Se I, Br II, Kr III, Rb IV,..) has a 4p$^4(^3\rm P)$ ground term with $J=2,1,0$ splitting. Potential IR lines include $^3$P$_1-^3$P$_2$ (2-1, 5.03 $\mu$m, $A=0.17$ s$^{-1}$), $^3$P$_0-^3$P$_1$ (3-2, 18.35 $\mu$m, $A=0.010$ s$^{-1}$), and (trans-term) $^1$D$_2-^3$P$_2$ (4-1, 1.04 $\mu$m, $A=0.63$ s$^{-1}$). 

\subsection{Se II}
\noindent The energies of levels 2-5 ($^2$D$^{\rm o}_{3/2}$, $^2$D$^{\rm o}_{5/2}$, $^2$P$^{\rm o}_{1/2}$, $^2$P$^{\rm o}_{3/2}$) were corrected to NIST values \citep{Moore1971}. Theoretical A-values for transitions between the first 5 levels are provided by \citet[][their Table 6]{Biemont1986-2}. 
The $^2$D$^{\rm o}_{5/2}-^2$D$^{\rm o}_{3/2}$ transition (within the second term) is at 16.22 \mum, with $A=2.9 \times 10^{-3}$ s$^{-1}$.

\subsection{Se III}
\noindent  The energies of levels 2-5 ($^3$P$_1$, $^3$P$_2$, $^1$D$_2$, $^1$S$_0$) were corrected to NIST values \citep{Moore1971}. 
Theoretical A-values for transitions between the first 5 levels are provided by \citet{Biemont1986-2} and \citet{Sterling2017}, we use the latter source as it also reports R-matrix calculations of collision strengths for these transitions, which we use. Potentially important IR lines include $^3$P$_2$-$^3$P$_1$  (3-2, 4.55 $\mu$m, $A=0.16$ s$^{-1}$), 
and $^3$P$_1$-$^3$P$_0$ (2-1, 5.74 $\mu$m, $A=0.082$ s$^{-1}$).

\subsection{Se IV}
\noindent Level 2 ($^2$P$^{\rm o}_{3/2}$) was corrected to the NIST value \citep{Moore1971}. Collision strengths for this transition were implemented \citep[][K. Butler, priv. comm.]{Sterling2007}. A potential IR line is $^2$P$^{\rm o}_{3/2}-^2$P$^{\rm o}_{1/2}$ (2-1, 2.29 $\mu$m, $A=0.75$ s$^{-1}$).  

\subsection{Br I}
\noindent  Level 2 ($^2$P$_{1/2}$) was corrected to the NIST value \citep{Tech1963}. The Br I isoelectronic sequence (Br I, Kr II, Rb III, Sr IV, ..) has a 4p$^5(^2\rm P)$ ground term with $J=3/2,1/2$ splitting. A potential IR line is $^2$P$_{1/2}-^2P_{3/2}$ (2-1, 2.71 $\mu$m, $A=0.90$ s$^{-1}$). 

\subsection{Br II}
\noindent The energies of levels 2-5 ($^3$P$_1$, $^3$P$_0$, $^1$D$_2$, $^1$S$_0$) were corrected to NIST values \citep{Moore1971}. 
Theoretical A-values for transitions between the first 5 levels are provided by \citet[][their Table 8]{Biemont1986-1}. Potential IR lines include $^3$P$_1-^3$P$\bm{_0}$ (2-1, 3.19 $\mu$m, $A=0.67$ s$^{-1}$) and $^3$P$_0-^3$P$_1$ (3-2, 14.26 $\mu$m, $A=0.022$ s$^{-1}$).

\subsection{Br III}
\noindent  The energies of levels 2-5 ($^2$D$^{\rm o}_{3/2}$, $^2$D$^{\rm o}_{5/2}$, $^2$P$^{\rm o}_{1/2}$, $^2$P$^{\rm o}_{3/2}$) were corrected to NIST values \citep{Moore1971}. A potential IR line is (second-term) $^2$D$^{\mbox{o}}_{5/2}-^2$D$^{\mbox{o}}_{3/2}$ (3-2, 7.94 $\mu$m, $A=2.0\times 10^{-2}$ s$^{-1}$). 

\subsection{Br IV}
\noindent The energies of levels 2-4 ($^3$P$_1$, $^3$P$_2$, $^1$D$_2$)  were corrected to NIST values \citep{Joshi1971}. 
Theoretical A-values for transitions between the first 5 levels are provided by \citet{Biemont1986-2}. Potential IR lines include $^3$P$_1-^3$P$_0$ (2-1, 3.81 $\mu$m, $A=0.32$ s$^{-1}$) and $^3$P$_2-^3$P$_1$ (3-2, 3.34 $\mu$m, $A=0.33$ s$^{-1}$).

\subsection{Kr I}
\noindent  The first excited state of Kr I lies high at 79,971 cm$^{-1}$, no calibrations were made. The Kr I isoelectronic sequence (Kr I, Rb II, Sr III, Y IV,...) has a 4p$^6(^1$S) ground term with no splitting.

\subsection{Kr II}
\noindent The energy of level 2 ($^2$P$_{1/2}$) was corrected to the NIST value \citep{Saloman2007}. The third state lies high at 109,000 cm$^{-1}$. A potential IR line is $^2$P$_{1/2}-^2$P$_{3/2}$ (2-1, 1.86 $\mu$m, $A=2.78$ s$^{-1}$).

Collision strengths were computed for the 2-1 transition, using a 70 state Breit-Pauli R-Matrix (BPRM) calculation.  This is similar to the calculations by \cite{Mulholland2024} for Sr~{\sc ii} and Y~{\sc ii} and we refer the interested reader to that paper for further details of and references to the methods.  The Kr~{\sc ii} target wavefunctions were computed using {\sc autostructure} \citep{Badnell1997,Badnell1986} with 7 configurations (4s$^2$4p$^5$, 4s4p$^6$, 4s$^2$4p$^4$4d, 4s$^2$4p$^4$4f, 4s$^2$4p$^4$5s, 4s$^2$4p$^4$5p, 4s$^2$4p$^4$5d) being included in the configuration expansion.  The radial wavefunctions were optimised by
minimising the sum of the energies of the lowest 30 LS-coupling terms which gave the following values for the parameters
of the Thomas-Fermi-Dirac-Amaldi potential:
 $\lambda_{\mathrm1s}=1.4240$, $\lambda_{\mathrm2s}=1.1393$, $\lambda_{\mathrm2p}=1.0856$
 $\lambda_{\mathrm3s}=1.0385$, $\lambda_{\mathrm3p}=1.0207$, $\lambda_{\mathrm3d}=1.0005$
 $\lambda_{\mathrm4s}=0.9913$, $\lambda_{\mathrm4p}=1.0039$, $\lambda_{\mathrm4d}=1.0424$
 $\lambda_{\mathrm4f}=1.0483$, $\lambda_{\mathrm5s}=1.0015$, $\lambda_{\mathrm5p}=0.9934$
 $\lambda_{\mathrm5d}=1.0062$. The BPRM calculation itself was performed using the Belfast codes \citep{Berrington1995} in versions maintained by Dr. Connor Ballance which are to be found at
\cite{connorb}. J values up to 37 were included explicitly, while larger values were estimated via a ``top-up'' procedure. The energy grid included of order 15000 points up to the highest threshold (approximately 1.554 Ryd) with a further 860 points to extend the grid up to roughly 10 Ryd. This was augmented with the infinity energy point provided by {\sc autostructure}.  The resulting cross sections were convolved with a Maxwellian distribution and are available in adf04 format.

\subsection{Kr III}
\noindent The energies of levels 2-5 ($^3$P$_1$, $^3$P$_0$, $^1$D$_2$, $^1$S$_0$) were corrected to the NIST values \citep{Saloman2007}.  
Theoretical A-values for transitions between the first 5 levels are provided by \citet[][their Table 8]{Biemont1986-1}. \citet{Schoning1997} provide R-matrix calculations of collision strengths. Potential IR lines include $^3$P$_1-^3$P$_2$ (2-1, 2.20 $\mu$m, $A=2.0$ s$^{-1}$) and $^3$P$_0-^3$P$_1$ (3-2, 13.08 $\mu$m, $A=0.029$ s$^{-1}$).

\subsection{Kr IV}
\noindent The energies of levels 2-5 ($^2$D$^{\rm o}_{3/2}$, $^2$D$^{\rm o}_{5/2}$, $^2$P$^{\rm o}_{1/2}$, $^2$P$^{\rm o}_{3/2}$) were corrected to NIST values \citep{Saloman2007}. 
\citet{Schoning1997} provide R-matrix calculations of collision strengths. A potential IR line is (second-term) $^2$D$^{\rm o}_{5/2}-^2$D$^{\rm o}_{3/2}$ (3-2, 6.01 $\mu$m, $A=0.046$ s$^{-1}$).  

\subsection{Rb I}
\noindent  The energies of levels 2-6 ($^2$P$^{\rm o}_{1/2}$, $^2$P$^{\rm o}_{3/2}$, $^2$D$_{5/2}$, $^2$D$_{3/2}$, $^2$S$_{1/2}$)  were corrected to NIST values \citep{Sansonetti2006}, which included a reordering of levels 4-6. Rb I has no expected strong IR lines. Rb I has a single-electron 5s($^2\rm S$) ground state.

\subsection{Rb II}
\noindent  The first excited state of Rb II lies high at 133,341 cm$^{-1}$, no calibrations were made.

\subsection{Rb III}
\noindent  The energy of level 2 ($^2$P$_{1/2}$) was corrected to the NIST value \citep{Sansonetti2006}. The third level lies high at 130,032 cm $^{-1}$. A potential IR line is $^2$P$_{1/2}-^2$P$_{3/2}$ (2-1, 1.36 $\mu$m, $A=7.21$ s$^{-1}$). 

\subsection{Rb IV}
\noindent The energies of levels 2-5 ($^3$P$_1$, $^3$P$_0$, $^1$D$_2$, $^1$S$_0$) were corrected to NIST values \citep{Sansonetti2006}. 
Theoretical A-values for transitions between the first 5 levels are provided by \citet[][their Table 8]{Biemont1986-1}. We use collision strengths for the first 5 levels from 
\citet{Sterling2016}\footnote{\citet{Sterling2016} also compute A-values, but list zero for several transitions that have non-zero entries in \citet{Biemont1986-1}. In their discussion they also use the  \citet{Biemont1986-1} values as benchmark.}. Potential IR lines include $^3$P$_1-^3$P$_2$ (2-1, 1.60 $\mu$m, $A=4.65$ s$^{-1}$) and $^3$P$_0-^3$P$_1$ (3-2, 14.54 $\mu$m, $A=2.3\times 10^{-2}$ s$^{-1}$). 

\subsection{Sr I}
\noindent  The energies of levels 2-9 ($^3$P$^{\rm o}_{0,1,2}$, $^3$D$_{1,2,3}$, $^1$D$_2$, $^1$P$^{\rm o}_1$) were corrected to NIST values \citep{Sansonetti2010}, which involved some reorderings. Sr I has a 5s$^2(^1\rm S)$ ground term with no splitting.

\subsection{Sr II}
\noindent  The energies of levels 2-17 ($^2$D$_{5/2,3/2}$, $^2$P$^{\rm o}_{1/2,3/2}$, $^2$S$_{1/2}$, 5d$(^2$D$_{5/2,3/2})$, 6p$(^2$P$^{\rm o}_{1/2,3/2}$), $^2$F$^{\rm o}_{7/2,5/2}$, 7s$(^2$S$_{1/2})$, 6d$(^2$D$_{3/2,5/2})$, 7p$(^2$P$^{\rm o}_{1/2,3/2}$)) were corrected to NIST values \citep{Moore1971}. Collision strengths from \citet{Mulholland2024} were used. 

\subsection{Sr III}
\noindent The first excited state of Sr III lies high at 176,434 cm$^{-1}$, no calibrations were made.

\subsection{Sr IV}
\noindent The energy of level 2 ($^2$P$_{1/2}$) was corrected to the NIST value \citep{Persson1978}. A potential IR line is $^2$P$_{1/2}-^2$P$_{3/2}$ (2-1, 1.028 $\mu$m, $A=15.8$ s$^{-1}$). 

\subsection{Y I}
\noindent  The energies of levels 2-8 (a$^2$D$_{5/2}$, z$^2$P$^{\rm o}_{1/2,3/2}$, a$^4$F$_{3/2,5/2,7/2,9/2}$) were corrected to NIST values \citep{Palmer1977}, which involved some reorderings. Because the \texttt{FAC} energy structure is significantly different to the NIST one for higher levels, the model atom was limited to these 8 states. Y I has a 4d5s$^2(^2\rm D)$ ground configuration with splitting into $J=3/2, 5/2$ (Zr II, Nb III, etc. have other ground configurations so the isoelectronic sequencing here breaks down). A potential IR line is $^2$D$_{5/2}-^2$D$_{3/2}$ (2-1, 18.85 $\mu$m, $A=5.5 \times 10^{-4}$ s$^{-1}$).

\subsection{Y II}
\noindent The energies of levels 2-12 (a$^3$D$_{1,2,3}$, a$^1$D$_{2}$, a$^3$F$_{2,3,4}$, a$^3$P$_{0,1,2}$, b$^1$D$_{2}$) were corrected to NIST values \citep{Nilsson1991}. We implemented collision strengths for transitions between the first 12 levels from \citet{Mulholland2024}. Y II has a 5s$^2(^1\rm S)$ ground term with no splitting. However, it has a rich level structure of quite low-energy metastable states, difficult to assess beforehand if IR lines from these may become strong. 

\subsection{Y III}
\noindent The energies of levels 2-17 were in \citet{Pognan2023} calibrated to NIST values. Here, we also add A-values for transitions between the first three states from \citet{Sahoo2008}. Y III has a 4p$^6(^2$D) ground term with $J=3/2,5/2$ splitting. Potential IR lines include $^2$D$_{5/2}-^2$D$_{3/2}$ (2-1, 13.81 $\mu$m, $A=4.1\times 10^{-3}$ s$^{-1}$), $5s^2$S$-4d^2$D$_{3/2}$ (3-1, 1.34 \mum,  $A=9.4\times 10^{-2}$ s$^{-1}$), and $5s^2$S$-4d^2$D$_{5/2}$ (3-2, 1.48 $\mu$m, $A=4.5\times 10^{-2}$ s$^{-1}$).

\subsection{Y IV}
\noindent The first excited state in Y IV lies high at 209,651 cm$^{-1}$, no calibrations were made.

\subsection{Zr I}
\noindent The energies of levels 2-12 (a$^3$F$_{3,4}$, a$^3$P$_{2,0,1}$, a$^5$F$_{1,2,3,4,5}$, a$^1$D$_{2}$, a$^1$G$_{4}$)  were corrected to NIST values \citep{Moore1971}.  
 Potential IR lines include $^3$F$_3-^3$F$_2$ (2-1, 17.53 $\mu$m, $A=3.2\times 10^{-3}$ s$^{-1}$), $^3$F$_4-^3$F$_3$ (3-2, 14.91 $\mu$m, $A=4.7\times 10^{-3}$ s$^{-1}$), and a blend of lines between $2-2.5$ $\mu$m. 

\subsection{Zr II}
\noindent The energies of levels 2-4 (a$^4$F$_{5/2,7/2,9/2}$) were corrected to NIST values \citep{Moore1971}. Potential MIR lines include $^4$F$_{9/2}-^4$F$_{7/2}$ (4-3, 17.87 \mum, $A=2.7 \times 10^{-3}$ s$^{-1}$) and  $^4$F$_{7/2}-^4$F$_{5/2}$ (3-2, 22.28 \mum, $A=2.1 \times 10^{-3}$ s$^{-1}$).

\subsection{Zr III}
\noindent The energies of levels 2-13 ($^3$F$_{3,4}$, $^1$D$_{2}$, $^3$P$_{0,1,2}$, $^1$G$_{4}$, $^3$D$_{1,2,3}$, $^1$S$_{0}$,  $^1$D$_{2}$) were corrected to NIST values \citep{Reader1997}. We also replaced the \texttt{FAC} A-values for transitions between these states with those of \citet[][Multi-Configuration Dirac-Hartree-Fock (MCDHF) and relativistic configuration interaction (RCI) calculations using GRASP2018]{Rynkun2020}. Potential IR lines include  $^3$F$_3-^3$F$_2$ (2-1, 14.67 $\mu$m, $A=7.5\times 10^{-3}$ s$^{-1}$) and $^3$F$_4-^3$F$_3$ (3-2, 12.42 $\mu$m, $A=1.0 \times 10^{-2}$ s$^{-1}$). 

\subsection{Zr IV}
\noindent The energies of levels 2-5 ($^2$D$_{5/2}$, $^2$S$_{1/2}$, $^2$P$^{\rm o}_{1/2,3/2}$)  were corrected to NIST values \citep{Reader1997}. Level 6 lies high at 137,413 cm$^{-1}$. 
Collision strengths for transitions between these first 5 levels were implemented \citep[][K. Butler, priv. comm.]{Dinerstein2006}. A potential IR line is $^2$D$_{5/2}-^2$D$_{3/2}$ (2-1, 8.00 $\mu$m, $A=0.012$ s$^{-1}$).

\subsection{Te I}
\noindent The energies of levels 2-10 ($^3$P$_{0,1}$, $^1$D$_{2}$, $^1$S$_{0}$, $^5$S$^{\rm o}_{2}$, $^3$S$^{\rm o}_{1}$, $^5$P$_{1,2,3}$) were calibrated to NIST values \citep{Morillon1975}. Because higher levels showed discrepancy between the \texttt{FAC} model and NIST, the model atom was capped at 10 levels - the eleventh level lies quite high at 54,683 cm$^{-1}$. We use collision strengths from \citet{Mulholland2024-Te}. A potential IR line is $^3$P$_1-^3$P$_2$ (3-1, 2.10 \mum, $A=2.27$ s$^{-1})$.

\subsection{Te II}
\noindent The energies of levels 2-8 ($^2$D$^{\rm o}_{3/2,5/2}$, $^2$P$^{\rm o}_{1/2,3/2}$, $^4$P$_{5/2,3/2,1/2}$) were calibrated to NIST values \citep{Eriksson1974}. Because higher levels showed discrepancy between the \texttt{FAC} model and NIST, the model atom was capped at 8 levels - the ninth level lies high at 78,448 cm$^{-1}$. A potential IR line is $^2$D$^{\rm o}_6-^2$D$^{\rm o}_4$ (3-2, 4.55 \mum, $A=0.098$ s$^{-1}$).

\subsection{Te III}
\noindent The energies of levels 2-4 ($^3$P$_{1,2}$, $^1$D$_{2}$) were calibrated to NIST values \citep{Moore1971}. Because higher levels showed discrepancy between the \texttt{FAC} model and NIST, the model atom was capped at 4 levels - the fifth level lies high at 82,885 cm$^{-1}$. A-values and collision strengths from \citet{Madonna2018} were used. Potential IR lines include $^3$P$_1-^3$P$_0$ (2-1, 2.10 \mum, $A=1.19$ s$^{-1}$) and $^3$P$_2-^3$P$_1$ (3-2, 2.92 \mum, $A=0.52$ s$^{-1}$).

\subsection{Te IV}
\noindent The energy of level 2 ($^2$P$^{\rm o}_{3/2}$) was calibrated to NIST values \citep{Moore1971}. Because higher levels showed discrepancy between the \texttt{FAC} model and NIST, the model atom was capped at 2 levels - the third level lies high at 92,772  cm$^{-1}$. A potential IR line is $^2$P$^{\rm o}_{3/2}-^2$P$^{\rm o}_{1/2}$ (2-1, 1.084 \mum, $A=7.1$ s$^{-1}$).

\subsection{Fe, Ni, Zn}
For Fe I-III, Ni I-Ni III we used the standard atomic data set in \texttt{SUMO}. For this study we added also a 2-level atom for Fe IV and a 7-level model atom for Ni IV \citep{Sugar1985}. For Zn I-IV we use the \texttt{FAC} model set, with energy calibrations done for the first few levels of each ion \citep{Sugar1995}. The only significant IR line candidate from zinc is from [Zn IV] 3.62 \mum.

\section{Sensitivity to deposition power}
\label{sec:sens_f}
\begin{figure}[htb]
\centering
\includegraphics[width=0.8\linewidth]{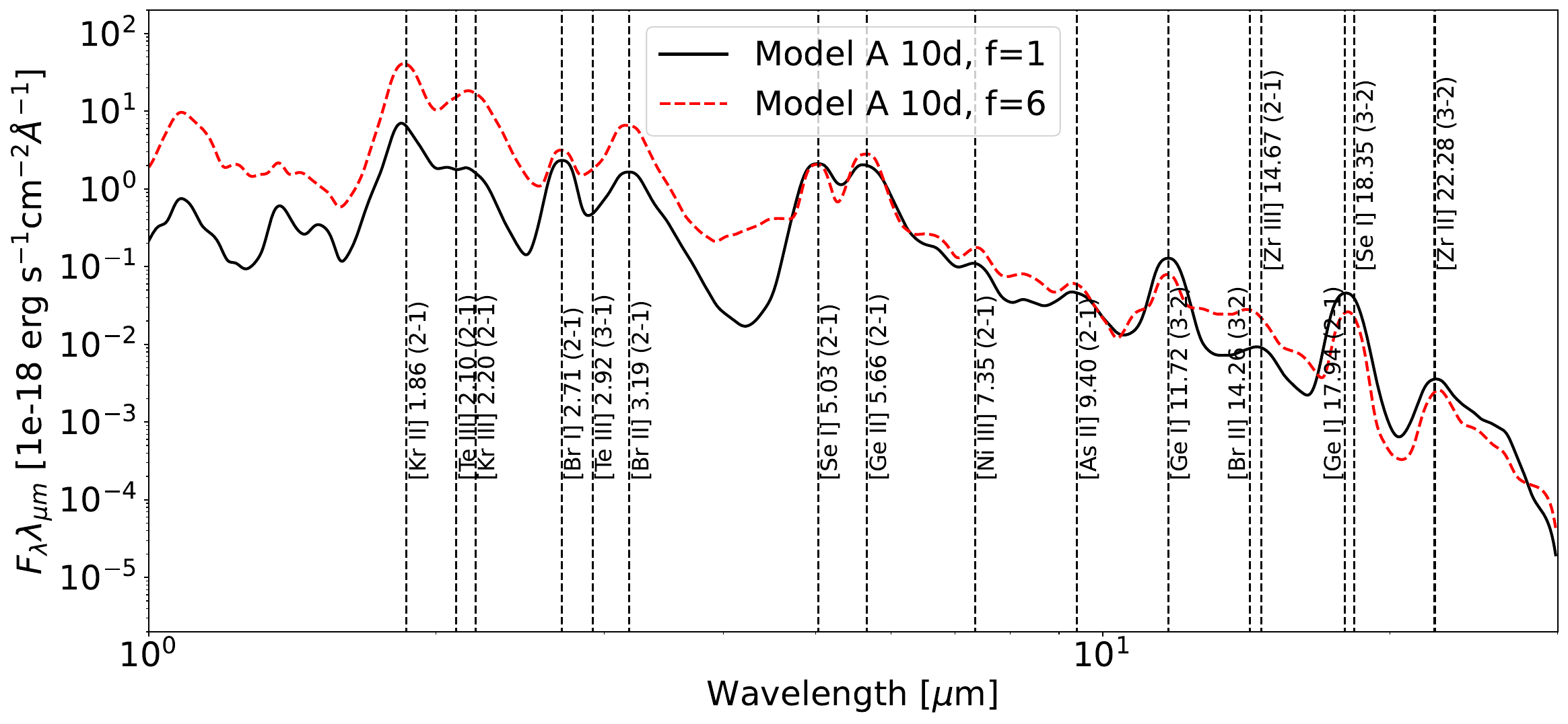}
\includegraphics[width=0.8\linewidth]{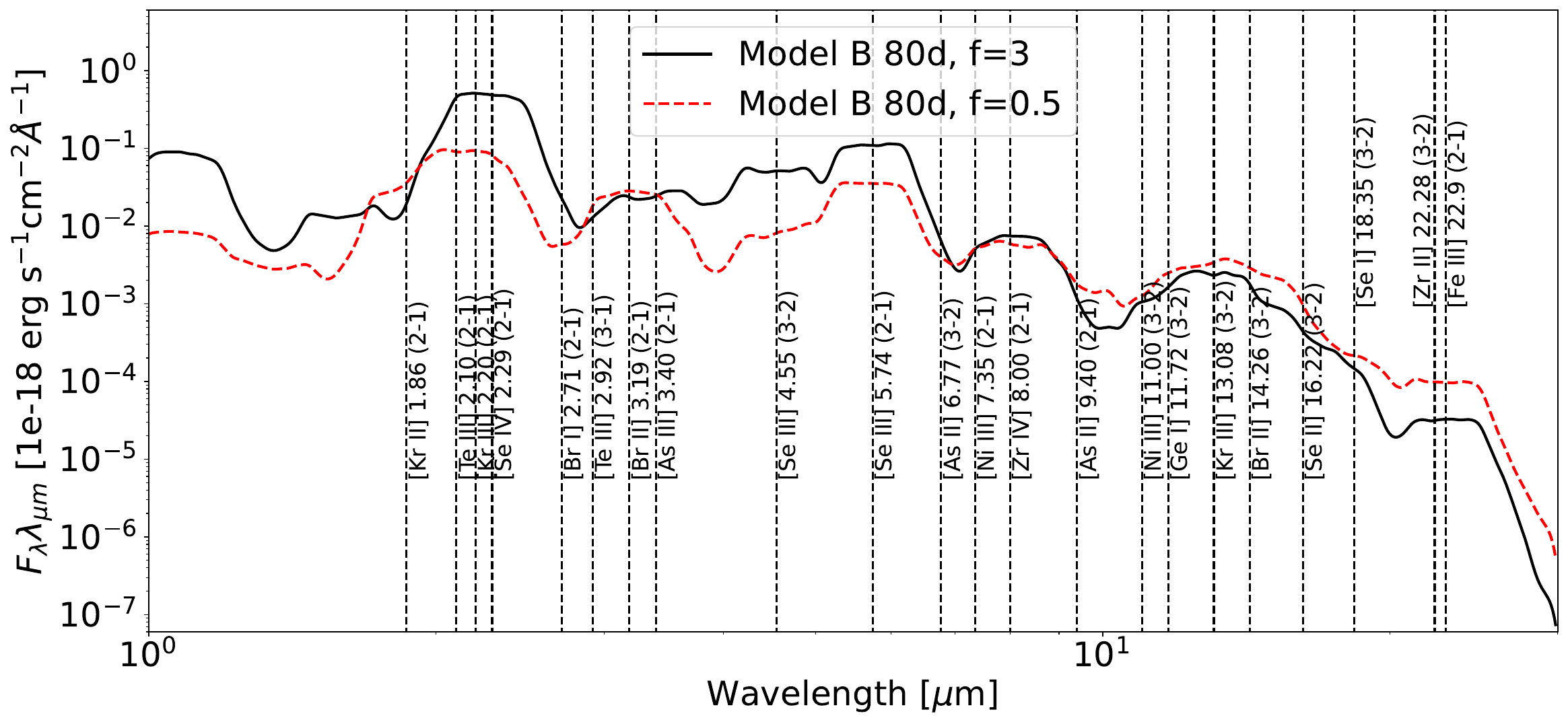}
\caption{Top: Comparison of model A with $f=1$ and $f=6$, at 10d. Bottom: Comparison of model B with $f=3$ and $f=0.5$, at 80d.}
\label{fig:compdep}
\end{figure}

In Fig. \ref{fig:compdep} we compare the model spectra when power levels change. In the top panel, $f$ changes from 1 to 6 in model A at 10d. The general temperature rise causes an increase in flux below $\sim$ 4 \mum. The changes after that are driven solely by ionization changes, as the lines become temperature-insensitive. These are however moderate enough that it is still largely the same ion and transitions that are dominant. This gives important robustness to the model predictions. In the bottom panel, model B at 80d is compared with $f=3$ and $f=0.5$.

\section{Ionization profiles}
\begin{figure*}[htb]
\includegraphics[width=0.49\linewidth]{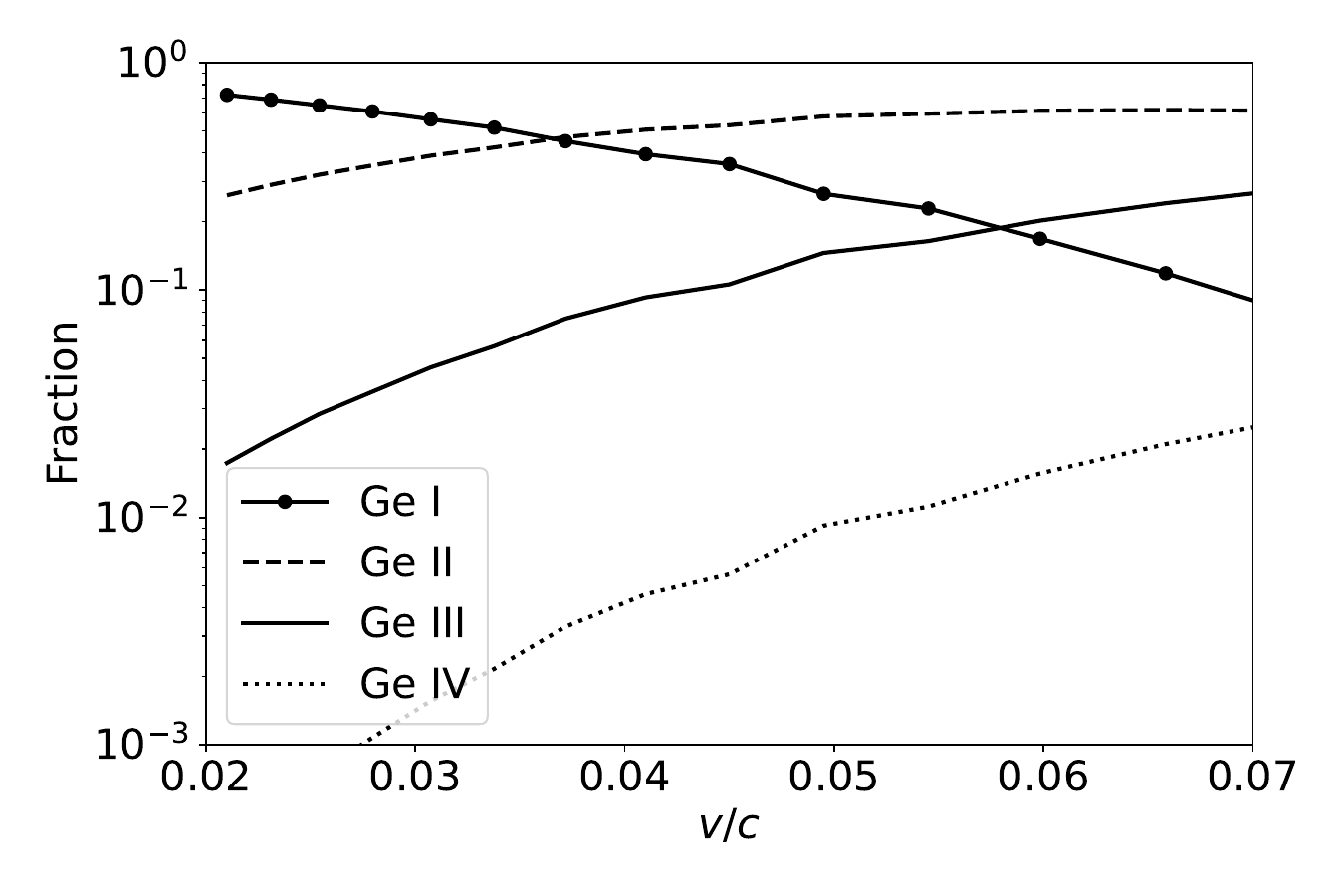}
\includegraphics[width=0.49\linewidth]{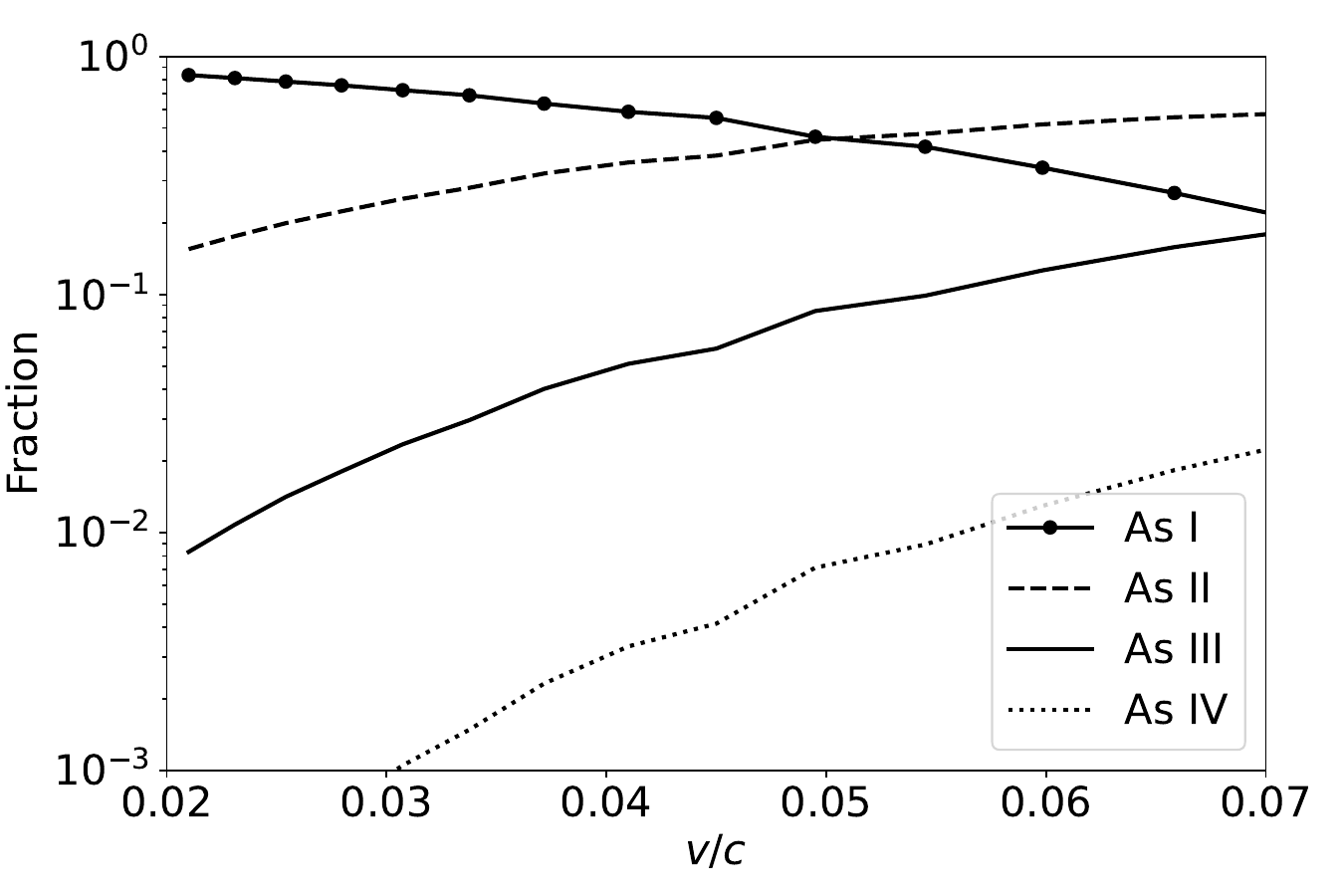}\\
\includegraphics[width=0.49\linewidth]{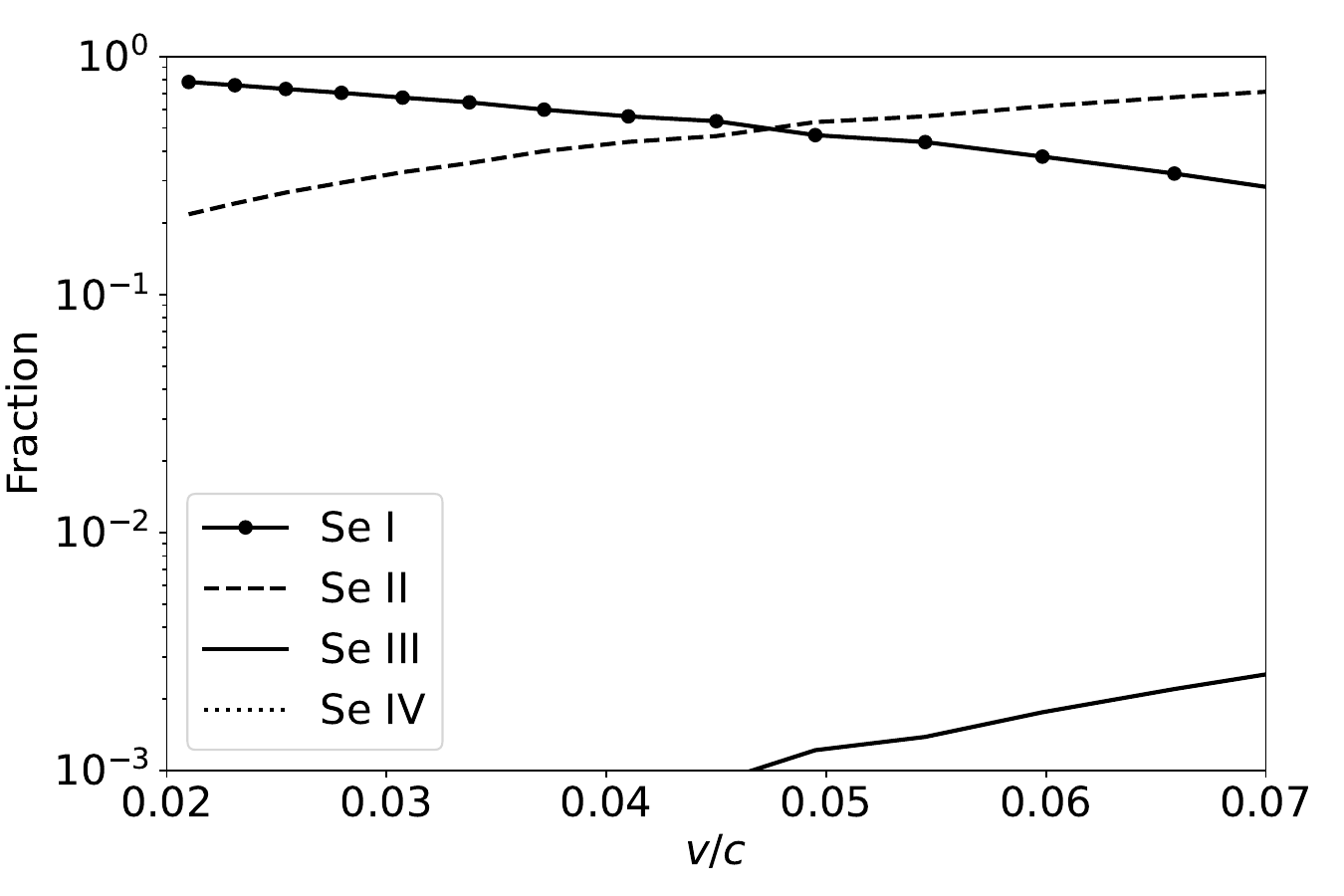}
\includegraphics[width=0.49\linewidth]{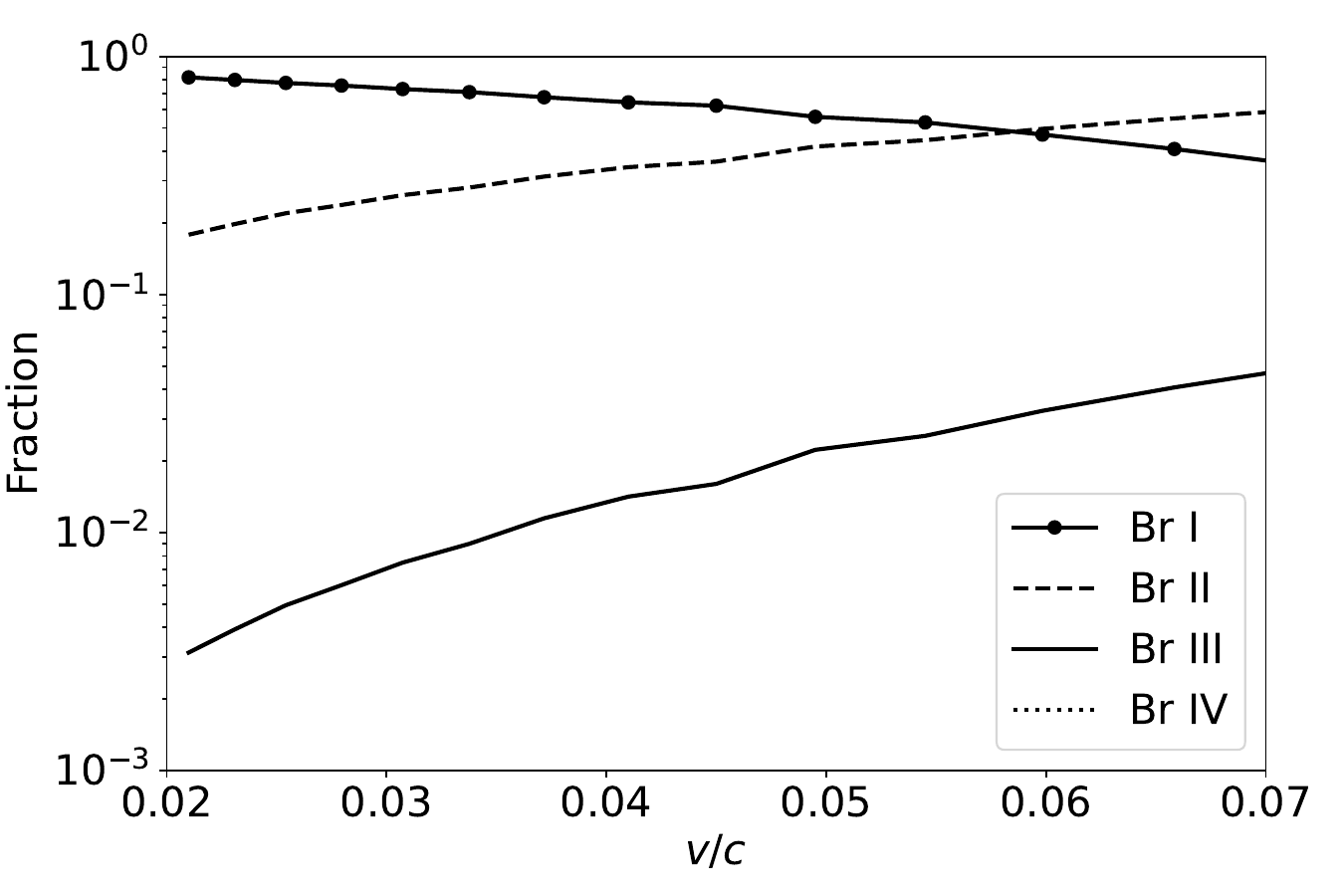}\\
\includegraphics[width=0.49\linewidth]{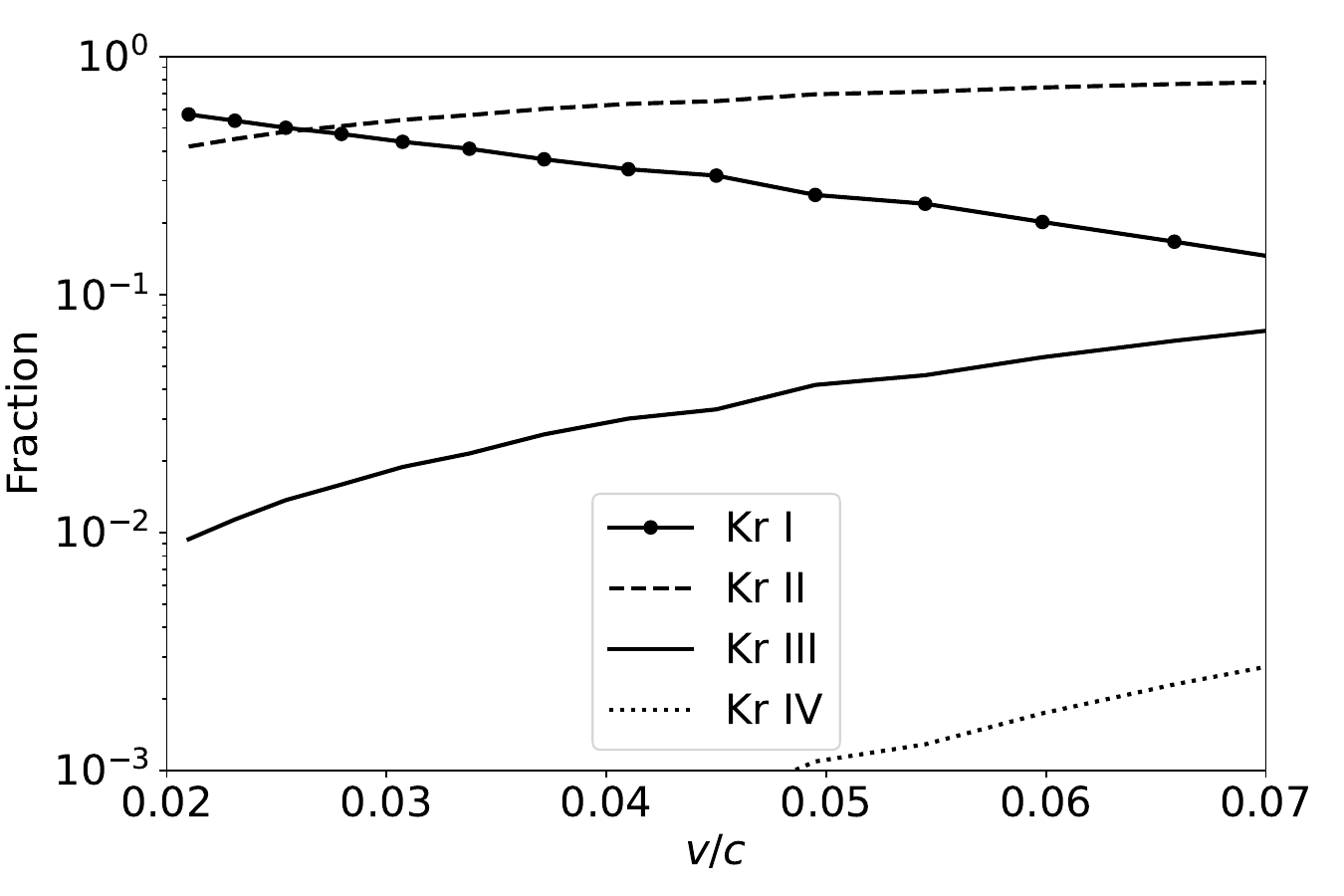}
\includegraphics[width=0.49\linewidth]{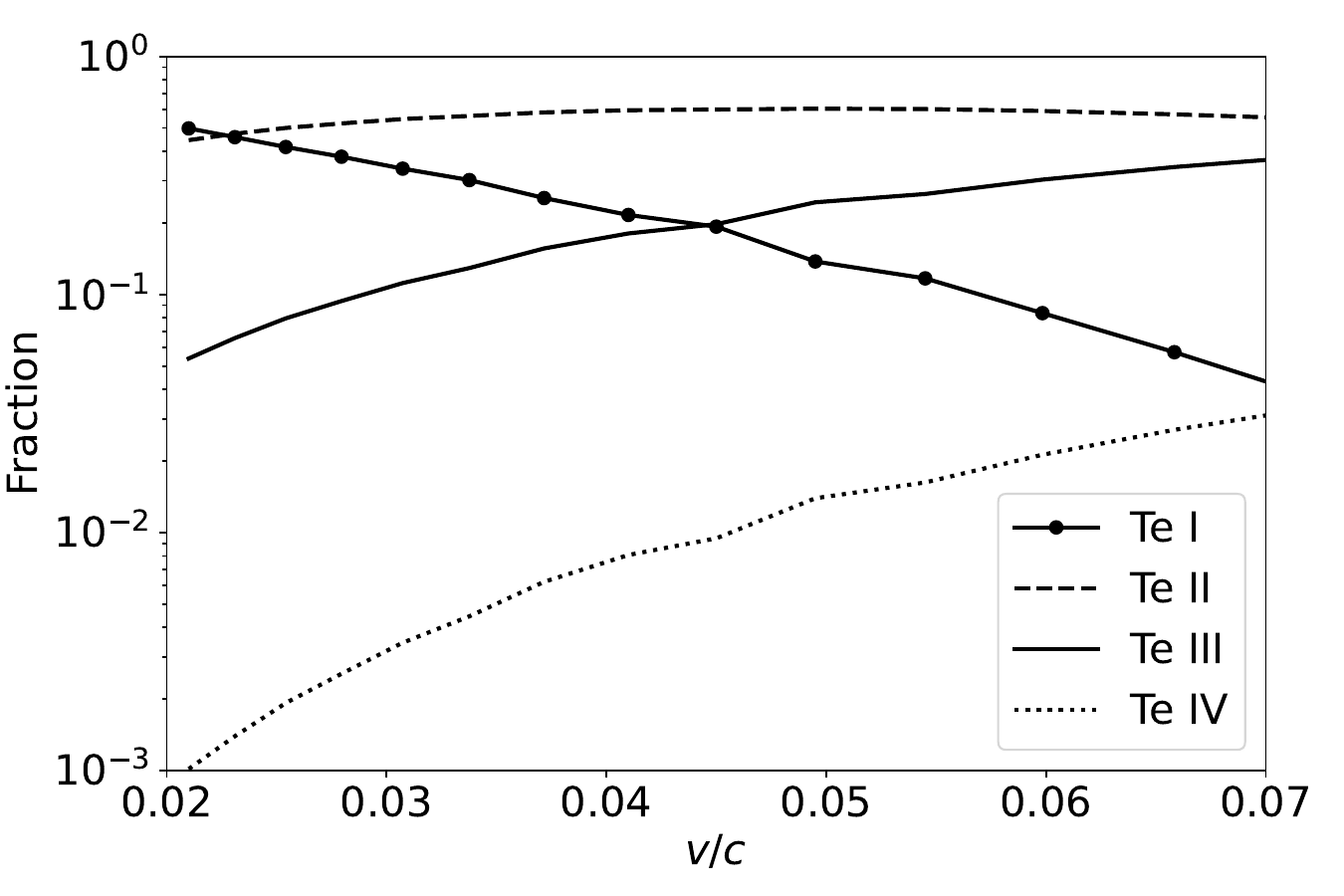}
\caption{Ionization profiles of selected important elements in model A, at 10d, zoomed in on the region $v/c \leq 0.07$ (which has $\gtrsim 80\%$ of the radioactive energy deposition). Neutral abundances are plotted with solid lines with dots, singly ionized with dashed lines, doubly ionized with solid lines, and triply ionized with dotted lines.}
\label{fig:modelA-low-10d-ionstructure}
\end{figure*}

\begin{figure*}[htb]
\includegraphics[width=0.49\linewidth]{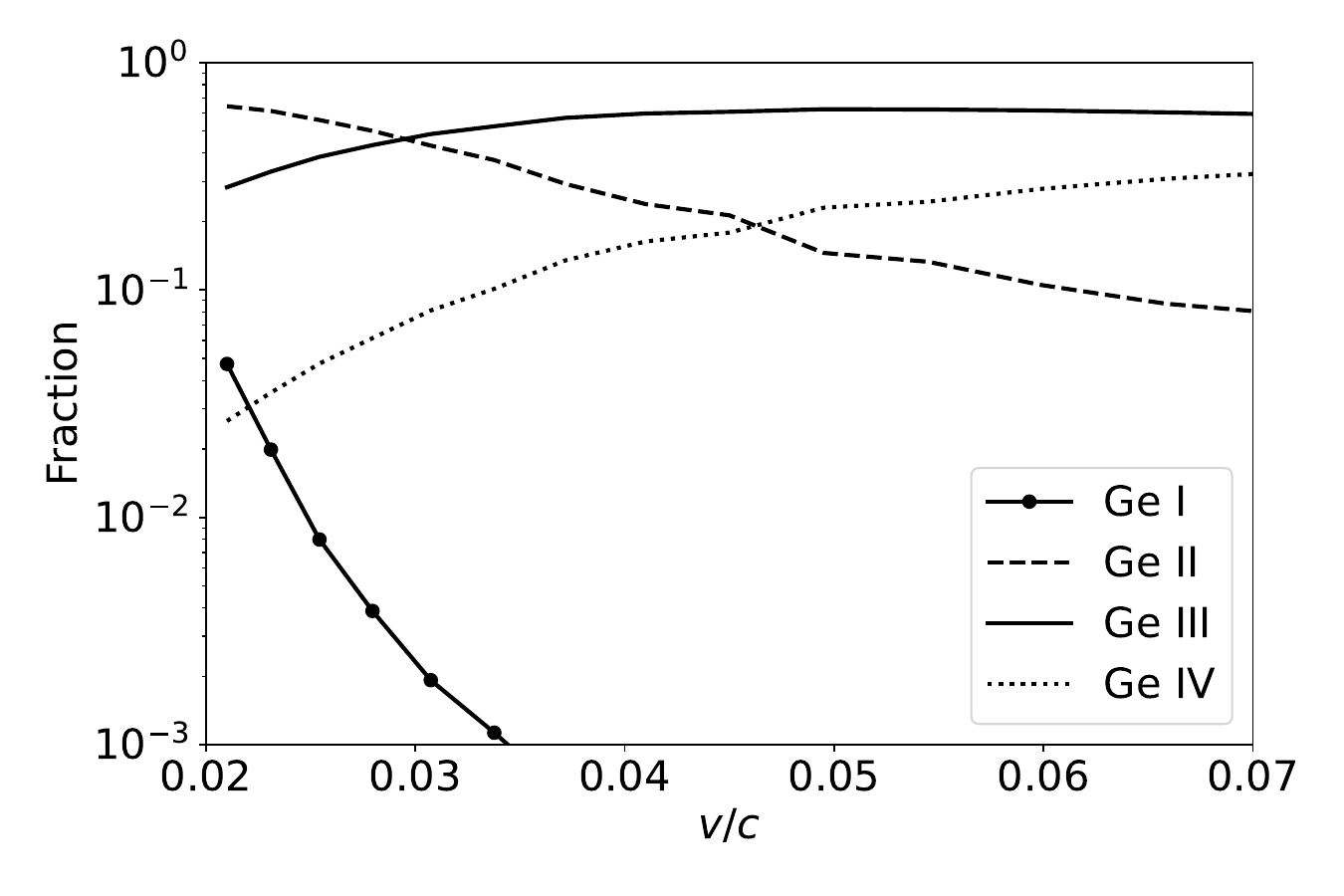}
\includegraphics[width=0.49\linewidth]{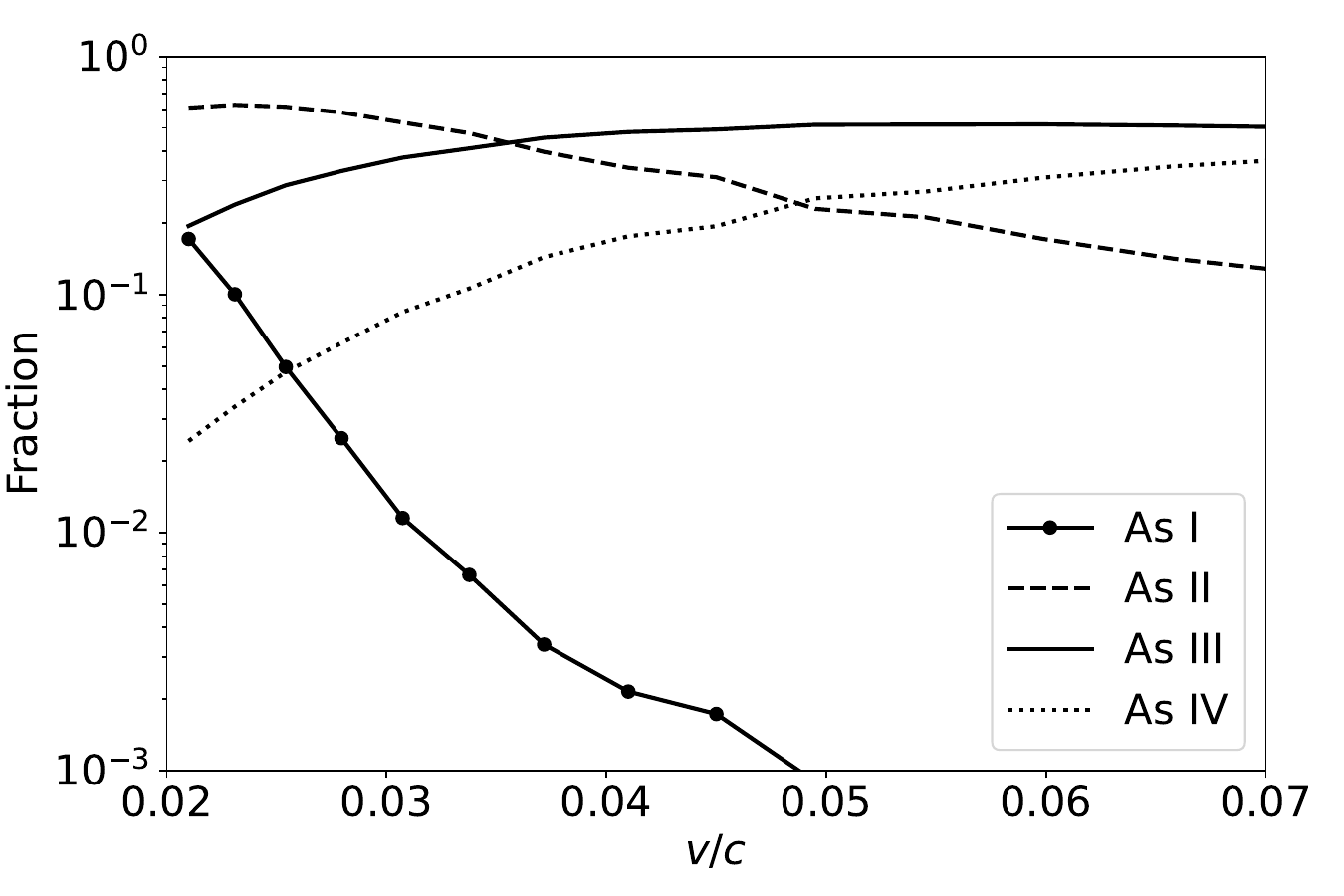}\\
\includegraphics[width=0.49\linewidth]{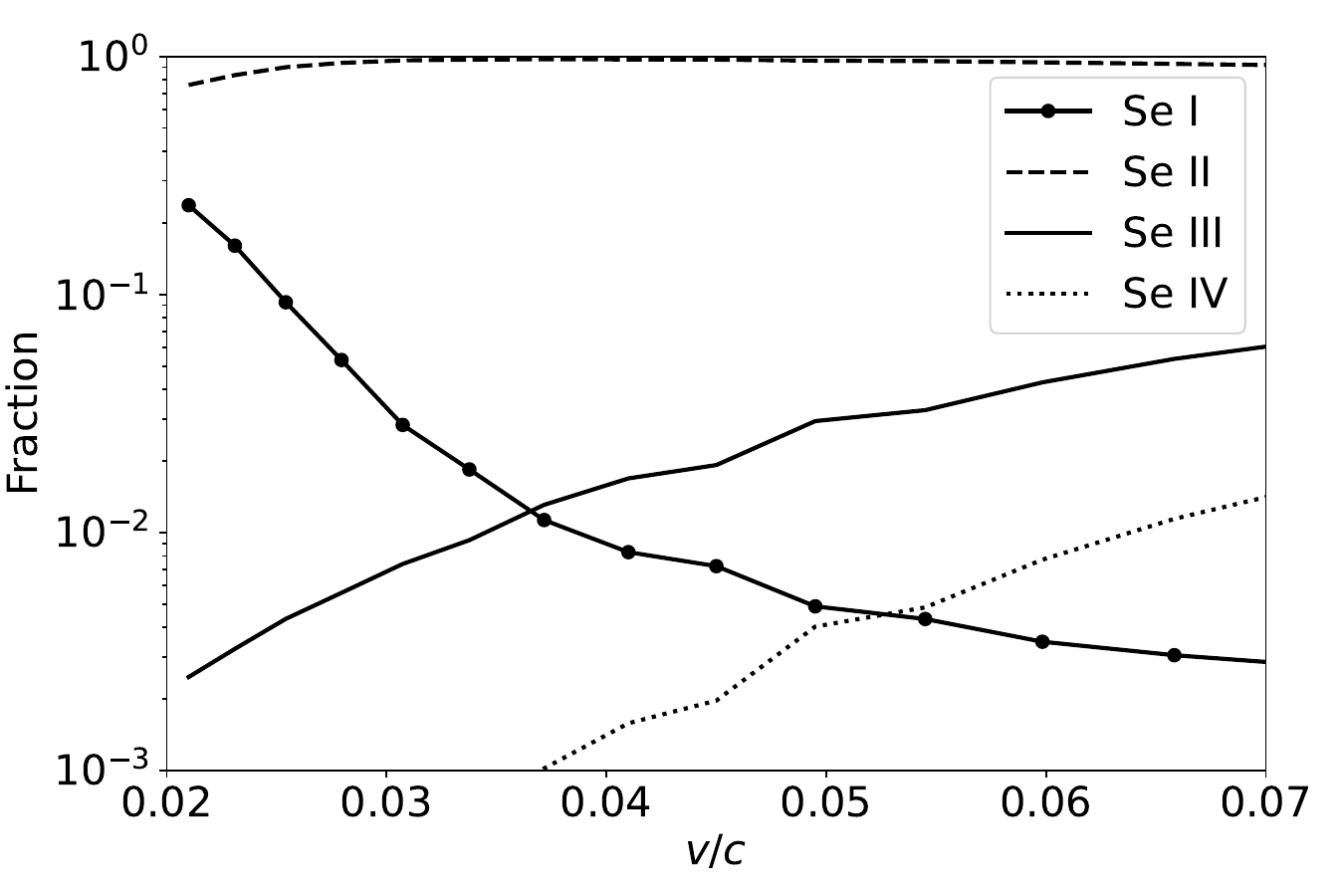}
\includegraphics[width=0.49\linewidth]{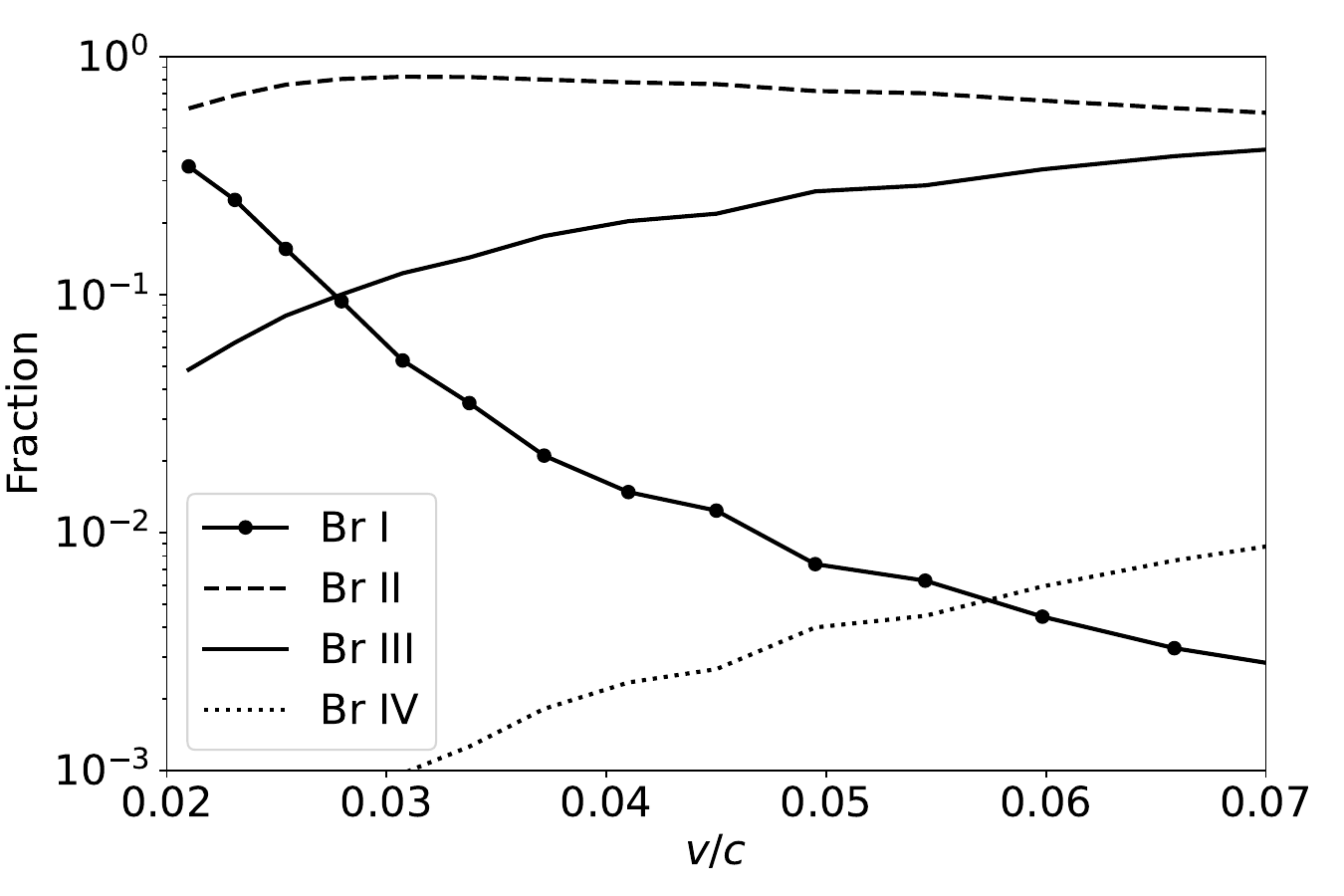}\\
\includegraphics[width=0.49\linewidth]{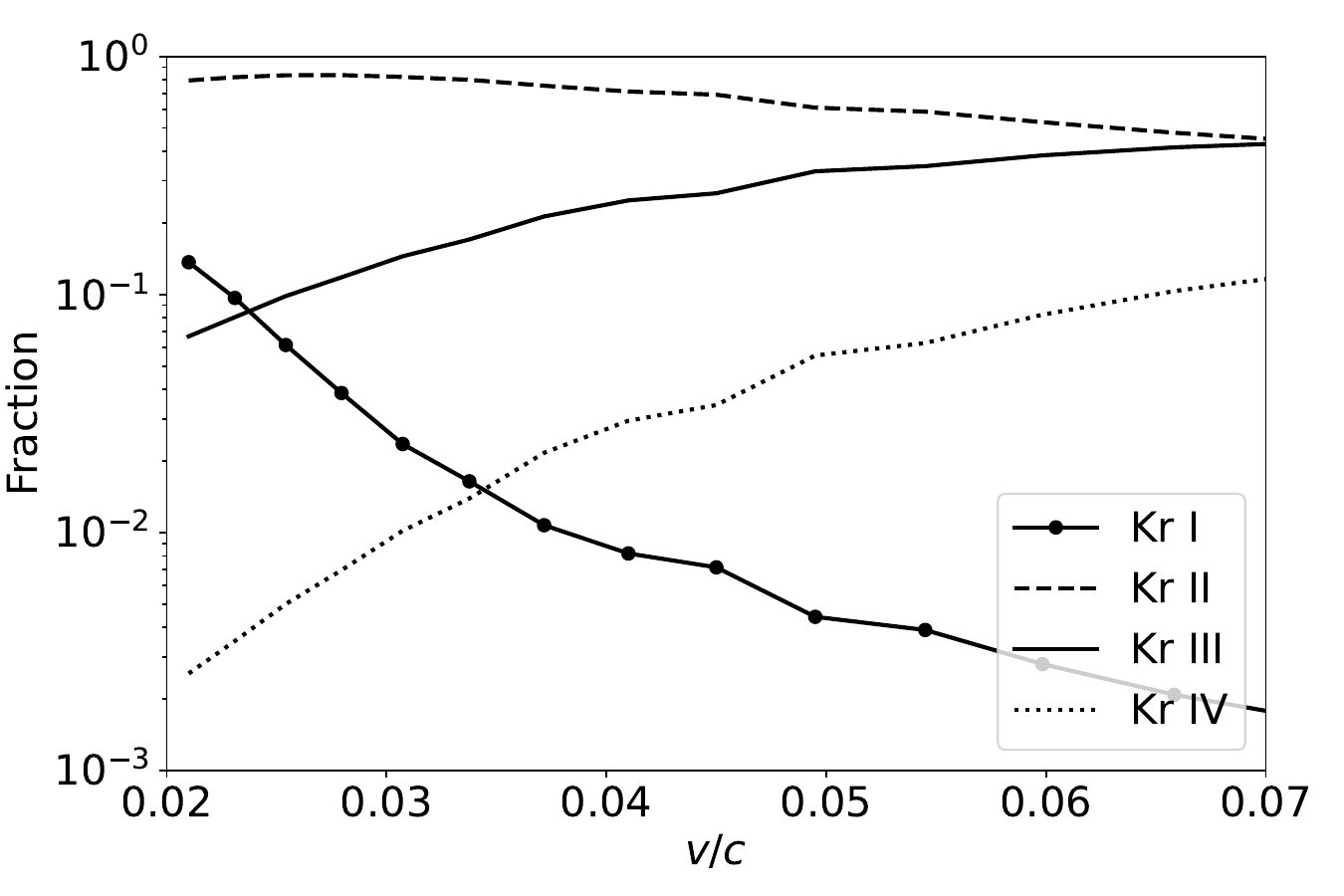}
\includegraphics[width=0.49\linewidth]{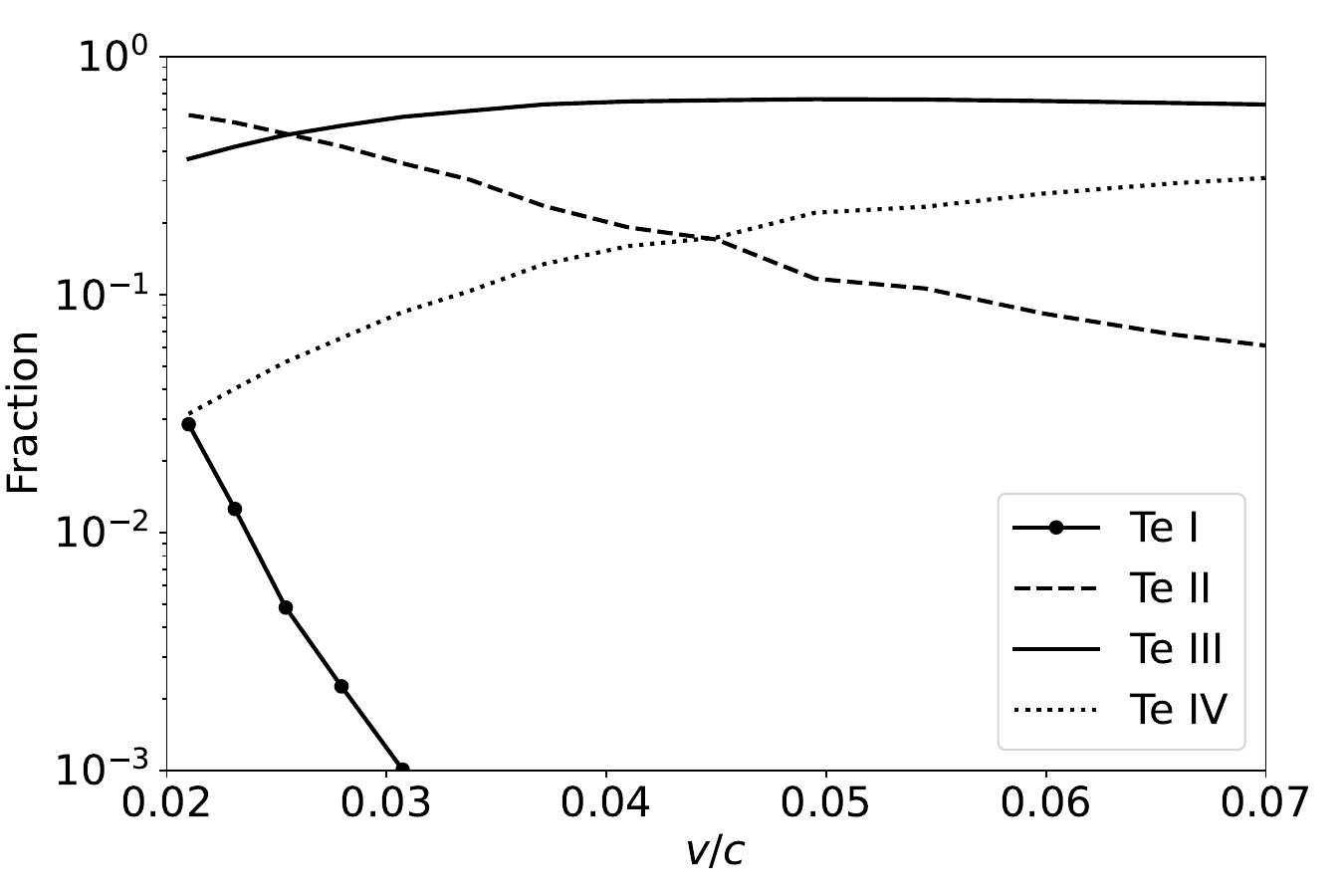}
\caption{Same as Fig. \ref{fig:modelA-low-10d-ionstructure}, at 40d.}
\label{fig:modelA-low-40d-ionstructure}
\end{figure*}

\begin{figure*}[htb]
\includegraphics[width=0.49\linewidth]{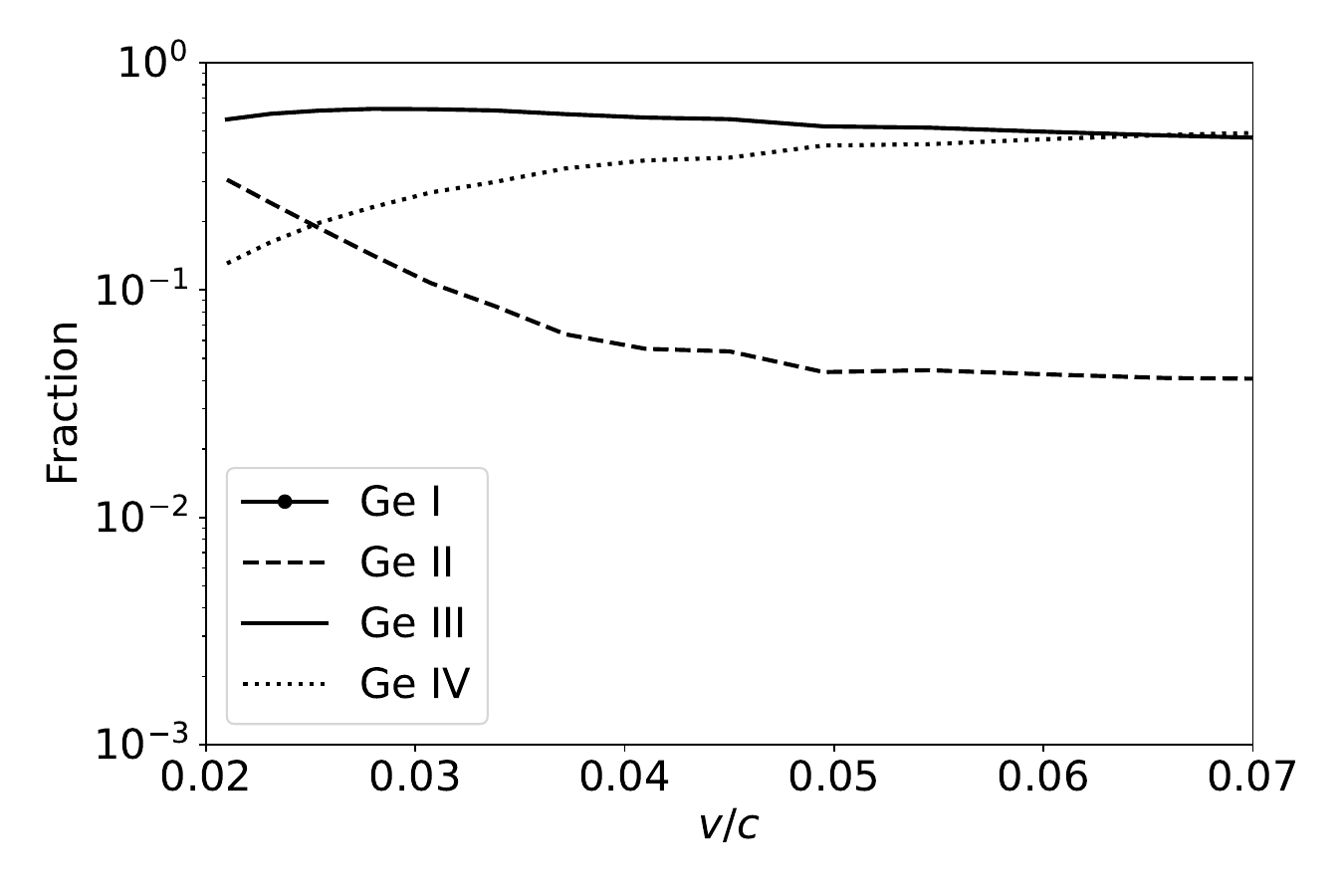}
\includegraphics[width=0.49\linewidth]{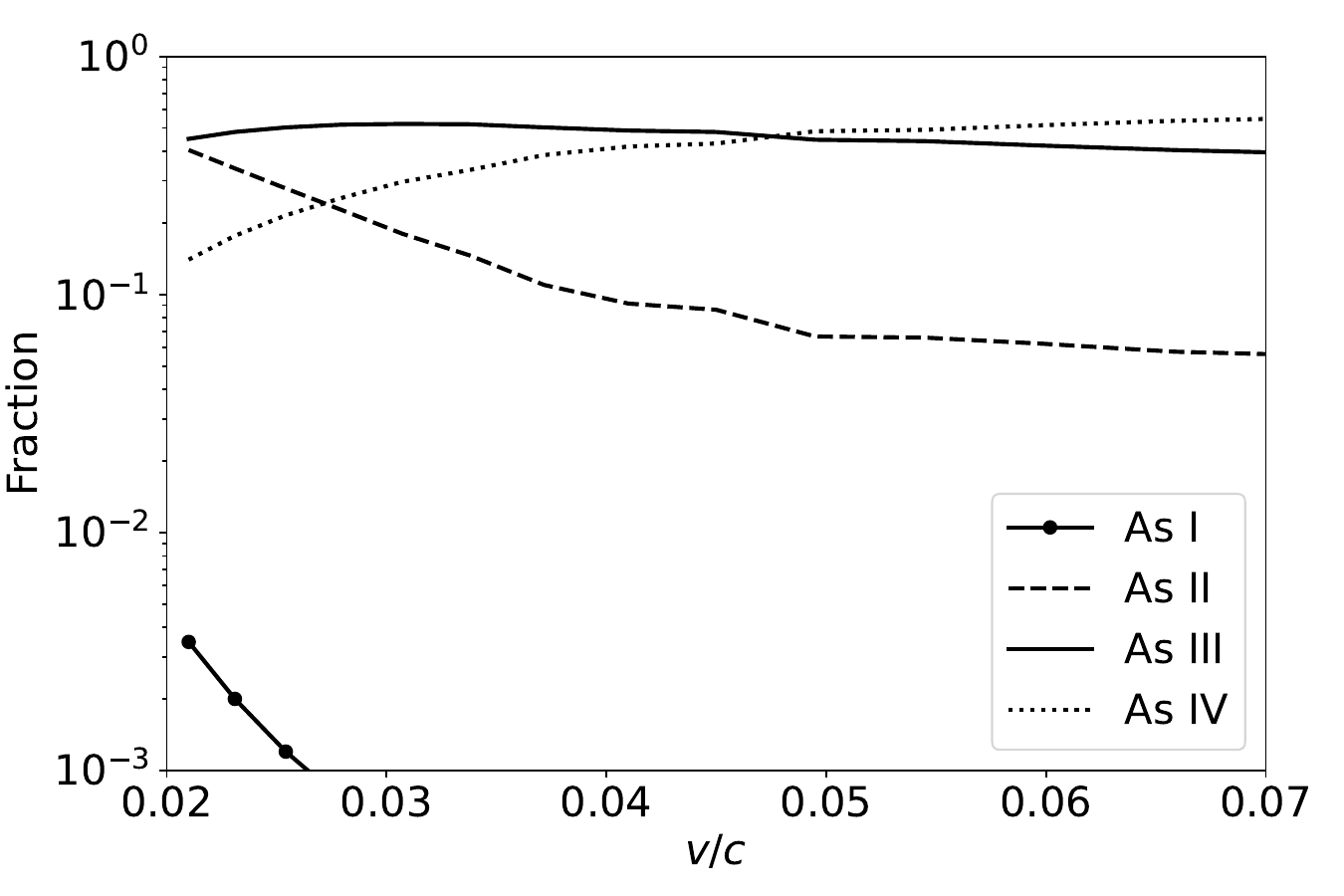}\\
\includegraphics[width=0.49\linewidth]{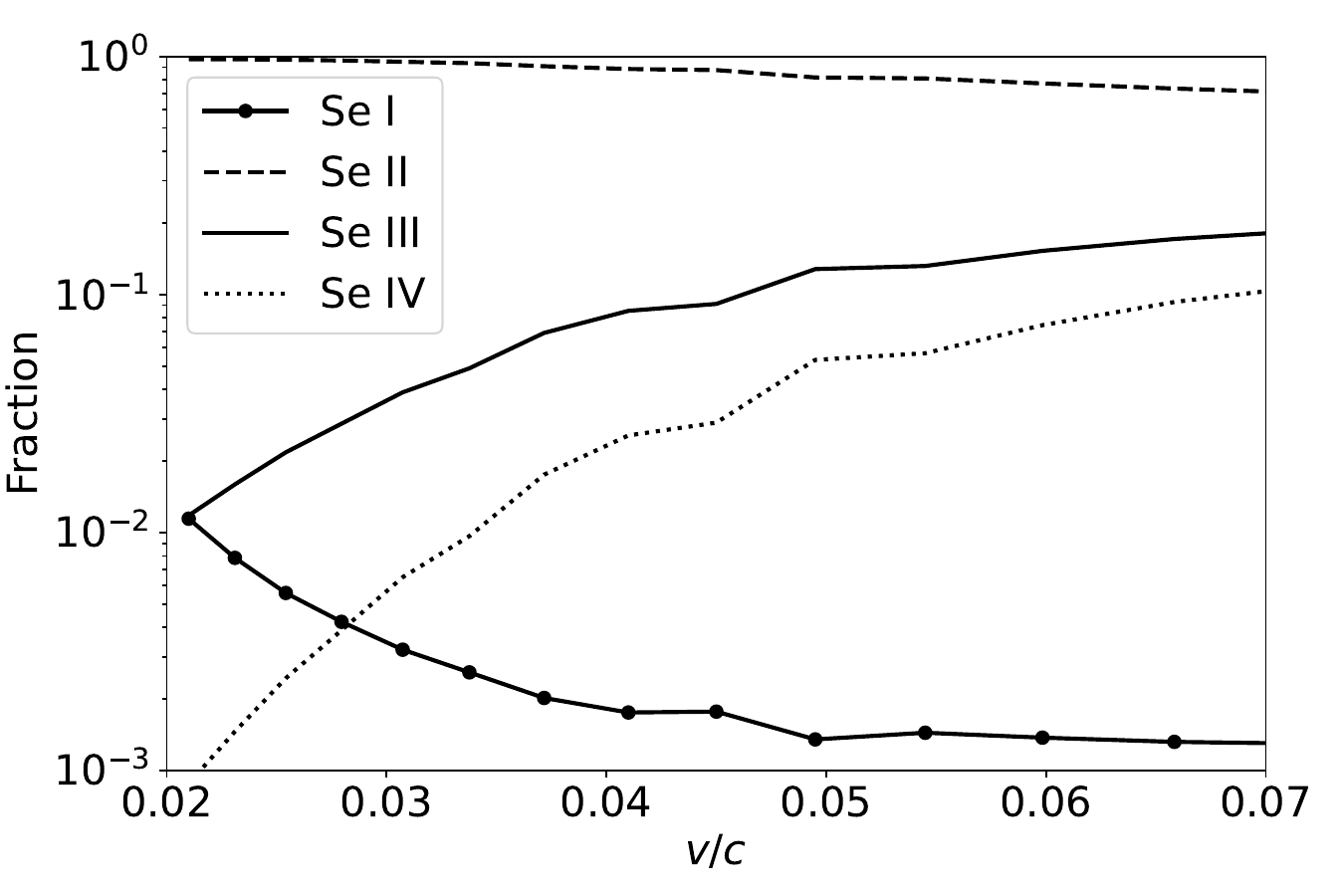}
\includegraphics[width=0.49\linewidth]{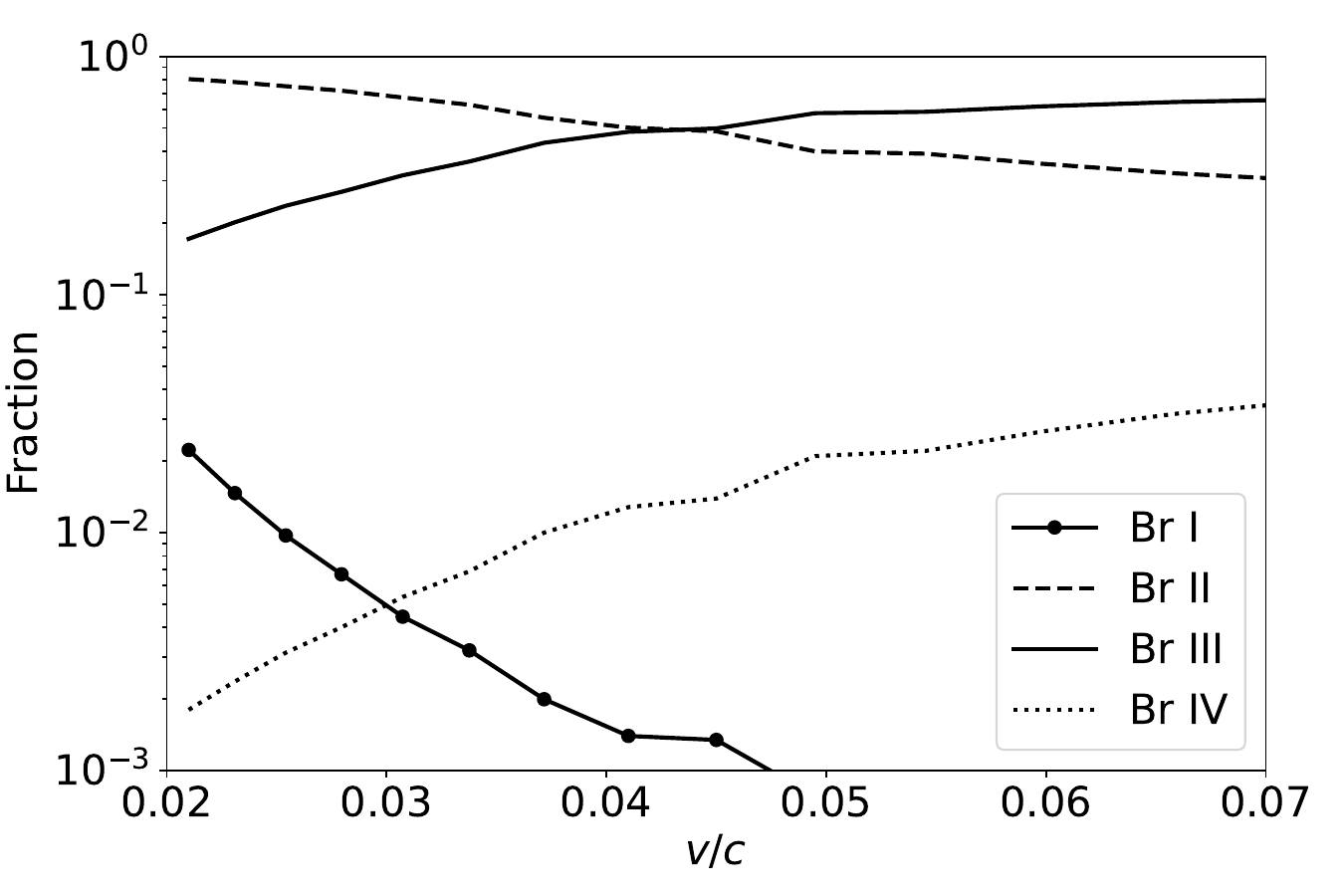}\\
\includegraphics[width=0.49\linewidth]{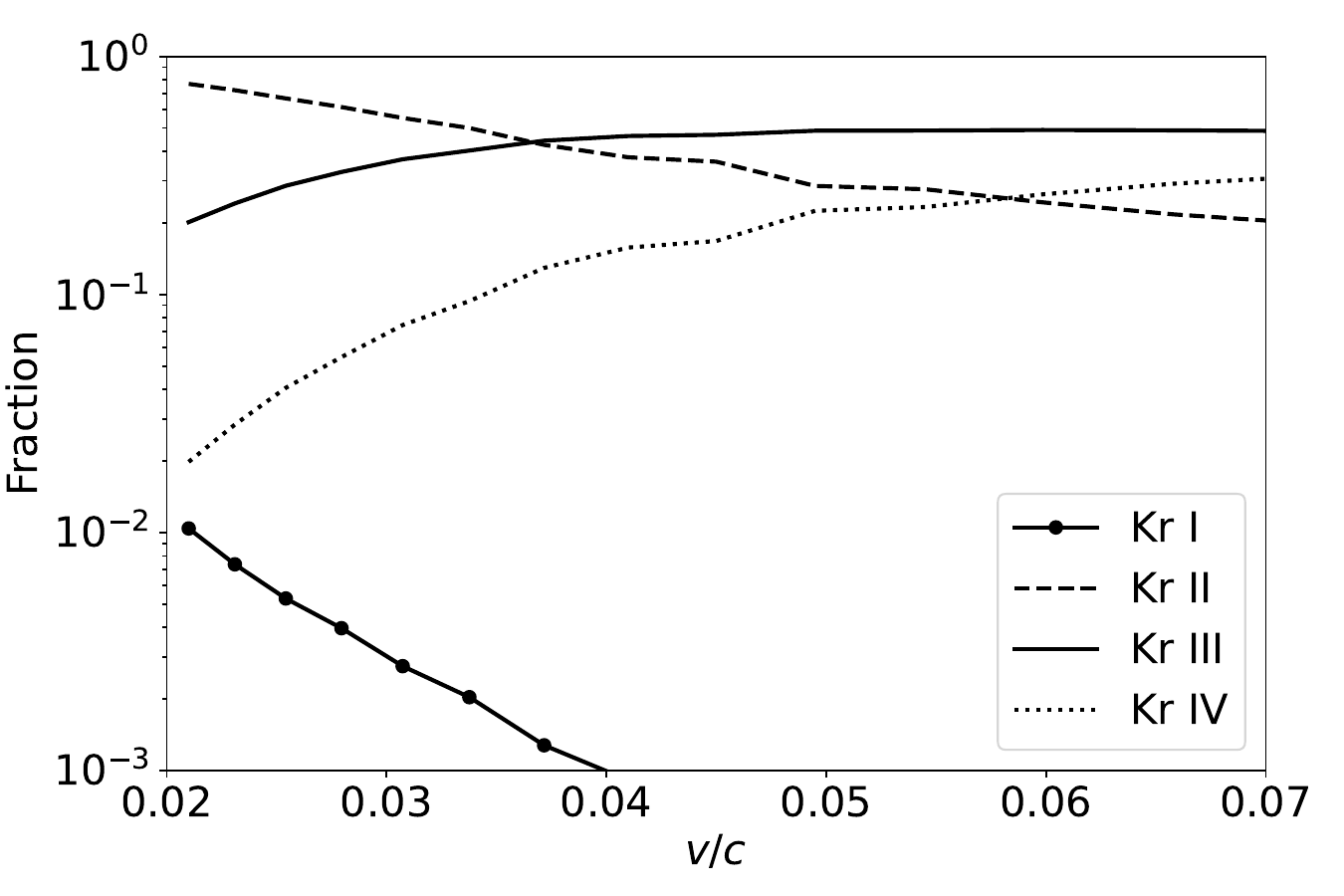}
\includegraphics[width=0.49\linewidth]{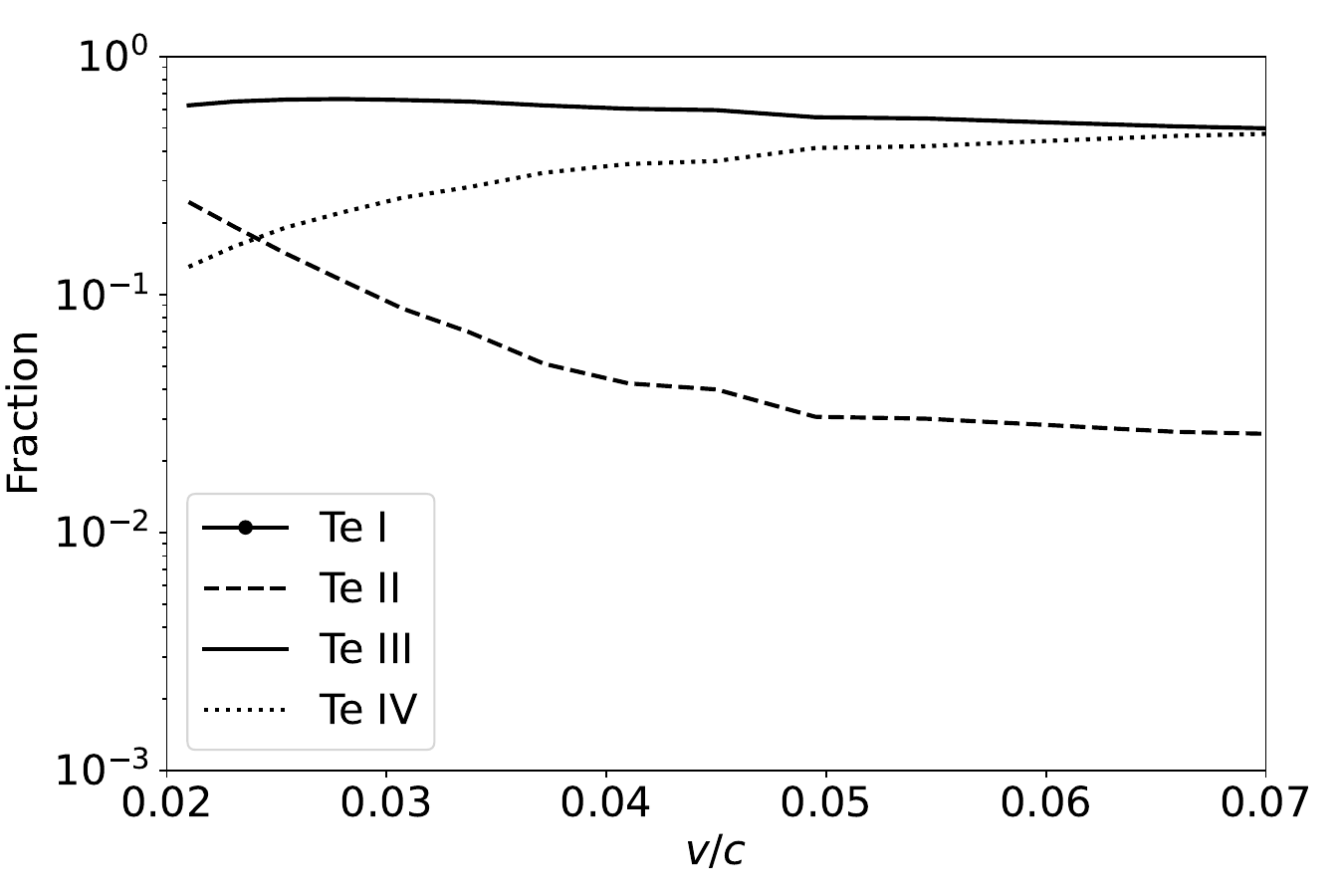}
\caption{Same as Fig. \ref{fig:modelA-low-10d-ionstructure}, at 80d.}
\label{fig:modelA-low-80d-ionstructure}
\end{figure*}

\begin{figure*}[htb]
\includegraphics[width=0.49\linewidth]{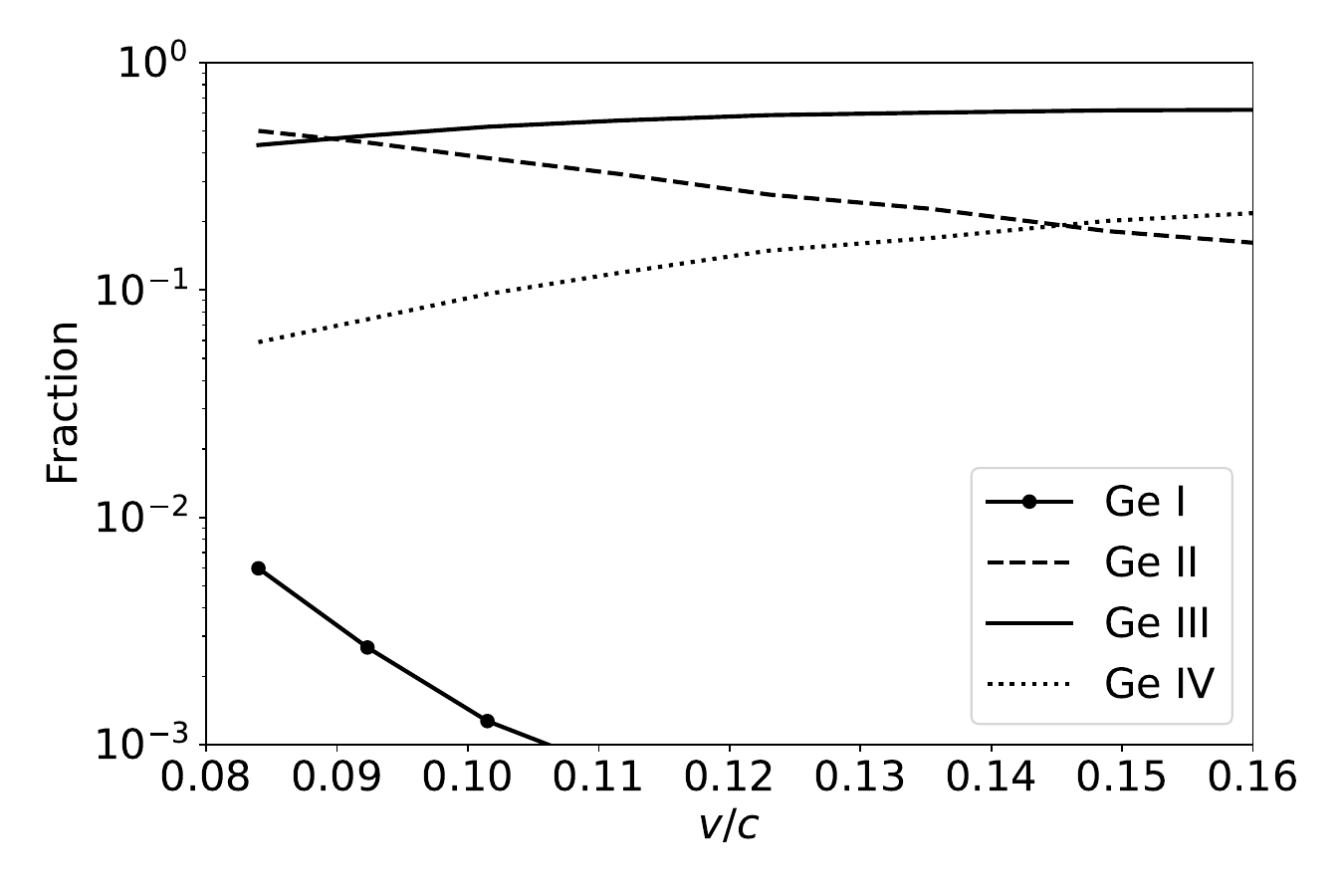}
\includegraphics[width=0.49\linewidth]{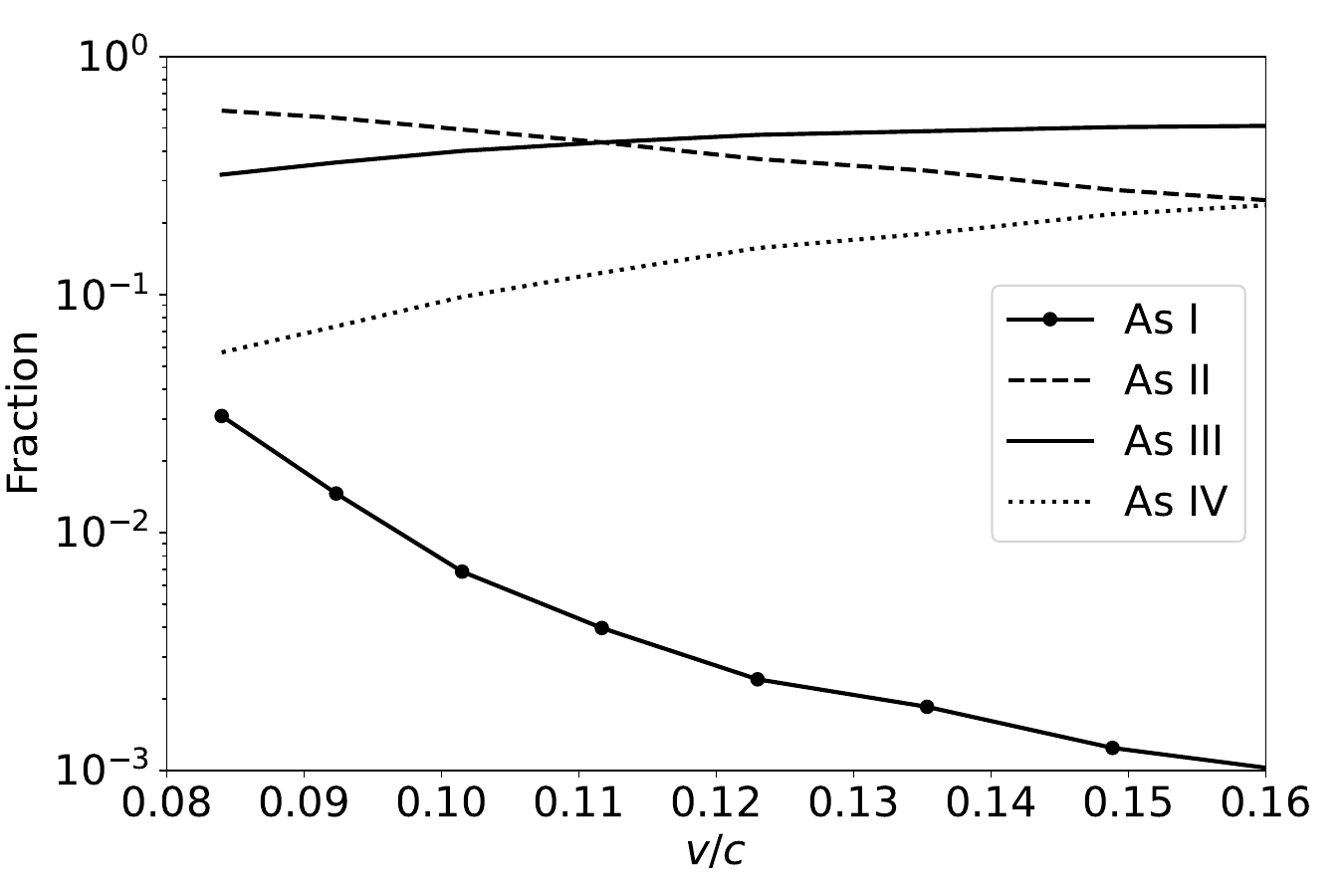}\\
\includegraphics[width=0.49\linewidth]{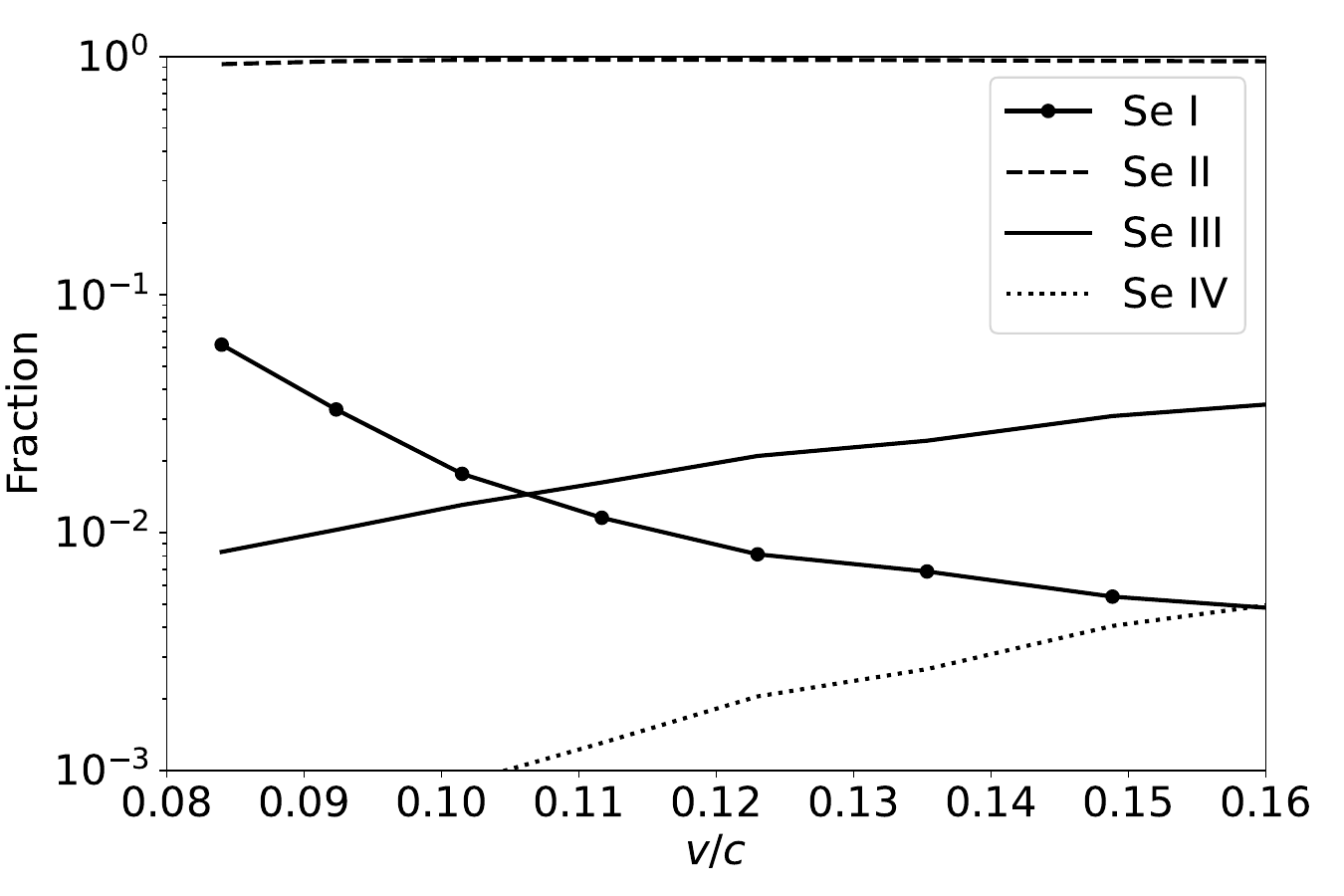}
\includegraphics[width=0.49\linewidth]{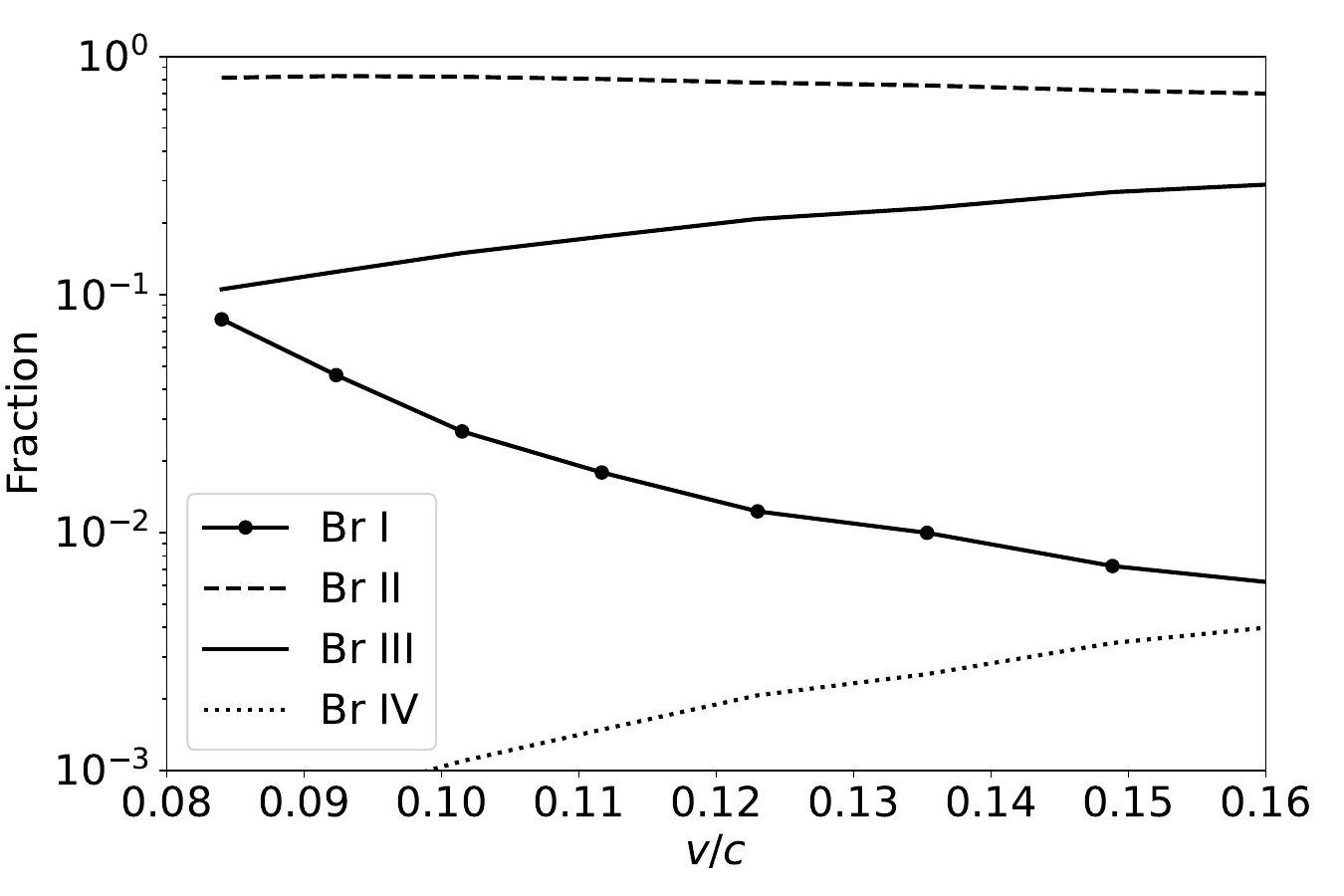}\\
\includegraphics[width=0.49\linewidth]{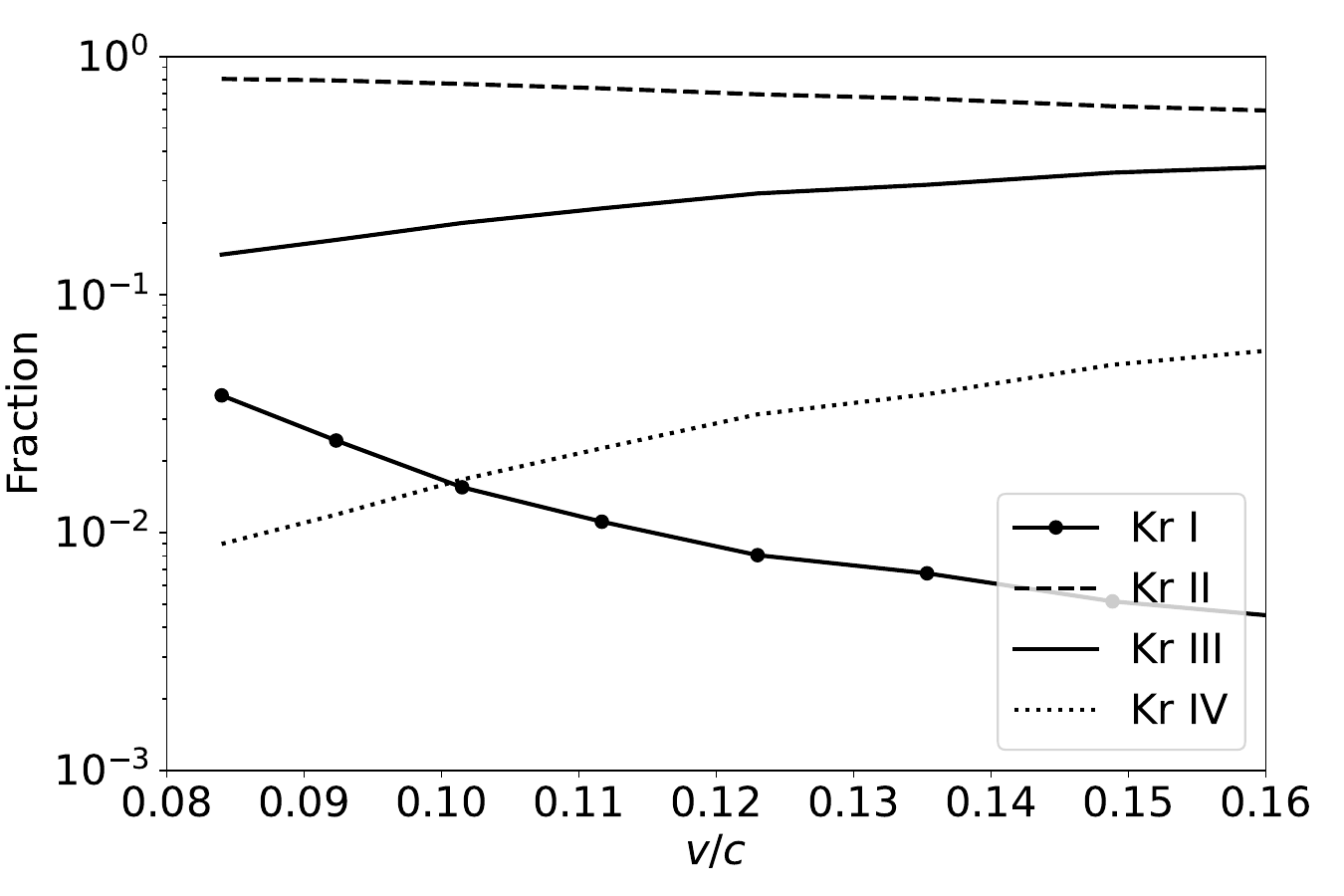}
\includegraphics[width=0.49\linewidth]{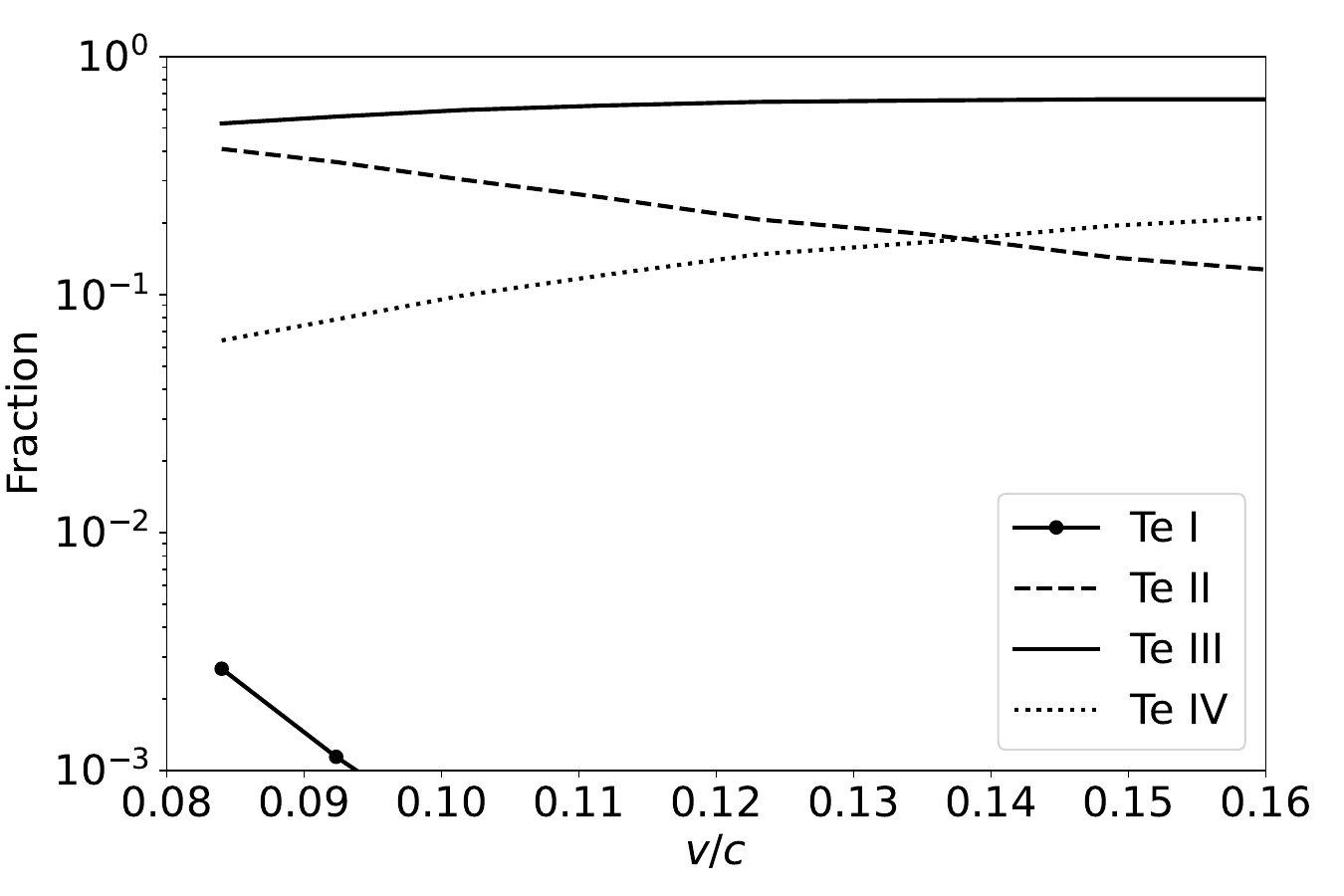}
\caption{Same as Fig. \ref{fig:modelA-low-10d-ionstructure}, but model B at 10d. The inner boundary is here at $v/c=0.08$ and the $\gtrsim 80\%$ radioactive deposition limit at $v/c=0.16$.}
\label{fig:modelB-high-10d-ionstructure}
\end{figure*}

\begin{figure*}[htb]
\includegraphics[width=0.49\linewidth]{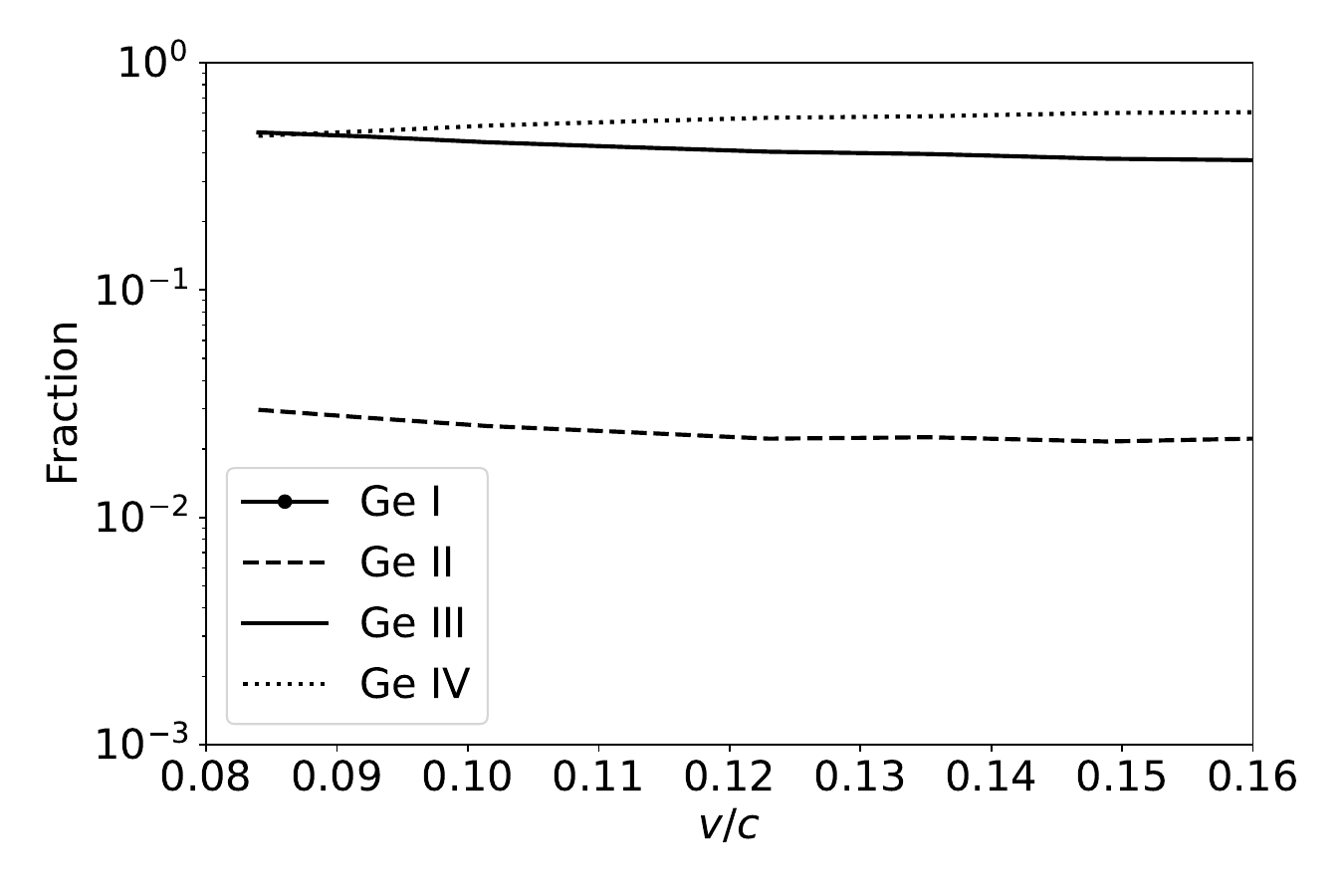}
\includegraphics[width=0.49\linewidth]{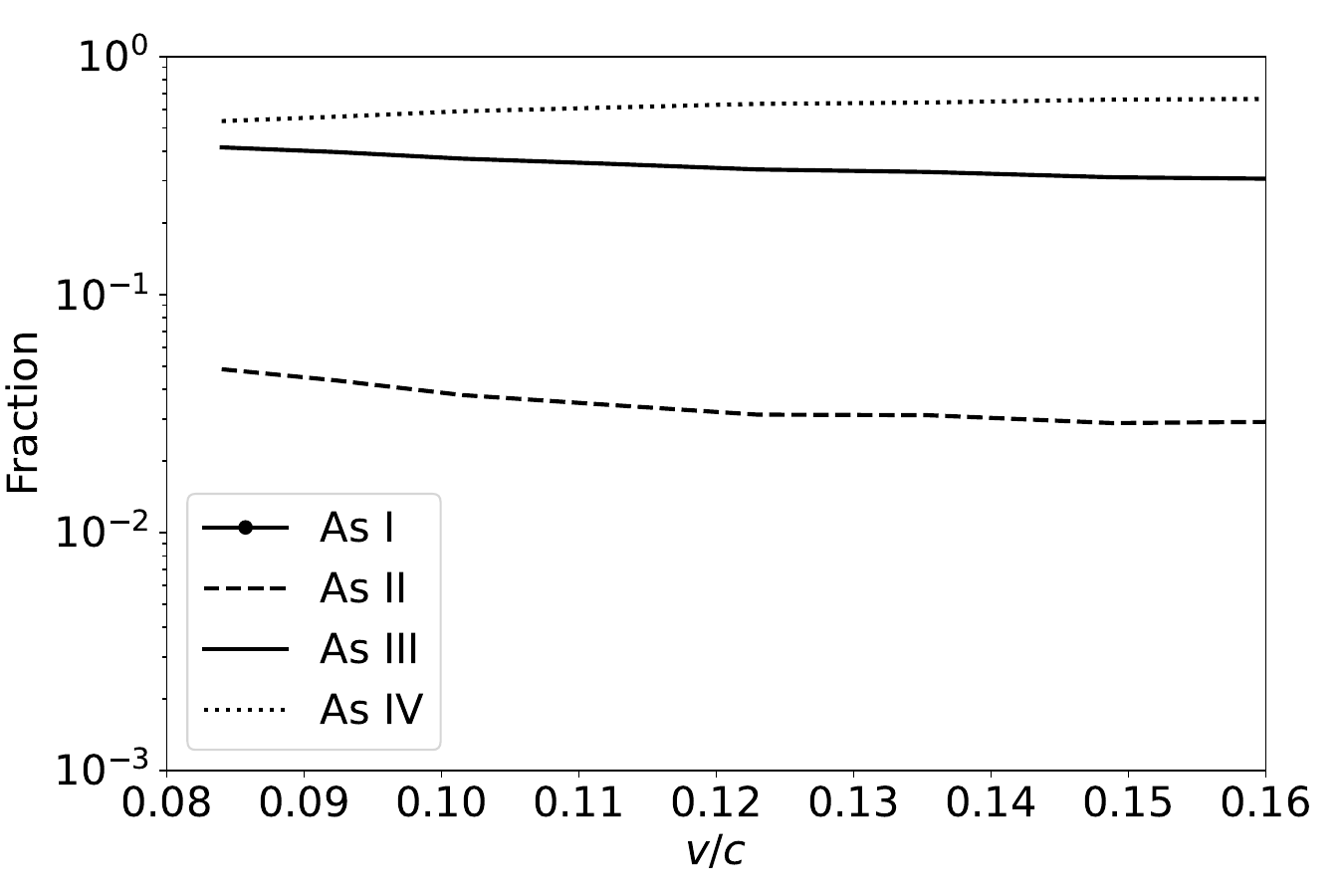}\\
\includegraphics[width=0.49\linewidth]{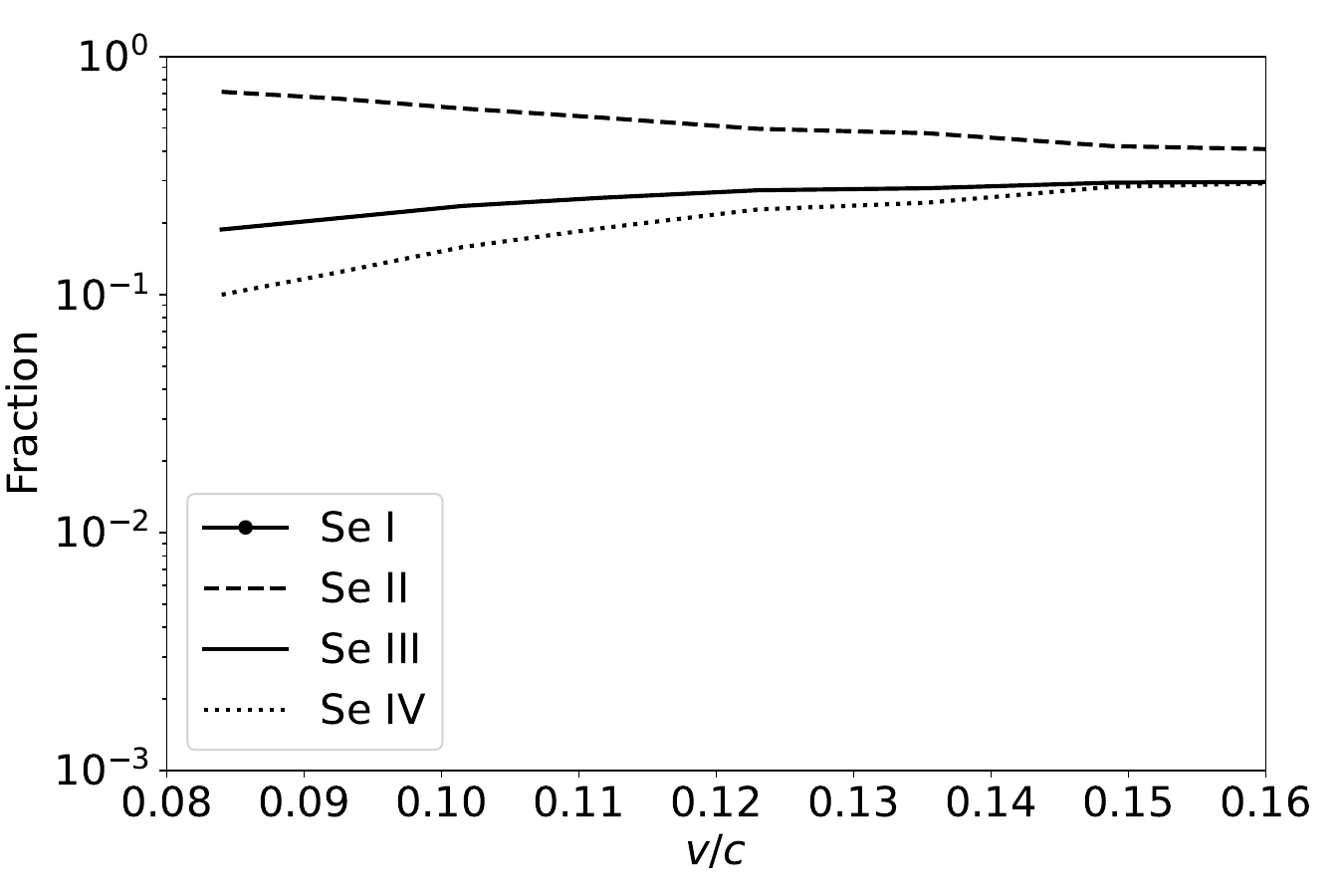}
\includegraphics[width=0.49\linewidth]{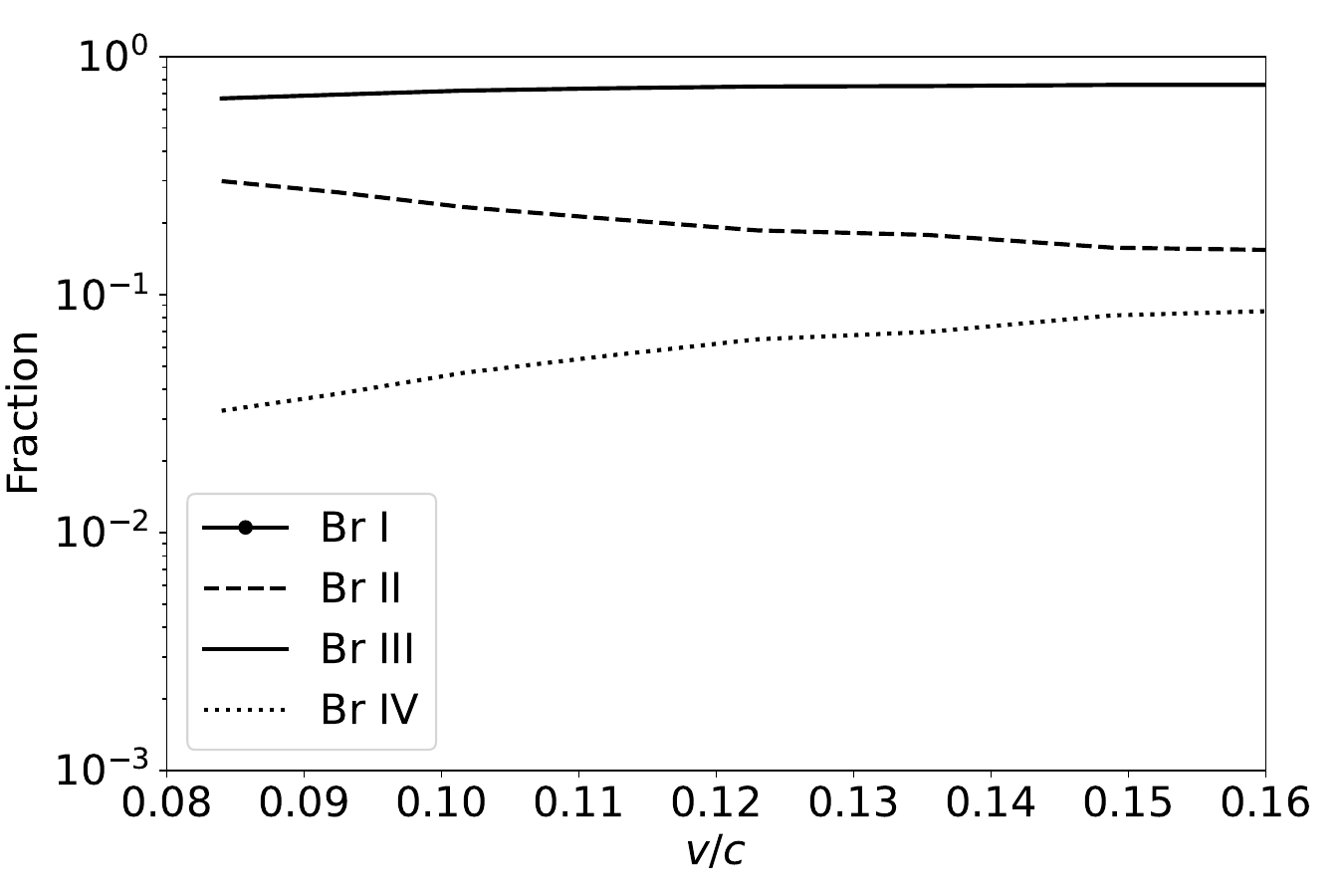}\\
\includegraphics[width=0.49\linewidth]{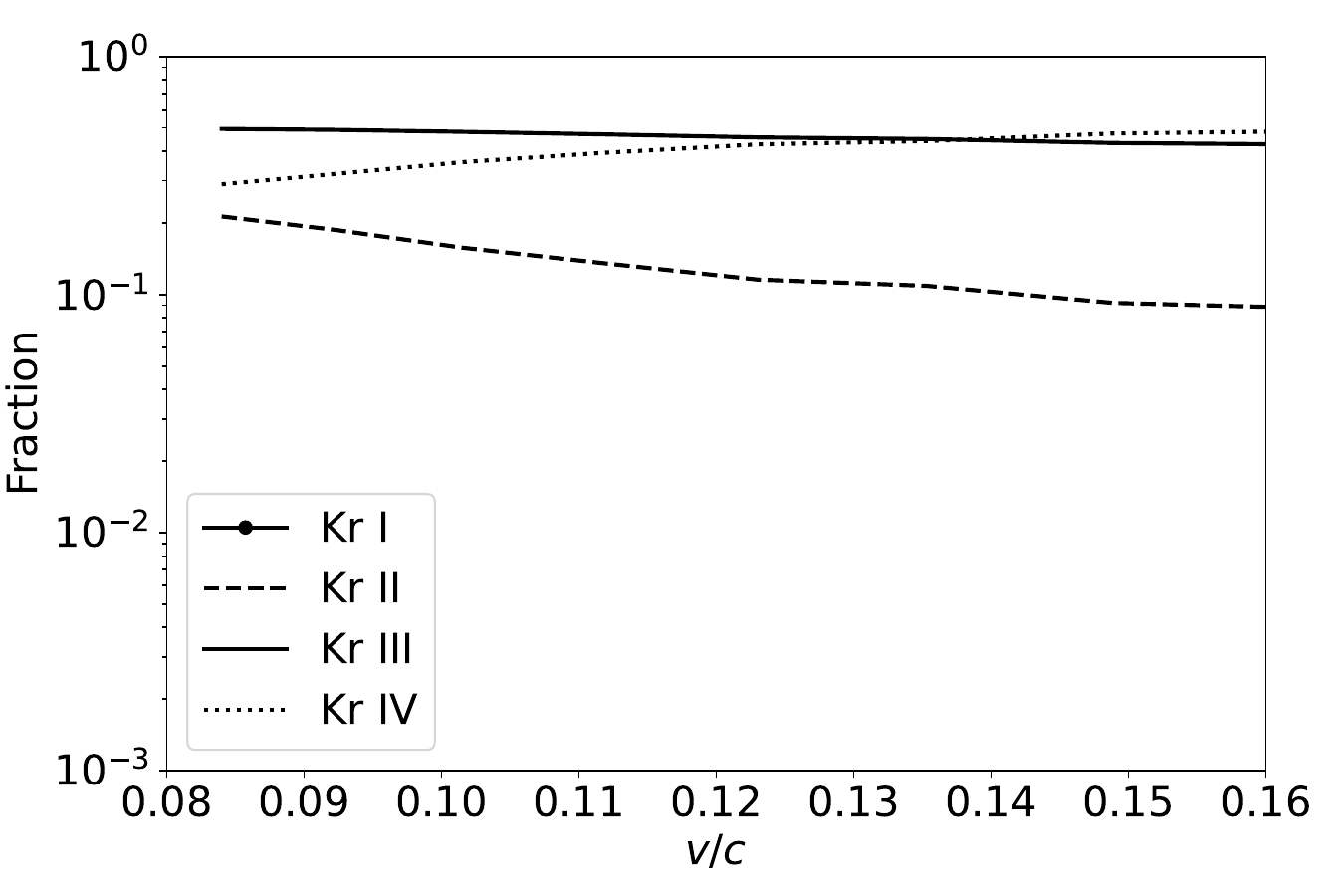}
\includegraphics[width=0.49\linewidth]{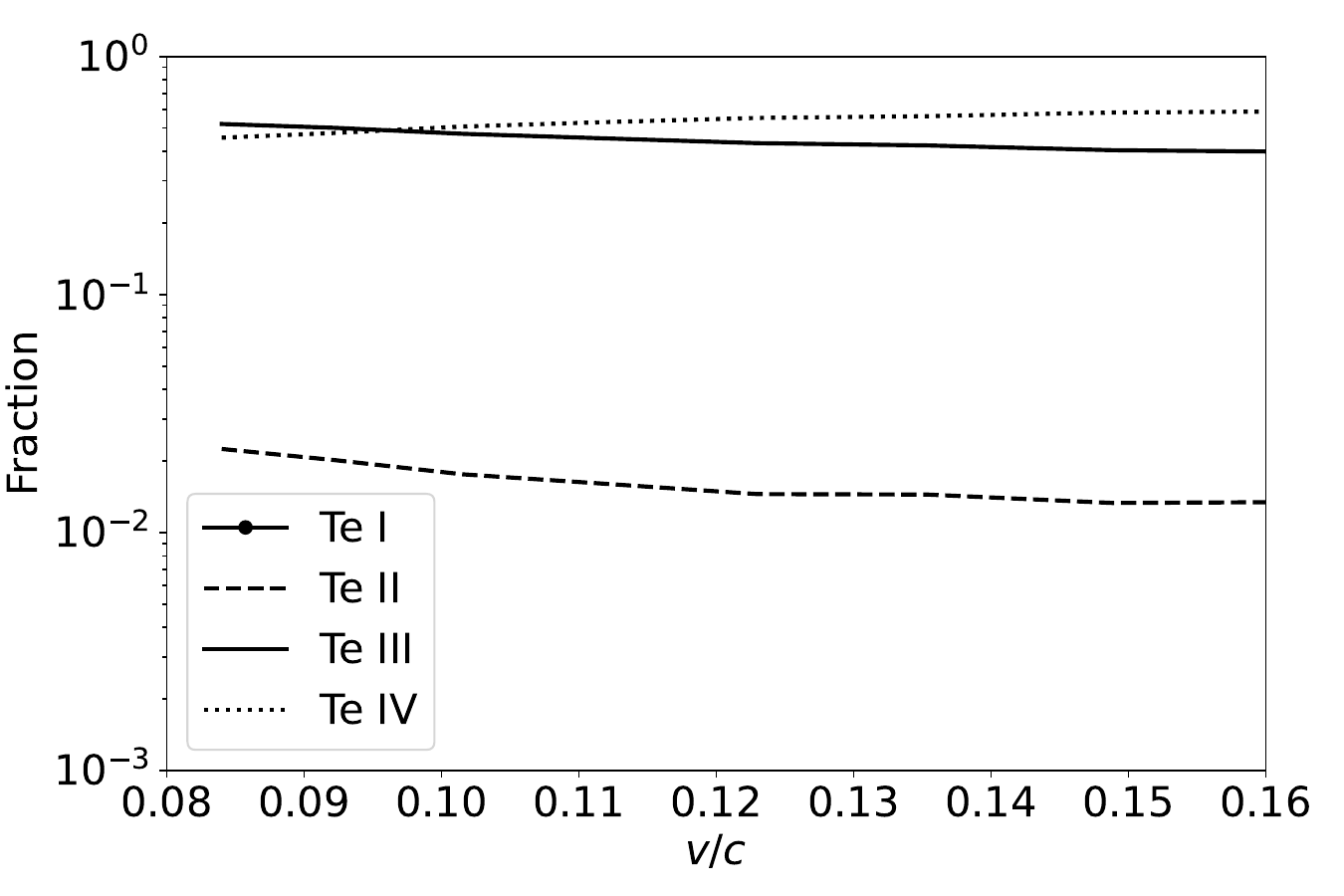}
\caption{Same as Fig. \ref{fig:modelB-high-10d-ionstructure}, at 40d.}
\label{fig:modelB-high-40d-ionstructure}
\end{figure*}

\begin{figure*}[htb]
\includegraphics[width=0.49\linewidth]{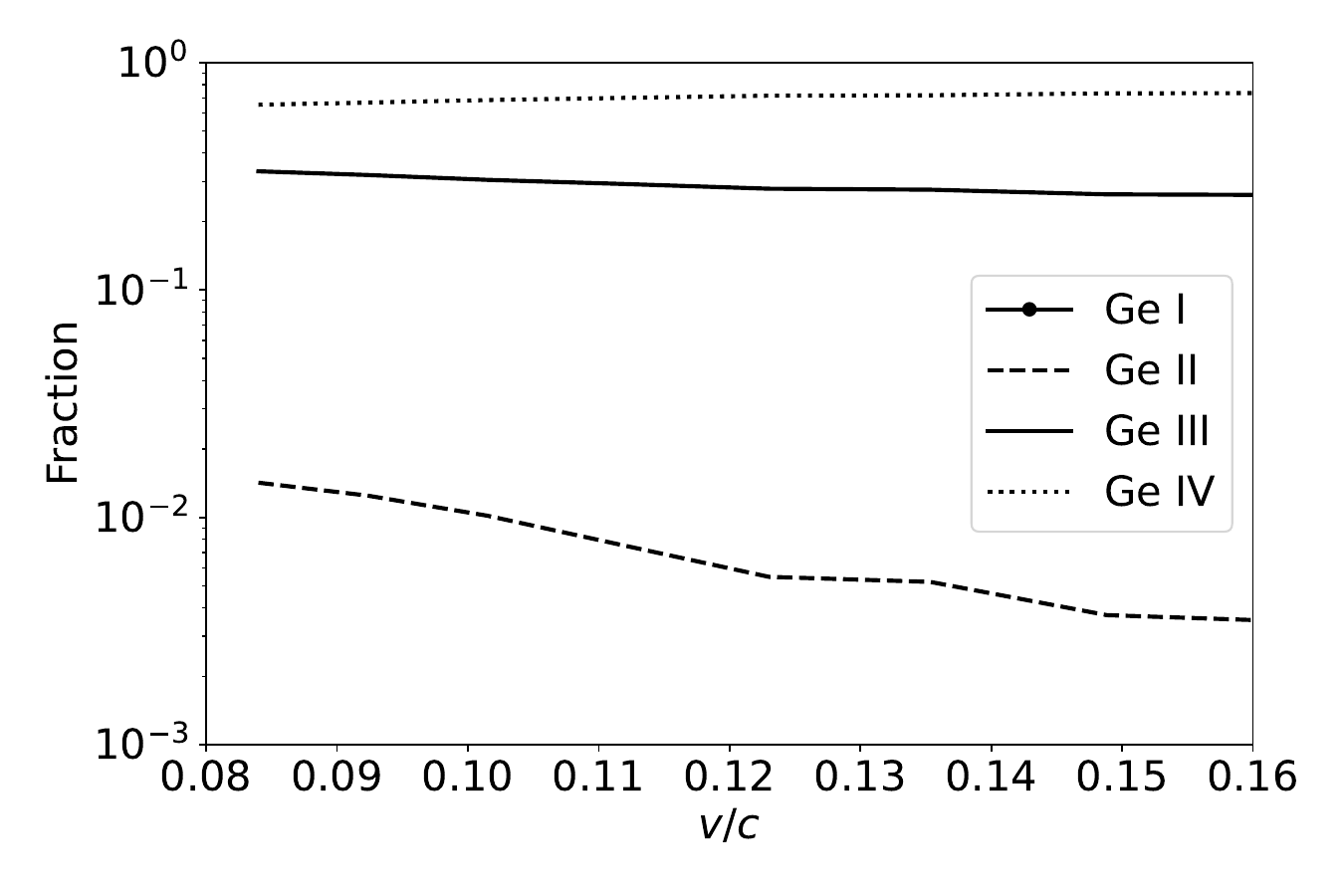}
\includegraphics[width=0.49\linewidth]{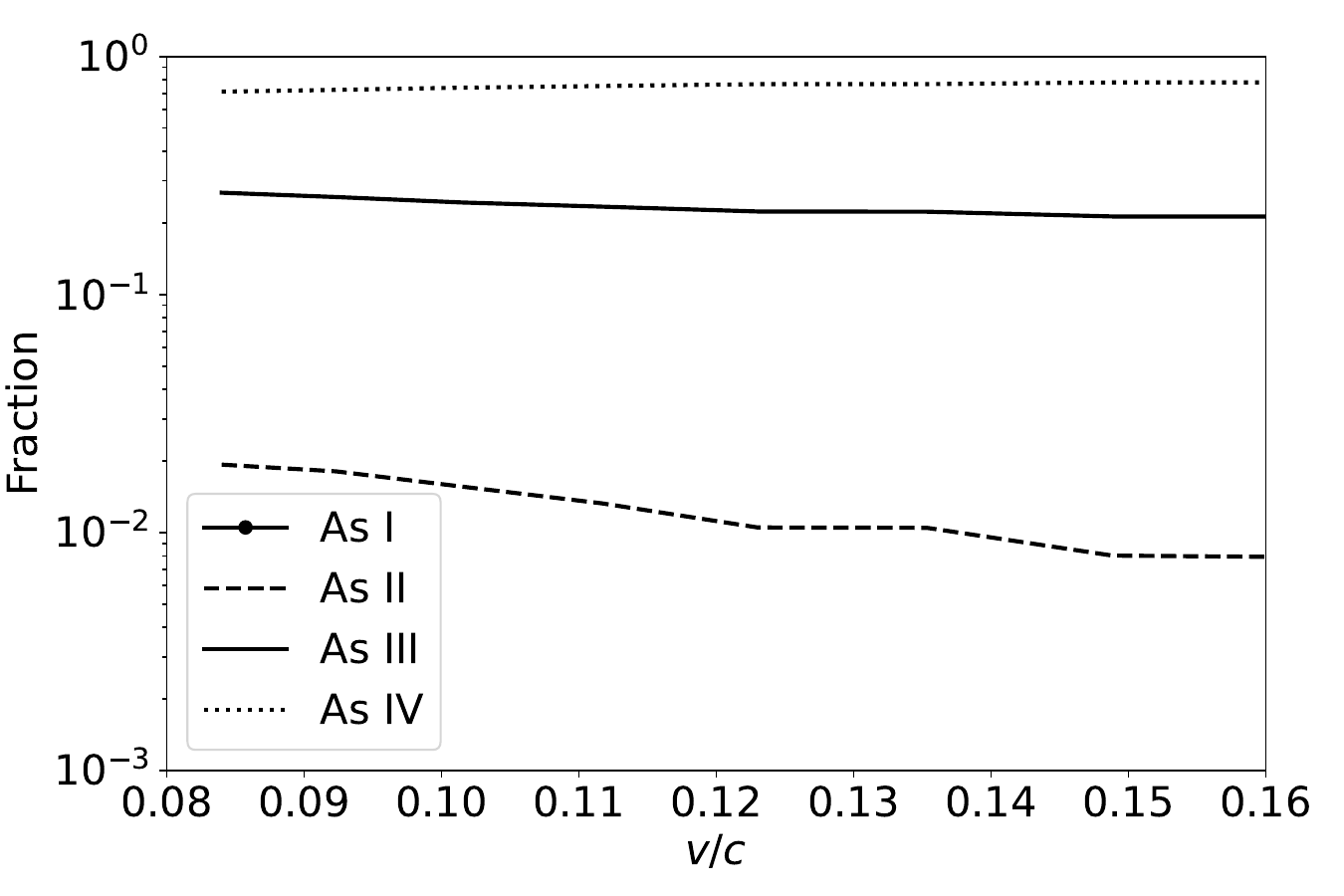}\\
\includegraphics[width=0.49\linewidth]{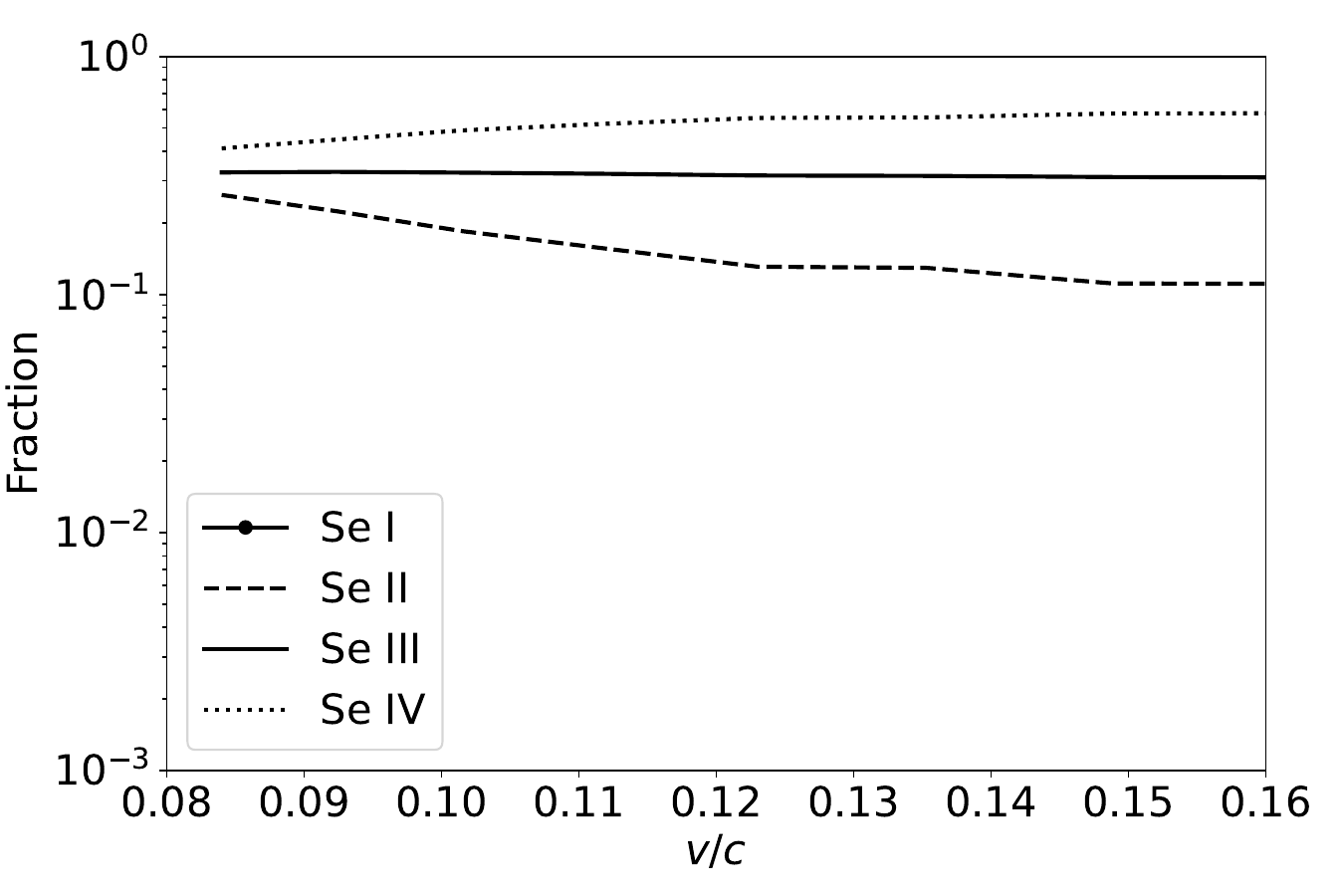}
\includegraphics[width=0.49\linewidth]{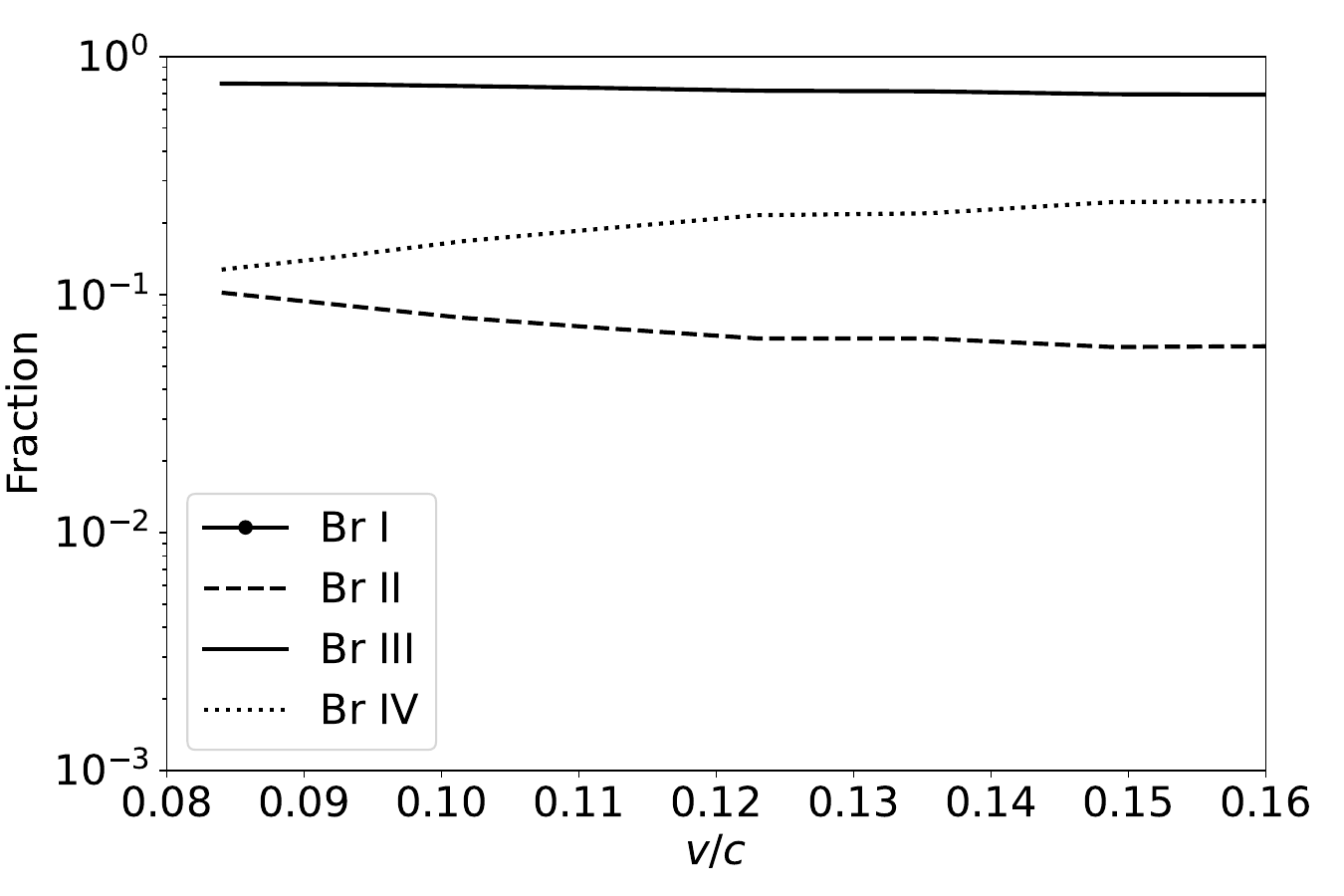}\\
\includegraphics[width=0.49\linewidth]{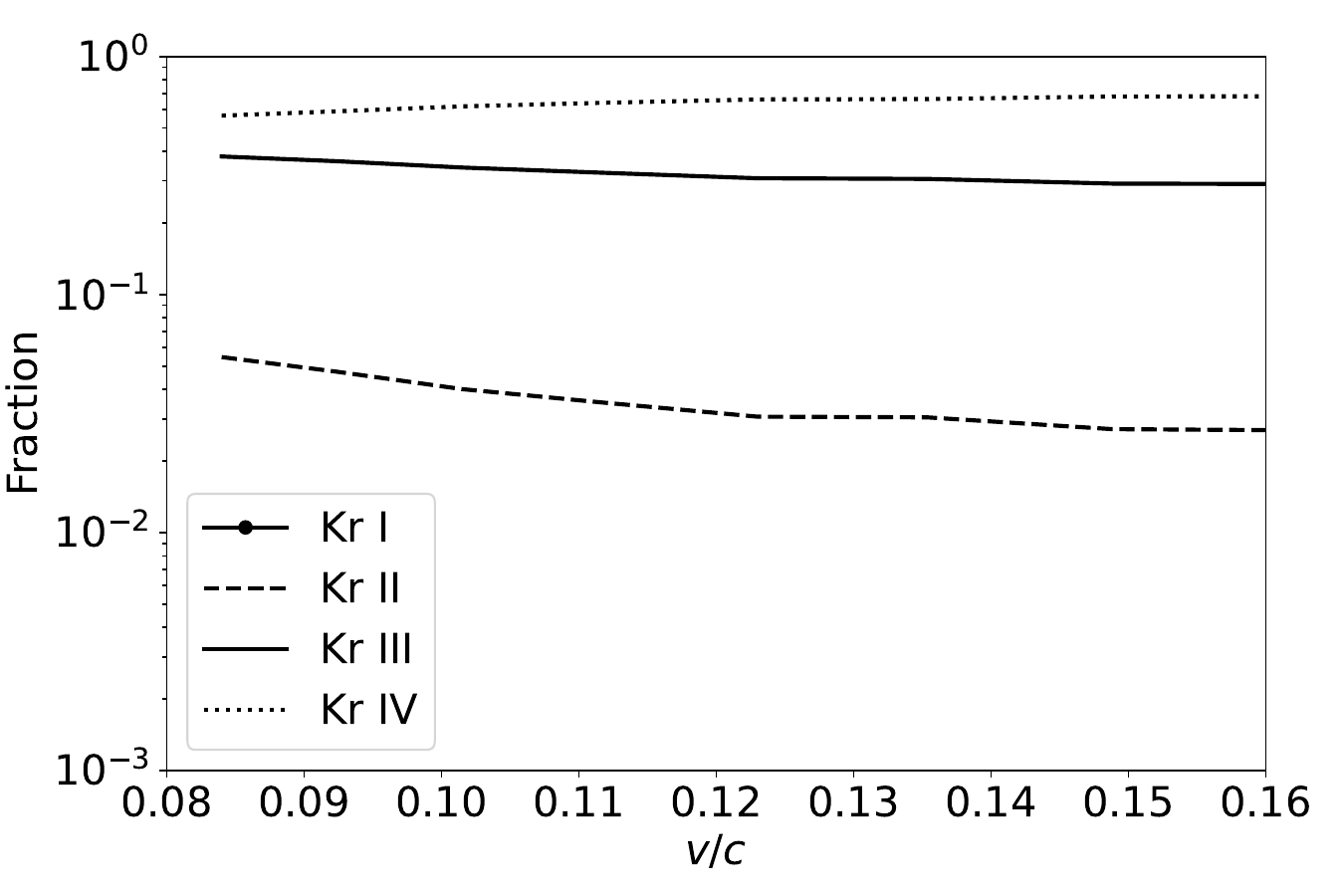}
\includegraphics[width=0.49\linewidth]{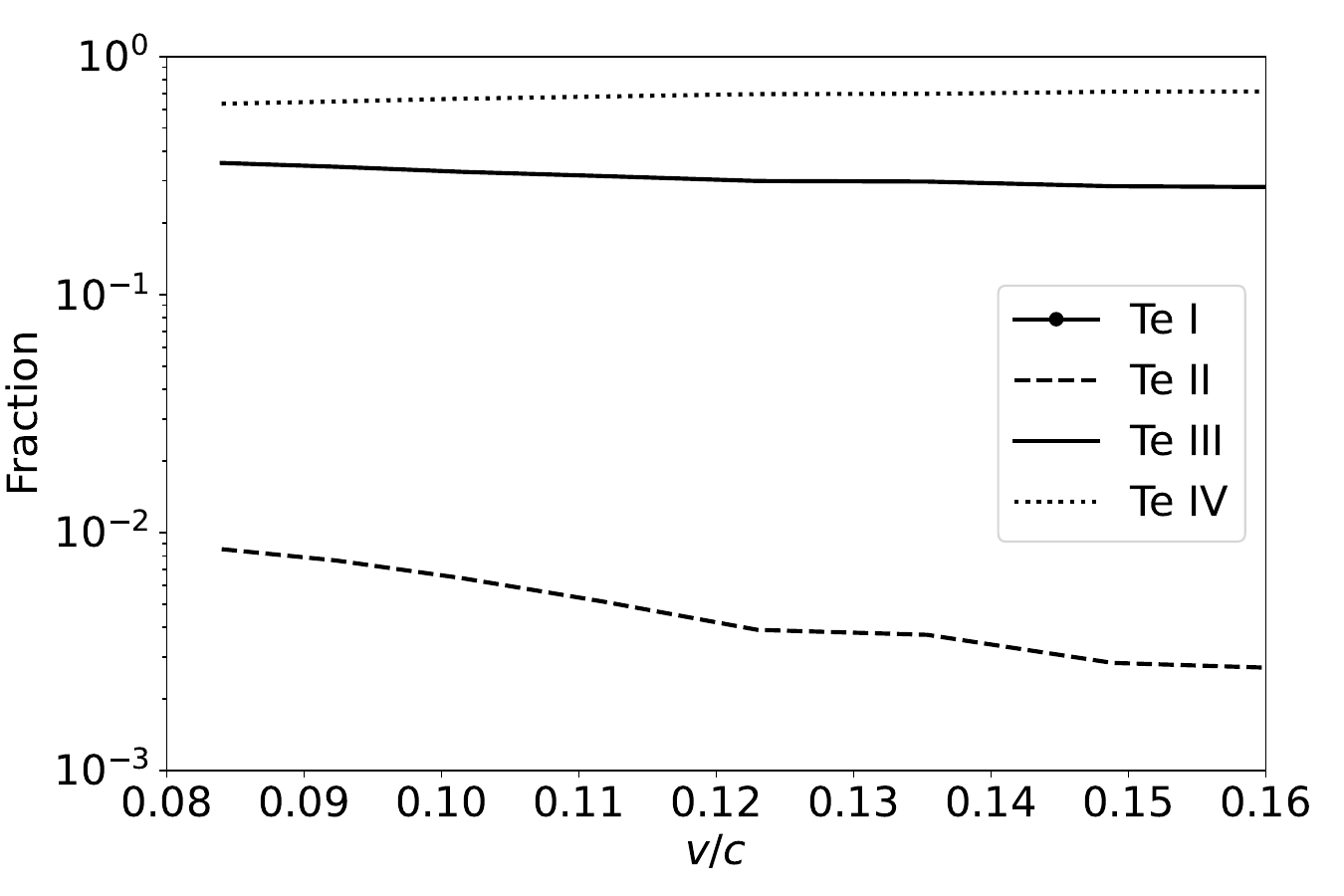}
\caption{Same as Fig. \ref{fig:modelB-high-10d-ionstructure}, at 80d.}
\label{fig:modelB-high-80d-ionstructure}
\end{figure*}

Figures \ref{fig:modelA-low-10d-ionstructure}-\ref{fig:modelB-high-80d-ionstructure} show the detailed ionization structures in the line-forming regions in the models, for elements with distinct lines.

\acknowledgements{We thank the referee for useful comments and suggestions for the manuscript. We acknowledge useful discussions with O. Just, S. Goriely and K. Hotokezaka. We acknowledge funding from the European Research Council
(ERC) under the European Union’s Horizon 2020 Research and
Innovation Program (ERC Starting Grant 803189 – SUPERSPEC). NCS acknowledges support from NSF award AST-2307116. JG thanks the Swedish Research Council for the
individual starting grant with contract no. 2020-05467. The com-
putations were enabled by resources provided by the National
Academic Infrastructure for Supercomputing in Sweden (NAISS),
and the Swedish National Infrastructure for Computing (SNIC),
at the Parallelldatorcentrum (PDC) Center for High Performance
Computing, Royal Institute of Technology (KTH), partially funded
by the Swedish Research Council through grant agreements no. 2022-
06725 and no. 2018-05973. We acknowledge also computing funding
by the Swedish Research Council (grant 2018-03799).}

\section*{Data availability}
All computed model spectra are available upon email request to the author, and will also be made available at the Zenodo archive.

\bibliography{./references.bib}{}

@ARTICLE{Gillanders2023_AT2023vfi,
       author = {{Gillanders}, James H. and {Troja}, Eleonora and {Fryer}, Chris L. and {Ristic}, Marko and {O'Connor}, Brendan and {Fontes}, Christopher J. and {Yang}, Yu-Han and {Domoto}, Nanae and {Rahmouni}, Salma and {Tanaka}, Masaomi and {Fox}, Ori D. and {Dichiara}, Simone},
        title = "{Heavy element nucleosynthesis associated with a gamma-ray burst}",
      journal = {arXiv e-prints},
         year = 2023,
        month = aug,
          eid = {arXiv:2308.00633},
        pages = {arXiv:2308.00633},
          doi = {10.48550/arXiv.2308.00633},
archivePrefix = {arXiv},
       eprint = {2308.00633},
 primaryClass = {astro-ph.HE},
       adsurl = {https://ui.adsabs.harvard.edu/abs/2023arXiv230800633G}
}

@article{Fujibayashi:2020qda,
    author = "Fujibayashi, Sho and Shibata, Masaru and Wanajo, Shinya and Kiuchi, Kenta and Kyutoku, Koutarou and Sekiguchi, Yuichiro",
    title = "{Mass ejection from disks surrounding a low-mass black hole: Viscous neutrino-radiation hydrodynamics simulation in full general relativity}",
    eprint = "2001.04467",
    archivePrefix = "arXiv",
    primaryClass = "astro-ph.HE",
    doi = "10.1103/PhysRevD.101.083029",
    journal = "Phys. Rev. D",
    volume = "101",
    number = "8",
    pages = "083029",
    year = "2020"
}

@ARTICLE{Sterling2007,
       author = {{Sterling}, N.~C. and {Dinerstein}, Harriet L. and {Kallman}, T.~R.},
        title = "{The Abundances of Light Neutron-Capture Elements in Planetary Nebulae. I. Photoionization Modeling and Ionization Corrections}",
      journal = {\apjs},
         year = 2007,
        month = mar,
       volume = {169},
       number = {1},
        pages = {37-61},
          doi = {10.1086/511031},
archivePrefix = {arXiv},
       eprint = {0706.1074},
 primaryClass = {astro-ph},
       adsurl = {https://ui.adsabs.harvard.edu/abs/2007ApJS..169...37S}
}

@ARTICLE{Pognan2026,
       author = {{Pognan}, Quentin and {Kawaguchi}, Kyohei and {Wanajo}, Shinya and {Fujibayashi}, Sho and {Jerkstrand}, Anders and {Grumer}, Jon},
        title = "{Lanthanide Impact on the Infra-Red Spectra of Nebular Phase Kilonovae}",
      journal = {\mnras},
         year = 2026,
        month = mar,
          doi = {10.1093/mnras/stag441},
archivePrefix = {arXiv},
       eprint = {2510.12413},
 primaryClass = {astro-ph.HE},
       adsurl = {https://ui.adsabs.harvard.edu/abs/2026MNRAS.tmp..416P}
}

@ARTICLE{Sneppen2024,
       author = {{Sneppen}, Albert and {Damgaard}, Rasmus and {Watson}, Darach and {Collins}, Christine E. and {Shingles}, Luke and {Sim}, Stuart A.},
        title = "{Helium features are inconsistent with the spectral evolution of the kilonova AT2017gfo}",
      journal = {\aap},
         year = 2024,
        month = dec,
       volume = {692},
          eid = {A134},
        pages = {A134},
          doi = {10.1051/0004-6361/202451450},
archivePrefix = {arXiv},
       eprint = {2407.12907},
 primaryClass = {astro-ph.HE},
       adsurl = {https://ui.adsabs.harvard.edu/abs/2024A&A...692A.134S}
}

@ARTICLE{Brethauer2026,
       author = {{Brethauer}, D. and {Kasen}, D. and {Margutti}, R. and {Chornock}, R.},
        title = "{Nonthermal Ionization of Kilonova Ejecta: Observable Impacts}",
      journal = {\apj},
         year = 2026,
        month = jan,
       volume = {996},
       number = {1},
          eid = {64},
        pages = {64},
          doi = {10.3847/1538-4357/ae1b8d},
archivePrefix = {arXiv},
       eprint = {2508.18364},
 primaryClass = {astro-ph.HE},
       adsurl = {https://ui.adsabs.harvard.edu/abs/2026ApJ...996...64B}
}

@ARTICLE{Sugar1995,
       author = {{Sugar}, Jack and {Musgrove}, Arlene},
        title = "{Energy Levels of Zinc, Zn I through Zn XXX}",
      journal = {Journal of Physical and Chemical Reference Data},
         year = 1995,
        month = nov,
       volume = {24},
       number = {6},
        pages = {1803-1872},
          doi = {10.1063/1.555971},
       adsurl = {https://ui.adsabs.harvard.edu/abs/1995JPCRD..24.1803S}
}

@BOOK{Sugar1985,
       author = {{Sugar}, Jack and {Corliss}, Charles},
        title = "{Atomic energy levels of the iron-period elements: Potassium through Nickel}",
         year = 1985,
       adsurl = {https://ui.adsabs.harvard.edu/abs/1985aeli.book.....S}
}

@online{connorb,
title = {Connor Ballance: Belfast atomic R-matrix codes},
year = {2026},
url = {https://connorb.freeshell.org},
urldate = {2026-03-08},
author = {Connor Ballance},
}

@article{Berrington1995,
title = {RMATRX1: Belfast atomic R-matrix codes},
journal = {Computer Physics Communications},
volume = {92},
number = {2},
pages = {290-420},
year = {1995},
issn = {0010-4655},
doi = {https://doi.org/10.1016/0010-4655(95)00123-8},
url = {https://www.sciencedirect.com/science/article/pii/0010465595001238},
author = {Keith A. Berrington and Werner B. Eissner and Patrick H. Norrington}
}

@ARTICLE{Badnell1986,
       author = {{Badnell}, N.~R.},
        title = "{Dielectronic recombination of Fe$^{22+}$ and Fe$^{21+}$}",
      journal = {Journal of Physics B Atomic Molecular Physics},
         year = 1986,
        month = nov,
       volume = {19},
       number = {22},
        pages = {3827-3835},
          doi = {10.1088/0022-3700/19/22/023},
       adsurl = {https://ui.adsabs.harvard.edu/abs/1986JPhB...19.3827B}
}

@ARTICLE{Badnell1997,
       author = {{Badnell}, N.~R.},
        title = "{On the effects of the two-body non-fine-structure operators of the Breit - Pauli Hamiltonian}",
      journal = {Journal of Physics B Atomic Molecular Physics},
         year = 1997,
        month = jan,
       volume = {30},
       number = {1},
        pages = {1-11},
          doi = {10.1088/0953-4075/30/1/005},
       adsurl = {https://ui.adsabs.harvard.edu/abs/1997JPhB...30....1B}
}

@ARTICLE{Rastinejad2025,
       author = {{Rastinejad}, J.~C. and {Fong}, W. and {Kilpatrick}, C.~D. and {Nicholl}, M. and {Metzger}, B.~D.},
        title = "{Uniform Modeling of Observed Kilonovae: Implications for Diversity and the Progenitors of Merger-driven Long Gamma-Ray Bursts}",
      journal = {\apj},
         year = 2025,
        month = feb,
       volume = {979},
       number = {2},
          eid = {190},
        pages = {190},
          doi = {10.3847/1538-4357/ad9c77},
archivePrefix = {arXiv},
       eprint = {2409.02158},
 primaryClass = {astro-ph.HE},
       adsurl = {https://ui.adsabs.harvard.edu/abs/2025ApJ...979..190R}
}

@ARTICLE{Just2015,
       author = {{Just}, O. and {Bauswein}, A. and {Ardevol Pulpillo}, R. and {Goriely}, S. and {Janka}, H.-T.},
        title = "{Comprehensive nucleosynthesis analysis for ejecta of compact binary mergers}",
      journal = {\mnras},
         year = 2015,
        month = mar,
       volume = {448},
       number = {1},
        pages = {541-567},
          doi = {10.1093/mnras/stv009},
archivePrefix = {arXiv},
       eprint = {1406.2687},
 primaryClass = {astro-ph.SR},
       adsurl = {https://ui.adsabs.harvard.edu/abs/2015MNRAS.448..541J}
}

@ARTICLE{Fernandez2013,
       author = {{Fern{\'a}ndez}, Rodrigo and {Metzger}, Brian D.},
        title = "{Delayed outflows from black hole accretion tori following neutron star binary coalescence}",
      journal = {\mnras},
         year = 2013,
        month = oct,
       volume = {435},
       number = {1},
        pages = {502-517},
          doi = {10.1093/mnras/stt1312},
archivePrefix = {arXiv},
       eprint = {1304.6720},
 primaryClass = {astro-ph.HE},
       adsurl = {https://ui.adsabs.harvard.edu/abs/2013MNRAS.435..502F}
}

@ARTICLE{Hjorth2017,
       author = {{Hjorth}, Jens and {Levan}, Andrew J. and {Tanvir}, Nial R. and {Lyman}, Joe D. and {Wojtak}, Rados{\l}aw and {Schr{\o}der}, Sophie L. and {Mandel}, Ilya and {Gall}, Christa and {Bruun}, Sofie H.},
        title = "{The Distance to NGC 4993: The Host Galaxy of the Gravitational-wave Event GW170817}",
      journal = {\apjl},
         year = 2017,
        month = oct,
       volume = {848},
       number = {2},
          eid = {L31},
        pages = {L31},
          doi = {10.3847/2041-8213/aa9110},
archivePrefix = {arXiv},
       eprint = {1710.05856},
 primaryClass = {astro-ph.GA},
       adsurl = {https://ui.adsabs.harvard.edu/abs/2017ApJ...848L..31H}
}

@ARTICLE{Yang2024,
       author = {{Yang}, Yu-Han and {Troja}, Eleonora and {O'Connor}, Brendan and {Fryer}, Chris L. and {Im}, Myungshin and {Durbak}, Joe and {Paek}, Gregory S.~H. and {Ricci}, Roberto and {Bom}, Cl{\'e}cio R. and {Gillanders}, James H. and {Castro-Tirado}, Alberto J. and {Peng}, Zong-Kai and {Dichiara}, Simone and {Ryan}, Geoffrey and {van Eerten}, Hendrik and {Dai}, Zi-Gao and {Chang}, Seo-Won and {Choi}, Hyeonho and {De}, Kishalay and {Hu}, Youdong and {Kilpatrick}, Charles D. and {Kutyrev}, Alexander and {Jeong}, Mankeun and {Lee}, Chung-Uk and {Makler}, Martin and {Navarete}, Felipe and {P{\'e}rez-Garc{\'\i}a}, Ignacio},
        title = "{A lanthanide-rich kilonova in the aftermath of a long gamma-ray burst}",
      journal = {\nat},
         year = 2024,
        month = feb,
       volume = {626},
       number = {8000},
        pages = {742-745},
          doi = {10.1038/s41586-023-06979-5},
archivePrefix = {arXiv},
       eprint = {2308.00638},
 primaryClass = {astro-ph.HE},
       adsurl = {https://ui.adsabs.harvard.edu/abs/2024Natur.626..742Y}
}

@ARTICLE{Wooden1993,
       author = {{Wooden}, Diane H. and {Rank}, David M. and {Bregman}, Jesse D. and {Witteborn}, Fred C. and {Tielens}, A.~G.~G.~M. and {Cohen}, Martin and {Pinto}, Philip A. and {Axelrod}, Timothy S.},
        title = "{Airborne Spectrophotometry of SN 1987A from 1.7 to 12.6 Microns: Time History of the Dust Continuum and Line Emission}",
      journal = {\apjs},
         year = 1993,
        month = oct,
       volume = {88},
        pages = {477},
          doi = {10.1086/191830},
       adsurl = {https://ui.adsabs.harvard.edu/abs/1993ApJS...88..477W}
}

@ARTICLE{Moseley1989,
       author = {{Moseley}, S.~H. and {Dwek}, E. and {Glaccum}, W. and {Graham}, J.~R. and {Loewenstein}, R.~F. and {Silverberg}, R.~F.},
        title = "{Far-Infrared Spectrophotometry of SN 1987A: Days 265 and 267}",
      journal = {\apj},
         year = 1989,
        month = dec,
       volume = {347},
        pages = {1119},
          doi = {10.1086/168201},
       adsurl = {https://ui.adsabs.harvard.edu/abs/1989ApJ...347.1119M}
}

@ARTICLE{Haas1990,
       author = {{Haas}, Michael R. and {Colgan}, Sean W.~J. and {Erickson}, Edwin F. and {Lord}, Steven D. and {Burton}, Michael G. and {Hollenbach}, David J.},
        title = "{Velocity-resolved Far-Infrared Spectra of Fe ii: Evidence for Mixing and Clumping in SN 1987A}",
      journal = {\apj},
         year = 1990,
        month = sep,
       volume = {360},
        pages = {257},
          doi = {10.1086/169115},
       adsurl = {https://ui.adsabs.harvard.edu/abs/1990ApJ...360..257H}
}

@ARTICLE{Sansonetti2010,
       author = {{Sansonetti}, J.~E. and {Nave}, G.},
        title = "{Wavelengths, Transition Probabilities, and Energy Levels for the Spectrum of Neutral Strontium (Sr I)}",
      journal = {Journal of Physical and Chemical Reference Data},
         year = 2010,
        month = sep,
       volume = {39},
       number = {3},
        pages = {033103-033103},
          doi = {10.1063/1.3449176},
       adsurl = {https://ui.adsabs.harvard.edu/abs/2010JPCRD..39c3103S}
}

@ARTICLE{Dougan2025,
       author = {{Dougan}, David J. and {McElroy}, Niall E. and {Ballance}, Connor P. and {Ramsbottom}, Catherine A.},
        title = "{Strontium I, III, IV, and V: electron impact excitation data for Kilonovae and white dwarf diagnostic applications}",
      journal = {\mnras},
         year = 2025,
        month = jul,
       volume = {541},
       number = {1},
        pages = {367-383},
          doi = {10.1093/mnras/staf1013},
archivePrefix = {arXiv},
       eprint = {2505.09788},
 primaryClass = {astro-ph.SR},
       adsurl = {https://ui.adsabs.harvard.edu/abs/2025MNRAS.541..367D}
}

@ARTICLE{Mulholland2024,
       author = {{Mulholland}, L.~P. and {McElroy}, N.~E. and {McNeill}, F.~L. and {Sim}, S.~A. and {Ballance}, C.~P. and {Ramsbottom}, C.~A.},
        title = "{New radiative and collisional atomic data for Sr II and Y II with application to Kilonova modelling}",
      journal = {\mnras},
         year = 2024,
        month = aug,
       volume = {532},
       number = {2},
        pages = {2289-2308},
          doi = {10.1093/mnras/stae1615},
archivePrefix = {arXiv},
       eprint = {2407.01398},
 primaryClass = {astro-ph.HE},
       adsurl = {https://ui.adsabs.harvard.edu/abs/2024MNRAS.532.2289M}
}

@ARTICLE{Travaglio2004,
       author = {{Travaglio}, Claudia and {Gallino}, Roberto and {Arnone}, Enrico and {Cowan}, John and {Jordan}, Faith and {Sneden}, Christopher},
        title = "{Galactic Evolution of Sr, Y, And Zr: A Multiplicity of Nucleosynthetic Processes}",
      journal = {\apj},
         year = 2004,
        month = feb,
       volume = {601},
       number = {2},
        pages = {864-884},
          doi = {10.1086/380507},
archivePrefix = {arXiv},
       eprint = {astro-ph/0310189},
 primaryClass = {astro-ph},
       adsurl = {https://ui.adsabs.harvard.edu/abs/2004ApJ...601..864T}
}

@ARTICLE{Cowan2021,
       author = {{Cowan}, John J. and {Sneden}, Christopher and {Lawler}, James E. and {Aprahamian}, Ani and {Wiescher}, Michael and {Langanke}, Karlheinz and {Mart{\'\i}nez-Pinedo}, Gabriel and {Thielemann}, Friedrich-Karl},
        title = "{Origin of the heaviest elements: The rapid neutron-capture process}",
      journal = {Reviews of Modern Physics},
         year = 2021,
        month = jan,
       volume = {93},
       number = {1},
          eid = {015002},
        pages = {015002},
          doi = {10.1103/RevModPhys.93.015002},
archivePrefix = {arXiv},
       eprint = {1901.01410},
 primaryClass = {astro-ph.HE},
       adsurl = {https://ui.adsabs.harvard.edu/abs/2021RvMP...93a5002C}
}

@ARTICLE{Arcones2023,
       author = {{Arcones}, Almudena and {Thielemann}, Friedrich-Karl},
        title = "{Origin of the elements}",
      journal = {\aapr},
         year = 2023,
        month = dec,
       volume = {31},
       number = {1},
          eid = {1},
        pages = {1},
          doi = {10.1007/s00159-022-00146-x},
       adsurl = {https://ui.adsabs.harvard.edu/abs/2023A&ARv..31....1A}
}

@ARTICLE{Curtis2023,
       author = {{Curtis}, Sanjana and {M{\"o}sta}, Philipp and {Wu}, Zhenyu and {Radice}, David and {Roberts}, Luke and {Ricigliano}, Giacomo and {Perego}, Albino},
        title = "{r-process nucleosynthesis and kilonovae from hypermassive neutron star post-merger remnants}",
      journal = {\mnras},
         year = 2023,
        month = feb,
       volume = {518},
       number = {4},
        pages = {5313-5322},
          doi = {10.1093/mnras/stac3128},
archivePrefix = {arXiv},
       eprint = {2112.00772},
 primaryClass = {astro-ph.HE},
       adsurl = {https://ui.adsabs.harvard.edu/abs/2023MNRAS.518.5313C}
}

@ARTICLE{Sneppen2023,
       author = {{Sneppen}, Albert and {Watson}, Darach},
        title = "{Discovery of a 760 nm P Cygni line in AT2017gfo: Identification of yttrium in the kilonova photosphere}",
      journal = {\aap},
         year = 2023,
        month = jul,
       volume = {675},
          eid = {A194},
        pages = {A194},
          doi = {10.1051/0004-6361/202346421},
archivePrefix = {arXiv},
       eprint = {2306.14942},
 primaryClass = {astro-ph.HE},
       adsurl = {https://ui.adsabs.harvard.edu/abs/2023A&A...675A.194S}
}

@ARTICLE{Tarumi2023,
       author = {{Tarumi}, Yuta and {Hotokezaka}, Kenta and {Domoto}, Nanae and {Tanaka}, Masaomi},
        title = "{Non-LTE analysis for Helium and Strontium lines in the kilonova AT2017gfo}",
      journal = {arXiv e-prints},
         year = 2023,
        month = feb,
          eid = {arXiv:2302.13061},
        pages = {arXiv:2302.13061},
          doi = {10.48550/arXiv.2302.13061},
archivePrefix = {arXiv},
       eprint = {2302.13061},
 primaryClass = {astro-ph.HE},
       adsurl = {https://ui.adsabs.harvard.edu/abs/2023arXiv230213061T}
}

@ARTICLE{Perego2022,
       author = {{Perego}, Albino and {Vescovi}, Diego and {Fiore}, Achille and {Chiesa}, Leonardo and {Vogl}, Christian and {Benetti}, Stefano and {Bernuzzi}, Sebastiano and {Branchesi}, Marica and {Cappellaro}, Enrico and {Cristallo}, Sergio and {Fl{\"o}rs}, Andreas and {Kerzendorf}, Wolfgang E. and {Radice}, David},
        title = "{Production of Very Light Elements and Strontium in the Early Ejecta of Neutron Star Mergers}",
      journal = {\apj},
         year = 2022,
        month = jan,
       volume = {925},
       number = {1},
          eid = {22},
        pages = {22},
          doi = {10.3847/1538-4357/ac3751},
archivePrefix = {arXiv},
       eprint = {2009.08988},
 primaryClass = {astro-ph.HE},
       adsurl = {https://ui.adsabs.harvard.edu/abs/2022ApJ...925...22P}
}

@ARTICLE{Domoto2021,
       author = {{Domoto}, Nanae and {Tanaka}, Masaomi and {Wanajo}, Shinya and {Kawaguchi}, Kyohei},
        title = "{Signatures of r-process Elements in Kilonova Spectra}",
      journal = {\apj},
         year = 2021,
        month = may,
       volume = {913},
       number = {1},
          eid = {26},
        pages = {26},
          doi = {10.3847/1538-4357/abf358},
archivePrefix = {arXiv},
       eprint = {2103.15284},
 primaryClass = {astro-ph.HE},
       adsurl = {https://ui.adsabs.harvard.edu/abs/2021ApJ...913...26D}
}

@ARTICLE{Watson2019,
       author = {{Watson}, Darach and {Hansen}, Camilla J. and {Selsing}, Jonatan and {Koch}, Andreas and {Malesani}, Daniele B. and {Andersen}, Anja C. and {Fynbo}, Johan P.~U. and {Arcones}, Almudena and {Bauswein}, Andreas and {Covino}, Stefano and {Grado}, Aniello and {Heintz}, Kasper E. and {Hunt}, Leslie and {Kouveliotou}, Chryssa and {Leloudas}, Giorgos and {Levan}, Andrew J. and {Mazzali}, Paolo and {Pian}, Elena},
        title = "{Identification of strontium in the merger of two neutron stars}",
      journal = {\nat},
         year = 2019,
        month = oct,
       volume = {574},
       number = {7779},
        pages = {497-500},
          doi = {10.1038/s41586-019-1676-3},
archivePrefix = {arXiv},
       eprint = {1910.10510},
 primaryClass = {astro-ph.HE},
       adsurl = {https://ui.adsabs.harvard.edu/abs/2019Natur.574..497W}
}

@ARTICLE{Vieria2025,
       author = {{Vieira}, Nicholas and {Ruan}, John J. and {Haggard}, Daryl and {Drout}, Maria R. and {Fern{\'a}ndez}, Rodrigo},
        title = "{Spectroscopic r-Process Abundance Retrieval for Kilonovae III: Self-Consistent Spectral and Light Curve Modeling of the GW170817 Kilonova}",
      journal = {arXiv e-prints},
         year = 2025,
        month = apr,
          eid = {arXiv:2504.10696},
        pages = {arXiv:2504.10696},
          doi = {10.48550/arXiv.2504.10696},
archivePrefix = {arXiv},
       eprint = {2504.10696},
 primaryClass = {astro-ph.HE},
       adsurl = {https://ui.adsabs.harvard.edu/abs/2025arXiv250410696V}
}

@ARTICLE{Wollaeger2018,
       author = {{Wollaeger}, Ryan T. and {Korobkin}, Oleg and {Fontes}, Christopher J. and {Rosswog}, Stephan K. and {Even}, Wesley P. and {Fryer}, Christopher L. and {Sollerman}, Jesper and {Hungerford}, Aimee L. and {van Rossum}, Daniel R. and {Wollaber}, Allan B.},
        title = "{Impact of ejecta morphology and composition on the electromagnetic signatures of neutron star mergers}",
      journal = {\mnras},
         year = 2018,
        month = aug,
       volume = {478},
       number = {3},
        pages = {3298-3334},
          doi = {10.1093/mnras/sty1018},
archivePrefix = {arXiv},
       eprint = {1705.07084},
 primaryClass = {astro-ph.HE},
       adsurl = {https://ui.adsabs.harvard.edu/abs/2018MNRAS.478.3298W}
}

@ARTICLE{Collins2023,
       author = {{Collins}, Christine E. and {Bauswein}, Andreas and {Sim}, Stuart A. and {Vijayan}, Vimal and {Mart{\'\i}nez-Pinedo}, Gabriel and {Just}, Oliver and {Shingles}, Luke J. and {Kromer}, Markus},
        title = "{3D radiative transfer kilonova modelling for binary neutron star merger simulations}",
      journal = {\mnras},
         year = 2023,
        month = may,
       volume = {521},
       number = {2},
        pages = {1858-1870},
          doi = {10.1093/mnras/stad606},
archivePrefix = {arXiv},
       eprint = {2209.05246},
 primaryClass = {astro-ph.HE},
       adsurl = {https://ui.adsabs.harvard.edu/abs/2023MNRAS.521.1858C}
}

@ARTICLE{Ricigliano2025,
       author = {{Ricigliano}, Giacomo and {Hotokezaka}, Kenta and {Arcones}, Almudena},
        title = "{Modelling the emission lines from r-process elements in supernova nebulae}",
      journal = {\mnras},
         year = 2025,
        month = nov,
       volume = {543},
       number = {3},
        pages = {2534-2552},
          doi = {10.1093/mnras/staf1577},
       adsurl = {https://ui.adsabs.harvard.edu/abs/2025MNRAS.543.2534R}
}

@ARTICLE{LRCA,
       author = {{Jerkstrand}, Anders},
        title = "{Spectral synthesis techniques for supernovae and kilonovae}",
      journal = {Living Reviews in Computational Astrophysics},
         year = 2025,
        month = aug,
       volume = {11},
       number = {1},
          eid = {1},
        pages = {1},
          doi = {10.1007/s41115-025-00022-2},
       adsurl = {https://ui.adsabs.harvard.edu/abs/2025LRCA...11....1J}
}

@ARTICLE{Kiuchi2023,
       author = {{Kiuchi}, Kenta and {Fujibayashi}, Sho and {Hayashi}, Kota and {Kyutoku}, Koutarou and {Sekiguchi}, Yuichiro and {Shibata}, Masaru},
        title = "{Self-Consistent Picture of the Mass Ejection from a One Second Long Binary Neutron Star Merger Leaving a Short-Lived Remnant in a General-Relativistic Neutrino-Radiation Magnetohydrodynamic Simulation}",
      journal = {\prl},
         year = 2023,
        month = jul,
       volume = {131},
       number = {1},
          eid = {011401},
        pages = {011401},
          doi = {10.1103/PhysRevLett.131.011401},
archivePrefix = {arXiv},
       eprint = {2211.07637},
 primaryClass = {astro-ph.HE},
       adsurl = {https://ui.adsabs.harvard.edu/abs/2023PhRvL.131a1401K}
}

@ARTICLE{Ferland2003,
       author = {{Ferland}, Gary J.},
        title = "{Quantitative Spectroscopy of Photoionized Clouds}",
      journal = {\araa},
         year = 2003,
        month = jan,
       volume = {41},
        pages = {517-554},
          doi = {10.1146/annurev.astro.41.011802.094836},
       adsurl = {https://ui.adsabs.harvard.edu/abs/2003ARA&A..41..517F}
}

@ARTICLE{Mulholland2025,
       author = {{Mulholland}, Leo P. and {Bromley}, Steven J. and {Ballance}, Connor P. and {Sim}, Stuart A. and {Ramsbottom}, Catherine A.},
        title = "{On the use of the Axelrod formula for thermal electron collisions in astrophysical modelling}",
      journal = {\jqsrt},
         year = 2025,
        month = nov,
       volume = {345},
          eid = {109545},
        pages = {109545},
          doi = {10.1016/j.jqsrt.2025.109545},
archivePrefix = {arXiv},
       eprint = {2503.05489},
 primaryClass = {astro-ph.HE},
       adsurl = {https://ui.adsabs.harvard.edu/abs/2025JQSRT.34509545M}
}

@ARTICLE{Wanajo2018,
       author = {{Wanajo}, Shinya and {M{\"u}ller}, Bernhard and {Janka}, Hans-Thomas and {Heger}, Alexander},
        title = "{Nucleosynthesis in the Innermost Ejecta of Neutrino-driven Supernova Explosions in Two Dimensions}",
      journal = {\apj},
         year = 2018,
        month = jan,
       volume = {852},
       number = {1},
          eid = {40},
        pages = {40},
          doi = {10.3847/1538-4357/aa9d97},
archivePrefix = {arXiv},
       eprint = {1701.06786},
 primaryClass = {astro-ph.SR},
       adsurl = {https://ui.adsabs.harvard.edu/abs/2018ApJ...852...40W}
}

@ARTICLE{Singh2025,
       author = {{Singh}, Suvam and {Harman}, Zolt{\'a}n and {Keitel}, Christoph H.},
        title = "{Dielectronic recombination studies of ions relevant to kilonovae and non-LTE plasma}",
      journal = {arXiv e-prints},
         year = 2025,
        month = apr,
          eid = {arXiv:2504.06639},
        pages = {arXiv:2504.06639},
          doi = {10.48550/arXiv.2504.06639},
archivePrefix = {arXiv},
       eprint = {2504.06639},
 primaryClass = {astro-ph.HE},
       adsurl = {https://ui.adsabs.harvard.edu/abs/2025arXiv250406639S}
}

@ARTICLE{McCann2025,
       author = {{McCann}, M. and {Ballance}, C.~P. and {McNeill}, F. and {Sim}, S.~A. and {Ramsbottom}, C.~A.},
        title = "{Electron-impact excitation of zirconium I-III in support of neutron star merger diagnostics}",
      journal = {\mnras},
         year = 2025,
        month = jul,
       volume = {540},
       number = {4},
        pages = {2923-2936},
          doi = {10.1093/mnras/staf866},
archivePrefix = {arXiv},
       eprint = {2503.16359},
 primaryClass = {physics.atom-ph},
       adsurl = {https://ui.adsabs.harvard.edu/abs/2025MNRAS.540.2923M}
}

@PHDTHESIS{Axelrod1980,
       author = {{Axelrod}, T.~S.},
        title = "{Late Time Optical Spectra from the Nickel(56) Model for Type I Supernovae.}",
       school = {University of California, Santa Cruz},
         year = 1980,
        month = jan,
       adsurl = {https://ui.adsabs.harvard.edu/abs/1980PhDT.........1A}
}

@ARTICLE{Jerkstrand2015,
       author = {{Jerkstrand}, A. and {Ergon}, M. and {Smartt}, S.~J. and {Fransson}, C. and {Sollerman}, J. and {Taubenberger}, S. and {Bersten}, M. and {Spyromilio}, J.},
        title = "{Late-time spectral line formation in Type IIb supernovae, with application to SN 1993J, SN 2008ax, and SN 2011dh}",
      journal = {\aap},
         year = 2015,
        month = jan,
       volume = {573},
          eid = {A12},
        pages = {A12},
          doi = {10.1051/0004-6361/201423983},
archivePrefix = {arXiv},
       eprint = {1408.0732},
 primaryClass = {astro-ph.HE},
       adsurl = {https://ui.adsabs.harvard.edu/abs/2015A&A...573A..12J}
}

@ARTICLE{Sukhbold2016,
       author = {{Sukhbold}, Tuguldur and {Ertl}, T. and {Woosley}, S.~E. and {Brown}, Justin M. and {Janka}, H. -T.},
        title = "{Core-collapse Supernovae from 9 to 120 Solar Masses Based on Neutrino-powered Explosions}",
      journal = {\apj},
         year = 2016,
        month = apr,
       volume = {821},
       number = {1},
          eid = {38},
        pages = {38},
          doi = {10.3847/0004-637X/821/1/38},
archivePrefix = {arXiv},
       eprint = {1510.04643},
 primaryClass = {astro-ph.HE},
       adsurl = {https://ui.adsabs.harvard.edu/abs/2016ApJ...821...38S}
}

@ARTICLE{Roederer2022,
       author = {{Roederer}, Ian U. and {Lawler}, James E. and {Den Hartog}, Elizabeth A. and {Placco}, Vinicius M. and {Surman}, Rebecca and {Beers}, Timothy C. and {Ezzeddine}, Rana and {Frebel}, Anna and {Hansen}, Terese T. and {Hattori}, Kohei and {Holmbeck}, Erika M. and {Sakari}, Charli M.},
        title = "{The R-process Alliance: A Nearly Complete R-process Abundance Template Derived from Ultraviolet Spectroscopy of the R-process-enhanced Metal-poor Star HD 222925}",
      journal = {\apjs},
         year = 2022,
        month = jun,
       volume = {260},
       number = {2},
          eid = {27},
        pages = {27},
          doi = {10.3847/1538-4365/ac5cbc},
archivePrefix = {arXiv},
       eprint = {2205.03426},
 primaryClass = {astro-ph.SR},
       adsurl = {https://ui.adsabs.harvard.edu/abs/2022ApJS..260...27R}
}

@ARTICLE{Roederer2012,
       author = {{Roederer}, Ian U. and {Lawler}, James E.},
        title = "{Detection of Elements at All Three r-process Peaks in the Metal-poor Star HD 160617}",
      journal = {\apj},
         year = 2012,
        month = may,
       volume = {750},
       number = {1},
          eid = {76},
        pages = {76},
          doi = {10.1088/0004-637X/750/1/76},
archivePrefix = {arXiv},
       eprint = {1204.3901},
 primaryClass = {astro-ph.SR},
       adsurl = {https://ui.adsabs.harvard.edu/abs/2012ApJ...750...76R}
}

@ARTICLE{Cowan2005,
       author = {{Cowan}, John J. and {Sneden}, Christopher and {Beers}, Timothy C. and {Lawler}, James E. and {Simmerer}, Jennifer and {Truran}, James W. and {Primas}, Francesca and {Collier}, Jason and {Burles}, Scott},
        title = "{Hubble Space Telescope Observations of Heavy Elements in Metal-Poor Galactic Halo Stars}",
      journal = {\apj},
         year = 2005,
        month = jul,
       volume = {627},
       number = {1},
        pages = {238-250},
          doi = {10.1086/429952},
archivePrefix = {arXiv},
       eprint = {astro-ph/0502591},
 primaryClass = {astro-ph},
       adsurl = {https://ui.adsabs.harvard.edu/abs/2005ApJ...627..238C}
}

@ARTICLE{Bromley2023,
       author = {{Bromley}, S.~J. and {McCann}, M. and {Loch}, S.~D. and {Ballance}, C.~P.},
        title = "{Electron-impact Excitation of Pt I-III: The Importance of Metastables and Collision Processes in Neutron Star Merger and Laboratory Plasmas}",
      journal = {\apjs},
         year = 2023,
        month = sep,
       volume = {268},
       number = {1},
          eid = {22},
        pages = {22},
          doi = {10.3847/1538-4365/ace5a1},
       adsurl = {https://ui.adsabs.harvard.edu/abs/2023ApJS..268...22B}
}

@ARTICLE{SterlingWitthoeft2011,
       author = {{Sterling}, N.~C. and {Witthoeft}, M.~C.},
        title = "{Atomic data for neutron-capture elements. I. Photoionization and recombination properties of low-charge selenium ions}",
      journal = {\aap},
         year = 2011,
        month = may,
       volume = {529},
          eid = {A147},
        pages = {A147},
          doi = {10.1051/0004-6361/201116718},
archivePrefix = {arXiv},
       eprint = {1102.5341},
 primaryClass = {astro-ph.SR},
}

@INPROCEEDINGS{Dinerstein2006,
       author = {{Dinerstein}, Harriet L. and {Lacy}, J.~H. and {Sellgren}, K. and {Sterling}, N.~C.},
        title = "{A Search for Gas-Phase Zirconium in s-process Enriched Planetary Nebulae}",
    booktitle = {American Astronomical Society Meeting Abstracts},
         year = 2006,
       series = {American Astronomical Society Meeting Abstracts},
       volume = {209},
        month = dec,
          eid = {156.09},
        pages = {156.09},
       adsurl = {https://ui.adsabs.harvard.edu/abs/2006AAS...20915609D}
}

@ARTICLE{Mulholland2024-Te,
       author = {{Mulholland}, L.~P. and {McNeill}, F. and {Sim}, S.~A. and {Ballance}, C.~P. and {Ramsbottom}, C.~A.},
        title = "{Collisional and radiative data for tellurium ions in kilonovae modelling and laboratory benchmarks}",
      journal = {\mnras},
         year = 2024,
        month = nov,
       volume = {534},
       number = {4},
        pages = {3423-3438},
          doi = {10.1093/mnras/stae2331},
archivePrefix = {arXiv},
       eprint = {2410.05958},
 primaryClass = {astro-ph.SR},
       adsurl = {https://ui.adsabs.harvard.edu/abs/2024MNRAS.534.3423M}
}

@INCOLLECTION{Jerkstrand2017,
       author = {{Jerkstrand}, Anders},
        title = "{Spectra of Supernovae in the Nebular Phase}",
    booktitle = {Handbook of Supernovae},
         year = 2017,
       editor = {{Alsabti}, Athem W. and {Murdin}, Paul},
        pages = {795},
          doi = {10.1007/978-3-319-21846-5_29},
       adsurl = {https://ui.adsabs.harvard.edu/abs/2017hsn..book..795J}
}

@ARTICLE{Eriksson1974,
       author = {{Eriksson}, K. B. S.},
        title = "{Transitions between the levels of the ground configuration in Te ii}",
      journal = {Journal of the Optical Society of America},
         year = 1974,
        month = sep,
       volume = {64},
       number = {9},
        pages = {1272-1272},
          doi = {10.1088/0031-8949/12/3/006},
       adsurl = {https://ui.adsabs.harvard.edu/abs/1975PhyS...12..145M}
}

@ARTICLE{Morillon1975,
       author = {{Morillon}, C. and {Verg{\`e}s}, J.},
        title = "{Etude des Configurations Fondamentales 5s$^{2}$5p$^{4}$ de Te I et I II par l'Observation des Raies Multipolaires Infrarouges}",
      journal = {\physscr},
         year = 1975,
        month = sep,
       volume = {12},
       number = {3},
        pages = {145-156},
          doi = {10.1088/0031-8949/12/3/006},
       adsurl = {https://ui.adsabs.harvard.edu/abs/1975PhyS...12..145M}
}

@ARTICLE{Gillanders2024,
       author = {{Gillanders}, J.~H. and {Sim}, S.~A. and {Smartt}, S.~J. and {Goriely}, S. and {Bauswein}, A.},
        title = "{Modelling the spectra of the kilonova AT2017gfo - II. Beyond the photospheric epochs}",
      journal = {\mnras},
         year = 2024,
        month = apr,
       volume = {529},
       number = {3},
        pages = {2918-2945},
          doi = {10.1093/mnras/stad3688},
archivePrefix = {arXiv},
       eprint = {2306.15055},
 primaryClass = {astro-ph.HE},
       adsurl = {https://ui.adsabs.harvard.edu/abs/2024MNRAS.529.2918G}
}

@ARTICLE{Gillanders2025,
       author = {{Gillanders}, J.~H. and {Smartt}, S.~J.},
        title = "{Analysis of the JWST spectra of the kilonova AT 2023vfi accompanying GRB 230307A}",
      journal = {\mnras},
         year = 2025,
        month = apr,
       volume = {538},
       number = {3},
        pages = {1663-1689},
          doi = {10.1093/mnras/staf287},
archivePrefix = {arXiv},
       eprint = {2408.11093},
 primaryClass = {astro-ph.HE},
       adsurl = {https://ui.adsabs.harvard.edu/abs/2025MNRAS.538.1663G}
}

@ARTICLE{Pian2017,
       author = {{Pian}, E. and {D'Avanzo}, P. and {Benetti}, S. and {Branchesi}, M. and {Brocato}, E. and {Campana}, S. and {Cappellaro}, E. and {Covino}, S. and {D'Elia}, V. and {Fynbo}, J.~P.~U. and {Getman}, F. and {Ghirlanda}, G. and {Ghisellini}, G. and {Grado}, A. and {Greco}, G. and {Hjorth}, J. and {Kouveliotou}, C. and {Levan}, A. and {Limatola}, L. and {Malesani}, D. and {Mazzali}, P.~A. and {Melandri}, A. and {M{\o}ller}, P. and {Nicastro}, L. and {Palazzi}, E. and {Piranomonte}, S. and {Rossi}, A. and {Salafia}, O.~S. and {Selsing}, J. and {Stratta}, G. and {Tanaka}, M. and {Tanvir}, N.~R. and {Tomasella}, L. and {Watson}, D. and {Yang}, S. and {Amati}, L. and {Antonelli}, L.~A. and {Ascenzi}, S. and {Bernardini}, M.~G. and {Bo{\"e}r}, M. and {Bufano}, F. and {Bulgarelli}, A. and {Capaccioli}, M. and {Casella}, P. and {Castro-Tirado}, A.~J. and {Chassande-Mottin}, E. and {Ciolfi}, R. and {Copperwheat}, C.~M. and {Dadina}, M. and {De Cesare}, G. and {di Paola}, A. and {Fan}, Y.~Z. and {Gendre}, B. and {Giuffrida}, G. and {Giunta}, A. and {Hunt}, L.~K. and {Israel}, G.~L. and {Jin}, Z. -P. and {Kasliwal}, M.~M. and {Klose}, S. and {Lisi}, M. and {Longo}, F. and {Maiorano}, E. and {Mapelli}, M. and {Masetti}, N. and {Nava}, L. and {Patricelli}, B. and {Perley}, D. and {Pescalli}, A. and {Piran}, T. and {Possenti}, A. and {Pulone}, L. and {Razzano}, M. and {Salvaterra}, R. and {Schipani}, P. and {Spera}, M. and {Stamerra}, A. and {Stella}, L. and {Tagliaferri}, G. and {Testa}, V. and {Troja}, E. and {Turatto}, M. and {Vergani}, S.~D. and {Vergani}, D.},
        title = "{Spectroscopic identification of r-process nucleosynthesis in a double neutron-star merger}",
      journal = {\nat},
         year = 2017,
        month = nov,
       volume = {551},
       number = {7678},
        pages = {67-70},
          doi = {10.1038/nature24298},
archivePrefix = {arXiv},
       eprint = {1710.05858},
 primaryClass = {astro-ph.HE},
       adsurl = {https://ui.adsabs.harvard.edu/abs/2017Natur.551...67P}
}

@ARTICLE{Hotokezaka2023,
       author = {{Hotokezaka}, Kenta and {Tanaka}, Masaomi and {Kato}, Daiji and {Gaigalas}, Gediminas},
        title = "{Tellurium emission line in kilonova AT 2017gfo}",
      journal = {\mnras},
         year = 2023,
        month = nov,
       volume = {526},
       number = {1},
        pages = {L155-L159},
          doi = {10.1093/mnrasl/slad128},
archivePrefix = {arXiv},
       eprint = {2307.00988},
 primaryClass = {astro-ph.HE},
       adsurl = {https://ui.adsabs.harvard.edu/abs/2023MNRAS.526L.155H}
}

@ARTICLE{Kotak2006,
       author = {{Kotak}, Rubina and {Meikle}, Peter and {Pozzo}, Monica and {van Dyk}, Schuyler D. and {Farrah}, Duncan and {Fesen}, Robert and {Filippenko}, Alexei V. and {Foley}, Ryan J. and {Fransson}, Claes and {Gerardy}, Christopher L. and {H{\"o}flich}, Peter A. and {Lundqvist}, Peter and {Mattila}, Seppo and {Sollerman}, Jesper and {Wheeler}, J. Craig},
        title = "{Spitzer Measurements of Atomic and Molecular Abundances in the Type IIP SN 2005af}",
      journal = {\apjl},
         year = 2006,
        month = nov,
       volume = {651},
       number = {2},
        pages = {L117-L120},
          doi = {10.1086/509655},
archivePrefix = {arXiv},
       eprint = {astro-ph/0609706},
 primaryClass = {astro-ph},
       adsurl = {https://ui.adsabs.harvard.edu/abs/2006ApJ...651L.117K}
}

@ARTICLE{Kotak2009,
       author = {{Kotak}, R. and {Meikle}, W.~P.~S. and {Farrah}, D. and {Gerardy}, C.~L. and {Foley}, R.~J. and {Van Dyk}, S.~D. and {Fransson}, C. and {Lundqvist}, P. and {Sollerman}, J. and {Fesen}, R. and {Filippenko}, A.~V. and {Mattila}, S. and {Silverman}, J.~M. and {Andersen}, A.~C. and {H{\"o}flich}, P.~A. and {Pozzo}, M. and {Wheeler}, J.~C.},
        title = "{Dust and The Type II-Plateau Supernova 2004et}",
      journal = {\apj},
         year = 2009,
        month = oct,
       volume = {704},
       number = {1},
        pages = {306-323},
          doi = {10.1088/0004-637X/704/1/306},
archivePrefix = {arXiv},
       eprint = {0904.3737},
 primaryClass = {astro-ph.SR},
       adsurl = {https://ui.adsabs.harvard.edu/abs/2009ApJ...704..306K}
}

@ARTICLE{Barnes2021,
       author = {{Barnes}, Jennifer and {Zhu}, Y.~L. and {Lund}, K.~A. and {Sprouse}, T.~M. and {Vassh}, N. and {McLaughlin}, G.~C. and {Mumpower}, M.~R. and {Surman}, R.},
        title = "{Kilonovae Across the Nuclear Physics Landscape: The Impact of Nuclear Physics Uncertainties on r-process-powered Emission}",
      journal = {\apj},
         year = 2021,
        month = sep,
       volume = {918},
       number = {2},
          eid = {44},
        pages = {44},
          doi = {10.3847/1538-4357/ac0aec},
archivePrefix = {arXiv},
       eprint = {2010.11182},
 primaryClass = {astro-ph.HE},
       adsurl = {https://ui.adsabs.harvard.edu/abs/2021ApJ...918...44B}
}

@ARTICLE{Metzger2014,
       author = {{Metzger}, Brian D. and {Fern{\'a}ndez}, Rodrigo},
        title = "{Red or blue? A potential kilonova imprint of the delay until black hole formation following a neutron star merger}",
      journal = {\mnras},
         year = 2014,
        month = jul,
       volume = {441},
       number = {4},
        pages = {3444-3453},
          doi = {10.1093/mnras/stu802},
archivePrefix = {arXiv},
       eprint = {1402.4803},
 primaryClass = {astro-ph.HE},
       adsurl = {https://ui.adsabs.harvard.edu/abs/2014MNRAS.441.3444M}
}

@ARTICLE{Goriely2015,
       author = {{Goriely}, S. and {Bauswein}, A. and {Just}, O. and {Pllumbi}, E. and {Janka}, H. -Th.},
        title = "{Impact of weak interactions of free nucleons on the r-process in dynamical ejecta from neutron star mergers}",
      journal = {\mnras},
         year = 2015,
        month = oct,
       volume = {452},
       number = {4},
        pages = {3894-3904},
          doi = {10.1093/mnras/stv1526},
archivePrefix = {arXiv},
       eprint = {1504.04377},
 primaryClass = {astro-ph.SR},
       adsurl = {https://ui.adsabs.harvard.edu/abs/2015MNRAS.452.3894G}
}

@ARTICLE{Smartt2017,
       author = {{Smartt}, S.~J. and {Chen}, T. -W. and {Jerkstrand}, A. and {Coughlin}, M. and {Kankare}, E. and {Sim}, S.~A. and {Fraser}, M. and {Inserra}, C. and {Maguire}, K. and {Chambers}, K.~C. and {Huber}, M.~E. and {Kr{\"u}hler}, T. and {Leloudas}, G. and {Magee}, M. and {Shingles}, L.~J. and {Smith}, K.~W. and {Young}, D.~R. and {Tonry}, J. and {Kotak}, R. and {Gal-Yam}, A. and {Lyman}, J.~D. and {Homan}, D.~S. and {Agliozzo}, C. and {Anderson}, J.~P. and {Angus}, C.~R. and {Ashall}, C. and {Barbarino}, C. and {Bauer}, F.~E. and {Berton}, M. and {Botticella}, M.~T. and {Bulla}, M. and {Bulger}, J. and {Cannizzaro}, G. and {Cano}, Z. and {Cartier}, R. and {Cikota}, A. and {Clark}, P. and {De Cia}, A. and {Della Valle}, M. and {Denneau}, L. and {Dennefeld}, M. and {Dessart}, L. and {Dimitriadis}, G. and {Elias-Rosa}, N. and {Firth}, R.~E. and {Flewelling}, H. and {Fl{\"o}rs}, A. and {Franckowiak}, A. and {Frohmaier}, C. and {Galbany}, L. and {Gonz{\'a}lez-Gait{\'a}n}, S. and {Greiner}, J. and {Gromadzki}, M. and {Guelbenzu}, A. Nicuesa and {Guti{\'e}rrez}, C.~P. and {Hamanowicz}, A. and {Hanlon}, L. and {Harmanen}, J. and {Heintz}, K.~E. and {Heinze}, A. and {Hernandez}, M. -S. and {Hodgkin}, S.~T. and {Hook}, I.~M. and {Izzo}, L. and {James}, P.~A. and {Jonker}, P.~G. and {Kerzendorf}, W.~E. and {Klose}, S. and {Kostrzewa-Rutkowska}, Z. and {Kowalski}, M. and {Kromer}, M. and {Kuncarayakti}, H. and {Lawrence}, A. and {Lowe}, T.~B. and {Magnier}, E.~A. and {Manulis}, I. and {Martin-Carrillo}, A. and {Mattila}, S. and {McBrien}, O. and {M{\"u}ller}, A. and {Nordin}, J. and {O'Neill}, D. and {Onori}, F. and {Palmerio}, J.~T. and {Pastorello}, A. and {Patat}, F. and {Pignata}, G. and {Podsiadlowski}, Ph. and {Pumo}, M.~L. and {Prentice}, S.~J. and {Rau}, A. and {Razza}, A. and {Rest}, A. and {Reynolds}, T. and {Roy}, R. and {Ruiter}, A.~J. and {Rybicki}, K.~A. and {Salmon}, L. and {Schady}, P. and {Schultz}, A.~S.~B. and {Schweyer}, T. and {Seitenzahl}, I.~R. and {Smith}, M. and {Sollerman}, J. and {Stalder}, B. and {Stubbs}, C.~W. and {Sullivan}, M. and {Szegedi}, H. and {Taddia}, F. and {Taubenberger}, S. and {Terreran}, G. and {van Soelen}, B. and {Vos}, J. and {Wainscoat}, R.~J. and {Walton}, N.~A. and {Waters}, C. and {Weiland}, H. and {Willman}, M. and {Wiseman}, P. and {Wright}, D.~E. and {Wyrzykowski}, {\L}. and {Yaron}, O.},
        title = "{A kilonova as the electromagnetic counterpart to a gravitational-wave source}",
      journal = {\nat},
         year = 2017,
        month = nov,
       volume = {551},
       number = {7678},
        pages = {75-79},
          doi = {10.1038/nature24303},
archivePrefix = {arXiv},
       eprint = {1710.05841},
 primaryClass = {astro-ph.HE},
       adsurl = {https://ui.adsabs.harvard.edu/abs/2017Natur.551...75S}
}

@ARTICLE{Macaluso2019,
       author = {{Macaluso}, D.~A. and {Aguilar}, A. and {Kilcoyne}, A.~L.~D. and {Bilodeau}, R.~C. and {Ju{\'a}rez}, A.~M. and {Dumitriu}, I. and {Hardy}, D. and {Sterling}, N.~C. and {Bautista}, M.},
        title = "{Absolute single photoionization cross-sections of Br$^{3+}$: experiment and theory}",
      journal = {Journal of Physics B Atomic Molecular Physics},
         year = 2019,
        month = jul,
       volume = {52},
       number = {14},
          eid = {145002},
        pages = {145002},
          doi = {10.1088/1361-6455/ab0e22},
       adsurl = {https://ui.adsabs.harvard.edu/abs/2019JPhB...52n5002M}
}

@ARTICLE{McLaughlin2019,
       author = {{McLaughlin}, Brendan M. and {Babb}, James F.},
        title = "{Photoionization of Rb$^{2+}$ ions in the valence energy region 37-44 eV}",
      journal = {Journal of Physics B Atomic Molecular Physics},
         year = 2019,
        month = jun,
       volume = {52},
       number = {12},
          eid = {125201},
        pages = {125201},
          doi = {10.1088/1361-6455/ab1e99},
archivePrefix = {arXiv},
       eprint = {1904.10770},
 primaryClass = {physics.atom-ph},
       adsurl = {https://ui.adsabs.harvard.edu/abs/2019JPhB...52l5201M}
}

@ARTICLE{Sterling2015,
       author = {{Sterling}, N.~C. and {Porter}, R.~L. and {Dinerstein}, Harriet L.},
        title = "{The Abundances of Light Neutron-capture Elements in Planetary Nebulae. III. The Impact of New Atomic Data on Nebular Selenium and Krypton Abundance Determinations}",
      journal = {\apjs},
         year = 2015,
        month = jun,
       volume = {218},
       number = {2},
          eid = {25},
        pages = {25},
          doi = {10.1088/0067-0049/218/2/25},
archivePrefix = {arXiv},
       eprint = {1505.01162},
 primaryClass = {astro-ph.SR},
       adsurl = {https://ui.adsabs.harvard.edu/abs/2015ApJS..218...25S}
}

@ARTICLE{Sterling2011,
       author = {{Sterling}, N.~C.},
        title = "{Atomic data for neutron-capture elements. II. Photoionization and recombination properties of low-charge krypton ions}",
      journal = {\aap},
         year = 2011,
        month = sep,
       volume = {533},
          eid = {A62},
        pages = {A62},
          doi = {10.1051/0004-6361/201117471},
archivePrefix = {arXiv},
       eprint = {1107.3843},
 primaryClass = {astro-ph.SR},
       adsurl = {https://ui.adsabs.harvard.edu/abs/2011A&A...533A..62S}
}

@ARTICLE{Banerjee2025,
       author = {{Banerjee}, Smaranika and {Jerkstrand}, Anders and {Badnell}, Nigel and {Pognan}, Quentin and {Ferguson}, Niamh and {Grumer}, Jon},
        title = "{Nebular Spectra of Kilonovae with Detailed Recombination Rates. I. Light r-process Composition}",
      journal = {\apj},
         year = 2025,
        month = oct,
       volume = {992},
       number = {1},
          eid = {19},
        pages = {19},
          doi = {10.3847/1538-4357/adf6ba},
archivePrefix = {arXiv},
       eprint = {2501.18345},
 primaryClass = {astro-ph.HE},
       adsurl = {https://ui.adsabs.harvard.edu/abs/2025ApJ...992...19B}
}

@ARTICLE{Levan2024,
       author = {{Levan}, Andrew J. and {Gompertz}, Benjamin P. and {Salafia}, Om Sharan and {Bulla}, Mattia and {Burns}, Eric and {Hotokezaka}, Kenta and {Izzo}, Luca and {Lamb}, Gavin P. and {Malesani}, Daniele B. and {Oates}, Samantha R. and {Ravasio}, Maria Edvige and {Rouco Escorial}, Alicia and {Schneider}, Benjamin and {Sarin}, Nikhil and {Schulze}, Steve and {Tanvir}, Nial R. and {Ackley}, Kendall and {Anderson}, Gemma and {Brammer}, Gabriel B. and {Christensen}, Lise and {Dhillon}, Vikram S. and {Evans}, Phil A. and {Fausnaugh}, Michael and {Fong}, Wen-fai and {Fruchter}, Andrew S. and {Fryer}, Chris and {Fynbo}, Johan P.~U. and {Gaspari}, Nicola and {Heintz}, Kasper E. and {Hjorth}, Jens and {Kennea}, Jamie A. and {Kennedy}, Mark R. and {Laskar}, Tanmoy and {Leloudas}, Giorgos and {Mandel}, Ilya and {Martin-Carrillo}, Antonio and {Metzger}, Brian D. and {Nicholl}, Matt and {Nugent}, Anya and {Palmerio}, Jesse T. and {Pugliese}, Giovanna and {Rastinejad}, Jillian and {Rhodes}, Lauren and {Rossi}, Andrea and {Saccardi}, Andrea and {Smartt}, Stephen J. and {Stevance}, Heloise F. and {Tohuvavohu}, Aaron and {van der Horst}, Alexander and {Vergani}, Susanna D. and {Watson}, Darach and {Barclay}, Thomas and {Bhirombhakdi}, Kornpob and {Breedt}, Elm{\'e} and {Breeveld}, Alice A. and {Brown}, Alexander J. and {Campana}, Sergio and {Chrimes}, Ashley A. and {D'Avanzo}, Paolo and {D'Elia}, Valerio and {De Pasquale}, Massimiliano and {Dyer}, Martin J. and {Galloway}, Duncan K. and {Garbutt}, James A. and {Green}, Matthew J. and {Hartmann}, Dieter H. and {Jakobsson}, P{\'a}ll and {Kerry}, Paul and {Kouveliotou}, Chryssa and {Langeroodi}, Danial and {Le Floc'h}, Emeric and {Leung}, James K. and {Littlefair}, Stuart P. and {Munday}, James and {O'Brien}, Paul and {Parsons}, Steven G. and {Pelisoli}, Ingrid and {Sahman}, David I. and {Salvaterra}, Ruben and {Sbarufatti}, Boris and {Steeghs}, Danny and {Tagliaferri}, Gianpiero and {Th{\"o}ne}, Christina C. and {de Ugarte Postigo}, Antonio and {Kann}, David Alexander},
        title = "{Heavy-element production in a compact object merger observed by JWST}",
      journal = {\nat},
         year = 2024,
        month = feb,
       volume = {626},
       number = {8000},
        pages = {737-741},
          doi = {10.1038/s41586-023-06759-1},
archivePrefix = {arXiv},
       eprint = {2307.02098},
 primaryClass = {astro-ph.HE},
       adsurl = {https://ui.adsabs.harvard.edu/abs/2024Natur.626..737L}
}

@ARTICLE{Jerkstrand2012,
       author = {{Jerkstrand}, A. and {Fransson}, C. and {Maguire}, K. and {Smartt}, S. and {Ergon}, M. and {Spyromilio}, J.},
        title = "{The progenitor mass of the Type IIP supernova SN 2004et from late-time spectral modeling}",
      journal = {\aap},
         year = 2012,
        month = oct,
       volume = {546},
          eid = {A28},
        pages = {A28},
          doi = {10.1051/0004-6361/201219528},
archivePrefix = {arXiv},
       eprint = {1208.2183},
 primaryClass = {astro-ph.HE},
       adsurl = {https://ui.adsabs.harvard.edu/abs/2012A&A...546A..28J}
}

@INPROCEEDINGS{Dinerstein2021,
       author = {{Dinerstein}, H.~L. and {Sterling}, N.~C. and {Vacca}, W.~D. and {Bautista}, M.~A.},
        title = "{New Infrared Emission Lines of the s-Process Enriched Neutron-Capture Elements Br and Rb in Planetary Nebulae}",
    booktitle = {American Astronomical Society Meeting Abstracts},
         year = 2021,
       series = {American Astronomical Society Meeting Abstracts},
       volume = {237},
        month = jan,
          eid = {548.13},
        pages = {548.13},
       adsurl = {https://ui.adsabs.harvard.edu/abs/2021AAS...23754813D}
}

@ARTICLE{Madonna2018,
       author = {{Madonna}, Simone and {Bautista}, Manuel and {Dinerstein}, Harriet L. and {Sterling}, N.~C. and {Garc{\'\i}a-Rojas}, Jorge and {Kaplan}, Kyle F. and {Rubio-D{\'\i}ez}, Maria M. and {Castro-Rodr{\'\i}guez}, Nieves and {Garz{\'o}n}, Francisco},
        title = "{Neutron-capture Elements in Planetary Nebulae: First Detections of Near-infrared [Te III] and [Br V] Emission Lines}",
      journal = {\apjl},
         year = 2018,
        month = jul,
       volume = {861},
       number = {1},
          eid = {L8},
        pages = {L8},
          doi = {10.3847/2041-8213/aaccef},
archivePrefix = {arXiv},
       eprint = {1806.05458},
 primaryClass = {astro-ph.GA},
       adsurl = {https://ui.adsabs.harvard.edu/abs/2018ApJ...861L...8M}
}

@ARTICLE{Dinerstein2001a,
       author = {{Dinerstein}, Harriet L.},
        title = "{Neutron-Capture Elements in Planetary Nebulae: Identification of Two Near-Infrared Emission Lines as [KR III] and [SE IV]}",
      journal = {\apjl},
         year = 2001,
        month = apr,
       volume = {550},
       number = {2},
        pages = {L223-L226},
          doi = {10.1086/319645},
       adsurl = {https://ui.adsabs.harvard.edu/abs/2001ApJ...550L.223D}
}

@ARTICLE{Dinerstein2001b,
       author = {{Dinerstein}, Harriet L. and {Geballe}, T.~R.},
        title = "{Detection and Significance of [Zn IV] 3.625 Microns in Planetary Nebulae}",
      journal = {\apj},
         year = 2001,
        month = nov,
       volume = {562},
       number = {1},
        pages = {515-520},
          doi = {10.1086/323469},
       adsurl = {https://ui.adsabs.harvard.edu/abs/2001ApJ...562..515D}
}

@ARTICLE{Pequignot1994,
       author = {{Pequignot}, D. and {Baluteau}, J. -P.},
        title = "{The identtification of krypton, xenon and other elements of rows 4, 5 and 6 of the periodic table in the planetary nebula NGC 7027.}",
      journal = {\aap},
         year = 1994,
        month = mar,
       volume = {283},
        pages = {593-625},
       adsurl = {https://ui.adsabs.harvard.edu/abs/1994A&A...283..593P}
}

@ARTICLE{Kawaguchi2021,
       author = {{Kawaguchi}, Kyohei and {Fujibayashi}, Sho and {Shibata}, Masaru and {Tanaka}, Masaomi and {Wanajo}, Shinya},
        title = "{A Low-mass Binary Neutron Star: Long-term Ejecta Evolution and Kilonovae with Weak Blue Emission}",
      journal = {\apj},
         year = 2021,
        month = jun,
       volume = {913},
       number = {2},
          eid = {100},
        pages = {100},
          doi = {10.3847/1538-4357/abf3bc},
archivePrefix = {arXiv},
       eprint = {2012.14711},
 primaryClass = {astro-ph.HE},
       adsurl = {https://ui.adsabs.harvard.edu/abs/2021ApJ...913..100K}
}

@ARTICLE{Goriely1999,
       author = {{Goriely}, S.},
        title = "{Uncertainties in the solar system r-abundance distribution}",
      journal = {\aap},
         year = 1999,
        month = feb,
       volume = {342},
        pages = {881-891},
       adsurl = {https://ui.adsabs.harvard.edu/abs/1999A&A...342..881G}
}

@ARTICLE{Fujibayashi2023a,
       author = {{Fujibayashi}, Sho and {Kiuchi}, Kenta and {Wanajo}, Shinya and {Kyutoku}, Koutarou and {Sekiguchi}, Yuichiro and {Shibata}, Masaru},
        title = "{Comprehensive Study of Mass Ejection and Nucleosynthesis in Binary Neutron Star Mergers Leaving Short-lived Massive Neutron Stars}",
      journal = {\apj},
         year = 2023,
        month = jan,
       volume = {942},
       number = {1},
          eid = {39},
        pages = {39},
          doi = {10.3847/1538-4357/ac9ce0},
archivePrefix = {arXiv},
       eprint = {2205.05557},
 primaryClass = {astro-ph.HE},
       adsurl = {https://ui.adsabs.harvard.edu/abs/2023ApJ...942...39F}
}

@ARTICLE{Just2023,
       author = {{Just}, O. and {Vijayan}, V. and {Xiong}, Z. and {Goriely}, S. and {Soultanis}, T. and {Bauswein}, A. and {Guilet}, J. and {Janka}, H. -Th. and {Mart{\'\i}nez-Pinedo}, G.},
        title = "{End-to-end Kilonova Models of Neutron Star Mergers with Delayed Black Hole Formation}",
      journal = {\apjl},
         year = 2023,
        month = jul,
       volume = {951},
       number = {1},
          eid = {L12},
        pages = {L12},
          doi = {10.3847/2041-8213/acdad2},
archivePrefix = {arXiv},
       eprint = {2302.10928},
 primaryClass = {astro-ph.HE},
       adsurl = {https://ui.adsabs.harvard.edu/abs/2023ApJ...951L..12J}
}

@ARTICLE{Fernandez2024,
       author = {{Fern{\'a}ndez}, Rodrigo and {Just}, Oliver and {Xiong}, Zewei and {Mart{\'\i}nez-Pinedo}, Gabriel},
        title = "{Viscous hydrodynamic evolution of neutron star merger accretion disks: A code comparison}",
      journal = {\prd},
         year = 2024,
        month = jul,
       volume = {110},
       number = {2},
          eid = {023001},
        pages = {023001},
          doi = {10.1103/PhysRevD.110.023001},
archivePrefix = {arXiv},
       eprint = {2307.02554},
 primaryClass = {astro-ph.HE},
       adsurl = {https://ui.adsabs.harvard.edu/abs/2024PhRvD.110b3001F}
}

@ARTICLE{Villar2018,
       author = {{Villar}, V.~A. and {Cowperthwaite}, P.~S. and {Berger}, E. and {Blanchard}, P.~K. and {Gomez}, S. and {Alexander}, K.~D. and {Margutti}, R. and {Chornock}, R. and {Eftekhari}, T. and {Fazio}, G.~G. and {Guillochon}, J. and {Hora}, J.~L. and {Nicholl}, M. and {Williams}, P.~K.~G.},
        title = "{Spitzer Space Telescope Infrared Observations of the Binary Neutron Star Merger GW170817}",
      journal = {\apjl},
         year = 2018,
        month = jul,
       volume = {862},
       number = {1},
          eid = {L11},
        pages = {L11},
          doi = {10.3847/2041-8213/aad281},
archivePrefix = {arXiv},
       eprint = {1805.08192},
 primaryClass = {astro-ph.HE},
       adsurl = {https://ui.adsabs.harvard.edu/abs/2018ApJ...862L..11V}
}

@ARTICLE{Rynkun2020,
       author = {{Rynkun}, P. and {Gaigalas}, G. and {J{\"o}nsson}, P.},
        title = "{Theoretical studies of energy levels and transition data for Zr III}",
      journal = {\aap},
         year = 2020,
        month = may,
       volume = {637},
          eid = {A10},
        pages = {A10},
          doi = {10.1051/0004-6361/201937243},
       adsurl = {https://ui.adsabs.harvard.edu/abs/2020A&A...637A..10R}
}

@ARTICLE{Persson1978,
       author = {{Persson}, Willy},
        title = "{An Extended Analysis of the Spectrum of Trebly-Ionized Strontium}",
      journal = {\physscr},
         year = 1978,
        month = apr,
       volume = {17},
       number = {4},
        pages = {387-410},
          doi = {10.1088/0031-8949/17/4/001},
       adsurl = {https://ui.adsabs.harvard.edu/abs/1978PhyS...17..387P}
}

@ARTICLE{Sahoo2008,
       author = {{Sahoo}, B.~K. and {Nataraj}, H.~S. and {Das}, B.~P. and {Chaudhuri}, R.~K. and {Mukherjee}, D.},
        title = "{Theoretical estimation of lifetimes of the lowest metastable states in Sc III and Y III}",
      journal = {Journal of Physics B Atomic Molecular Physics},
         year = 2008,
        month = mar,
       volume = {41},
       number = {5},
          eid = {055702},
        pages = {055702},
          doi = {10.1088/0953-4075/41/5/055702},
archivePrefix = {arXiv},
       eprint = {0707.3282},
 primaryClass = {physics.atom-ph},
       adsurl = {https://ui.adsabs.harvard.edu/abs/2008JPhB...41e5702S}
}

@ARTICLE{Nilsson1991,
       author = {{Nilsson}, A.~E. and {Johansson}, S. and {Kurucz}, R.~L.},
        title = "{The spectrum of singly ionized yttrium, Y II}",
      journal = {\physscr},
         year = 1991,
        month = sep,
       volume = {44},
       number = {3},
        pages = {226-257},
          doi = {10.1088/0031-8949/44/3/003},
       adsurl = {https://ui.adsabs.harvard.edu/abs/1991PhyS...44..226N}
}

@ARTICLE{Sterling2016,
       author = {{Sterling}, N.~C. and {Dinerstein}, Harriet L. and {Kaplan}, Kyle F. and {Bautista}, Manuel A.},
        title = "{Discovery of Rubidium, Cadmium, and Germanium Emission Lines in the Near-infrared Spectra of Planetary Nebulae}",
      journal = {\apjl},
         year = 2016,
        month = mar,
       volume = {819},
       number = {1},
          eid = {L9},
        pages = {L9},
          doi = {10.3847/2041-8205/819/1/L9},
archivePrefix = {arXiv},
       eprint = {1602.03188},
 primaryClass = {astro-ph.SR},
}

@ARTICLE{Sansonetti2006,
       author = {{Sansonetti}, J.~E.},
        title = "{Wavelengths, Transition Probabilities, and Energy Levels for the Spectra of Rubidium (Rb I through Rb XXXVII)}",
      journal = {Journal of Physical and Chemical Reference Data},
         year = 2006,
        month = mar,
       volume = {35},
       number = {1},
        pages = {301-421},
          doi = {10.1063/1.2035727},
       adsurl = {https://ui.adsabs.harvard.edu/abs/2006JPCRD..35..301S}
}

@ARTICLE{Shirai2007,
       author = {{Shirai}, T. and {Reader}, J. and {Kramida}, A.~E. and {Sugar}, J.},
        title = "{Spectral Data for Gallium: Ga I through Ga XXXI}",
      journal = {Journal of Physical and Chemical Reference Data},
         year = 2007,
        month = jun,
       volume = {36},
       number = {2},
        pages = {509-615},
          doi = {10.1063/1.2207144},
       adsurl = {https://ui.adsabs.harvard.edu/abs/2007JPCRD..36..509S}
}

@ARTICLE{Joshi1971,
       author = {{Joshi}, Y. N. and {Budhiraja}, C. J.},
        title = "{Spectrum of Trebly Ionized Bromine}",
      journal = {Can. J. Phys.},
         year = 1971,
       volume = {49},
        pages = {670–677},
          doi = {10.1139/p71-084},
}

@ARTICLE{Tech1963,
       author = {{Tech}, J. L.},
        title = "{Analysis of the Spectrum of Neutral Atomic Bromine (Br I)}",
      journal = {J. Res. Natl. Bur. Stand. (U.S.), Sect. A},
         year = 1963,
       volume = {67},
        pages = {505-554},
          doi = {10.6028/jres.067A.051},
}

@ARTICLE{Biemont1986-1,
       author = {{Bi{\'e}mont}, E. and {Hansen}, J.~E.},
        title = "{Forbidden Transitions in 3p$^{4}$ and 4p$^{4}$ Configurations}",
      journal = {\physscr},
         year = 1986,
        month = aug,
       volume = {34},
       number = {2},
        pages = {116-130},
          doi = {10.1088/0031-8949/34/2/005},
       adsurl = {https://ui.adsabs.harvard.edu/abs/1986PhyS...34..116B}
}

@ARTICLE{Biemont1986-2,
       author = {{Bi{\'e}mont}, E. and {Hansen}, J.~E.},
        title = "{Magnetic-Dipole and Electric-Quadrupole Transitions in the Ground-State Configurations of the Germanium and Arsenic Isoelectronic Sequences}",
      journal = {\physscr},
         year = 1986,
        month = feb,
       volume = {33},
       number = {2},
        pages = {117-134},
          doi = {10.1088/0031-8949/33/2/006},
       adsurl = {https://ui.adsabs.harvard.edu/abs/1986PhyS...33..117B}
}

@ARTICLE{Sugar1993,
       author = {{Sugar}, Jack and {Musgrove}, Arlene},
        title = "{Energy Levels of Germanium, Ge I through Ge XXXII}",
      journal = {Journal of Physical and Chemical Reference Data},
         year = 1993,
        month = sep,
       volume = {22},
       number = {5},
        pages = {1213-1278},
          doi = {10.1063/1.555929},
       adsurl = {https://ui.adsabs.harvard.edu/abs/1993JPCRD..22.1213S}
}

@ARTICLE{Schoning1997,
       author = {{Schöning}, T.},
        title = "{Effective collision strengths for transitions in the 4p\^k (k = 2-4) ground configurations of KR III, KR IV and KR V}",
      journal = {\aaps},
         year = 1997,
        month = apr,
       volume = {122},
        pages = {277-283},
          doi = {10.1051/aas:1997133},
       adsurl = {https://ui.adsabs.harvard.edu/abs/1997A&AS..122..277S}
}

@ARTICLE{Sterling2017,
       author = {{Sterling}, N.~C. and {Madonna}, S. and {Butler}, K. and {Garc{\'\i}a-Rojas}, J. and {Mashburn}, A.~L. and {Morisset}, C. and {Luridiana}, V. and {Roederer}, I.~U.},
        title = "{Identification of Near-infrared [Se III] and [Kr VI] Emission Lines in Planetary Nebulae}",
      journal = {\apj},
         year = 2017,
        month = may,
       volume = {840},
       number = {2},
          eid = {80},
        pages = {80},
          doi = {10.3847/1538-4357/aa6c28},
archivePrefix = {arXiv},
       eprint = {1704.00741},
 primaryClass = {astro-ph.SR},
       adsurl = {https://ui.adsabs.harvard.edu/abs/2017ApJ...840...80S}
}

@ARTICLE{Saloman2007,
       author = {{Saloman}, E.~B.},
        title = "{Energy Levels and Observed Spectral Lines of Krypton, Kr I through Kr XXXVI}",
      journal = {Journal of Physical and Chemical Reference Data},
         year = 2007,
        month = mar,
       volume = {36},
       number = {1},
        pages = {215-386},
          doi = {10.1063/1.2227036},
       adsurl = {https://ui.adsabs.harvard.edu/abs/2007JPCRD..36..215S}
}

@ARTICLE{Prantzos2020,
       author = {{Prantzos}, N. and {Abia}, C. and {Cristallo}, S. and {Limongi}, M. and {Chieffi}, A.},
        title = "{Chemical evolution with rotating massive star yields II. A new assessment of the solar s- and r-process components}",
      journal = {\mnras},
         year = 2020,
        month = jan,
       volume = {491},
       number = {2},
        pages = {1832-1850},
          doi = {10.1093/mnras/stz3154},
archivePrefix = {arXiv},
       eprint = {1911.02545},
 primaryClass = {astro-ph.GA},
       adsurl = {https://ui.adsabs.harvard.edu/abs/2020MNRAS.491.1832P}
}

@BOOK{Moore1971,
       author = {{Moore}, C. E. },
        title = "{Atomic Energy Levels as Derived from the Analyses of Optical Spectra – Chromium through Niobium}",
         year = 1971,
       adsurl = {}
}

@ARTICLE{Reader1997,
       author = {{Reader}, Joseph and {Acquista}, Nicolo},
        title = "{Spectrum and energy levels of doubly-ionized zirconium (Zr III)}",
      journal = {\physscr},
         year = 1997,
        month = mar,
       volume = {55},
       number = {3},
        pages = {310-329},
          doi = {10.1088/0031-8949/55/3/009},
       adsurl = {https://ui.adsabs.harvard.edu/abs/1997PhyS...55..310R}
}

@BOOK{Palmer1977,
       author = {{Palmer}, B. A.},
        title = "{The First Spectrum of Yttrium and an Automatic Comparator for Its Measurement}",
         year = 1977,
       adsurl = {}
}

@ARTICLE{Waxman2019,
       author = {{Waxman}, Eli and {Ofek}, Eran O. and {Kushnir}, Doron},
        title = "{Late-time Kilonova Light Curves and Implications to GW170817}",
      journal = {\apj},
         year = 2019,
        month = jun,
       volume = {878},
       number = {2},
          eid = {93},
        pages = {93},
          doi = {10.3847/1538-4357/ab1f71},
archivePrefix = {arXiv},
       eprint = {1902.01197},
 primaryClass = {astro-ph.HE},
       adsurl = {https://ui.adsabs.harvard.edu/abs/2019ApJ...878...93W}
}

@ARTICLE{Kasen2019,
       author = {{Kasen}, Daniel and {Barnes}, Jennifer},
        title = "{Radioactive Heating and Late Time Kilonova Light Curves}",
      journal = {\apj},
         year = 2019,
        month = may,
       volume = {876},
       number = {2},
          eid = {128},
        pages = {128},
          doi = {10.3847/1538-4357/ab06c2},
archivePrefix = {arXiv},
       eprint = {1807.03319},
 primaryClass = {astro-ph.HE},
       adsurl = {https://ui.adsabs.harvard.edu/abs/2019ApJ...876..128K}
}

@ARTICLE{Kasliwal2022,
       author = {{Kasliwal}, Mansi M. and {Kasen}, Daniel and {Lau}, Ryan M. and {Perley}, Daniel A. and {Rosswog}, Stephan and {Ofek}, Eran O. and {Hotokezaka}, Kenta and {Chary}, Ranga-Ram and {Sollerman}, Jesper and {Goobar}, Ariel and {Kaplan}, David L.},
        title = "{Spitzer mid-infrared detections of neutron star merger GW170817 suggests synthesis of the heaviest elements}",
      journal = {\mnras},
         year = 2022,
        month = feb,
       volume = {510},
       number = {1},
        pages = {L7-L12},
          doi = {10.1093/mnrasl/slz007},
archivePrefix = {arXiv},
       eprint = {1812.08708},
 primaryClass = {astro-ph.HE},
       adsurl = {https://ui.adsabs.harvard.edu/abs/2022MNRAS.510L...7K}
}

@ARTICLE{Lodders2003,
       author = {{Lodders}, Katharina},
        title = "{Solar System Abundances and Condensation Temperatures of the Elements}",
      journal = {\apj},
         year = 2003,
        month = jul,
       volume = {591},
       number = {2},
        pages = {1220-1247},
          doi = {10.1086/375492},
       adsurl = {https://ui.adsabs.harvard.edu/abs/2003ApJ...591.1220L}
}

@ARTICLE{Tanaka17,
       author = {{Tanaka}, Masaomi and {Utsumi}, Yousuke and {Mazzali}, Paolo A. and {Tominaga}, Nozomu and {Yoshida}, Michitoshi and {Sekiguchi}, Yuichiro and {Morokuma}, Tomoki and {Motohara}, Kentaro and {Ohta}, Kouji and {Kawabata}, Koji S. and {Abe}, Fumio and {Aoki}, Kentaro and {Asakura}, Yuichiro and {Baar}, Stefan and {Barway}, Sudhanshu and {Bond}, Ian A. and {Doi}, Mamoru and {Fujiyoshi}, Takuya and {Furusawa}, Hisanori and {Honda}, Satoshi and {Itoh}, Yoichi and {Kawabata}, Miho and {Kawai}, Nobuyuki and {Kim}, Ji Hoon and {Lee}, Chien-Hsiu and {Miyazaki}, Shota and {Morihana}, Kumiko and {Nagashima}, Hiroki and {Nagayama}, Takahiro and {Nakaoka}, Tatsuya and {Nakata}, Fumiaki and {Ohsawa}, Ryou and {Ohshima}, Tomohito and {Okita}, Hirofumi and {Saito}, Tomoki and {Sumi}, Takahiro and {Tajitsu}, Akito and {Takahashi}, Jun and {Takayama}, Masaki and {Tamura}, Yoichi and {Tanaka}, Ichi and {Terai}, Tsuyoshi and {Tristram}, Paul J. and {Yasuda}, Naoki and {Zenko}, Tetsuya},
        title = "{Kilonova from post-merger ejecta as an optical and near-Infrared counterpart of GW170817}",
      journal = {\pasj},
         year = 2017,
        month = dec,
       volume = {69},
       number = {6},
          eid = {102},
        pages = {102},
          doi = {10.1093/pasj/psx121},
archivePrefix = {arXiv},
       eprint = {1710.05850},
 primaryClass = {astro-ph.HE},
       adsurl = {https://ui.adsabs.harvard.edu/abs/2017PASJ...69..102T}
}

@ARTICLE{Wanajo2014,
       author = {{Wanajo}, Shinya and {Sekiguchi}, Yuichiro and {Nishimura}, Nobuya and
         et al.},
        title = "{Production of All the r-process Nuclides in the Dynamical Ejecta of Neutron Star Mergers}",
      journal = {\apjl},
         year = 2014,
        month = jul,
       volume = {789},
       number = {2},
          eid = {L39},
        pages = {L39},
          doi = {10.1088/2041-8205/789/2/L39},
archivePrefix = {arXiv},
       eprint = {1402.7317},
 primaryClass = {astro-ph.SR},
       adsurl = {https://ui.adsabs.harvard.edu/abs/2014ApJ...789L..39W}
}

@ARTICLE{Hotokezaka2021,
       author = {{Hotokezaka}, Kenta and {Tanaka}, Masaomi and {Kato}, Daiji and {Gaigalas}, Gediminas},
        title = "{Nebular emission from lanthanide-rich ejecta of neutron star merger}",
      journal = {\mnras},
         year = 2021,
        month = oct,
       volume = {506},
       number = {4},
        pages = {5863-5877},
          doi = {10.1093/mnras/stab1975},
archivePrefix = {arXiv},
       eprint = {2102.07879},
 primaryClass = {astro-ph.HE},
       adsurl = {https://ui.adsabs.harvard.edu/abs/2021MNRAS.506.5863H}
}

@ARTICLE{Hotokezaka2022,
       author = {{Hotokezaka}, Kenta and {Tanaka}, Masaomi and {Kato}, Daiji and {Gaigalas}, Gediminas},
        title = "{Tungsten versus Selenium as a potential source of kilonova nebular emission observed by Spitzer}",
      journal = {\mnras},
         year = 2022,
        month = sep,
       volume = {515},
       number = {1},
        pages = {L89-L93},
          doi = {10.1093/mnrasl/slac071},
archivePrefix = {arXiv},
       eprint = {2204.00737},
 primaryClass = {astro-ph.HE},
       adsurl = {https://ui.adsabs.harvard.edu/abs/2022MNRAS.515L..89H}
}

@ARTICLE{Pognan2023,
       author = {{Pognan}, Quentin and {Grumer}, Jon and {Jerkstrand}, Anders and {Wanajo}, Shinya},
        title = "{NLTE spectra of kilonovae}",
      journal = {\mnras},
         year = 2023,
        month = dec,
       volume = {526},
       number = {4},
        pages = {5220-5248},
          doi = {10.1093/mnras/stad3106},
archivePrefix = {arXiv},
       eprint = {2309.01134},
 primaryClass = {astro-ph.HE},
       adsurl = {https://ui.adsabs.harvard.edu/abs/2023MNRAS.526.5220P}
}

@ARTICLE{Bulla2019,
       author = {{Bulla}, M. and {Covino}, S. and {Kyutoku}, K. and {Tanaka}, M. and {Maund}, J.~R. and {Patat}, F. and {Toma}, K. and {Wiersema}, K. and {Bruten}, J. and {Jin}, Z.~P. and {Testa}, V.},
        title = "{The origin of polarization in kilonovae and the case of the gravitational-wave counterpart AT 2017gfo}",
      journal = {Nature Astronomy},
         year = 2019,
        month = mar,
       volume = {3},
        pages = {99-106},
          doi = {10.1038/s41550-018-0593-y},
archivePrefix = {arXiv},
       eprint = {1809.04078},
 primaryClass = {astro-ph.HE},
       adsurl = {https://ui.adsabs.harvard.edu/abs/2019NatAs...3...99B}
}

@ARTICLE{Pognan2025,
       author = {{Pognan}, Quentin and {Wu}, Meng-Ru and {Mart{\'\i}nez-Pinedo}, Gabriel and {da Silva}, Ricardo Ferreira and {Jerkstrand}, Anders and {Grumer}, Jon and {Fl{\"o}rs}, Andreas},
        title = "{Actinide signatures in low electron fraction kilonova ejecta}",
      journal = {\mnras},
         year = 2025,
        month = jan,
       volume = {536},
       number = {3},
        pages = {2973-2992},
          doi = {10.1093/mnras/stae2778},
archivePrefix = {arXiv},
       eprint = {2409.16210},
 primaryClass = {astro-ph.HE},
       adsurl = {https://ui.adsabs.harvard.edu/abs/2025MNRAS.536.2973P}
}

@ARTICLE{Pognan2022a,
       author = {{Pognan}, Quentin and {Jerkstrand}, Anders and {Grumer}, Jon},
        title = "{On the validity of steady-state for nebular phase kilonovae}",
      journal = {\mnras},
         year = 2022,
        month = mar,
       volume = {510},
       number = {3},
        pages = {3806-3837},
          doi = {10.1093/mnras/stab3674},
archivePrefix = {arXiv},
       eprint = {2112.07484},
 primaryClass = {astro-ph.HE},
       adsurl = {https://ui.adsabs.harvard.edu/abs/2022MNRAS.510.3806P}
}

@ARTICLE{Pognan2022b,
       author = {{Pognan}, Quentin and {Jerkstrand}, Anders and {Grumer}, Jon},
        title = "{NLTE effects on kilonova expansion opacities}",
      journal = {\mnras},
         year = 2022,
        month = jul,
       volume = {513},
       number = {4},
        pages = {5174-5197},
          doi = {10.1093/mnras/stac1253},
archivePrefix = {arXiv},
       eprint = {2202.09245},
 primaryClass = {astro-ph.HE},
       adsurl = {https://ui.adsabs.harvard.edu/abs/2022MNRAS.513.5174P}
}

@ARTICLE{Jerkstrand11,
       author = {{Jerkstrand}, A. and {Fransson}, C. and {Kozma}, C.},
        title = "{The $^{44}$Ti-powered spectrum of SN 1987A}",
      journal = {\aap},
         year = 2011,
        month = jun,
       volume = {530},
          eid = {A45},
        pages = {A45},
          doi = {10.1051/0004-6361/201015937},
archivePrefix = {arXiv},
       eprint = {1103.3653},
 primaryClass = {astro-ph.HE},
       adsurl = {https://ui.adsabs.harvard.edu/abs/2011A&A...530A..45J}
}

@ARTICLE{Villar17,
       author = {{Villar}, V.~A. and {Guillochon}, J. and {Berger}, E. and
         {Metzger}, B.~D. and {Cowperthwaite}, P.~S. and {Nicholl}, M. and {Alexand
        er}, K.~D. and {Blanchard}, P.~K. and {Chornock}, R. and
         {Eftekhari}, T. and {Fong}, W. and {Margutti}, R. and
         {Williams}, P.~K.~G.},
        title = "{The Combined Ultraviolet, Optical, and Near-infrared Light Curves of the Kilonova Associated with the Binary Neutron Star Merger GW170817: Unified Data Set, Analytic Models, and Physical Implications}",
      journal = {\apjl},
         year = 2017,
        month = dec,
       volume = {851},
       number = {1},
          eid = {L21},
        pages = {L21},
          doi = {10.3847/2041-8213/aa9c84},
archivePrefix = {arXiv},
       eprint = {1710.11576},
 primaryClass = {astro-ph.HE},
       adsurl = {https://ui.adsabs.harvard.edu/abs/2017ApJ...851L..21V}
}

@ARTICLE{Just2022b,
       author = {{Just}, O. and {Goriely}, S. and {Janka}, H. -Th and {Nagataki}, S. and {Bauswein}, A.},
        title = "{Neutrino absorption and other physics dependencies in neutrino-cooled black hole accretion discs}",
      journal = {\mnras},
         year = 2022,
        month = jan,
       volume = {509},
       number = {1},
        pages = {1377-1412},
          doi = {10.1093/mnras/stab2861},
archivePrefix = {arXiv},
       eprint = {2102.08387},
 primaryClass = {astro-ph.HE},
       adsurl = {https://ui.adsabs.harvard.edu/abs/2022MNRAS.509.1377J}
}

@ARTICLE{Fujibayashi2020b,
       author = {{Fujibayashi}, Sho and {Wanajo}, Shinya and {Kiuchi}, Kenta and {Kyutoku}, Koutarou and {Sekiguchi}, Yuichiro and {Shibata}, Masaru},
        title = "{Postmerger Mass Ejection of Low-mass Binary Neutron Stars}",
      journal = {\apj},
         year = 2020,
        month = oct,
       volume = {901},
       number = {2},
          eid = {122},
        pages = {122},
          doi = {10.3847/1538-4357/abafc2},
archivePrefix = {arXiv},
       eprint = {2007.00474},
 primaryClass = {astro-ph.HE},
       adsurl = {https://ui.adsabs.harvard.edu/abs/2020ApJ...901..122F}
}
\bibliographystyle{aasjournal}

\end{document}